\documentclass[10pt]{article} 
\usepackage{indentfirst,amsmath,latexsym,amssymb,amsbsy,amscd}  

\newcommand{\be}{\begin{equation}}
\newcommand{\ee}{\end{equation}}
\newcommand{\ra}{\rightarrow}
\newcommand{\la}{\leftarrow}
\newcommand{\ua}{\uparrow}
\newcommand{\da}{\downarrow}
\newcommand{\lra}{\leftrightarrow}
\newcommand{\uda}{\updownarrow}

\newcommand{\qed}{\hfill $\bullet$}
\newcommand{\cast}{\circledast}
\newcommand{\ccirc}{\circledcirc}

\newcommand{\shrtpll}{\shortparallel}

\newtheorem{prop}{Proposition}
\newtheorem{thm}[prop]{Theorem}
\newtheorem{alg}[prop]{Algorithm}
\newtheorem{lem}[prop]{Lemma}
\newtheorem{cor}[prop]{Corollary}

\newenvironment{ex}[1]{\trivlist \item[\hskip \labelsep{\bf Example {#1}}]}{\qed \endtrivlist}
\newenvironment{prf}{\trivlist \item[\hskip \labelsep{\bf Proof.}]}{\qed \endtrivlist}

\begin{document}


\newcommand{\je}[1]{j={#1},\ldots,\ell}
\newcommand{\ifof}{if and only if}
\newcommand{\bone}{\mathbf{1}}
\newcommand{\bzero}{\mathbf{0}}

\newcommand{\bsigma}{{\boldsymbol{\sigma}}}

\newcommand{\msfcpg}{{\mathsf{G}}}
\newcommand{\msfg}{{\mathsf{g}}}
\newcommand{\msfgdt}{{{\dot{\mathsf{g}}}}}
\newcommand{\msfgddt}{{{\ddot{\mathsf{g}}}}}

\newcommand{\bmcpz}{{\mathbf{Z}}}
\newcommand{\bmcpx}{{\mathbf{X}}}
\newcommand{\bmcpy}{{\mathbf{Y}}}
\newcommand{\mbbc}{{\mathbb{C}}}
\newcommand{\bmcpr}{{\mathbf{R}}}
\newcommand{\bmcpu}{{\mathbf{U}}}
\newcommand{\bmcpi}{{\mathbf{I}}}

\newcommand{\bmd}{{\mathbf{d}}}
\newcommand{\bmdbr}{{{\bar{\bmd}}}}
\newcommand{\bmb}{{\mathbf{b}}}
\newcommand{\bma}{{\mathbf{a}}}

\newcommand{\bmy}{{\mathbf{y}}}
\newcommand{\bmm}{{\mathbf{m}}}
\newcommand{\bmz}{{\mathbf{z}}}
\newcommand{\bms}{{\mathbf{s}}}
\newcommand{\bmt}{{\mathbf{t}}}
\newcommand{\bmcpw}{{\mathbf{W}}}

\newcommand{\bmf}{{\mathbf{f}}}

\newcommand{\bmc}{{\mathbf{c}}}
\newcommand{\bmu}{{\mathbf{u}}}
\newcommand{\bmr}{{\mathbf{r}}}
\newcommand{\bmg}{{\mathbf{g}}}
\newcommand{\bmi}{{\mathbf{i}}}
\newcommand{\bme}{{\mathbf{e}}}
\newcommand{\bmh}{{\mathbf{h}}}
\newcommand{\bmk}{{\mathbf{k}}}
\newcommand{\bmp}{{\mathbf{p}}}
\newcommand{\bmn}{{\mathbf{n}}}

\newcommand{\bmw}{{\mathbf{w}}}
\newcommand{\bml}{{\mathbf{l}}}
\newcommand{\bmx}{{\mathbf{x}}}
\newcommand{\bmcpq}{{\mathbf{Q}}}
\newcommand{\bmq}{{\mathbf{q}}}
\newcommand{\bmv}{{\mathbf{v}}}
\newcommand{\bmcpv}{{\mathbf{V}}}

\newcommand{\bmbht}{{\hat{\bmb}}}
\newcommand{\bmcht}{{\hat{\bmc}}}
\newcommand{\bmsht}{{\hat{\bms}}}
\newcommand{\bmrht}{{\hat{\bmr}}}
\newcommand{\bmght}{{\hat{\bmg}}}
\newcommand{\bmhht}{{\hat{\bmh}}}
\newcommand{\bmeht}{{\hat{\bme}}}

\newcommand{\bmubv}{{\breve{\bmu}}}
\newcommand{\bmuinv}{{\bmu^{-1}}}

\newcommand{\bmcck}{{\check{\bmc}}}
\newcommand{\bmuck}{{\check{\bmu}}}
\newcommand{\bmrck}{{\check{\bmr}}}
\newcommand{\bmgck}{{\check{\bmg}}}

\newcommand{\bmabr}{{\bar{\bma}}}
\newcommand{\bmbbr}{{\bar{\bmb}}}
\newcommand{\bmcbr}{{\bar{\bmc}}}
\newcommand{\bmubr}{{\bar{\bmu}}}
\newcommand{\bmrbr}{{\bar{\bmr}}}
\newcommand{\bmsbr}{{\bar{\bms}}}
\newcommand{\bmgbr}{{\bar{\bmg}}}
\newcommand{\bmhbr}{{\bar{\bmh}}}
\newcommand{\bmebr}{{\bar{\bme}}}
\newcommand{\bmvbr}{{\bar{\bmv}}}
\newcommand{\bmwbr}{{\bar{\bmw}}}

\newcommand{\bmbddt}{{\ddot{\bmb}}}
\newcommand{\bmaddt}{{\ddot{\bma}}}
\newcommand{\bmcddt}{{\ddot{\bmc}}}
\newcommand{\bmuddt}{{\ddot{\bmu}}}
\newcommand{\bmvddt}{{\ddot{\bmv}}}
\newcommand{\bmwddt}{{\ddot{\bmw}}}
\newcommand{\bmrddt}{{\ddot{\bmr}}}
\newcommand{\bmsddt}{{\ddot{\bms}}}
\newcommand{\bmgddt}{{\ddot{\bmg}}}
\newcommand{\bmeddt}{{\ddot{\bme}}}

\newcommand{\bmbdt}{{\dot{\bmb}}}
\newcommand{\bmadt}{{\dot{\bma}}}
\newcommand{\bmcdt}{{\dot{\bmc}}}
\newcommand{\bmudt}{{\dot{\bmu}}}
\newcommand{\bmvdt}{{\dot{\bmv}}}
\newcommand{\bmwdt}{{\dot{\bmw}}}
\newcommand{\bmrdt}{{\dot{\bmr}}}
\newcommand{\bmsdt}{{\dot{\bms}}}
\newcommand{\bmgdt}{{\dot{\bmg}}}
\newcommand{\bmxdt}{{\dot{\bmx}}}
\newcommand{\bmydt}{{\dot{\bmy}}}
\newcommand{\bmedt}{{\dot{\bme}}}

\newcommand{\bmaac}{{\acute{\bma}}}
\newcommand{\bmbac}{{\acute{\bmb}}}
\newcommand{\bmcac}{{\acute{\bmc}}}
\newcommand{\bmsac}{{\acute{\bms}}}
\newcommand{\bmuac}{{\acute{\bmu}}}
\newcommand{\bmvac}{{\acute{\bmv}}}
\newcommand{\bmrac}{{\acute{\bmr}}}
\newcommand{\bmgac}{{\acute{\bmg}}}

\newcommand{\bmagr}{{\grave{\bma}}}
\newcommand{\bmbgr}{{\grave{\bmb}}}
\newcommand{\bmcgr}{{\grave{\bmc}}}
\newcommand{\bmsgr}{{\grave{\bms}}}
\newcommand{\bmugr}{{\grave{\bmu}}}
\newcommand{\bmvgr}{{\grave{\bmv}}}
\newcommand{\bmrgr}{{\grave{\bmr}}}
\newcommand{\bmggr}{{\grave{\bmg}}}

\newcommand{\aht}{{\hat{a}}}
\newcommand{\aac}{{\acute{a}}}
\newcommand{\agr}{{\grave{a}}}
\newcommand{\addt}{{\ddot{a}}}
\newcommand{\adt}{{\dot{a}}}
\newcommand{\abr}{{\bar{a}}}

\newcommand{\bht}{{\hat{b}}}
\newcommand{\bac}{{\acute{b}}}
\newcommand{\bgr}{{\grave{b}}}
\newcommand{\bddt}{{\ddot{b}}}
\newcommand{\bdt}{{\dot{b}}}
\newcommand{\bbr}{{\bar{b}}}

\newcommand{\cht}{{\hat{c}}}
\newcommand{\cac}{{\acute{c}}}
\newcommand{\cgr}{{\grave{c}}}
\newcommand{\cddt}{{\ddot{c}}}
\newcommand{\cdt}{{\dot{c}}}
\newcommand{\cbr}{{\bar{c}}}

\newcommand{\rht}{{\hat{r}}}
\newcommand{\rac}{{\acute{r}}}
\newcommand{\rgr}{{\grave{r}}}
\newcommand{\rddt}{{\ddot{r}}}
\newcommand{\rdt}{{\dot{r}}}
\newcommand{\rck}{{\check{r}}}
\newcommand{\rbr}{{\bar{r}}}

\newcommand{\vbr}{{\bar{v}}}
\newcommand{\ebr}{{\bar{e}}}

\newcommand{\fbr}{{\bar{f}}}
\newcommand{\gdt}{{\dot{g}}}
\newcommand{\ght}{{\hat{g}}}
\newcommand{\gddt}{{\ddot{g}}}
\newcommand{\gbr}{{\bar{g}}}

\newcommand{\xdt}{{\dot{x}}}
\newcommand{\ydt}{{\dot{y}}}
\newcommand{\fdt}{{\dot{f}}}
\newcommand{\kdt}{{\dot{k}}}
\newcommand{\kddt}{{\ddot{k}}}

\newcommand{\odotht}{{\hat{\odot}}}
\newcommand{\cdotht}{{\hat{\cdot}}}
\newcommand{\bmuht}{{\hat{\bmu}}}
\newcommand{\bmudtht}{{\hat{\bmudt}}}
\newcommand{\bmuddtht}{{\hat{\bmuddt}}}
\newcommand{\bmvht}{{\hat{\bmv}}}
\newcommand{\bmvdtht}{{\hat{\bmvdt}}}
\newcommand{\bmvddtht}{{\hat{\bmvddt}}}
\newcommand{\caleht}{{\hat{\cale}}}
\newcommand{\calvht}{{\hat{\calv}}}
\newcommand{\callht}{{\hat{\call}}}

\newcommand{\bmghat}{{\hat{\mathbf{g}}}}
\newcommand{\bmgbar}{{\bar{\mathbf{g}}}}
\newcommand{\bmehat}{{\hat{\mathbf{e}}}}

\newcommand{\ssr}{Schreier series}
\newcommand{\fssr}{forward Schreier series}
\newcommand{\bssr}{backward Schreier series}
\newcommand{\gm}{generator matrix}
\newcommand{\gms}{generator matrices}
\newcommand{\stm}{static matrix}
\newcommand{\stms}{static matrices}
\newcommand{\inpm}{input matrix}
\newcommand{\inpms}{input matrices}
\newcommand{\sgen}{shift generator}
\newcommand{\sgens}{shift generators}
\newcommand{\glab}{generator label}
\newcommand{\glabs}{generator labels}
\newcommand{\mchn}{matrix chain}

\newcommand{\ellctl}{$\ell$-controllable}
\newcommand{\ellpctl}{$\ell'$-controllable}
\newcommand{\cdc}{coset decomposition chain}
\newcommand{\creps}{coset representatives}
\newcommand{\crep}{coset representative}
\newcommand{\crepc}{coset representative chain}
\newcommand{\compset}{complete set of coset representatives}
\newcommand{\fss}{full symmetry system}
\newcommand{\consub}{consistent subsystem}
\newcommand{\nss}{natural symmetry system}
\newcommand{\nsss}{natural symmetry systems}
\newcommand{\scgs}{strongly controllable group system}
\newcommand{\sctigs}{strongly controllable time invariant group system}
\newcommand{\sctigss}{strongly controllable time invariant group systems}
\newcommand{\sctigc}{strongly controllable time invariant group code}
\newcommand{\fhgs}{first homomorphism theorem for group systems}

\newcommand{\xj}{{\{X_j\}}}
\newcommand{\yi}{{\{Y_i\}}}
\newcommand{\xjt}{{\{X_j^t\}}}
\newcommand{\yit}{{\{Y_i^t\}}}
\newcommand{\yj}{{\{Y_j\}}}
\newcommand{\yk}{{\{Y_k\}}}
\newcommand{\bi}{{\{B^{(i)}\}}}

\newcommand{\calf}{{\mathcal{F}}}
\newcommand{\calb}{{\mathcal{B}}}
\newcommand{\calh}{{\mathcal{H}}}
\newcommand{\call}{{\mathcal{L}}}
\newcommand{\calt}{{\mathcal{T}}}

\newcommand{\calq}{{\mathcal{Q}}}
\newcommand{\calr}{{\mathcal{R}}}
\newcommand{\calu}{{\mathcal{U}}}
\newcommand{\cale}{{\mathcal{E}}}
\newcommand{\cals}{{\mathcal{S}}}
\newcommand{\calv}{{\mathcal{V}}}
\newcommand{\calw}{{\mathcal{W}}}
\newcommand{\cali}{{\mathcal{I}}}

\newcommand{\mfrkcpu}{{\mathfrak{U}}}
\newcommand{\mfrkcpr}{{\mathfrak{R}}}
\newcommand{\mfrkr}{{\mathfrak{r}}}
\newcommand{\mfrkrdt}{{\dot{\mfrkr}}}
\newcommand{\mfrkrddt}{{\ddot{\mfrkr}}}
\newcommand{\mfrku}{{\mathfrak{u}}}
\newcommand{\mfrkudt}{{\dot{\mfrku}}}
\newcommand{\mfrkuddt}{{\ddot{\mfrku}}}
\newcommand{\mfrkubr}{{\bar{\mfrku}}}

\newcommand{\cpaht}{{\hat{A}}}
\newcommand{\cpvht}{{\hat{V}}}
\newcommand{\cpuht}{{\hat{U}}}
\newcommand{\cprht}{{\hat{R}}}
\newcommand{\calbht}{{\hat{\calb}}}
\newcommand{\caluht}{{\hat{\calu}}}
\newcommand{\circht}{{\hat{\circ}}}
\newcommand{\ccircht}{{\hat{\ccirc}}}
\newcommand{\starht}{{\hat{\star}}}

\newcommand{\rmdef}{\stackrel{\rm def}{=}}
\newcommand{\imp}{{\rm im}\,p}
\newcommand{\imtheta}{{\rm im}\,\theta}
\newcommand{\imfr}{{\rm im}\,f_r}
\newcommand{\imfu}{{\rm im}\,f_u}
\newcommand{\imhu}{{\rm im}\,h_u}
\newcommand{\imhv}{{\rm im}\,h_v}
\newcommand{\plus}{{+}}
\newcommand{\dimnd}{{\diamond}}
\newcommand{\dggr}{{\dagger}}
\newcommand{\ddggr}{{\ddagger}}
\newcommand{\vtri}{{\vartriangle}}

\newcommand{\fracfjk}[2]
{{
\frac{\calf^{#1}(\Delta_{#2}^t)}{\calf^{#1}(\Delta_{{#2}-1}^t)}
}}

\newcommand{\ssl}{{/ \!\! /}} 

\newcommand{\argu}{{\,\cdot\,}}

\newcommand{\ovast}{{\overline{*}}}
\newcommand{\ovtimes}{{\overline{\times}}}
\newcommand{\ovcirc}{{\overline{\circ}}}
\newcommand{\ovstar}{{\overline{\star}}}
\newcommand{\ovcdot}{{\overline{\cdot}}}
\newcommand{\ovphi}{{\overline{\phi}}}
\newcommand{\ovtheta}{{\overline{\theta}}}
\newcommand{\ovomega}{{\overline{\omega}}}
\newcommand{\ovalpha}{{\overline{\alpha}}}

\newcommand{\eid}{{{\stackrel{e}{=}}}}

\newcommand{\tridnjk}[2]{{\triangledown_{#1,#2}}}
\newcommand{\tridnjkt}[3]{{\triangledown_{#1,#2}^{#3}}}
\newcommand{\tridnjktarg}[4]{{\tridnjkt{#1}{#2}{#3}(#4)}}
\newcommand{\prodjktarg}[4]{{{#4}_{#1,#2}^{#3}}}
\newcommand{\grpjktarg}[5]{{(\tridnjktarg{#1}{#2}{#3}{#4},\prodjktarg{#1}{#2}{#3}{#5})}}

\newcommand{\proditarg}[3]{{{#3}_{#1}^{#2}}}

\newcommand{\squarit}[2]{{\Box_{#1}^{#2}}}
\newcommand{\squaritarg}[3]{{\squarit{#1}{#2}(#3)}}
\newcommand{\squargrparg}[4]{{(\squaritarg{#1}{#2}{#3},\proditarg{#1}{#2}{#4})}}

\newcommand{\capkcalu}[1]{{K_{\bmk_{q+}^{#1}}^{\bmt_q^{#1}}(\calu)}}



\title{Any strongly controllable group system or group shift or any linear block code
is a linear system whose input is a generator group}

\author{Kenneth M. Mackenthun Jr. (email:  {\tt ken1212576@gmail.com})}
\maketitle

\vspace{3mm}
{\bf ABSTRACT}
\vspace{3mm}

Consider any sequence of finite groups $A^t$, where $t$ takes values in an integer index 
set $\mathbf{Z}$.  A group system $A$ is a set of sequences with components in
$A^t$ that forms a group under componentwise addition in $A^t$, for each $t\in\mathbf{Z}$.
In the setting of group systems, a natural definition of a linear system is a homomorphism from a group
of inputs to an output group system $A$.  We show that any group can be the input group
of a linear system and some group system.  In general the kernel of the homomorphism is nontrivial.
We show that any $\ell$-controllable complete group system $A$ is a linear system whose
input group is a generator group $({\mathcal{U}},\circ)$, deduced from $A$, and then the kernel 
is always trivial.  The input set ${\mathcal{U}}$ is a set of tensors, a double 
Cartesian product space of sets $R_{0,k}^t$, with indices $k$, for $0\le k\le\ell$, and time $t$, for
$t\in\mathbf{Z}$.  $R_{0,k}^t$ is a set of unique generator labels for the
generators in $A$ with nontrivial span for the time interval $[t,t+k]$.  We show
the generator group contains an elementary system, an infinite collection of elementary groups, one for 
each $k$ and $t$, defined on small subsets of ${\mathcal{U}}$, in the shape of triangles,
which form a tile like structure over ${\mathcal{U}}$.  There is a homomorphism from 
each elementary group to any elementary group defined on smaller tiles of the former group.
Any elementary system is sufficient to define a unique generator group up to isomorphism,
and therefore is sufficient to construct a linear system and group system as well.
Any linear block code is a 
strongly controllable group system which is nontrivial on a finite time interval.
Therefore it is a linear system whose input is a generator group, and we use the generator group
to obtain new results on the structure of block codes.  We give a new definition of group system
isomorphism using the generator group.
There is a harmonic theory of group systems:  a spectral domain expansion of $A$ has
been previously given by Forney and Trott, and we give a time domain expansion here.
We show these two expansions can be more simply obtained using the generator group, 
along with many other expansions as well.

{\bf Comment:}  Final editing.

\newpage

\topmargin      =-0.25in
\evensidemargin =0.10in 
\oddsidemargin  =-0.90in
\textheight     = 9.5in
\headheight     = 0.0in
\headsep        = 0.0in

\newpage
\vspace{3mm}
{\bf 1.  INTRODUCTION}
\vspace{3mm}

Consider any sequence of finite groups $A^t$, where $t$ takes values in an integer index 
set $\bmcpz$.  A group system $A$ is a set of sequences with components in
$A^t$ that forms a group under componentwise addition in $A^t$, for each $t\in\bmcpz$.
We will prove the following simple extension of the
first homomorphism theorem to show that it is possible to use a homomorphism to construct
a group system from any input group.  We can think of the following construction theorem 
as a {\it first homomorphism theorem for group systems}.

\begin{thm}
\label{thmone}
Consider any group $\msfcpg$.  Suppose there is a homomorphism $p^t:  \msfcpg\ra \msfcpg^t$ from $\msfcpg$
to a group $\msfcpg^t$ for each $t\in\bmcpz$.  In general group $\msfcpg^t$ may be different for each 
$t\in\bmcpz$.  Define the direct product group $(\msfcpg_\amalg,\plus)$ 
by $(\msfcpg_\amalg,\plus)\rmdef\cdots\times \msfcpg^t\times \msfcpg^{t+1}\times\cdots$.
There is a homomorphism $p$ from $\msfcpg$ to the direct product group $(\msfcpg_\amalg,\plus)$
defined by $p:  \msfcpg\ra \msfcpg_\amalg$ with assignment $p:  \msfg\mapsto\msfg_\amalg$, 
where sequence $\msfg_\amalg$ is defined by 
$$
\msfg_\amalg\rmdef\ldots,p^t(\msfg),p^{t+1}(\msfg),\ldots.
$$
Then
$$
\msfcpg/\msfcpg_K\simeq\imp,
$$
where group $\imp$ is the image of the homomorphism $p$, and where 
$\msfcpg_K$ is the kernel of the homomorphism $p$.  We have $\imp$ is a group system 
defined by a componentwise operation in $\msfcpg^t$ for each $t\in\bmcpz$.  Lastly we have 
$\msfcpg\simeq\imp$ \ifof\ the kernel $\msfcpg_K$ of the homomorphism $p$ is the identity.
\end{thm}

In an engineering sense, we can think of group $\msfcpg$ as an input group and
group system $\imp$ as an output group.
In the setting of groups, it is natural to define a linear system as a homomorphism from a group
of inputs to a group of outputs.
Therefore we can regard any group system constructed by the above theorem as a linear system.

In general, if we use any group as an input, the kernel of the homomorphism is nontrivial,
and the linear system may not be very interesting.
However, in the case of any $\ell$-controllable complete group system $A$, we show that 
there is a unique input group up to isomorphism, 
the $(\ell+1)$-depth generator group $(\calu,\circ)$ of $A$,
which gives a homomorphism $f_u$ with a trivial kernel, as shown in (\ref{chain1a}).
This shows that any $\ell$-controllable complete group system $A$
is a generalization of the classical linear system.
Since the kernel is trival, the generator group is isomorphic to the group system $A$, 
and we can use the generator group to study the structure of the group system $A$.
\be
\label{chain1a}
\begin{array}{lll}
\da & {\stackrel{=}{\la}}        & \imfu         \\
\da &                            & \ua\,f_u      \\
A   &                            & (\calu,\circ) 
\end{array}
\ee

Furthermore, the input group is derived from the output group; in other words,
the generator group $(\calu,\circ)$ is derived from the group system $A$.
We obtain the set $\calu$ in the following way.
Any group system $A$ has a description as a canonical realization, 
group system $B$ \cite{FT}.  A generator $\bmg^{[t,t+k]}$ is a sequence in $B$ which 
is the identity except for a nontrivial span on the time interval $[t,t+k]$ \cite{FT}. 
We can think of all the generators as a basis of $B$ (or of $A$).
To obtain a particular output sequence $\bmb$ in $B$, we select a generator from the basis for each
time $t$, $t\in\bmcpz$, for $0\le k\le\ell$.  This list of generators forms a tensor $\bmr$.
An output sequence $\bmb\in B$ or $\bma\in A$ can be calculated directly from a tensor $\bmr$.
The set of tensors $\bmr$ forms set $\calr$.  Any tensor $\bmr\in\calr$ can be converted
to another tensor $\bmu$ by replacing the generators $\bmg^{[t,t+k]}$ in $\bmr$ with 
unique \glabs\ in $\bmu$.  The set of tensors $\bmu$ forms set $\calu$.  
The set $\calu$ is a double 
Cartesian product space of sets $R_{0,k}^t$, with indices $k$, for $0\le k\le\ell$, and time $t$, for
$t\in\bmcpz$, where $R_{0,k}^t$ is a set of unique generator labels $r_{0,k}^t$.  The set of tensors
$\bmu\in\calu$ is a set of inputs for the linear system (\ref{chain1a}).
Since the homomorphism $f_u$ in (\ref{chain1a}) is a bijection, we can find a
unique input $\bmu\in\calu$ for each output $\bma$ in $A$, and we say the linear system
is invertible.  An invertible linear system is usually the case of interest for
a communications engineer.

Because of (\ref{chain1a}), the study and construction of generator groups $(\calu,\circ)$
is essentially the study and construction of \ellctl\ complete group systems.  Further,
we show that any $(\ell+1)$-depth generator group $(\calu,\circ)$ of a group system $A$ contains
an $(\ell+1)$-depth elementary system $\cale_A$, which is not a group but a collection of elementary groups.
Given an $(\ell+1)$-depth elementary system $\cale_A$ of a group system $A$, 
we can always recover the $(\ell+1)$-depth generator group $(\calu,\circ)$ of $A$,
up to an isomorphic and essentially identical group.
Therefore any \ellctl\ complete group system $A$ 
can be constructed from either the generator group $(\calu,\circ)$ of $A$ or
the $(\ell+1)$-depth elementary system $\cale_A$ of $A$,
using the \fhgs, and the homomorphism $f_u$ is a bijection (see (\ref{chain1b})).  
Moreover, starting from any constructed $(\ell+1)$-depth elementary system $\cale$, 
the \fhgs\ can always construct an \ellctl\ complete group system.
Therefore the study and construction of \ellctl\ complete group systems
is also essentially the study and construction of elementary systems.
\be
\label{chain1b}
\begin{array}{lllll}
\da & {\stackrel{=}{\la}}        & \imfu         &                           &         \\
\da &                            & \ua\,f_u      &                           &         \\
A   &                            & (\calu,\circ) & {\stackrel{\simeq}{\lra}} & \cale_A
\end{array}
\ee

An elementary system is an infinite collection of elementary groups, one for 
each $k$ and $t$, defined on small subsets of $\calu$, in the shape of triangles,
which form a tile like structure, or snakeskin structure,
over $\calu$.  There is a homomorphism from 
each elementary group to any elementary group wholely contained in the former group.
The elementary system is nested, since the elementary groups in each row are one step
larger than elementary groups in the preceding row.  Because of nesting and the homomorphism
property, it is easy to construct an elementary system row by row and from this any
$\ell$-controllable complete group system.

A special case of the \ellctl\ complete group system is a 
group system which is only nontrivial on a finite time interval $[t,t+\ell]$.
We call this a block group system, often called a block code over a
group, a linear block code over a group, or a linear block code \cite{MS}.  
Since a block group system is a group system,
any block group system can be decomposed into a set of generators \cite{FT}.
Knowing the decomposition into generators, we can describe the form of the set $\calu$.
For the block group system, the set $\calu$ has a particularly simple form:
the sets of \glabs\ which may be nontrivial form a pyramid shape in $\calu$, or
a Pascal like triangle in which each row is nested in the row below it.  Because 
the nontrivial \glabs\ have a pyramid shape, the nontrivial elementary groups
have a pyramid shape as well.
Since the nontrivial \glabs\ and elementary groups have a pyramid shape, the 
construction of a block group system using its elementary system
is particularly simple.  We give an example showing the construction
of the binary $(8,4,4)$ extended Hamming code or first order Reed-Muller code
as a linear system.

We use the generator group to describe the additive structure of the codewords and
generators which comprise the block group system.  Since a linear block code over
a group is embedded in any linear block code over a field, vector space, ring, or
module, these new results on additive structure of the generators apply to 
these cases as well.

The study of group shifts started with the work of Kitchens \cite{KIT} in symbolic
dynamics theory; the study of group systems started with the work of Willems \cite{JW} in 
linear system theory, and the study of group codes started with the work of 
Forney and Trott \cite{FT}, and Loeliger and Mittelholzer \cite{LM} in coding theory.
Kitchens \cite{KIT} introduced the idea of a group shift \cite{LMr}
and showed that a group shift has finite memory, i.e., 
it is a shift of finite type \cite{LMr}.  Willems reenvisaged linear system theory
by starting with a set and group of sequences and deriving all properties of the
system from this set.  Following the work of Willems, Forney and Trott \cite{FT} 
take the view that all properties of a group code $A$ (we use notation
$A$ in place of $C$ in \cite{FT}) can be obtained by just starting 
with a set of sequences that has a group property.  
They describe the state group and state code of $A$.
A time invariant group code is essentially a group shift.
They show that any group code that is complete
(the group code can be characterized by its local behavior, see \cite{FT})
can be wholely specified by a sequence of connected labeled group trellis sections 
(which may vary in time), a trellis.
They explain the important idea of ``shortest length code sequences''
or generators.  A generator is a code sequence which is not a combination
of shorter sequences.  In a strongly controllable group code, the 
nontrivial portions of all generators have a bounded length.  They
give an encoder whose inputs are generators and whose outputs are codewords 
in the group code.  At each time $t$, a finite set of generators is used to 
give an output letter in the codeword.

Loeliger and Mittelholzer \cite{LM} obtain an analog of the derivation
of Forney and Trott starting with a trellis having group properties instead 
of the group of sequences $A$.  To derive their encoder, 
they use an intersection of paths which split and merge to the 
identity path in the trellis, an analog
of the quotient group of code sequences (granule) in $A$
used in \cite{FT}.

Forney and Trott suggest the term group system as an alternative to
group code.  In this paper, we principally use the term 
group system rather than group code because some results have analogues 
in classical systems theory and harmonic analysis.  
We consider the general time varying group system.  
Time invariant group systems are the same as
group shifts \cite{LMr}; therefore the results here also apply 
to group shifts.  A general time varying group system can have nontrivial values
over a finite interval, in which case it is a block group system, often called
a block code over a group \cite{FT} or linear block code.
Therefore the results here also apply to block codes.

Both Willems and Forney and Trott have taken the approach to systems theory that
all properties of the system can be derived starting from the sequences in the system themselves.
Willems has shown for linear systems that this approach gives a notion of an input sequence.
Similarly Forney and Trott have shown that there is a notion of input for a group system,
the sequence of first components of the generators.  However as discussed in \cite{FT},
they do not show there is a homomorphic map from the 
input sequence space to the output sequence space.  Therefore, they do not show
a group system is a linear system.   Likewise, as also discussed in \cite{FT}, the work of
Brockett and Willsky \cite{BW} on ``finite group homomorphic sequential systems''
does not suffice to realize general linear systems over groups.
One of the main aims of this paper is to show that the generators themselves 
form a group, and there is a homomorphism from this group, the input group, to the group system,
the output group.  Therefore a group system is a linear system.  This 
fills a gap in the theory of Willems and Forney and Trott.  The generator group
is very revealing of the structure of group systems
and seems to give a new approach to understanding group systems and related studies
such as linear systems, group codes, group shifts, and linear block codes.

The Forney-Trott encoder (and Loeliger-Mittelholzer analog) is
a global approach to finding an encoder of $A$.  They use a \cdc\ of
infinite extent.  The group code $A$ has a normal chain
$A_0\subset A_1\subset\cdots\subset A_k\subset\cdots\subset A_\ell=A$,
where each $A_k$ is a $k$-controllable subcode of $A$ \cite{FT}.
The $k$-controllable subcode $A_k$ is a sequence of quotient groups (granules)  
$\Gamma^{[t,t+k]}$.  The \creps\ of the granule $\Gamma^{[t,t+k]}$
are generators $\bmg_a^{[t,t+k]}$ which are
trivial except for the time interval $[t,t+k]$.  The product $\prod_{t\in\bmcpz}\bmg_a^{[t,t+k]}$
of the generators is a \crep\ of subcode $A_k$.  The product over all the subcodes,
$\prod_{0\le k\le\ell}\prod_{t\in\bmcpz}\bmg_a^{[t,t+k]}$, is an output sequence in $A$.

We develop a different encoder which also uses a \cdc\ of infinite extent.
We apply the Schreier refinement theorem and Zassenhaus lemma used to prove 
the Jordan-Holder theorem \cite{ROT}
to two different normal chains.  The refinement of the Zassenhaus lemma
is the \cdc\ we use.  The sequence of quotient groups in the refinement
normal chain have representatives which include the Forney-Trott generators.
The resulting encoder is time reversed and product interchanged compared
to the Forney-Trott encoder \cite{KM5}.  The encoder we find has the form of
a time convolution, but the encoders of Forney and Trott and Loeliger and Mittelholzer
do not have the form of a time convolution.  For this reason, we call the encoder 
found here \cite{KM5} a time domain encoder, 
and the Forney and Trott encoder a spectral domain encoder.  
These results show there is a time and harmonic theory of group systems \cite{KM5}.
The time domain and spectral domain encoders are derived from first principles.
In this paper we show the time and spectral domain encoders of $A$ can be more easily obtained
as the \crepc s of two different \cdc s of the generator group of $A$.  We show the generator
group has many many other \cdc s which give other encoders and expansions.

Forney gives a fast soft decision decoding algorithm
for the Reed-Muller codes, the Golay code, and the Leech lattice \cite{FY0}.
On one level, the algorithm can be interpreted as an efficient walk through the 
pyramid shape or Pascal triangle of \glabs\ in $\calu$ or elementary groups in
$(\calu,\circ)$ or $\cale_A$.  The \crepc s\ found for different \cdc s of the
generator group also amount to walks through the generator group.  With this interpretation, 
it may be possible to use these \crepc s to find efficient soft decision decoding algorithms for
other linear block codes, or to construct new linear block codes which have
efficient soft decision decoding algorithms.  This topic is not explored
further here.

We now discuss the organization of the paper in more detail.
A group system $A$ is a set of sequences $\bma$ of letters $a^t$ in alphabet $A^t$,
for each $t\in\bmcpz$, that forms a group under componentwise group addition \cite{FT}.
We only consider strongly controllable group systems $A$, in which there is
a fixed least integer $\ell$ such that for any time $t$, for any pair of sequences
$\bma,\bmadt\in A$, there is a sequence $\bmaddt\in A$ that agrees with $\bma$ on
$(-\infty,t]$ and agrees with $\bmadt$ on $[t+\ell,\infty)$ \cite{FT}.
In this case we say $A$ is \ellctl.  

Forney and Trott show that any group system $A$ has a canonical realization
$(A,\bsigma(A))$, where $\bsigma(A)$ is the state code of $A$, the sequences
of states of $A$.  The canonical realization is a group system $B$ and a minimal
description of $A$.  The canonical realization has an equivalent modeling
as a trellis diagram, group trellis $B$.
The vertices of $B$ form the state code of $A$.  The labels of branches in $B$
form sequences in $A$.  We can regard group trellis $B$ as a connected sequence 
of trellis sections $B^t$, also called branch
groups, where each branch $b^t$ in $B^t$ is an ordered triple $(s^t,a^t,s^{t+1})$
of state, alphabet letter, and next state.  We only study group systems $A$ that are complete.  
These are group systems which can
be completely characterized by local behavior, i.e., no global constraints are
needed.  If $A$ is complete, then $A$ is completely specified by its
trellis diagram $B$ \cite{JW}.  This material is reviewed in Section 2.

In group trellis $B$, the sequence of branches that split from the 
identity path for each $t\in\bmcpz$, and merge to the identity path for each 
$t\in\bmcpz$, form two normal chains of $B$.
The Schreier refinement theorem \cite{ROT} can be applied to these two normal
chains to obtain another normal chain, a refinement of the two
chains.  A \crep\ of a quotient group of the refinement chain
is a generator $\bmg^{[t,t+k]}$.
The \creps\ of the quotient groups of the refinement chain reduce
to a sequence of generators which form a tensor $\bmr$.
The set of all possible tensors $\bmr$ is a set $\calr$.  At each time $t$, 
a component of the tensor $\bmr$ is a triangular matrix, the \stm\
$\tridnjktarg{0}{0}{t}{\bmr}$.  Each \stm\ $\tridnjktarg{0}{0}{t}{\bmr}$
is formed by branches of generators $\bmg^{[t,t+k]}$ at times $t-j$, for $j=0,1,\ldots,\ell$,
for $j\le k\le\ell$.  Each path 
$\bmr\in\calr$ is encoded into a path $\bmb\in B$, and any $\bmb\in B$
is the encoding of some path $\bmr\in\calr$.  This gives a bijection $\calr\ra B$.
It is easy to find the encoder for $\bma$ from the encoder for $\bmb$, and
then compare with the Forney-Trott encoder.  This is discussed in Section 3.

In Section 4, we study the time $t$ component of the refinement chain of $B$.
We show that the branches in generators in $\bmr$ at time $t$, e.g., the branches in \stm\
$\tridnjktarg{0}{0}{t}{\bmr}$, form a \crepc\ of a \cdc\ of the branch group $B^t$.

If $\bmbdt,\bmbddt$ are two paths in $B$, then their product $\bmbdt\bmbddt$ 
is another path in $B$.  Let $\bmrdt\ra\bmbdt$ and $\bmrddt\ra\bmbddt$ under the
bijection $\calr\ra B$.  We define an operation $*$ on $\calr$ by 
$\bmrdt*\bmrddt\rmdef\bmrbr$ if $\bmrbr\mapsto\bmbdt\bmbddt$ under the
bijection $\calr\ra B$.  The set of tensors $\calr$ with operation $*$ forms a 
group $(\calr,*)$ called a decomposition group of $A$, and $(\calr,*)\simeq B$.
So far we have shown
$$
A\,\,{\stackrel{\simeq}{\ra}}\,\,B\,\,{\stackrel{\simeq}{\ra}}\,\,(\calr,*).
$$
This is discussed in Subsection 5.1.

Let $\tridnjktarg{0}{0}{t}{\calr}$ be the set of triangular forms
$\{\tridnjktarg{0}{0}{t}{\bmr}:  \bmr\in\calr\}$. 
Each triangular form $\tridnjktarg{0}{0}{t}{\bmr}$ has 
representatives $r_{j,k}^{t-j}$ as entries, for
columns $j$ such that $0\le j\le\ell$, for rows $k$ such that $j\le k\le\ell$.
The entry $r_{j,k}^{t-j}$ is the component at time $t$ in generator $\bmg^{[t-j,t-j+k]}$.
Since the entries in $\tridnjktarg{0}{0}{t}{\bmr}$ form a \crepc\ of $B^t$, then $B^t$ induces a
group $\grpjktarg{0}{0}{t}{\calr}{\cast}$ on the set of triangular forms 
$\tridnjktarg{0}{0}{t}{\calr}$.  The upper coordinates of the set 
$\tridnjktarg{0}{0}{t}{\calr}$ are a set of subtriangles $\tridnjktarg{k}{k}{t}{\calr}$ 
of $\tridnjktarg{0}{0}{t}{\calr}$.  In any finite group $G$, the upper
coordinates of a \compset\ form a group induced by group multiplication
in $G$.  This means the set of subtriangles of upper coordinates 
$\tridnjktarg{k}{k}{t}{\calr}$ forms a group $\grpjktarg{k}{k}{t}{\calr}{\cast}$
under multiplication induced by $B^t$, for $k$ such that $0\le k\le\ell$.
Shifts $\tridnjktarg{k-j}{k}{t-j}{\calr}$ of subtriangles $\tridnjktarg{k}{k}{t}{\calr}$, 
for $0\le j\le k$, contain representatives from the same set of generators
as representatives of $\tridnjktarg{k}{k}{t}{\calr}$.  
This means that a representative in $\tridnjktarg{k-j}{k}{t-j}{\calr}$ can be put
in 1-1 corrrespondence with a representative in $\tridnjktarg{k}{k}{t}{\calr}$
if they belong to the same generator.  Under this 
1-1 correspondence, the set of subtriangles $\tridnjktarg{k-j}{k}{t-j}{\calr}$ forms a group
$\grpjktarg{k-j}{k}{t-j}{\calr}{\cast}$ isomorphic to 
$\grpjktarg{k}{k}{t}{\calr}{\cast}$, for $0\le j\le k$.
We call groups $\grpjktarg{j}{k}{t}{\calr}{\cast}$,
for $t\in\bmcpz$, for $0\le k\le\ell$, for $0\le j\le k$, 
the {\it elementary groups} of $(\calr,*)$.  We have shown that any decomposition 
group $(\calr,*)$ can be reduced to an infinite collection of elementary groups.
The elementary groups have a homomorphism property:  if any set
$\tridnjktarg{m}{n}{t}{\calr}$ forms subtriangles in another set $\tridnjktarg{j}{k}{t}{\calr}$,
then the projection map $\tridnjktarg{j}{k}{t}{\bmr}\mapsto\tridnjktarg{m}{n}{t}{\bmr}$
defines a homomorphism from $\grpjktarg{j}{k}{t}{\calr}{\cast}$ to
$\grpjktarg{m}{n}{t}{\calr}{\cast}$.  This is discussed in Subsection 5.2.

An element $\bmr\in\calr$ is equivalent to a sequence of generators $\bmg^{[t,t+k]}$ for 
$0\le k\le\ell$, for each $t\in\bmcpz$.  We replace each
generator $\bmg^{[t,t+k]}$ in $\bmr\in\calr$ with a \glab\ $r_{0,k}^t$.
Under the assignment $\bmg^{[t,t+k]}\mapsto r_{0,k}^t$
for $0\le k\le\ell$, for each $t\in\bmcpz$, a tensor $\bmr\in\calr$
becomes a tensor $\bmu$, $\bmr\mapsto\bmu$.  Let $\calu$ be the set
of tensors $\bmu$ obtained from $\calr$ this way.
The operation $*$ in $(\calr,*)$ determines an operation $\circ$ on
$\calu$, and this gives a group $(\calu,\circ)$ isomorphic to $(\calr,*)$, called a generator group.
The group $(\calu,\circ)$ collapses the sequence of generators 
in $\bmr\in\calr$ to a sequence of \glabs\ in $\bmu\in\calu$.
The group $(\calu,\circ)$ collapses any sequence of isomorphic elementary groups
$\grpjktarg{j}{k}{t}{\calr}{\cast}$ in $(\calr,*)$, for $0\le j\le k$, to a single
isomorphic elementary group $\grpjktarg{0}{k}{t}{\calu}{\ccirc}$ in $(\calu,\circ)$.
There is a homomorphism among elementary groups of $(\calu,\circ)$ in the same way as
for $(\calr,*)$.  We use the \fhgs\ to recover group system $A$ from 
generator group $(\calu,\circ)$.  Essentially, the \fhgs\ constructs the generators of $A$ from 
the \glabs\ of $(\calu,\circ)$.  Thus we have shown the following more detailed version
of (\ref{chain1a}), where $f_u$ is a homomorphism.
\be
\begin{array}{lllllll}
 \da & \la                        & \la   & \la                        & \la         & {\stackrel{=}{\la}}      & \imfu         \\
 \da &                            &       &                            &             &                          & \ua f_u       \\
 A   & {\stackrel{\simeq}{\ra}}   & B     & {\stackrel{\simeq}{\ra}}   & (\calr,*)   & {\stackrel{\simeq}{\ra}} & (\calu,\circ)           
\end{array}
\ee
This shows that a strongly controllable complete group system
$A$ is a linear system whose input group is a generator group.
This is discussed in Section 6.  In Subsection 6.6, we give an efficient
algorithm to construct all \ellctl\ complete group systems up to isomorphism from the set of
all generator groups of \ellctl\ complete group systems.  

In Subsection 6.4, we give a harmonic theory of group systems.  We show the spectral domain
encoder and time domain encoder are related at a deeper level, and that in general there are
many other encoders and expansions besides the spectral and time domain.

The infinite collection of elementary groups $\grpjktarg{0}{k}{t}{\calu}{\ccirc}$
of $(\calu,\circ)$, for $0\le k\le\ell$, for $t\in\bmcpz$, forms a 
nested tile pattern on the double Cartesian product space $\calu$.
Since $0\le k\le\ell$, and $t\in\bmcpz$, the product space has finite depth and  
infinite length.  We call the double Cartesian product space an elementary set.
We call the infinite collection of elementary groups an elementary list.
The elementary set and elementary list, together with a homomorphism
condition, form an $(\ell+1)$-depth {\it elementary system} $\cale_A$.  
The homomorphism condition is that for each $k$ such that $0\le k<\ell$, for each
$t\in\bmcpz$, there is a homomorphism from 
elementary group $\grpjktarg{0}{k}{t}{\calu}{\ccirc}$ to
the next two largest elementary groups nested in $\grpjktarg{0}{k}{t}{\calu}{\ccirc}$. 
Then we have shown that the generator group of any \ellctl\ complete group system $A$ can be reduced
to an elementary system $\cale_A$.  This is discussed in Section 7.

In the remainder of Section 7, we show that this process can be reversed.
We can use the infinite collection of elementary groups of $\cale_A$ to 
define a global operation on the elementary set $\calu$.
This gives an $(\ell+1)$-depth global group
$(\calu,\star)$ which is isomorphic and essentially identical
to generator group $(\calu,\circ)$.  Then again we can
use the \fhgs\ to recover $A$ from global group $(\calu,\star)$; the
mapping $f_u$ remains the same since $(\calu,\circ)$ and $(\calu,\star)$
share the same group elements.  
Thus we have shown the following more detailed version of (\ref{chain1b}).
\be
\begin{array}{lllllllll}
 \da & \la                        & \la   & \la                        & \la         & {\stackrel{=}{\la}}      & \imfu         &     &             \\
 \da &                            &       &                            &             &                          & \ua\,f_u      &     &             \\
 A   & {\stackrel{\simeq}{\ra}}   & B     & {\stackrel{\simeq}{\ra}}   & (\calr,*)   & {\stackrel{\simeq}{\ra}} & (\calu,\circ) & \ra & \cale_A     \\       
     &                            &       &                            &             &                          & \uda\,\eid    &     & \uda =      \\
     &                            &       &                            &             &                          & (\calu,\star) & \la & \cale_A
\end{array}
\ee
In Subsection 7.3, we give a new notion of isomorphism for group systems which appears to be
more appropriate than the definition of isomorphism for finite groups.  We give an efficient
algorithm to construct all \ellctl\ complete group systems up to this new notion of
isomorphism from the set of all elementary systems $\cale_A$.

Moreover, if we construct any $(\ell+1)$-depth elementary system $\cale$, we can use
this same process to construct an \ellctl\ complete group system from $\cale$.  Therefore the
construction of an \ellctl\ complete group system amounts to the
construction of any $(\ell+1)$-depth elementary system $\cale$.
Since the $(\ell+1)$-depth elementary system has finite depth and is nested by depth, the construction
of any $(\ell+1)$-depth elementary system is very simple.  The algorithm is essentially a construction of 
the generators of the group system $B$ and $A$.  This appears to be the 
first construction algorithm given for a general time varying group system.
This is discussed in Subsection 7.4.
We show in v7 and earlier versions of \cite{KM6} that the sequence of quotient groups
used here in construction is the dual of the sequence of $k$-controllable
subcodes of $A$ used to construct the encoder in \cite{FT}.

Since a group system is a linear system, we can use the generator group to 
study the structure of the group system.  The additive structue of the \glabs\
in the generator group describes the additive structure of the generators in
the group system.  Moreover, finite or infinite collections of elementary
groups of $(\calu,\circ)$ also form a group, as shown in Subsection 6.3.  
Then we can also use collections of elementary groups to parse out the structure of the group system.

In Section 8, we study a special case of the general time varying \ellctl\ group system
which is only nontrivial on a finite time interval $[t,t+\ell]$, called
a block group system.  For a block group system, the set $\calu$ has a particularly simple form:
the sets of \glabs\ which may be nontrivial form a pyramid shape in $\calu$; the nontrivial
elementary groups in $(\calu,\circ)$ also form a pyramid shape.
In Section 8.1, we use all possible combinations of
elementary groups in the pyramid shape in $(\calu,\circ)$ that form a group 
to study the structure of all block codes with $\ell=2$.
In Section 8.2, we give an example of a homomorphism between the generator group and group system 
of the $(8,4,4)$ extended Hamming code, where $\ell=3$.
Finally, since the nontrivial elementary groups in $(\calu,\circ)$ have a pyramid shape, the 
construction of a block group system using its elementary system
is particularly simple.  In Section 8.3, we give an example of this
construction for the $(8,4,4)$ extended Hamming code.

\newpage
\vspace{3mm}
{\bf 2.  GROUP SYSTEMS}
\vspace{3mm}

This section gives a very brief review of some fundamental concepts in 
\cite{FT}, and introduces some definitions used here.
We follow the notation of Forney and Trott as closely as possible.
One significant difference is that subscript $k$ in \cite{FT} denotes
time; we use $t$ (an integer) in place of $k$.  In any notation, a
superscript is used exclusively to indicate time; thus $t$ always 
appears as a superscript in any notation.  We use notation $\bone$
in a generic sense as the identity of any group of sequences; the particular group
should be clear by context.  Similarly, we use notation $1^t$
in a generic sense as the identity of any group defined for time $t$.

Forney and Trott study a collection of sequences with time axis defined on the
set of integers $\bmcpz$, whose components $a^t$ are taken from 
an {\it alphabet group} or {\it alphabet} $A^t$ at each time $t$, $t\in\bmcpz$.  
The set of sequences is a group under componentwise addition in $A^t$ \cite{FT}.  
We call this a {\it group system} $A$.
A sequence $\bma$ in $A$ is given by
\be
\label{c9}
\bma=\ldots,a^{t-1},a^t,a^{t+1},\ldots,
\ee
where $a^t\in A^t$ is the component at time $t$.
We assume that all elements of $A^t$ are represented in $A$ at time $t$.

A group system $A$ can be isomorphic to another group system or another group.
Let $G$ be a group or group system.  Then $A\simeq G$ \ifof\ there is a bijection
$A\ra G$ such that if $\bmadt,\bmaddt\in A$ and $\bmadt\mapsto\gdt$,
$\bmaddt\mapsto\gddt$, then $\bmadt\bmaddt\mapsto\gdt\gddt$.

In this paper, we study {\it complete} group systems  \cite{JW,FT}.  A complete
group system can be characterized by its local behavior; in particular complete
systems can be generated by their trellis diagrams \cite{FT}.
Forney and Trott construct their canonical encoder for a complete strongly 
controllable group system.  Completeness is the same as closure in symbolic 
dynamics \cite{LMr}.   Therefore a time invariant complete group system 
$A$ is the same thing as a {\it group shift} in symbolic dynamics.  
Kitchens \cite{KIT} shows that any time invariant complete group system 
(group shift) must have finite state spaces.  In this paper, we use the
language associated with group systems \cite{FT} rather than group shifts
\cite{LMr}.  For an incomplete group system, a global constraint is 
required to fully specify the group system.  Some examples of group systems 
that require a global constraint are given in \cite{FT}.

As in \cite{FT}, we use conventional notation for time intervals.
If $m\le n$, the time interval $[m,n]$ starts at time $m$, ends at time $n$, and
has {\it length} length $n-m+1$.  We also write time interval $[m,n]$ as
$[m,n+1)$.  The time interval $[m,m]$ or $[m,m+1)$ has length 1 and is also written just $m$.

Let $A$ be a group system, and let $\bma$ be a sequence in $A$.
Using (\ref{c9}), define the projection map at time $t$, 
$\chi^t:  A\ra A^t$, by the assignment $\bma\mapsto a^t$.
Define the projection map $\chi^{[t_1,t_2]}:  A\ra A^{t_1}\times\cdots\times A^{t_2}$ 
by the assignment $\bma\mapsto (a^{t_1},\ldots,a^{t_2})$.  In general,
we say that sequence $\bma$ has {\it span} $t_2-t_1+1$ if $\bma$ is the same as the identity
sequence except for a finite segment $(a^{t_1},\ldots,a^{t_2})$ of length $t_2-t_1+1$,
where $a^{t_1}\ne1^{t_1}$, $a^{t_2}\ne1^{t_2}$, and $1^t$ is the identity of $A^t$.  
We define $A^{[t_1,t_2]}$ to be the sequences in $A$ which are the identity outside time 
interval $[t_1,t_2]$.

A group system $A$ is {\it $[m,n)$-controllable} if for any
$\bmadt,\bmaddt\in A$, there exists a sequence $\bma\in A$ with
$\chi^{(-\infty,m)}(\bma)=\chi^{(-\infty,m)}(\bmadt)$ and 
$\chi^{[n,+\infty)}(\bma)=\chi^{[n,+\infty)}(\bmaddt)$.  
Then the finite segment $\chi^{[m,n)}(\bma)$ of length $n-m$ in $\bma$ 
connects the past $\chi^{(-\infty,m)}(\bmadt)$ of $\bmadt$ to
the future $\chi^{[n,+\infty)}(\bmaddt)$ of $\bmaddt$ \cite{FT}.  
A group system $A$ is {\it $l$-controllable} if there is an integer $l>0$ 
such that $A$ is $[t,t+l)$-controllable for all $t\in\bmcpz$.  
A group system $A$ is 
{\it strongly controllable} if it is $l$-controllable for some integer $l$.  
The least integer $l$ for which a group system
$A$ is strongly controllable is denoted as $\ell$.
In this paper we study strongly controllable group systems.

For each $t\in\bmcpz$, define $X_a^t$ to be the set of all sequences $\bma$ in $A$ for which 
$a^n=1^n$ for $n<t$, where $1^n$ is the identity of $A^n$ at time $n$ (see Figure
\ref{statedefn}).  For each $t\in\bmcpz$, define $Y_a^t$ to be the set of all sequences $\bma$ 
in $A$ for which $a^n=1^n$ for $n>t$.
The {\it canonic state space} $\Sigma_a^t$ of $A$ at time $t$ is defined to be 
$$
\Sigma_a^t\rmdef\frac{A}{Y_a^{t-1}X_a^t}.
$$
(Note that $A$ is the same as $C$ in \cite{FT}, $Y_a^{t-1}$ is the same as $C^{t^-}$, 
and $X_a^t$ is the same as $C^{t^+}$.)
From Figure \ref{statedefn}, it is evident the definition of the state at time $t$
involves a split between time $t-1$ and time $t$ \cite{FT}.  The canonic state space 
is unique.  The group system satisfies the {\it axiom of state}:  whenever
two sequences pass through the same state at a given time,
the concatenation of the past of either with the future of the other
is a valid sequence \cite{FT}.

\begin{figure}[h]
\centering

\begin{picture}(100,80)(0,-10)

\put(0,0){\line(1,0){100}}
\put(0,2){\line(1,0){40}}
\put(40,4){\line(1,0){50}}
\put(40,2){\line(1,1){50}}
\put(40,4){\line(-1,1){50}}

\put(30,-9){\makebox(0,0)[t]{$t-1$}}
\put(50,-9){\makebox(0,0)[t]{$t$}}
\put(70,-9){\makebox(0,0)[t]{$t+1$}}
\put(30,-5){\line(0,1){5}}
\put(50,-5){\line(0,1){5}}
\put(70,-5){\line(0,1){5}}
\put(3,50){\makebox(0,0)[l]{$Y_a^{t-1}$}}
\put(80,50){\makebox(0,0)[r]{$X_a^t$}}

\end{picture}

\caption{Definition of $Y_a^{t-1}$ and $X_a^t$.}
\label{statedefn}

\end{figure}

The state $\sigma^t(\bma)$ of a system sequence $\bma$ at time $t$
is determined by the natural map
$$
\sigma^t:  A\ra A/(Y_a^{t-1}X_a^t)=\Sigma_a^t,
$$
a homomorphism.  There is therefore a well defined state 
sequence $\bsigma(\bma)=\{\sigma^t(\bma): t\in\bmcpz\}$
associated with each $\bma\in A$,
\be
\label{c9a}
\bsigma(\bma)=\ldots,s^{t-1},s^t,s^{t+1},\ldots,
\ee
where $s^t\in\Sigma_a^t$ for each $t\in\bmcpz$.
The set of sequences $\{\bsigma(\bma): \bma\in A\}$ is the {\it state code}
$\bsigma(A)$ of $A$.  The {\it canonical realization} $(A,\bsigma(A))$ of $A$
is the set of all pairs of sequences $(\bma,\bsigma(\bma))$,
\be
\label{s90}
\{(\bma,\bsigma(\bma)):  \bma\in A\}.
\ee
The canonical realization $(A,\bsigma(A))$ is a group system isomorphic to $A$ under the assignment
$\bma\mapsto(\bma,\bsigma(\bma))$.  The canonical realization of $A$ is a {\it minimal} 
realization of $A$ \cite{FT}; a minimal realization of $A$ is a state description of $A$
that uses the minimum number of states.

It is useful to write the canonical realization in a different form.
We write an element $(\bma,\bsigma(\bma))$ of the canonic realization
as a sequence $\bmb$,
\be
\label{eqn0}
\bmb=\ldots,b^{t-1},b^t,b^{t+1},\ldots,
\ee
where component $b^t$ is given by an ordered triple $b^t=(s^t,a^t,s^{t+1})$, where 
$s^t\in\Sigma^t$ is the state in $\bsigma(\bma)$ at time $t$, $a^t$ is
the component of $\bma$ at time $t$, and $s^{t+1}\in\Sigma^{t+1}$ is the state 
in $\bsigma(\bma)$ at time $t+1$.  At each time $t$, the set of components $b^t$ 
forms a subdirect product group $B^t$, a subgroup of the direct 
product group $\Sigma^t\times A^t\times\Sigma^{t+1}$.  
The set of sequences $\bmb$ forms a group system $B$ with componentwise
addition in group $B^t$.  Group system $B$ is just an isomorphic copy of
the canonical realization under the assignment $(\bma,\bsigma(\bma))\mapsto\bmb$.

$B$ is a group system.  Therefore we can find its canonic state space.  
For each $t\in\bmcpz$, define $X^t$ to be the set of all sequences $\bmb$ in $B$ for which 
$b^n=1^n$ for $n<t$, where $1^n$ is the identity of $B^n$ at time $n$.  For each $t\in\bmcpz$, 
define $Y^t$ to be the set of all sequences $\bmb$ in $B$ for which $b^n=1^n$ for $n>t$.
Then the canonic state space $\Sigma_b^t$ of $B$ at time $t$ is
$$
\Sigma_b^t\rmdef\frac{B}{Y^{t-1} X^t}.
$$
It is clear that $\Sigma_a^t\simeq\Sigma_b^t$.  For example, if $1_{\Sigma_a}^t$ is
the identity state of $\Sigma_a^t$, then the identity state of $\Sigma_b^t$ is 
formed by sequences $Y^{t-1} X^t$ of the form
$$
\ldots,(s^{t-1},a^{t-1},1_{\Sigma_a}^t),(1_{\Sigma_a}^t,a^t,s^{t+1}),\ldots.
$$

$B$ can be used to find a {\it trellis diagram} of $A$ \cite{FT}.
The components of $B^t$ can be arranged in a bipartite graph, a {\it trellis section} $T^t$,
where the left vertices are states in $\Sigma^t$, the right vertices
are states in $\Sigma^{t+1}$, and the label of a {\it branch} 
$b^t=(s^t,a^t,s^{t+1})$ between state $s^t$ and state $s^{t+1}$ is $a^t\in A^t$.  
We think of $b^t=(s^t,a^t,s^{t+1})$ as a branch in {\it branch group} $B^t$.  
Then the canonic realization $(A,\bsigma(A))$ of $A$ can be described by a {\it trellis} graph,
a connected sequence of trellis sections, where $T^t$ and $T^{t+1}$ are joined
together using their common states in $\Sigma^{t+1}$ \cite{FT}.  Then
we can also think of group system $B$ as a {\it group trellis}, a graphical model
of $(A,\bsigma(A))$ and $A$.

For the purposes of this paper, we summarize these results \cite{FT} as follows.

\begin{thm}
\label{thm1}
We have an isomorphism $A\simeq B$ given by the bijection $\alpha:  B\ra A$ with assignment
$\alpha:  \bmb\mapsto\bma$.  If $\bmb\in B$, where $\bmb$ is given by
$$
\bmb=\ldots,b^{t-1},b^t,b^{t+1},\ldots,
$$
and where for each time $t$, $b^t=(s^t,a^t,s^{t+1})$, then $\bma\in A$ is given by
$$
\bma=\ldots,a^{t-1},a^t,a^{t+1},\ldots,
$$
where the sequence of states $\bsigma(\bma)$ of $\bma$ is given by
$$
\bsigma(\bma)=\ldots,s^{t-1},s^t,s^{t+1},\ldots.
$$
The bijection $\alpha:  B\ra A$ with assignment $\alpha:  \bmb\mapsto\bma$
is given by a map $\alpha^t:  B^t\ra A^t$ with assignment $b^t=(s^t,a^t,s^{t+1})\mapsto a^t$
for each $t\in\bmcpz$, where $\alpha^t$ is a homorphism from branch group $B^t$ to alphabet 
group $A^t$.
\end{thm}
Note that we can go in reverse as well: there is a bijection $\alpha^{-1}:  A\ra B$
given by the assignment $\alpha^{-1}:  \bma\mapsto\bmb$, where $\bmb$ can be determined from
$\bma$ and $\bsigma(\bma)$.  We combine both results by using the notation
$A\,\,{\stackrel{\simeq}{\lra}}\,\,B$.

We now review some results from \cite{FT} on the construction of $A$ from its
fundamental components, the generators.  Assume a group system $A$ is
\ellctl\ and complete.  Forney and Trott \cite{FT} 
define the $k$-controllable subcode $A_k$ of an \ellctl\ group code $A$, for $0\le k\le\ell$.  
The $k$-{\it controllable subcode} $A_k$ of a group code $A$ is defined as the set of 
combinations of sequences of span $k+1$ or less:
$$
A_k=\prod_tA^{[t,t+k]}.
$$
They show
\be
\label{norm1}
A_0\subset A_1\subset\ldots A_{k-1}\subset A_k\subset\ldots A_\ell=A
\ee
is a normal series.  A chain coset decomposition yields a one-to-one correspondence
\be
\label{norm2}
A\lra A_0\times (A_1/A_0)\times\cdots\times (A_k/A_{k-1})\times\cdots\times (A_\ell/A_{\ell-1}).
\ee

For $1\le k\le\ell$, the quotient groups $(A_k/A_{k-1})$ may be further evaluated as follows.
In their Code Granule Theorem \cite{FT}, they show
$A_k/A_{k-1}$ is isomorphic to a direct product of quotient groups $\Gamma^{[t,t+k]}$,
\be
\label{norm3}
A_k/A_{k-1}\simeq\prod_t\Gamma^{[t,t+k]},
\ee
where $\Gamma^{[t,t+k]}$ is defined by
$$  
\Gamma^{[t,t+k]}\rmdef\frac{A^{[t,t+k]}}{A^{[t,t+k)}A^{(t,t+k]}}.
$$
$\Gamma^{[t,t+k]}$ is called a {\it granule}.
A \crep\ of $\Gamma^{[t,t+k]}$ is called a {\it Forney-Trott generator} $\bmg_a^{[t,t+k]}$
of $A$.  The \crep\ of $A^{[t,t+k)}A^{(t,t+k]}$ is always taken
to be the identity sequence.  In case
$\Gamma^{[t,t+k]}$ is isomorphic to the identity group, the identity
sequence is the only coset representative.  
A nonidentity generator is an element of
$A^{[t,t+k]}$ but not of $A^{[t,t+k)}$ or of $A^{(t,t+k]}$, so its span
is exactly $k+1$.  Thus every nonidentity generator
is a codeword that cannot be expressed as a combination of shorter
codewords \cite{FT}.

If $Q$ is any quotient group, let $[Q]$ denote a set of \creps\
of $Q$, or a transversal of $Q$.  Let $[\Gamma^{[t,t+k]}]$ be a transversal of $\Gamma^{[t,t+k]}$.
It follows from (\ref{norm3}) that the set $\prod_t \left[\Gamma^{[t,t+k]}\right]$
is a set of coset representatives for the cosets of $A_{k-1}$ in $A_k$.
We know that the set of \creps\ of the granule $\Gamma^{[t,t+k]}$, or of transversal 
$[\Gamma^{[t,t+k]}]$, is a set of generators $\{\bmg_a^{[t,t+k]}\}$.
Then from (\ref{norm2}), any sequence $\bma$ 
can be uniquely evaluated as a product
$$
\bma=A_0\prod_{k=1}^{\ell} \prod_{t=-\infty}^{+\infty} \bmg_a^{[t,t+k]}.
$$
Any element of $A^{[t,t]}$ is a \crep\ $\bmg_a^{[t,t]}$.
Then $A_0=\prod_t \{\bmg_a^{[t,t]}\}$.  It follows that 
(Generator Theorem \cite{FT}) every sequence $\bma$ 
can be uniquely expressed as a product
\be
\label{encftc}
\bma=\prod_{k=0}^{\ell} \prod_{t=-\infty}^{+\infty} \bmg_a^{[t,t+k]}
\ee
of generators $\bmg_a^{[t,t+k]}$.  Thus every sequence $\bma$
is a product of some sequence of generators, and conversely, every
sequence of generators corresponds to some sequence $\bma$.
Then a component $a^t$ of $\bma$ is given by
\be
\label{encftd}
a^t=\prod_{k=0}^\ell \left(\prod_{j=0}^k \chi^t(\bmg_a^{[t-j,t-j+k]})\right).
\ee

Forney and Trott show there is a notion of input associated with a group system $A$.
Forney and Trott define the {\it input group} $F^t$ of a group system $A$ at time $t$
to be the projection $F^t\rmdef\chi^t(X^t)$.  In their Input Granule 
Theoerem \cite{FT}, they show $F^t$ has a normal chain decomposition
$$
F_0^t\subset F_1^t\subset\cdots\subset F_{k-1}^t\subset F_k^t\subset\cdots\subset F_\ell^t=F^t,
$$
where each quotient group $F_k^t/F_{k-1}^t$ is isomorphic to a granule,
$$
F_k^t/F_{k-1}^t\simeq\Gamma^{[t,t+k]},
$$
for $0<k\le\ell$, and $F_0^t\rmdef A^{[t,t]}$.
An {\it input} $\bmi^t$ of $A$ at time $t$ is the selection of one representative
from each quotient group $F_k^t/F_{k-1}^t$ of $F^t$, for $0\le k\le\ell$.
Equivalently, from the Input Granule Theorem we can take an input $\bmi^t$
at time $t$ as the first component of a generator $\bmg_a^{[t,t+k]}$
of granule $\Gamma^{[t,t+k]}$, for $0\le k\le\ell$.  The set of all inputs $\bmi^t$
of $A$ that can occur at time $t$ is $\bmcpi^t$.  This is just the set of 
all possible ways of selecting one representative from each quotient group 
$F_k^t/F_{k-1}^t$, for $0\le k\le\ell$.  An {\it input sequence} $\bmi$
of $A$ is a sequence of inputs $\ldots,\bmi^t,\bmi^{t+1},\ldots$ that can occur 
in $A$.  In general, the set of input sequences that can occur in $A$ is a subset 
of the Cartesian product $\bigotimes_{-\infty<t<+\infty} \bmcpi^t$ \cite{FT}.
If $A$ is complete, then the set of input sequences is the full Cartesian product
$\bigotimes_{-\infty<t<+\infty} \bmcpi^t$ \cite{FT}.

A {\it basis} $\calb$ of $A$ is a smallest set of shortest length generators 
that is sufficient to generate the group system $A$ \cite{FY1}.
It follows from the encoder in \cite{FT} that a basis is a set of \creps\ 
of $\Gamma^{[t,t+k]}$, for $0\le k\le\ell$, for each $t\in\bmcpz$.
At each time $t\in\bmcpz$, the \creps\ of $\Gamma^{[t,t+k]}$ form set $\calb^t$, where
$$
\calb^t\rmdef\cup_{\{0\le k\le\ell\}} [\Gamma^{[t,t+k]}].
$$
Then basis $\calb$ is the set $\cup_{t\in\bmcpz} \calb^t$.  

The encoder in (\ref{encftc}) forms output $\bma$ from a list of generators selected from
basis $\calb$.  For each time $t\in\bmcpz$, for each $k$ such that $0\le k\le\ell$, 
a single generator $\bmg_a^{[t,t+k]}$ is selected from set $[\Gamma^{[t,t+k]}]$.
For this paper, the way in which we order the selected generators in $\calb$ is important.
The inner product $\prod_t \bmg_a^{[t,t+k]}$ in (\ref{encftc}) uses the selected 
generators in the order
\be
\label{eqncalg}
\{\bmg_a^{[t,t+k]}:  -\infty<t<+\infty,0\le k\le\ell\},
\ee
meaning the list
\begin{multline*}
\ldots,\bmg_a^{[t,t]},\bmg_a^{[t+1,t+1]},\ldots,
\ldots,\bmg_a^{[t,t+1]},\bmg_a^{[t+1,t+2]},\ldots,
\ldots,\bmg_a^{[t,t+k]},\bmg_a^{[t+1,t+1+k]},\ldots, \\
\ldots,\bmg_a^{[t,t+\ell]},\bmg_a^{[t+1,t+1+\ell]},\ldots.
\end{multline*}
We call this the {\it spectral domain order} and the corresponding encoder in 
(\ref{encftc}) using this order the {\it spectral domain encoder}.

There is another way to order the selected generators in $\calb$, as the list
\be
\label{eqncalg1}
\{\bmg_a^{[t,t+k]}:  0\le k\le\ell,-\infty<t<+\infty\},
\ee
in which the time index $t$ and $k$ index in (\ref{eqncalg}) are reversed.
We call this the {\it time domain order}.  In the next section, 
we show there is a corresponding encoder using this order,
in which the inner and outer product in (\ref{encftc}) are reversed,
the {\it time domain encoder}.

\newpage
\vspace{3mm}
{\bf 3.  THE NORMAL CHAIN OF $B$}
\vspace{3mm}

The group system $B$ is more suited to the development in this paper, and henceforth we
study $B$ instead of $A$.
Recall that for each $t\in\bmcpz$, we have defined $X^t$ to be the set of all sequences $\bmb$ in $B$ 
for which $b^n=1^n$ for $n<t$, where $1^n$ is the identity of $B^n$ at time $n$.
And for each $t\in\bmcpz$, we have defined $Y^t$ to be the set of all sequences $\bmb$ 
in $B$ for which $b^n=1^n$ for $n>t$.

It is clear that $X^t\lhd B$ and $Y^t\lhd B$ for each $t\in\bmcpz$. 
Then the group $B$ has two normal series (and chief series)
\be
\label{chnx}
\bone\cdots\subset X^{t+1}\subset X^t\subset X^{t-1}\subset\cdots\subset X^{t-j+1}\subset X^{t-j}\subset\cdots\subset X^{t-\ell+1}\subset X^{t-\ell}\subset\cdots B,
\ee
and
\be
\label{chny}
\bone\cdots\subset Y^{t-1}\subset Y^t\subset Y^{t+1}\subset\cdots\subset Y^{t+k-1}\subset Y^{t+k}\subset\cdots\subset Y^{t+\ell-1}\subset Y^{t+\ell}\subset\cdots B.
\ee
In (\ref{chnx}) we have chosen index $j$ such that $0\le j\le\ell$ and in (\ref{chny})
we have chosen index $k$ such that $0\le k\le\ell$.
The Schreier refinement theorem used to prove the Jordan-H\"{o}lder 
theorem \cite{ROT} shows how to obtain a refinement of two normal series
by inserting one into the other.  We can obtain a refinement of (\ref{chnx})
by inserting (\ref{chny}) between each two successive terms of (\ref{chnx}).
This gives the normal series shown in (\ref{sminf}).  The normal series
is given by the $\ell+3$ rows in the middle of (\ref{sminf}) and the rows denoted by the two
vertical ellipses near the top and bottom.  The bottom row and top row in (\ref{sminf}) are
limiting groups, explained further below.  The normal series is
an infinite series of columns, with each column an infinite series of groups.
We have only shown $\ell+1$ columns of the infinite series in (\ref{sminf}). 
Since (\ref{chnx}) and (\ref{chny}) are chief series, the normal chain 
(\ref{sminf}) is a chief series.

\begin{figure}

\be
\label{sminf}
\begin{array}{cccccccc}
& \shrtpll & \shrtpll && \shrtpll && \shrtpll & \\

& X^{t+1}(X^t) & X^t(X^{t-1}) & \cdots & X^{t-j+1}(X^{t-j}) & \cdots & X^{t-\ell+1}(X^{t-\ell}) & \\

& \cup & \cup && \cup && \cup & \\

& \vdots & \vdots & \vdots & \vdots & \vdots & \vdots & \\

& \vdots & \vdots & \vdots & \vdots & \vdots & \vdots & \\

& \cup & \cup && \cup && \cup & \\

& X^{t+1}(X^t\cap Y^{t+\ell+1}) & X^t(X^{t-1}\cap Y^{t+\ell}) & \cdots & X^{t-j+1}(X^{t-j}\cap Y^{t+\ell-j+1}) & \cdots & X^{t-\ell+1}(X^{t-\ell}\cap Y^{t+1}) & \\

& \cup & \cup && \cup && \cup & \\

& X^{t+1}(X^t\cap Y^{t+\ell}) & X^t(X^{t-1}\cap Y^{t+\ell-1}) & \cdots & X^{t-j+1}(X^{t-j}\cap Y^{t+\ell-j}) & \cdots & X^{t-\ell+1}(X^{t-\ell}\cap Y^{t}) & \\

& \cup & \cup && \cup && \cup & \\

& X^{t+1}(X^t\cap Y^{t+\ell-1}) & X^t(X^{t-1}\cap Y^{t+\ell-2}) & \cdots & X^{t-j+1}(X^{t-j}\cap Y^{t+\ell-j-1}) & \cdots & X^{t-\ell+1}(X^{t-\ell}\cap Y^{t-1}) & \\

& \cup & \cup && \cup && \cup & \\

& \cdots & \cdots & \cdots & \cdots & \cdots & \cdots & \\

& \cup & \cup && \cup && \cup & \\

& X^{t+1}(X^t\cap Y^{t+k}) & X^t(X^{t-1}\cap Y^{t+k-1}) & \cdots & X^{t-j+1}(X^{t-j}\cap Y^{t+k-j}) & \cdots & X^{t-\ell+1}(X^{t-\ell}\cap Y^{t+k-\ell}) & \\

& \cup & \cup && \cup && \cup & \\

& X^{t+1}(X^t\cap Y^{t+k-1}) & X^t(X^{t-1}\cap Y^{t+k-2}) & \cdots & X^{t-j+1}(X^{t-j}\cap Y^{t+k-j-1}) & \cdots & X^{t-\ell+1}(X^{t-\ell}\cap Y^{t+k-\ell-1}) & \\

& \cup & \cup && \cup && \cup & \\

\cdots & \cdots & \cdots & \cdots & \cdots & \cdots & \cdots & \cdots \\

& \cup & \cup && \cup && \cup & \\

& X^{t+1}(X^t\cap Y^{t+j}) & X^t(X^{t-1}\cap Y^{t+j-1}) & \cdots & X^{t-j+1}(X^{t-j}\cap Y^{t}) & \cdots & X^{t-\ell+1}(X^{t-\ell}\cap Y^{t+j-\ell}) & \\

& \cup & \cup && \cup && \cup & \\

& X^{t+1}(X^t\cap Y^{t+j-1}) & X^t(X^{t-1}\cap Y^{t+j-2}) & \cdots & X^{t-j+1}(X^{t-j}\cap Y^{t-1}) & \cdots & X^{t-\ell+1}(X^{t-\ell}\cap Y^{t+j-\ell-1}) & \\

& \cup & \cup && \cup && \cup & \\

& \cdots & \cdots & \cdots & \cdots & \cdots & \cdots & \\

& \cup & \cup && \cup && \cup & \\

& X^{t+1}(X^t\cap Y^{t+1}) & X^t(X^{t-1}\cap Y^{t}) & \cdots & X^{t-j+1}(X^{t-j}\cap Y^{t-j+1}) & \cdots & X^{t-\ell+1}(X^{t-\ell}\cap Y^{t-\ell+1}) & \\

& \cup & \cup && \cup && \cup & \\

& X^{t+1}(X^t\cap Y^t) & X^t(X^{t-1}\cap Y^{t-1}) & \cdots & X^{t-j+1}(X^{t-j}\cap Y^{t-j}) & \cdots & X^{t-\ell+1}(X^{t-\ell}\cap Y^{t-\ell}) & \\

& \cup & \cup && \cup && \cup & \\

& X^{t+1}(X^t\cap Y^{t-1}) & X^t(X^{t-1}\cap Y^{t-2}) & \cdots & X^{t-j+1}(X^{t-j}\cap Y^{t-j-1}) & \cdots & X^{t-\ell+1}(X^{t-\ell}\cap Y^{t-\ell-1}) & \\

& \cup & \cup && \cup && \cup & \\

& \vdots & \vdots & \vdots & \vdots & \vdots & \vdots & \\

& \vdots & \vdots & \vdots & \vdots & \vdots & \vdots & \\

& \cup & \cup && \cup && \cup & \\

& X^{t+1} & X^t & \cdots & X^{t-j+1} & \cdots & X^{t-\ell+1} & \\

& \shrtpll & \shrtpll && \shrtpll && \shrtpll &
\end{array}
\ee
\end{figure}

We now show that (\ref{sminf}) is indeed a refinement of (\ref{chnx}).
First we show that in each column the group in the bottom row is contained
in all groups in the infinite column of groups, and the group in the top row
contains all groups in the infinite column of groups.
Any term in (\ref{sminf}) is of the form
$$
X^{i+1}(X^i\cap Y^{i+m})
$$
for some integer pair $i,m\in\bmcpz$.  For example, term $X^{t-j+1}(X^{t-j}\cap Y^{t+k-j})$
is of the form $X^{i+1}(X^i\cap Y^{i+m})$ for $i=t-j$ and $m=k$.   Fix $i\in\bmcpz$.  
For any $m\in\bmcpz$, we have $X^{i+1}\subset X^{i+1}(X^i\cap Y^{i+m})$
Therefore the group in the bottom row is contained in all groups 
in the infinite column of groups.  For any $m\in\bmcpz$,
it is clear that $X^{i+1}(X^i\cap Y^{i+m})\subset X^{i+1}(X^i)$.
Therefore the group in the top row contains all groups in the 
infinite column of groups.  Now note that $X^{i+1}(X^i)=X^i$.
This means that a group in the top row is the same as the group
in the bottom row in the next column.  But now note the groups in
the bottom row form the same sequence as (\ref{chnx}).  Therefore (\ref{sminf}) is 
indeed a refinement of (\ref{chnx}).

The factor $X^{t-j+1}$ in term $X^{t-j+1}(X^{t-j}\cap Y^{t+k-j})$ in (\ref{sminf})
is called an {\it integration factor}, and the factor $(X^{t-j}\cap Y^{t+k-j})$
in term $X^{t-j+1}(X^{t-j}\cap Y^{t+k-j})$ is called a
{\it derivative factor}.  The derivative factors in each row of a finite band of rows
in (\ref{sminf}), absent the normal chain given here, are essentially what is used in 
\cite{FT} to find a \cdc\ for the spectral domain (see also \cite{LM}).  
The product of the derivative factors in each row
of the finite band corresponds to a $k$-controllable subcode of \cite{FT}.

We now show the normal chain (\ref{sminf}) contains all the sequences $\bmb\in B$.  Since $B$
is complete, any $\bmb\in B$ can be specified by its projection over finite time intervals
\cite{FT}.  Then we just have to show that the normal chain contains the projection of $\bmb$
over any finite time interval $[t,t+n]$, $\chi^{[t,t+n]}(\bmb)$.  But since $B$ is 
\ellctl, the group $(X^i\cap Y^{i+m})$ contains $\chi^{[t,t+n]}(\bmb)$
for $i\le t-\ell$ and $i+m\ge t+n+\ell$.  Therefore the group
$X^{i+1}(X^i\cap Y^{i+m})$ in (\ref{sminf}) contains $\chi^{[t,t+n]}(\bmb)$
for $i\le t-\ell$ and $i+m\ge t+n+\ell$.  This gives the following.

\begin{thm}
Any sequence $\bmb\in B$ is completely specified by the normal chain (\ref{sminf}).
\end{thm}

Since any sequence $\bmb\in B$ is specified by the normal chain (\ref{sminf}), we can use the
\cdc\ of (\ref{sminf}) to find any sequence $\bmb$ in $B$.
In any normal chain, we may form the quotient group of two successive groups in the chain.
Then a normal chain of groups, as in (\ref{sminf}), gives a series of quotient groups.
A general quotient group obtained from (\ref{sminf}) is of the form
\be
\label{qgx}
\frac{X^{i+1}(X^i\cap Y^{i+m})}{X^{i+1}(X^i\cap Y^{i+m-1})}
\ee
for any integer pair $i,m\in\bmcpz$.  We call this the {\it time domain granule}.
Note that the time domain granule has half infinite extent while the granule
$\Gamma^{[t,t+k]}$ of \cite{FT} has finite extent.
We wish to find a transversal of the time domain granule.  
The \creps\ of the time domain granule are called {\it generators}.
Since $(X^{i+1}\cap Y^{i+m})\subset X^{i+1}$, we have
$X^{i+1}(X^i\cap Y^{i+m-1})=X^{i+1}(X^i\cap Y^{i+m-1})(X^{i+1}\cap Y^{i+m})$,
and we may rewrite (\ref{qgx}) as
\be
\label{qgx0}
\frac{X^{i+1}(X^i\cap Y^{i+m})}{X^{i+1}(X^i\cap Y^{i+m-1})(X^{i+1}\cap Y^{i+m})}.
\ee

We now find a transversal of (\ref{qgx0}).  An element of the numerator group is of 
the form $\bmx\bmy$, where $\bmx\in X^{i+1}$ and $\bmy\in (X^i\cap Y^{i+m})$.
Then a coset of normal subgroup $X^{i+1}(X^i\cap Y^{i+m-1})(X^{i+1}\cap Y^{i+m})$ is 
\begin{align*}
\bmx\bmy X^{i+1}(X^i\cap Y^{i+m-1})(X^{i+1}\cap Y^{i+m}) 
&=\bmx X^{i+1}\bmy(X^i\cap Y^{i+m-1})(X^{i+1}\cap Y^{i+m}) \\
&=X^{i+1}\bmy(X^i\cap Y^{i+m-1})(X^{i+1}\cap Y^{i+m}).
\end{align*}

\begin{thm}
A coset of normal subgroup $X^{i+1}(X^i\cap Y^{i+m-1})(X^{i+1}\cap Y^{i+m})$ is 
$X^{i+1}\bmy(X^i\cap Y^{i+m-1})(X^{i+1}\cap Y^{i+m})$ where $\bmy\in (X^i\cap Y^{i+m})$.
A \crep\ of coset $X^{i+1}\bmy(X^i\cap Y^{i+m-1})(X^{i+1}\cap Y^{i+m})$ is
$\bmxdt\bmydt$ where $\bmxdt$ is in $X^{i+1}$ and $\bmydt$ is in
$\bmy(X^i\cap Y^{i+m-1})(X^{i+1}\cap Y^{i+m})$.
A transversal of the quotient group (\ref{qgx0}) is a selection of one \crep\
$\bmxdt\bmydt$ from each coset $X^{i+1}\bmy(X^i\cap Y^{i+m-1})(X^{i+1}\cap Y^{i+m})$.
\end{thm}
We may always select a \crep\ $\bmxdt\bmydt$ of coset $X^{i+1}\bmy(X^i\cap Y^{i+m-1})(X^{i+1}\cap Y^{i+m})$
to be $\bone\bmydt$ where $\bone$ is the identity of $X^{i+1}$.
But note that $\bmy(X^i\cap Y^{i+m-1})(X^{i+1}\cap Y^{i+m})$ is a coset of
normal subgroup $(X^i\cap Y^{i+m-1})(X^{i+1}\cap Y^{i+m})$ in quotient group
\be
\label{qgx1}
\frac{(X^i\cap Y^{i+m})}{(X^i\cap Y^{i+m-1})(X^{i+1}\cap Y^{i+m})}.
\ee
Then $\bmydt$ is a \crep\ of (\ref{qgx1}).  This gives the following.

\begin{cor}
A transversal of the quotient group (\ref{qgx1}) is a transversal
of the quotient group (\ref{qgx0}), but the reverse is only true if $\bmxdt\bmydt=\bone\bmydt$ 
or $\bmxdt=\bone$ for each \crep\ $\bmxdt\bmydt$ of (\ref{qgx0}).
\end{cor}

We have just shown that a transversal of (\ref{qgx1}) is a transversal of (\ref{qgx0}).
It is not surprising then that there is a homomorphism from (\ref{qgx0}) to
(\ref{qgx1}).  In fact this result is an application of the Zassenhaus lemma
used in the proof of the Schreier refinement theorem \cite{ROT}.  We 
first restate the following lemma, excised from the proof of the Zassenhaus 
lemma (see p.\ 100 of \cite{ROT}).

\begin{lem}
\label{lem26xx}
{\bf (from proof of Zassenhaus lemma)}
Let $U\lhd U^*$ and $V\lhd V^*$ be four subgroups of a group $G$.
Then $D=(U^*\cap V)(U\cap V^*)$ is a normal subgroup of $U^*\cap V^*$.
If $g\in U(U^*\cap V^*)$, then $g=uu^*$ for $u\in U$ and $u^*\in U^*\cap V^*$.
Define function $f:  U(U^*\cap V^*)\ra (U^*\cap V^*)/D$ by $f(g)=f(uu^*)=Du^*$.
Then $f$ is a well defined homomorphism with kernel $U(U^*\cap V)$ and
$$
\frac{U(U^*\cap V^*)}{U(U^*\cap V)}\simeq\frac{U^*\cap V^*}{D}.
$$
\end{lem}

Note that (\ref{qgx0}) is equivalent to (\ref{qgx}).
We now use Lemma \ref{lem26xx} to show there is a homomorphism from (\ref{qgx}) to
(\ref{qgx1}).  Let $U=X^{i+1}$ and $U^*=X^i$.  Let $V=Y^{i+m-1}$ and $V^*=Y^{i+m}$.
Note that $U\lhd U^*$ and $V\lhd V^*$.  Then
\begin{align*}
\frac{U(U^*\cap V^*)}{U(U^*\cap V)} &=
\frac{X^{i+1}(X^i\cap Y^{i+m})}{X^{i+1}(X^i\cap Y^{i+m-1})},
\end{align*}
and
$$
\frac{U^*\cap V^*}{D}=\frac{U^*\cap V^*}{(U^*\cap V)(U\cap V^*)}
=\frac{(X^i\cap Y^{i+m})}{(X^i\cap Y^{i+m-1})(X^{i+1}\cap Y^{i+m})}.
$$
Now the function $f$ of Lemma \ref{lem26xx} is a homomorphism from (\ref{qgx}) to
(\ref{qgx1}).

\begin{thm}
\label{dmnh}
There is a homomorphism from the quotient group (\ref{qgx}), or equivalently (\ref{qgx0}),
to the quotient group (\ref{qgx1}), given by the function $f$ of the Zassenhaus lemma.
\end{thm}

Choose a basis $\calb$.  We now choose a \crep\ or generator in the time domain granules
(\ref{qgx}) formed by (\ref{sminf}) for any integer pair $i,m\in\bmcpz$.  Fix any $i\in\bmcpz$.
There are three cases to consider for (\ref{qgx}), depending on the value of $m$.
For $m<0$, we have $(X^i\cap Y^{i+m})=\bone$ and so (\ref{qgx}) reduces to $X^{i+1}(\bone)/X^{i+1}(\bone)=
X^{i+1}/X^{i+1}$.  Then \crep\ $\bmxdt\bmydt$ is $\bmxdt\bone$.  We may always select $\bmxdt$
to be $\bone$.  Then for $m<0$, we can choose the \crep\ or generator of (\ref{qgx}) to be $\bone$.

We next consider the case $0\le m\le\ell$.  If we select a \crep\ $\bmxdt\bmydt$ of 
coset $X^{i+1}\bmy(X^i\cap Y^{i+m-1})(X^{i+1}\cap Y^{i+m})$
to be $\bone\bmydt$ where $\bone$ is the identity of $X^{i+1}$, then a transversal 
of the quotient group (\ref{qgx1}) is a transversal of the quotient group (\ref{qgx0}).
But for $0\le m\le\ell$ note that (\ref{qgx1}) is the same as the Forney-Trott granule, 
or {\it spectral domain granule}, for $B$,
\be
\label{spctrl}
\frac{B^{[i,i+m]}}{B^{[i,i+m-1]}B^{[i+1,i+m]}}.
\ee
This gives the following.

\begin{cor}
\label{cor10}
For $0\le m\le\ell$, a transversal of the spectral domain granule (\ref{spctrl}) is a transversal of 
the time domain granule (\ref{qgx}), but the reverse is only true if $\bmxdt\bmydt=\bone\bmydt$ 
or $\bmxdt=\bone$ for each \crep\ $\bmxdt\bmydt$ of the time domain granule.
\end{cor}
Then for $0\le m\le\ell$, we can choose the \crep\ or generator of (\ref{qgx}) to be $\bone\bmydt$.
In this case, the \crep\ or generator of (\ref{qgx}) is the same as the \crep\ $\bmg^{[i,i+m]}$
of the spectral domain granule (\ref{spctrl}) of $B$.  Therefore the \crep\ of the time domain
granule can be chosen to be the same as the Forney-Trott generator of $B$.  Note that for $0\le m\le\ell$, 
Theorem \ref{dmnh} shows there is a homomorphism from the time domain granule to the spectral 
domain granule given by the function $f$ of the Zassenhaus lemma.

Lastly we consider the case $m>\ell$.  Since $B$ is \ellctl, in (\ref{qgx1}) there can be no
elements of $(X^i\cap Y^{i+m})$ that are not elements of $(X^i\cap Y^{i+m-1})(X^{i+1}\cap Y^{i+m})$.
Then
$$
X^i\cap Y^{i+m}=(X^i\cap Y^{i+m-1})(X^{i+1}\cap Y^{i+m}),
$$
and quotient group (\ref{qgx1}) is trivial.  Then  quotient group (\ref{qgx0}) and 
(\ref{qgx}) is trivial.  Then we can choose the \crep\ $\bmxdt\bmydt$ or generator of (\ref{qgx}) 
to be $\bmxdt\bmydt=\bone\bone$ or $\bone$.

\begin{figure}

\be
\label{smgen}
\begin{array}{llllllll}
& \vdots & \vdots & \vdots & \vdots & \vdots & \vdots & \\
& \vdots & \vdots & \vdots & \vdots & \vdots & \vdots & \\
& \bone & \bone & \cdots & \bone & \cdots & \bone & \\
& \bmg^{[t,t+\ell]}   & \bmg^{[t-1,t-1+\ell]}   & \cdots & \bmg^{[t-j,t-j+\ell]}   & \cdots & \bmg^{[t-\ell,t]} & \\
& \bmg^{[t,t+\ell-1]} & \bmg^{[t-1,t-1+\ell-1]} & \cdots & \bmg^{[t-j,t-j+\ell-1]} & \cdots & \bmg^{[t-\ell,t-1]} & \\
& \cdots              & \cdots                & \cdots & \cdots                  & \cdots & \cdots & \\
& \bmg^{[t,t+k]}      & \bmg^{[t-1,t-1+k]}    & \cdots & \bmg^{[t-j,t-j+k]}      & \cdots & \bmg^{[t-\ell,t-\ell+k]} & \\
& \bmg^{[t,t+k-1]}    & \bmg^{[t-1,t-1+k-1]}  & \cdots & \bmg^{[t-j,t-j+k-1]}    & \cdots & \bmg^{[t-\ell,t-\ell+k-1]} & \\
\cdots & \cdots       & \cdots                & \cdots & \cdots                  & \cdots & \cdots & \cdots \\
& \bmg^{[t,t+j]}      & \bmg^{[t-1,t-1+j]}    & \cdots & \bmg^{[t-j,t]}          & \cdots & \bmg^{[t-\ell,t-\ell+j]} & \\
& \bmg^{[t,t+j-1]}    & \bmg^{[t-1,t-1+j-1]}  & \cdots & \bmg^{[t-j,t-1]}        & \cdots & \bmg^{[t-\ell,t-\ell+j-1]} & \\
& \cdots              & \cdots                & \cdots & \cdots                  & \cdots & \cdots & \\
& \bmg^{[t,t+1]}      & \bmg^{[t-1,t]}        & \cdots & \bmg^{[t-j,t-j+1]}      & \cdots & \bmg^{[t-\ell,t-\ell+1]} & \\
& \bmg^{[t,t]}        & \bmg^{[t-1,t-1]}      & \cdots & \bmg^{[t-j,t-j]}        & \cdots & \bmg^{[t-\ell,t-\ell]} & \\
& \bone & \bone & \cdots & \bone & \cdots & \bone & \\
& \vdots & \vdots & \vdots & \vdots & \vdots & \vdots & \\
& \vdots & \vdots & \vdots & \vdots & \vdots & \vdots &
\end{array}
\ee
\end{figure}

We have found generators of the time domain granule for each $i\in\bmcpz$ and
ranges $m<0$, $0\le m\le\ell$, and $m>\ell$.  
We can arrange the selected set of generators in the same order as (\ref{sminf}), as
shown in (\ref{smgen}).  For $m<0$, there are an infinite number of rows at the bottom 
of (\ref{smgen}) which are filled with the identity generator; for $m>\ell$ there are 
an infinite number of rows at the top of (\ref{smgen}) which are filled with the identity generator.
There are a finite number of rows, $\ell+1$, which can have nontrivial generators $\bmg^{[i,i+m]}$
for $0\le m\le\ell$.  Let $\bmr$ be the infinite series of finite columns given by generators in
the finite band in (\ref{smgen}), i.e., the generators $\bmg^{[i,i+m]}$ for $0\le m\le\ell$,
for each $i\in\bmcpz$.
Note that $\bmr$ is a tensor.  Let $\calr$ be the set of all possible tensors $\bmr$.

We can arrange the sequence of generators in the finite band in (\ref{smgen}) into the product
\be
\label{prod}
\cdots\bmg^{[i+1,i+1+\ell]}\bmg^{[i,i]}\bmg^{[i,i+1]}\cdots\bmg^{[i,i+\ell]}
\bmg^{[i-1,i-1]}\bmg^{[i-1,i]}\cdots\bmg^{[i-1,i-1+\ell]}\bmg^{[i-2,i-2]}\cdots.
\ee
Each generator $\bmg^{[i,i+m]}$ is a path in $B$.  Then the product (\ref{prod}) 
is understood to mean a multiplication of paths in $B$.

\begin{thm}
The sequence (\ref{prod}) is a well defined product in $B$.  Therefore (\ref{prod}) 
is a well defined path $\bmb$ in $B$, where $\bmb$ is given by
\be
\label{enctdx}
\bmb=\prod_{i=+\infty}^{-\infty} \left(\prod_{m=0}^\ell \bmg^{[i,i+m]})\right).
\ee
\end{thm}

\begin{prf}
Each term in (\ref{prod}) is a path in $B$.  The paths in sequence (\ref{prod})
are multiplied according to the definition of product in $B$, which means
component by component multiplication for $-\infty<i<+\infty$.  The infinite
product (\ref{prod}) is well defined if for each $i\in\bmcpz$, there are only 
finitely many terms $\bmg^{[i,i+m]}$ having a nontrivial component \cite{FT}.
Therefore the final result is a well defined path in $B$.
\end{prf}

Choose a basis $\calb$.  If a basis $\calb$ of $B$ is chosen, then $\calr$ is fixed.

\begin{thm}
\label{thm9}
Fix basis $\calb$ of $B$.  Then $\calr$ is fixed, and there is a 
bijection $\eta:  \calr\ra B$ given by assignment $\eta:  \bmr\mapsto\bmb$, 
where $\bmb$ is an encoding of $\bmr$ using product (\ref{enctdx}) on the 
generators in $\bmr$.
\end{thm}

\begin{prf}
Since (\ref{sminf}) forms a \cdc\ of $B$, and since (\ref{enctdx}) is a selection of
one representative from each nontrivial quotient group in the chain, then any $\bmb\in B$ can
be obtained by the composition (\ref{enctdx}).  The assignment $\eta:  \bmr\mapsto\bmb$
is a bijection since any unique selection of \creps\ gives a unique $\bmb\in B$.
\end{prf}

A component $b^t$ of $\bmb$ at any time $t$ can be obtained by the 
composition of the time $t$ component $\chi^t$ of generators in (\ref{smgen}).
There are only a finite number of generators in (\ref{smgen}) with nontrivial components
at time $t$.  These are generators $\bmg^{[t-j,t-j+k]}$, for $0\le j\le\ell$, for $j\le k\le\ell$,
shown in the triangular matrix (\ref{smgent}).
The components at time $t$ of all other generators
$\bmg^{[i,i+m]}$ in (\ref{smgen}) are $\chi^t(\bmg^{[i,i+m]})=\chi^t(\bone)$, 
where $\bone$ is the identity of $B$.  Then a component $b^t$ of $\bmb$ 
can be obtained by the composition of terms
$\chi^t(\bmg^{[t-j,t-j+k]})$ for $0\le j\le\ell$, for $j\le k\le\ell$, as
\be
\label{enctdc}
b^t=\prod_{j=0}^\ell \left(\prod_{k=j}^\ell \chi^t(\bmg^{[t-j,t-j+k]})\right).
\ee
Developed in another way \cite{KM6}, the encoder (\ref{enctdc}) is an estimator 
which corrects its path with a new estimate at each time $t$.

We can obtain a corresponding result for $A$ immediately from (\ref{enctdc}) as
\be
\label{enctdc1}
a^t=\prod_{j=0}^\ell \left(\prod_{k=j}^\ell \chi^t(\bmg_A^{[t-j,t-j+k]})\right),
\ee
where the generator $\bmg^{[t-j,t-j+k]}$ in $B$ is replaced with the generator
$\bmg_A^{[t-j,t-j+k]}$ in $A$ using the 1-1 correspondence $A\lra B$ in the isomosphism $A\simeq B$.
In other words, the generator $\bmg_A^{[t-j,t-j+k]}$ in (\ref{enctdc1}) is just the sequence of 
alphabet letters in $\bmg^{[t-j,t-j+k]}$ in (\ref{enctdc}).
The encoder in (\ref{enctdc1}) can be written as $a^t=\prod_{j=0}^\ell h_j^{t-j}$, where
the inner term in parentheses is some function of time, say $h_j^{t-j}$.
Thus the encoder has the form of a time convolution, reminiscent of a linear system.  
The encoder (\ref{encftd}) in \cite{FT} does not have a convolution
property.  For this reason, we say the encoder (\ref{enctdx}) or (\ref{enctdc}) 
or (\ref{enctdc1}) is a time domain encoder,
and the encoder (\ref{encftc}) or (\ref{encftd}) in \cite{FT} a spectral domain encoder.
The time domain encoder has a granule based construction of input granules and 
state granules in the same way as the spectral domain encoder in \cite{FT}.

We have seen from Corollary \ref{cor10} (see also \cite{KM5}) that the \crep\ of the time domain
granule of $B$ can be chosen to be the same as the Forney-Trott generator of the spectral domain granule
of $B$.  In this case, the generator $\bmg_A^{[t-j,t-j+k]}$ in (\ref{enctdc1}) can be the same
as the generator $\bmg_a^{[t-j,t-j+k]}$ in (\ref{encftd}).
Then comparing (\ref{encftd}) and (\ref{enctdc1}), we see that the time domain encoder
is just an interchange of the double product in the spectral domain encoder \cite{KM5}.  
If both (\ref{encftd}) and (\ref{enctdc1}) use the same
generators and the alphabet group $A^t$ is abelian, then the interchange
does not matter and both encoders give the same output at time $t$.
If $A^t$ is nonabelian, this may not be the case.

The spectral domain encoder is a {\it minimal} encoder because there is
a bijection between the state granules at any time $t$ and the states of the canonical
realization \cite{FT}.  Since the time domain encoder can use the same generators as the
spectral domain encoder, the state granules of the time domain and spectral domain
encoder can be the same.  Therefore the time domain encoder is also a minimal encoder.

\begin{figure}

\be
\label{smgent}
\begin{array}{llllll}
\bmg^{[t,t+\ell]}   & \bmg^{[t-1,t-1+\ell]}   & \cdots & \bmg^{[t-j,t-j+\ell]}   & \cdots & \bmg^{[t-\ell,t]} \\
\bmg^{[t,t+\ell-1]} & \bmg^{[t-1,t-1+\ell-1]} & \cdots & \bmg^{[t-j,t-j+\ell-1]} & \cdots & \\      
\cdots              & \cdots                  & \cdots & \cdots                  & \cdots & \\      
\bmg^{[t,t+k]}      & \bmg^{[t-1,t-1+k]}      & \cdots & \bmg^{[t-j,t-j+k]}      &        & \\      
\bmg^{[t,t+k-1]}    & \bmg^{[t-1,t-1+k-1]}    & \cdots & \bmg^{[t-j,t-j+k-1]}    &        & \\      
\cdots              & \cdots                  & \cdots & \cdots                  &        & \\      
\bmg^{[t,t+j]}      & \bmg^{[t-1,t-1+j]}      & \cdots & \bmg^{[t-j,t]}          &        & \\     
\bmg^{[t,t+j-1]}    & \bmg^{[t-1,t-1+j-1]}    & \cdots &                         &        & \\      
\cdots              & \cdots                  & \cdots &                         &        & \\      
\bmg^{[t,t+1]}      & \bmg^{[t-1,t]}          &        &                         &        & \\      
\bmg^{[t,t]}        &                         &        &                         &        &         
\end{array}
\ee
\end{figure}

Since the only generators in (\ref{smgen}) with nontrivial components at time $t$ are generators
$\bmg^{[t-j,t-j+k]}$, for $0\le j\le\ell$, for $j\le k\le\ell$, we have the following.

\begin{lem}
\label{lem9}
The only quotient groups, or time domain granules, formed from (\ref{sminf}) that have transversals 
with nontrivial components at time $t$ are of the form
\be
\label{qgqg}
\Lambda_{j,k}^t\rmdef\frac{X^{t-j+1}(X^{t-j}\cap Y^{t+k-j})}{X^{t-j+1}(X^{t-j}\cap Y^{t+k-j-1})}
\ee
for $0\le j\le\ell$, for $j\le k\le\ell$.
\end{lem}

Fix time $t\in\bmcpz$.  For $0\le j\le\ell$, for $j\le k\le\ell$, let generator $\bmg^{[t-j,t-j+k]}$ 
be a representative in quotient group $\Lambda_{j,k}^t$.  We assume $\bmg^{[t-j,t-j+k]}$ has the
sequence of components
\be
\label{tps0}
\bmg^{[t-j,t-j+k]}=
\ldots,1^{t-j-2},1^{t-j-1},r_{0,k}^{t-j},r_{1,k}^{t-j},\ldots,r_{m,k}^{t-j},\ldots,r_{k,k}^{t-j},1^{t-j+k+1},1^{t-j+k+2},\ldots,
\ee
where $0\le m\le k$.  For $0\le m\le k$, the superscript $t-j$ in $r_{m,k}^{t-j}$ indicates 
that $r_{m,k}^{t-j}$ is in a generator $\bmg^{[t-j,t-j+k]}$ whose nontrivial components 
start at time $t-j$.  The component at time $t-j$ in $\bmg^{[t-j,t-j+k]}$ is $r_{0,k}^t$.
In general, the component $r_{m,k}^{t-j}$ is at time $t-j+m$.  Note that the component at time $t$ in 
generator $\bmg^{[t-j,t-j+k]}$ is $r_{j,k}^{t-j}$.
Then component $r_{j,k}^{t-j}$ is some branch $b^t\in B^t$.
The quotient group $\Lambda_{j,k}^t$ can be the identity.  In this case the transversal
has only one representative and the generator $\bmg^{[t-j,t-j+k]}$ is the identity
sequence $\bone$ of $B$, where component $r_{m,k}^{t-j}$ 
in (\ref{tps0}) is the identity branch of $B^t$, for $0\le m\le k$.

Note that $\chi^t(\bmg^{[t-j,t-j+k]})=r_{j,k}^{t-j}$.  Then we can write the time $t$
components of the generators in (\ref{smgent}) as the triangular matrix (\ref{rttf}).
\be
\label{rttf}
\begin{array}{llllllllll}
  r_{0,\ell}^t   & r_{1,\ell}^{t-1}   & \cdots & \cdots & r_{j,\ell}^{t-j}   & \cdots & \cdots & \cdots & r_{\ell-1,\ell}^{t-\ell+1}   & r_{\ell,\ell}^{t-\ell} \\
  r_{0,\ell-1}^t & r_{1,\ell-1}^{t-1} & \cdots & \cdots & r_{j,\ell-1}^{t-j} & \cdots & \cdots & \cdots & r_{\ell-1,\ell-1}^{t-\ell+1} & \\
  \vdots & \vdots & \vdots & \vdots & \vdots & \vdots & \vdots & \vdots && \\
  r_{0,k}^t & r_{1,k}^{t-1} & \cdots & \cdots & r_{j,k}^{t-j} & \cdots & r_{k,k}^{t-k} &&& \\
  \vdots & \vdots & \vdots & \vdots & \vdots & \vdots &&&& \\
  \cdots & \cdots & \cdots & \cdots & r_{j,j}^{t-j} &&&&& \\
  \vdots & \vdots & \vdots &&&&&&& \\
  r_{0,2}^t   & r_{1,2}^{t-1} & r_{2,2}^{t-2} &&&&&&& \\
  r_{0,1}^t   & r_{1,1}^{t-1} &&&&&&&& \\
  r_{0,0}^t   &&&&&&&&&
\end{array}
\ee
We call (\ref{rttf}) the {\it static matrix} 
$\tridnjktarg{0}{0}{t}{\bmr}$ since all entries are at the same time epoch $t$.
Then we see that each sequence $\bmr\in\calr$ gives a sequence of \stms\
\be
\label{seqstms}
\ldots,\tridnjktarg{0}{0}{t}{\bmr},\tridnjktarg{0}{0}{t-1}{\bmr},\ldots.
\ee
Let $\tridnjktarg{0}{0}{t}{\calr}$ be the set of all \stms\ $\tridnjktarg{0}{0}{t}{\bmr}$.
In other words, $\tridnjktarg{0}{0}{t}{\calr}\rmdef\{\tridnjktarg{0}{0}{t}{\bmr}:  \bmr\in\calr\}$.

We have written a sequence $\bma\in A$ as $\ldots,a^{t-1},a^t,\ldots$ and a sequence
$\bmb\in B$ as $\ldots,b^{t-1},b^t,\ldots$.
We have chosen to write the sequence of \stms\ (\ref{seqstms}) in reverse time order
since it reflects the order of indices in (\ref{smgen}), 
which reflects the order of indices in (\ref{sminf}).  It is a natural ordering
for this problem.  But writing the sequence in
reverse time order on a piece of paper does not change the physics of time:
the component at time epoch $t-1$ still occurs before the component at time epoch $t$,
as it does for $\bma$ and $\bmb$.

Using the \stm\ $\tridnjktarg{0}{0}{t}{\bmr}$, we can rewrite (\ref{enctdc}) in the equivalent form as
\begin{align}
b^t
\label{enctda}
&=\prod_{j=0}^\ell \left(\prod_{k=j}^\ell \chi^t(\bmg^{[t-j,t-j+k]})\right) \\
\label{enctd}
&=\prod_{j=0}^\ell \left(\prod_{k=j}^\ell r_{j,k}^{t-j}\right),
\end{align} 
where the inner product in parentheses in (\ref{enctd})  
is just the product of terms in the $j$-th column of $\tridnjktarg{0}{0}{t}{\bmr}$.
By the convention used here, equation (\ref{enctd})
is evaluated as
\be
\label{enctd1}
b^t=r_{0,0}^tr_{0,1}^tr_{0,2}^t\cdots r_{0,\ell}^tr_{1,1}^{t-1}\cdots r_{1,\ell}^{t-1}r_{2,2}^{t-2}
\cdots r_{j,j}^{t-j}\cdots r_{j,k}^{t-j}\cdots r_{j,\ell}^{t-j}\cdots 
r_{\ell-1,\ell-1}^{t-\ell+1} r_{\ell-1,\ell}^{t-\ell+1} r_{\ell,\ell}^{t-\ell}.
\ee

Using (\ref{enctd}) we can give the following enhancement of Theorem \ref{thm9}.

\begin{thm}
\label{thm11}
Fix basis $\calb$ of $B$.  Then $\calr$ is fixed, and there is a 
bijection $\eta:  \calr\ra B$ given by assignment $\eta:  \bmr\mapsto\bmb$, 
where component $b^t$ of $\bmb$ is an encoding of the representatives in
$\tridnjktarg{0}{0}{t}{\bmr}$ of $\bmr$ using (\ref{enctd}), for each $t\in\bmcpz$.
\end{thm}

\begin{cor}
\label{cor12}
For each $t\in\bmcpz$, for each $\tridnjktarg{0}{0}{t}{\bmr}$ in $\tridnjktarg{0}{0}{t}{\calr}$,
there is some $b^t$ in $B^t$ which is an encoding of the representatives in
$\tridnjktarg{0}{0}{t}{\bmr}$ using (\ref{enctd}).
\end{cor}

\begin{prf}
Fix $t\in\bmcpz$.  For each $\tridnjktarg{0}{0}{t}{\bmr}$ in $\tridnjktarg{0}{0}{t}{\calr}$, 
let $\eta:  \bmr\mapsto\bmb$.
Then from Theorem \ref{thm11}, $b^t$ of $\bmb$ is an encoding of the representatives in
$\tridnjktarg{0}{0}{t}{\bmr}$ using (\ref{enctd}).
\end{prf}

\begin{cor}
\label{cor13}
For each $t\in\bmcpz$, for each $b^t$ in $B^t$, $b^t$ is an encoding of the representatives in
some $\tridnjktarg{0}{0}{t}{\bmr}$ in $\tridnjktarg{0}{0}{t}{\calr}$ using (\ref{enctd}).
\end{cor}

\begin{prf}
Fix $t\in\bmcpz$.  For each $b^t$ in $B^t$, there is a $\bmb$ in $B$ with component $b^t$.
Let $\eta:  \bmr\mapsto\bmb$.
Then from Theorem \ref{thm11}, $b^t$ of $\bmb$ is an encoding of the representatives in
$\tridnjktarg{0}{0}{t}{\bmr}$ using (\ref{enctd}).
\end{prf}

\newpage
\vspace{3mm}
{\bf 4.  THE NORMAL CHAIN OF $B^t$}
\vspace{3mm}

\begin{figure}

\be
\label{sminfatt}
\begin{array}{cccccccc}
& \shrtpll & \shrtpll && \shrtpll && \shrtpll & \\

& \chi^n(X^{t+1}(X^t)) & \chi^n(X^t(X^{t-1})) & \cdots & \chi^n(X^{t-j+1}(X^{t-j})) & \cdots & \chi^n(X^{t-\ell+1}(X^{t-\ell})) & \\

& \cup & \cup && \cup && \cup & \\

& \vdots & \vdots & \vdots & \vdots & \vdots & \vdots & \\

& \vdots & \vdots & \vdots & \vdots & \vdots & \vdots & \\

& \cup & \cup && \cup && \cup & \\

& \chi^n(X^{t+1}(X^t\cap Y^{t+\ell+1})) & \chi^n(X^t(X^{t-1}\cap Y^{t+\ell})) & \cdots & \chi^n(X^{t-j+1}(X^{t-j}\cap Y^{t+\ell-j+1})) & \cdots & \chi^n(X^{t-\ell+1}(X^{t-\ell}\cap Y^{t+1})) & \\

& \cup & \cup && \cup && \cup & \\

& \chi^n(X^{t+1}(X^t\cap Y^{t+\ell})) & \chi^n(X^t(X^{t-1}\cap Y^{t+\ell-1})) & \cdots & \chi^n(X^{t-j+1}(X^{t-j}\cap Y^{t+\ell-j})) & \cdots & \chi^n(X^{t-\ell+1}(X^{t-\ell}\cap Y^{t})) & \\

& \cup & \cup && \cup && \cup & \\

& \chi^n(X^{t+1}(X^t\cap Y^{t+\ell-1})) & \chi^n(X^t(X^{t-1}\cap Y^{t+\ell-2})) & \cdots & \chi^n(X^{t-j+1}(X^{t-j}\cap Y^{t+\ell-j-1})) & \cdots & \chi^n(X^{t-\ell+1}(X^{t-\ell}\cap Y^{t-1})) & \\

& \cup & \cup && \cup && \cup & \\

& \cdots & \cdots & \cdots & \cdots & \cdots & \cdots & \\

& \cup & \cup && \cup && \cup & \\

& \chi^n(X^{t+1}(X^t\cap Y^{t+k})) & \chi^n(X^t(X^{t-1}\cap Y^{t+k-1})) & \cdots & \chi^n(X^{t-j+1}(X^{t-j}\cap Y^{t+k-j})) & \cdots & \chi^n(X^{t-\ell+1}(X^{t-\ell}\cap Y^{t+k-\ell})) & \\

& \cup & \cup && \cup && \cup & \\

& \chi^n(X^{t+1}(X^t\cap Y^{t+k-1})) & \chi^n(X^t(X^{t-1}\cap Y^{t+k-2})) & \cdots & \chi^n(X^{t-j+1}(X^{t-j}\cap Y^{t+k-j-1})) & \cdots & \chi^n(X^{t-\ell+1}(X^{t-\ell}\cap Y^{t+k-\ell-1})) & \\

& \cup & \cup && \cup && \cup & \\

\cdots & \cdots & \cdots & \cdots & \cdots & \cdots & \cdots & \cdots \\

& \cup & \cup && \cup && \cup & \\

& \chi^n(X^{t+1}(X^t\cap Y^{t+j})) & \chi^n(X^t(X^{t-1}\cap Y^{t+j-1})) & \cdots & \chi^n(X^{t-j+1}(X^{t-j}\cap Y^{t})) & \cdots & \chi^n(X^{t-\ell+1}(X^{t-\ell}\cap Y^{t+j-\ell})) & \\

& \cup & \cup && \cup && \cup & \\

& \chi^n(X^{t+1}(X^t\cap Y^{t+j-1})) & \chi^n(X^t(X^{t-1}\cap Y^{t+j-2})) & \cdots & \chi^n(X^{t-j+1}(X^{t-j}\cap Y^{t-1})) & \cdots & \chi^n(X^{t-\ell+1}(X^{t-\ell}\cap Y^{t+j-\ell-1})) & \\

& \cup & \cup && \cup && \cup & \\

& \cdots & \cdots & \cdots & \cdots & \cdots & \cdots & \\

& \cup & \cup && \cup && \cup & \\

& \chi^n(X^{t+1}(X^t\cap Y^{t+1})) & \chi^n(X^t(X^{t-1}\cap Y^{t})) & \cdots & \chi^n(X^{t-j+1}(X^{t-j}\cap Y^{t-j+1})) & \cdots & \chi^n(X^{t-\ell+1}(X^{t-\ell}\cap Y^{t-\ell+1})) & \\

& \cup & \cup && \cup && \cup & \\

& \chi^n(X^{t+1}(X^t\cap Y^t)) & \chi^n(X^t(X^{t-1}\cap Y^{t-1})) & \cdots & \chi^n(X^{t-j+1}(X^{t-j}\cap Y^{t-j})) & \cdots & \chi^n(X^{t-\ell+1}(X^{t-\ell}\cap Y^{t-\ell})) & \\

& \cup & \cup && \cup && \cup & \\

& \chi^n(X^{t+1}(X^t\cap Y^{t-1})) & \chi^n(X^t(X^{t-1}\cap Y^{t-2})) & \cdots & \chi^n(X^{t-j+1}(X^{t-j}\cap Y^{t-j-1})) & \cdots & \chi^n(X^{t-\ell+1}(X^{t-\ell}\cap Y^{t-\ell-1})) & \\

& \cup & \cup && \cup && \cup & \\

& \vdots & \vdots & \vdots & \vdots & \vdots & \vdots & \\

& \vdots & \vdots & \vdots & \vdots & \vdots & \vdots & \\

& \cup & \cup && \cup && \cup & \\

& \chi^n(X^{t+1}) & \chi^n(X^t) & \cdots & \chi^n(X^{t-j+1}) & \cdots & \chi^n(X^{t-\ell+1}) & \\

& \shrtpll & \shrtpll && \shrtpll && \shrtpll &
\end{array}
\ee
\end{figure}

Any sequence $\bmb$ in $B$ has a well defined component at time $n$, $b^n$.  The set of
all components at time $n$ forms group $B^n$.  Then
the projection of group system $B$ on the component at time $n$ is group $B^n\rmdef\chi^n(B)$.
There is a homomorphism from the group system $B$ to group $B^n$
given by the projection $\chi^n$.
\begin{thm}
\label{homoa}
Let $\chi^n:  B\ra B^n$ be the projection map given by assignment $\chi^n:  \bmb\mapsto b^n$.  
There is a homomorphism from $B$ to $B^n$ given by map $\chi^n$, for each $n\in\bmcpz$.
\end{thm}

\begin{prf}
Let $\chi^n:  \bmbdt\mapsto \bdt^n$ and $\chi^n:  \bmbddt\mapsto \bddt^n$.  Then
$\chi^n(\bmbdt\bmbddt)=\bdt^n\bddt^n=\chi^n(\bmbdt)\chi^n(\bmbddt)$.
\end{prf}

The projection of the normal series of $B$, (\ref{sminf}), is a normal series of group $B^n$, 
as shown in (\ref{sminfatt}).  Any term in (\ref{sminfatt}) is of the form
$$
\chi^n(X^{i+1}(X^i\cap Y^{i+m}))
$$
for some integer pair $i,m\in\bmcpz$.
In any normal chain, we may form the quotient group of two successive groups in the chain.
Then a normal chain of groups, as in (\ref{sminfatt}), gives a series of quotient groups.
A general quotient group obtained from (\ref{sminfatt}) is of the form
\be
\label{qgxf}
\frac{\chi^n(X^{i+1}(X^i\cap Y^{i+m}))}{\chi^n(X^{i+1}(X^i\cap Y^{i+m-1}))}
\ee
As noted before we may rewrite (\ref{qgxf}) as
\begin{multline}
\frac{\chi^n(X^{i+1}(X^i\cap Y^{i+m}))}{\chi^n(X^{i+1}(X^i\cap Y^{i+m-1})(X^{i+1}\cap Y^{i+m}))} \\
\label{qgx0f}
=\frac{\chi^n(X^{i+1})\chi^n(X^i\cap Y^{i+m})}{\chi^n(X^{i+1})\chi^n((X^i\cap Y^{i+m-1})(X^{i+1}\cap Y^{i+m}))} \\
\end{multline}

We now find a transversal of (\ref{qgx0f}), in similar manner to finding a transversal
of (\ref{qgx0}).  An element of the numerator group is of 
the form $x^ny^n$, where $x^n\in\chi^n(X^{i+1})$ and $y^n\in\chi^n(X^i\cap Y^{i+m})$.
Then a coset of normal subgroup $\chi^n(X^{i+1})\chi^n((X^i\cap Y^{i+m-1})(X^{i+1}\cap Y^{i+m}))$ is 
\begin{align*}
x^ny^n\chi^n(X^{i+1})\chi^n((X^i\cap Y^{i+m-1})(X^{i+1}\cap Y^{i+m})) 
&=x^n\chi^n(X^{i+1}) y^n\chi^n((X^i\cap Y^{i+m-1})(X^{i+1}\cap Y^{i+m})) \\
&=\chi^n(X^{i+1}) y^n\chi^n((X^i\cap Y^{i+m-1})(X^{i+1}\cap Y^{i+m})).
\end{align*}

\begin{thm}
A coset of normal subgroup $\chi^n(X^{i+1})\chi^n((X^i\cap Y^{i+m-1})(X^{i+1}\cap Y^{i+m}))$ is 
$\chi^n(X^{i+1})y^n\chi^n((X^i\cap Y^{i+m-1})(X^{i+1}\cap Y^{i+m}))$ where $y^n\in\chi^n(X^i\cap Y^{i+m})$.
A \crep\ of coset $\chi^n(X^{i+1})y^n\chi^n((X^i\cap Y^{i+m-1})(X^{i+1}\cap Y^{i+m}))$ is
$\xdt^n\ydt^n$ where $\xdt^n$ is in $\chi^n(X^{i+1})$ and $\ydt^n$ is in
$y^n\chi^n((X^i\cap Y^{i+m-1})(X^{i+1}\cap Y^{i+m}))$.
A transversal of the quotient group (\ref{qgx0f}) is a selection of one \crep\
$\xdt^n\ydt^n$ from each coset $\chi^n(X^{i+1})y^n\chi^n((X^i\cap Y^{i+m-1})(X^{i+1}\cap Y^{i+m}))$.
\end{thm}
We may always select a \crep\ $\xdt^n\ydt^n$ of coset 
$\chi^n(X^{i+1})y^n\chi^n((X^i\cap Y^{i+m-1})(X^{i+1}\cap Y^{i+m}))$
to be $1^n \ydt^n$ where $1^n$ is the identity of $\chi^n(X^{i+1})$.
But note that $y^n\chi^n((X^i\cap Y^{i+m-1})(X^{i+1}\cap Y^{i+m}))$ is a coset of
normal subgroup $\chi^n((X^i\cap Y^{i+m-1})(X^{i+1}\cap Y^{i+m}))$ in quotient group
\be
\label{qgx1f}
\frac{\chi^n(X^i\cap Y^{i+m})}{\chi^n((X^i\cap Y^{i+m-1})(X^{i+1}\cap Y^{i+m}))}.
\ee
Then $\ydt^n$ is a \crep\ of (\ref{qgx1f}).  This gives the following.

\begin{cor}
A transversal of the quotient group (\ref{qgx1f}) is a transversal
of the quotient group (\ref{qgx0f}), but the reverse is only true if $\xdt^n\ydt^n=1^n \ydt^n$ 
or $\xdt^n=1^n$ for each \crep\ $\xdt^n\ydt^n$ of (\ref{qgx0f}).
\end{cor}

We have just shown that a transversal of (\ref{qgx1f}) is a transversal of (\ref{qgx0f}).
It is not surprising then that there is a homomorphism from (\ref{qgx0f}) to (\ref{qgx1f}).
Again this result is an application of the Zassenhaus lemma, Lemma \ref{lem26xx}.

Note that (\ref{qgx0f}) is equivalent to (\ref{qgxf}), and (\ref{qgxf}) is equivalent to
\be
\label{qgxff}
\frac{\chi^n(X^{i+1})\chi^n(X^i\cap Y^{i+m})}{\chi^n(X^{i+1})\chi^n(X^i\cap Y^{i+m-1})}.
\ee
We now use Lemma \ref{lem26xx} to show there is a homomorphism from (\ref{qgxff})
to (\ref{qgx1f}).  Let $U=\chi^n(X^{i+1})$ and $U^*=\chi^n(X^i)$.
Let $V=\chi^n(Y^{i+m-1})$ and $V^*=\chi^n(Y^{i+m})$.
Note that $U\lhd U^*$ and $V\lhd V^*$.  Then
\begin{align*}
\frac{U(U^*\cap V^*)}{U(U^*\cap V)}
&=\frac{\chi^n(X^{i+1})(\chi^n(X^i)\cap\chi^n(Y^{i+m}))}{\chi^n(X^{i+1})(\chi^n(X^i)\cap\chi^n(Y^{i+m-1}))} \\
&=\frac{\chi^n(X^{i+1})\chi^n(X^i\cap Y^{i+m})}{\chi^n(X^{i+1})\chi^n(X^i\cap Y^{i+m-1})},
\end{align*}
and
\begin{align*}
\frac{U^*\cap V^*}{D}
&=\frac{U^*\cap V^*}{(U^*\cap V)(U\cap V^*)} \\
&=\frac{\chi^n(X^i)\cap \chi^n(Y^{i+m})}{(\chi^n(X^i)\cap \chi^n(Y^{i+m-1}))(\chi^n(X^{i+1})\cap\chi^n(Y^{i+m}))} \\
&=\frac{\chi^n(X^i\cap Y^{i+m})}{\chi^n(X^i\cap Y^{i+m-1})\chi^n(X^{i+1}\cap Y^{i+m})} \\
&=\frac{\chi^n(X^i\cap Y^{i+m})}{\chi^n((X^i\cap Y^{i+m-1})(X^{i+1}\cap Y^{i+m}))}.
\end{align*}
Now the function $f$ of Lemma \ref{lem26xx} is a homomorphism from (\ref{qgxff}) to
(\ref{qgx1f}).

\begin{thm}
There is a homomorphism from quotient group (\ref{qgxff}),
or equivalently (\ref{qgxf}) and (\ref{qgx0f}), to quotient group (\ref{qgx1f}),
given by the function $f$ of the Zassenhaus lemma.
\end{thm}

We now show the time $n$ components $\chi^n$ of a transversal of the 
quotient group (\ref{qgx}) are a transversal of (\ref{qgxf}).

\begin{lem}
Fix $i\in\bmcpz$.  Let $0\le m\le\ell$.  For any $n$ such that $i\le n\le i+m$,
there is a homomorphism from $X^{i+1}(X^i\cap Y^{i+m})$ to
$\chi^n(X^{i+1}(X^i\cap Y^{i+m}))$ given by the projection map $\chi^n$.
The kernel of the homomorphism is $K_{i,m}$, the paths in $X^{i+1}(X^i\cap Y^{i+m})$
that are the identity at time $n$, and
$$
\frac{X^{i+1}(X^i\cap Y^{i+m})}{K_{i,m}}\simeq\chi^n(X^{i+1}(X^i\cap Y^{i+m})).
$$
\end{lem}

\begin{prf}
This is just the first homomorphism theorem.
\end{prf}

\begin{thm}
\label{thm17}
Fix $i\in\bmcpz$.  Let $0\le m\le\ell$.  For any $n$ such that $i\le n\le i+m$, 
there is a homomorphism from quotient group (\ref{qgx}) to quotient group (\ref{qgxf})
given by the projection map $\chi^n$, where a coset of (\ref{qgx}) is mapped to a coset
of (\ref{qgxf}), and with this mapping we have
\be
\label{qgqgt}
\frac{X^{i+1}(X^i\cap Y^{i+m})}{X^{i+1}(X^i\cap Y^{i+m-1})}
 \simeq\frac{\chi^n(X^{i+1}(X^i\cap Y^{i+m}))}{\chi^n(X^{i+1}(X^i\cap Y^{i+m-1}))}.
\ee
\end{thm}

\begin{prf}
$K_{i,m}$ are the paths in $X^{i+1}(X^i\cap Y^{i+m})$ that are the identity at time $n$.
We now show $K_{i,m}\subset X^{i+1}(X^i\cap Y^{i+m-1})$ for $i\le n\le i+m$.
First consider the case $n=i$.  The paths in $X^{i+1}(X^i\cap Y^{i+m})$ that are the 
identity at time $n$ are the paths in $X^{i+1}(X^{i+1}\cap Y^{i+m})=X^{i+1}$.
Then $K_{i,m}\subset X^{i+1}(X^i\cap Y^{i+m-1})$ for $n=i$.  Now consider the case
$i<n\le i+m$.  The paths in $X^{i+1}(X^i\cap Y^{i+m})$ split from the identity
path at time $i$ or later.  The paths in $X^{i+1}(X^i\cap Y^{i+m})$ that are the 
identity at time $n$ merge to the identity path at time 
$n$ or earlier.  Then the paths in $K_{i,m}$ are
in $(X^i\cap Y^{n-1})$.  Since $n\le i+m$, $(X^i\cap Y^{n-1})\subset(X^i\cap Y^{i+m-1})$.  
Then $K_{i,m}\subset X^{i+1}(X^i\cap Y^{i+m-1})$ for $i<n\le i+m$.
Then $K_{i,m}\subset X^{i+1}(X^i\cap Y^{i+m-1})$ for $i\le n\le i+m$.

The image of $X^{i+1}(X^i\cap Y^{i+m})$ under the projection $\chi^n$ is the group
$\chi^n(X^{i+1}(X^i\cap Y^{i+m}))$.  Since $X^{i+1}(X^i\cap Y^{i+m-1})$
contains the normal subgroup $K_{i,m}$ for $i\le n\le i+m$, then by the correspondence theorem, 
the image of $X^{i+1}(X^i\cap Y^{i+m-1})$ under projection $\chi^n$ is a normal 
subgroup of image $\chi^n(X^{i+1}(X^i\cap Y^{i+m}))$, and (\ref{qgqgt}) holds for $i\le n\le i+m$.
\end{prf}

\begin{cor}
\label{cor16}
For any $i,m\in\bmcpz$, for any $n\in\bmcpz$, the time $n$ components $\chi^n$ 
of a transversal of the quotient group (\ref{qgx}) are a transversal of (\ref{qgxf}).
\end{cor}

\begin{prf}
Fix $i\in\bmcpz$.  First consider the case $m<0$.  Then the quotient group (\ref{qgx}) is
$X^{i+1}/X^{i+1}$.  Then a \crep\ $\bmxdt\bmydt$ is $\bmxdt\bone$.  But then $\chi^n(\bmxdt)$ is 
a \crep\ of $\chi^n(X^{i+1})/\chi^n(X^{i+1})$ for any $n\in\bmcpz$.  Then the time $n$ components $\chi^n$ 
of a transversal of (\ref{qgx}) are a transversal of (\ref{qgxf}), and so Corollary \ref{cor16}
is true.  Next consider the case $m>\ell$.  Then the quotient group
(\ref{qgx}) is trivial.  Then any element of the numerator (\ref{qgx}) is a \crep\ $\bmxdt\bmydt$.
But then $\chi^n(\bmxdt\bmydt)$ is a \crep\ of (\ref{qgxf}) for any $n\in\bmcpz$.  Then Corollary
\ref{cor16} is true.  Lastly consider the case $0\le m\le\ell$.
Select a \crep\ $\bmxdt\bmydt$ of quotient group (\ref{qgx}).  For $n<i$, we have 
$\chi^n(\bmxdt\bmydt)=\chi^n(\bone\bone)=\chi^n(\bone)$ and Corollary \ref{cor16} holds.
For $n>i+m$, we have $\chi^n(\bmxdt\bmydt)=\chi^n(\bmxdt\bone)=\chi^n(\bmxdt)$ and 
Corollary \ref{cor16} holds.  Lastly consider the case $i\le n\le i+m$.  But in this case,
Theorem \ref{thm17} shows that the projection $\chi^n$ gives a bijection between cosets 
of (\ref{qgx}) and cosets of (\ref{qgxf}).  Therefore the projection $\chi^n$ of a transversal 
of (\ref{qgx}) is a transversal of (\ref{qgxf}).
\end{prf}

We now examine the normal series (\ref{sminfatt}) for time $n=t$.  First we reduce the normal series
(\ref{sminfatt}) to a finite number of columns.  We use the following lemma.

\begin{lem}
\label{cntrl}
We have $\chi^{[t,\infty)}(X^{t-\ell})=\chi^{[t,\infty)}(B)$ and
$\chi^{(-\infty,t]}(Y^{t+\ell})=\chi^{(-\infty,t]}(B)$. 
\end{lem}

\begin{prf}
Fix $t\in\bmcpz$.  Let $\bmb$ be an arbritrary path in $B$.  Since $B$ is \ellctl, there is a
$\bmbdt\in X^{t-\ell}$ such that $\chi^{[t,\infty)}(\bmbdt)=\chi^{[t,\infty)}(\bmb)$.
In other words, $\chi^{[t,\infty)}(X^{t-\ell})=\chi^{[t,\infty)}(B)$.
\end{prf}

There is a single ellipsis in the first column of (\ref{sminfatt}).  This denotes all columns 
to the left of the first column shown in (\ref{sminfatt}).  For $n=t$, these columns contain 
terms of the form $\chi^t(X^{i+1}(X^i\cap Y^{i+m}))$ for $i>t$ and $m\in\bmcpz$.  If $m\ge 0$,
then $\chi^t(X^{i+1}(X^i\cap Y^{i+m}))=\chi^t(\bone)=1^t$.  If $m<0$,
then $(X^i\cap Y^{i+m})=\bone$ and $\chi^t(X^{i+1}(X^i\cap Y^{i+m}))=1^t$.
Therefore all terms in the half infinite sequence of columns denoted by the
single ellipsis in the first column of (\ref{sminfatt}) reduce to $1^t$.
Similarly the first term $\chi^t(X^{t+1})$ in the bottom row of (\ref{sminfatt}) for $n=t$
reduces to $1^t$.

There is a single ellipsis in the last column of (\ref{sminfatt}).  This denotes all columns 
to the right of the last column shown in (\ref{sminfatt}).  For $n=t$, these columns contain 
terms of the form $\chi^t(X^{i+1}(X^i\cap Y^{i+m}))$ for $i+1\le t-\ell$ and $m\in\bmcpz$.  But 
for $i+1\le t-\ell$ and $m\in\bmcpz$,
\begin{align*}
\chi^t(X^{i+1}(X^i\cap Y^{i+m})) &=\chi^t(X^{i+1})\chi^t(X^i\cap Y^{i+m}) \\
 &=\chi^t(X^{i+1})\chi^t(X^i\cap Y^{i+m}) \\
 &=B^t\chi^t(X^i\cap Y^{i+m}) \\
 &=B^t,
\end{align*}
where $\chi^t(X^{i+1})=B^t$ by Lemma \ref{cntrl} since $i+1\le t-\ell$.
Therefore all terms in the half infinite sequence of columns denoted by the
single ellipsis in the last column of (\ref{sminfatt}) reduce to $B^t$.
Similarly the last term $\chi^t(X^{t-\ell+1}(X^{t-\ell}))$ in the top row of 
(\ref{sminfatt}) for $n=t$ reduces to $\chi^t(X^{t-\ell+1})\chi^t(X^{t-\ell})=B^t$ since
$\chi^t(X^{t-\ell})=B^t$ by Lemma \ref{cntrl}.
Therefore we have just reduced the infinite sequence of columns in (\ref{sminfatt}) to
the finite number $\ell+1$ of columns as shown in (\ref{sminfatt1a}).

The results $1^t=\chi^t(X^{t+1})$ and $\chi^t(X^{t-\ell+1}(X^{t-\ell}))=B^t$, together 
with the groups in the top and bottom rows of (\ref{sminfatt1a}), show that (\ref{sminfatt1a}) 
is a refinement of the normal series
$$
1^t=\chi^t(X^{t+1})\subset\chi^t(X^t)\subset\cdots\subset\chi^t(X^{t-j+1})\subset\cdots\subset\chi^t(X^{t-\ell+1})\subset\chi^t(X^{t-\ell})=B^t
$$
of $B^t$ by inserting the normal series
$$
1^t=\chi^t(Y^{t-1})\subset\chi^t(Y^t)\subset\cdots\subset\chi^t(Y^{t+k-1})\subset\cdots\subset\chi^t(Y^{t+\ell-1})\subset\chi^t(Y^{t+\ell})=B^t,
$$
where we have used $X^{i+1}(X^i)=X^i$ in the top row of (\ref{sminfatt1a}).

We now reduce the infinite number of groups in each column of (\ref{sminfatt1a}) to a 
finite number of groups.  We first consider
the half infinite column of groups in the bottom of each of the $\ell+1$ columns in 
(\ref{sminfatt1a}) of the form
$$
\chi^t(X^{t-j+1}(X^{t-j}\cap Y^{t-s}))
$$
for $0\le j\le\ell$ and $s\le 1$.  For these groups, we have
\begin{align*}
\chi^t(X^{t-j+1}(X^{t-j}\cap Y^{t-s})) &=\chi^t(X^{t-j+1})\chi^t(X^{t-j}\cap Y^{t-s}) \\
 &=\chi^t(X^{t-j+1})(\chi^t(X^{t-j})\cap\chi^t(Y^{t-s})) \\
 &=\chi^t(X^{t-j+1})(\chi^t(X^{t-j})\cap1^t) \\
 &=\chi^t(X^{t-j+1}).
\end{align*}
Therefore in each column of (\ref{sminfatt1a}), groups in the bottom half infinite column 
are all the same and equal the group in the bottom row.  This gives the 
infinite chain of equalities shown in the bottom half of each column of (\ref{sminfatt1}).

We now show a similar chain of equalities holds for groups in the half infinite 
column of groups at the top of each column in (\ref{sminfatt1a}).
We consider terms of the form 
$$
\chi^t(X^{t-j+1}(X^{t-j}\cap Y^{t-j+s}))
$$
for $0\le j\le\ell$ and $s\ge\ell$.
For these groups, we have
\begin{align*}
\chi^t(X^{t-j+1}(X^{t-j}\cap Y^{t-j+s})) &=\chi^t(X^{t-j+1})\chi^t(X^{t-j}\cap Y^{t-j+s}) \\
 &=\chi^t(X^{t-j+1})(\chi^t(X^{t-j})\cap\chi^t(Y^{t-j+s})) \\
 &=\chi^t(X^{t-j+1})(\chi^t(X^{t-j})\cap B^t) \\
 &=\chi^t(X^{t-j+1}(X^{t-j})),
\end{align*}
where $\chi^t(Y^{t-j+s})=B^t$ by Lemma \ref{cntrl}.  This collapses all groups in the top 
half infinite column of (\ref{sminfatt1a}) to the group in the top row, as shown
in (\ref{sminfatt1}).
Then the normal chain (\ref{sminfatt1a}) can be reduced to the normal chain (\ref{sminfatt1}).
The normal chain (\ref{sminfatt1}) collapses to a finite number of groups in each column after eliminating
duplicate groups, as shown in (\ref{sminfatt2}).  The normal chain (\ref{sminfatt2}) is a \cdc\
of $B^t$.  Then we have reduced the normal series (\ref{sminfatt1a}) to a finite number 
of groups in (\ref{sminfatt2}).

As a check on (\ref{sminfatt2}),
recall from Corollary \ref{cor16} that the time $n$ components $\chi^n$ of a transversal 
of the quotient group (\ref{qgx}) are a transversal of the quotient group (\ref{qgxf}) 
for any $n\in\bmcpz$ and $i,m\in\bmcpz$.  Then the time $t$ components $\chi^t$ of a 
transversal $[\Lambda_{j,k}^t]$ of the quotient group
\be
\label{qgqgx}
\Lambda_{j,k}^t=\frac{X^{t-j+1}(X^{t-j}\cap Y^{t+k-j})}{X^{t-j+1}(X^{t-j}\cap Y^{t+k-j-1})},
\ee
written as $\chi^t([\Lambda_{j,k}^t])$, are a transversal of the quotient group
\be
\label{qgxfx}
\frac{\chi^t(X^{t-j+1}(X^{t-j}\cap Y^{t+k-j}))}{\chi^t(X^{t-j+1}(X^{t-j}\cap Y^{t+k-j-1}))}.
\ee
But from Lemma \ref{lem9}, the only quotient groups formed from (\ref{sminf}) that have 
transversals with nontrivial components at time $t$ are of the form (\ref{qgqgx})
for $0\le j\le\ell$, for $j\le k\le\ell$.  Then we have the following.

\begin{lem}
The only nontrivial quotient groups at time $t$ of the form (\ref{qgxfx}) formed from (\ref{sminf})
are those for $0\le j\le\ell$, for $j\le k\le\ell$.
\end{lem}
But these quotient groups are exactly those formed from the normal chain (\ref{sminfatt2}).

Now, since there is a set of generators $\{\bmg^{[t-j,t-j+k]}\}$ which is a transversal $[\Lambda_{j,k}^t]$ 
of (\ref{qgqgx}), or
$$
[\Lambda_{j,k}^t]=\{\bmg^{[t-j,t-j+k]}\},
$$
for $0\le j\le\ell$, for $j\le k\le\ell$,
then the time $t$ components $\chi^t$ of transversal $[\Lambda_{j,k}^t]$, that is 
\be
\label{keyeqf}
\chi^t([\Lambda_{j,k}^t])=\{\chi^t(\bmg^{[t-j,t-j+k]})\}=\{r_{j,k}^{t-j}\},
\ee
are a transversal of (\ref{qgxfx}), for $0\le j\le\ell$, for $j\le k\le\ell$.
In this way we obtain a 
{\it complete set of coset representatives} for the normal chain of $B^t$ given 
by (\ref{sminfatt2}).  This gives the following result.

\begin{thm}
\label{thm25}
Fix time $t$.  For $0\le j\le\ell$, for $j\le k\le\ell$, let $\{\bmg^{[t-j,t-j+k]}\}$
be a set of generators which is a transversal $[\Lambda_{j,k}^t]$ of (\ref{qgqgx}).
For $0\le j\le\ell$, for $j\le k\le\ell$, 
the time $t$ components $\chi^t$ of transversal $[\Lambda_{j,k}^t]$
form a transversal $\chi^t([\Lambda_{j,k}^t])$ of (\ref{qgxfx}).  The union set of transversals,
\be
\label{unionset}
\cup_{\{0\le j\le\ell,j\le k\le\ell\}} \, \chi^t([\Lambda_{j,k}^t]),
\ee
forms a \compset\ for the normal chain of $B^t$ given by (\ref{sminfatt2}).
\end{thm}
Note that the components $\chi^t(\bmg^{[t-j,t-j+k]})$ used in the evaluation of $b^t$ in (\ref{enctdc}) 
and (\ref{enctda}), for each $b^t$ in $B^t$, are taken from the union set (\ref{unionset}),
a complete set of coset representatives for the normal chain of $B^t$ in (\ref{sminfatt2}).
Similarly the \creps\ $r_{j,k}^{t-j}$ used in the evaluation of $b^t$ in (\ref{enctd}), 
for each $b^t$ in $B^t$, are taken from the union set (\ref{unionset}).

By selecting one \crep\ $\chi^t(\bmg^{[t-j,t-j+k]})$ or $r_{j,k}^{t-j}$ from each transversal 
$\chi^t([\Lambda_{j,k}^t])$, for $0\le j\le\ell$, for $j\le k\le\ell$, 
we obtain a {\it coset representative chain}
for the normal chain of $B^t$ given by (\ref{sminfatt2}).
For each $t\in\bmcpz$, define a map $\eta^t:  \tridnjktarg{0}{0}{t}{\calr}\ra B^t$
given by assignment $\eta^t:  \tridnjktarg{0}{0}{t}{\bmr}\mapsto b^t$ if
$b^t$ is an encoding of the representatives in $\tridnjktarg{0}{0}{t}{\bmr}$
using (\ref{enctd}).

\begin{thm}
\label{thm29}
For each $t\in\bmcpz$, the map $\eta^t:  \tridnjktarg{0}{0}{t}{\calr}\ra B^t$ 
given by assignment $\eta^t:  \tridnjktarg{0}{0}{t}{\bmr}\mapsto b^t$ is a bijection,
where the representatives in $\tridnjktarg{0}{0}{t}{\bmr}$ form a \crepc\ of $b^t$ using (\ref{enctd}).
\end{thm}

\begin{prf}
We have just seen the representatives in $\tridnjktarg{0}{0}{t}{\bmr}$ form a \crepc\ of $b^t$
using (\ref{enctd}).  From Corollary \ref{cor12}, the encoding of the representatives 
in each $\tridnjktarg{0}{0}{t}{\bmr}$ in $\tridnjktarg{0}{0}{t}{\calr}$ 
using (\ref{enctd}) is into $B^t$.  From Corollary \ref{cor13},
it is onto $B^t$.  But since the \crepc\ $\tridnjktarg{0}{0}{t}{\bmr}$ of $b^t$ in $B^t$ is unique, 
the encoding is one to one.  Then $\eta^t$ is a bijection from $\tridnjktarg{0}{0}{t}{\calr}$ to $B^t$.
\end{prf} 

This gives the following corollary of Theorem \ref{thm11} and \ref{thm29}.

\begin{cor}
\label{cor26}
Fix basis $\calb$ of $B$.  Then $\calr$ is fixed, and there is a 
bijection $\eta:  \calr\ra B$ given by assignment $\eta:  \bmr\mapsto\bmb$, 
and for each $t\in\bmcpz$, we have $\eta^t:  \tridnjktarg{0}{0}{t}{\bmr}\mapsto\chi^t(\bmb)=b^t$ 
in the bijection $\eta^t:  \tridnjktarg{0}{0}{t}{\calr}\ra B^t$, where the representatives in
$\tridnjktarg{0}{0}{t}{\bmr}$ form a \crepc\ of $b^t$ using (\ref{enctd}). 
\end{cor}

\begin{prf}
From Theorem \ref{thm11}, we have $\eta^t:  \tridnjktarg{0}{0}{t}{\bmr}\mapsto\chi^t(\bmb)=b^t$ 
if $\eta:  \bmr\mapsto\bmb$ in bijection $\eta:  \calr\ra B$.  Now apply Theorem \ref{thm29}.
\end{prf}

\begin{lem}
\label{lem47}
Fix time $t$.  Let $\bmrdt,\bmrddt\in\calr$.  In the product
$\bmrdt*\bmrddt$, the representatives in 
$\tridnjktarg{0}{0}{t}{\bmrdt*\bmrddt}$ are uniquely determined by the representatives in
$\tridnjktarg{0}{0}{t}{\bmrdt}$ and $\tridnjktarg{0}{0}{t}{\bmrddt}$.
\end{lem}

\begin{prf}
Fix time $t$.  Let $\bmrdt,\bmrddt\in\calr$.  Suppose $\eta:  \bmrdt\mapsto\bmbdt$ and
$\eta:  \bmrddt\mapsto\bmbddt$.  Then from Corollary \ref{cor26}, we have
$\eta^t:  \tridnjktarg{0}{0}{t}{\bmrdt}\mapsto\chi^t(\bmbdt)\rmdef\bdt^t$ where the representatives in
$\tridnjktarg{0}{0}{t}{\bmrdt}$ form a \crepc\ of $\bdt^t$ using (\ref{enctd}).  Then from (\ref{enctd1}), 
\be
\label{exp1}
\bdt^t=\rdt_{0,0}^t\rdt_{0,1}^t\rdt_{0,2}^t
\cdots \rdt_{j,j}^{t-j}\cdots \rdt_{j,k}^{t-j}\cdots \rdt_{j,\ell}^{t-j} 
\cdots \rdt_{\ell-1,\ell-1}^{t-\ell+1} \rdt_{\ell-1,\ell}^{t-\ell+1} \rdt_{\ell,\ell}^{t-\ell}.
\ee
Also $\eta^t:  \tridnjktarg{0}{0}{t}{\bmrddt}\mapsto\chi^t(\bmbddt)\rmdef\bddt^t$ where the representatives in
$\tridnjktarg{0}{0}{t}{\bmrddt}$ form a \crepc\ of $\bddt^t$ using (\ref{enctd}).  Then from (\ref{enctd1}), 
\be
\label{exp2}
\bddt^t=\rddt_{0,0}^t\rddt_{0,1}^t\rddt_{0,2}^t
\cdots \rddt_{j,j}^{t-j}\cdots \rddt_{j,k}^{t-j}\cdots \rddt_{j,\ell}^{t-j} 
\cdots \rddt_{\ell-1,\ell-1}^{t-\ell+1} \rddt_{\ell-1,\ell}^{t-\ell+1} \rddt_{\ell,\ell}^{t-\ell}.
\ee
Since $\eta:  \bmrdt\mapsto\bmbdt$ and $\eta:  \bmrddt\mapsto\bmbddt$, then 
$\eta:  \bmrdt*\bmrddt\mapsto\bmbdt\bmbddt$.  Then from Corollary \ref{cor26},
$\eta^t:  \tridnjktarg{0}{0}{t}{\bmrdt*\bmrddt}\mapsto\chi^t(\bmbdt\bmbddt)=\bdt^t\bddt^t$
where the representatives in $\tridnjktarg{0}{0}{t}{\bmrdt*\bmrddt}$ form a \crepc\ of
$\bdt^t\bddt^t$ using (\ref{enctd}).  But from (\ref{exp1})-(\ref{exp2}), we have
\begin{multline}
\label{exp4} 
\bdt^t\bddt^t=
\left(
\rdt_{0,0}^t\rdt_{0,1}^t\rdt_{0,2}^t
\cdots \rdt_{j,j}^{t-j}\cdots \rdt_{j,k}^{t-j}\cdots \rdt_{j,\ell}^{t-j} 
\cdots \rdt_{\ell-1,\ell-1}^{t-\ell+1} \rdt_{\ell-1,\ell}^{t-\ell+1} \rdt_{\ell,\ell}^{t-\ell}
\right) \\
\left(
\rddt_{0,0}^t\rddt_{0,1}^t\rddt_{0,2}^t
\cdots \rddt_{j,j}^{t-j}\cdots \rddt_{j,k}^{t-j}\cdots \rddt_{j,\ell}^{t-j} 
\cdots \rddt_{\ell-1,\ell-1}^{t-\ell+1} \rddt_{\ell-1,\ell}^{t-\ell+1} \rddt_{\ell,\ell}^{t-\ell}
\right).
\end{multline}
Equation (\ref{exp4}) shows that the \crepc s in (\ref{exp1}) and (\ref{exp2})
form a unique product $\bdt^t\bddt^t$.  But the product $\bdt^t\bddt^t$ has a
unique \crepc.  Therefore the \crepc\ of $\bdt^t\bddt^t$ is uniquely determined by the product 
of the \crepc s of $\bdt^t$ and $\bddt^t$.
\end{prf}

From Theorem \ref{thm25}, the union set of transversals,
\be
\label{unionset1}
\cup_{\{0\le j\le\ell,j\le k\le\ell\}} \, \{r_{j,k}^{t-j}\},
\ee
where each representative $r_{j,k}^{t-j}$ is at time $t$,
forms a \compset\ for the normal chain of $B^t$ given by (\ref{sminfatt2}).
The representatives in a \compset\ are distinct.  This gives the following.

\begin{lem}
\label{lem25}
Fix time $t$.  The representatives in union set (\ref{unionset1}) are distinct.  Then
for $0\le j\le\ell$, for $j\le k\le\ell$, the representatives in set $\{r_{j,k}^{t-j}\}$
are distinct.
\end{lem}

\begin{thm}
\label{thm25a}
Fix time $t$.  For $0\le j\le\ell$, for $j\le k\le\ell$, let $\{\bmg^{[t-j,t-j+k]}\}$
be a set of generators which is a transversal $[\Lambda_{j,k}^t]$ of (\ref{qgqgx}).
We assume each $\bmg^{[t-j,t-j+k]}$ has the sequence of components
\be
\bmg^{[t-j,t-j+k]}=
\ldots,1^{t-j-2},1^{t-j-1},r_{0,k}^{t-j},r_{1,k}^{t-j},\ldots,r_{m,k}^{t-j},\ldots,r_{k,k}^{t-j},1^{t-j+k+1},1^{t-j+k+2},\ldots,
\ee
where $0\le m\le k$.  For each $m$ such that $0\le m\le k$, there is a bijection
$[\Lambda_{j,k}^t]\ra \, \{r_{m,k}^{t-j}\}$.
\end{thm}

\begin{prf}
Fix time $t'$.  From Theorem \ref{thm25}, the union set of transversals,
\be
\cup_{\{0\le j'\le\ell,j'\le k'\le\ell\}} \, \{r_{j',k'}^{t'-j'}\},
\ee
where each representative $r_{j',k'}^{t'-j'}$ is at time $t'$,
forms a \compset\ for the normal chain of $B^{t'}$ given by (\ref{sminfatt2}).
From Lemma \ref{lem25}, the representatives in set $\{r_{j',k'}^{t'-j'}\}$ are distinct
for $j',k'$ in the indicated ranges.  Now apply this result for $t'=t-j+j'$,
for $j'=0,1,\ldots,k$, and for $k'=k$.  Then the representatives in sets
$\{r_{j',k'}^{t'-j'}\}=\{r_{j',k}^{t-j}\}$ are distinct for $j'=0,1,\ldots,k$.
But these are sets of representatives in generators of $[\Lambda_{j,k}^t]$.
It follows that there is a bijection $[\Lambda_{j,k}^t]\ra \, \{r_{m,k}^{t-j}\}$, 
for $0\le m\le k$.
\end{prf}

For $0\le k\le\ell$, let $R_{0,k}^t$ be the set of first components $\{r_{0,k}^t\}$
of generators $\bmg^{[t,t+k]}$ which are a transversal of granule $\Lambda_{0,k}^t$.  
From Theorem \ref{thm25a},
we know there is a bijection $[\Lambda_{0,k}^t]\mapsto R_{0,k}^t$.  Then an input
$\bmi^t$ of $B$ at time $t$ is a selection of one representative $r_{0,k}^t$ from each set
$R_{0,k}^t$, for $0\le k\le\ell$.  Since the first column $\bmr_0^t$ of $\tridnjktarg{0}{0}{t}{\bmr}$ in 
(\ref{rttf}) are the first components of generators $\bmg^{[t,t+k]}$ at time $t$, for $0\le k\le\ell$,
the first column $\bmr_0^t$ of $\tridnjktarg{0}{0}{t}{\bmr}$ is an input $\bmi^t$ of $B$ at time $t$.
The set of all inputs $\bmi^t$ of $B$ that can occur at time $t$, or $\bmcpi^t$, 
is just the set of all possible ways of selecting one representative from each set
$R_{0,k}^t$, for $0\le k\le\ell$.  Then $\bmcpi^t$ is just 
$\bigotimes_{0\le k\le\ell} R_{0,k}^t$.  An input sequence $\bmi$
of $B$ is a sequence of inputs $\ldots,\bmi^t,\bmi^{t+1},\ldots$ that can occur 
in $B$.  If $B$ is complete, then the set of input sequences 
is the full Cartesian product $\bigotimes_{-\infty<t<+\infty} \bmcpi^t$.
If $\bmr\in\calr$, then the sequence
\be
\label{input1}
\ldots,\bmr_0^t,\bmr_0^{t-1},\ldots
\ee
of first columns of the sequence $\ldots,\tridnjktarg{0}{0}{t}{\bmr},\tridnjktarg{0}{0}{t-1}{\bmr},\ldots$ 
of \stms\ of $\bmr$ is an input sequence $\bmi$ of $B$ in reverse time order.  This gives the following.

\begin{lem}
\label{nmbrlem}
If $B$ is complete, the set of sequences (\ref{input1}) of first columns
of the sequence $\ldots,\tridnjktarg{0}{0}{t}{\bmr},\tridnjktarg{0}{0}{t-1}{\bmr},\ldots$ of \stms\ 
of $\bmr\in\calr$ is just the Cartesian product
\be
\label{nmbrin}
\bigotimes_{t=+\infty}^{-\infty} \bigotimes_{0\le k\le\ell} R_{0,k}^t.
\ee
\end{lem}

\newpage

\begin{figure}[tbp]

\be
\label{sminfatt1a}
\begin{array}{cccccc}
         &          &&          && B^t      \\

\shrtpll & \shrtpll && \shrtpll && \shrtpll \\

\chi^t(X^{t+1}(X^t)) & \chi^t(X^t(X^{t-1})) & \cdots & \chi^t(X^{t-j+1}(X^{t-j})) & \cdots & \chi^t(X^{t-\ell+1}(X^{t-\ell})) \\

\cup & \cup && \cup && \cup \\

\vdots & \vdots & \vdots & \vdots & \vdots & \vdots \\

\vdots & \vdots & \vdots & \vdots & \vdots & \vdots \\

\cup & \cup && \cup && \cup \\

\chi^t(X^{t+1}(X^t\cap Y^{t+\ell+1})) & \chi^t(X^t(X^{t-1}\cap Y^{t+\ell})) & \cdots & \chi^t(X^{t-j+1}(X^{t-j}\cap Y^{t+\ell-j+1})) & \cdots & \chi^t(X^{t-\ell+1}(X^{t-\ell}\cap Y^{t+1})) \\

\cup & \cup && \cup && \cup \\

\chi^t(X^{t+1}(X^t\cap Y^{t+\ell})) & \chi^t(X^t(X^{t-1}\cap Y^{t+\ell-1})) & \cdots & \chi^t(X^{t-j+1}(X^{t-j}\cap Y^{t+\ell-j})) & \cdots & \chi^t(X^{t-\ell+1}(X^{t-\ell}\cap Y^{t})) \\

\cup & \cup && \cup && \cup \\

\chi^t(X^{t+1}(X^t\cap Y^{t+\ell-1})) & \chi^t(X^t(X^{t-1}\cap Y^{t+\ell-2})) & \cdots & \chi^t(X^{t-j+1}(X^{t-j}\cap Y^{t+\ell-j-1})) & \cdots & \chi^t(X^{t-\ell+1}(X^{t-\ell}\cap Y^{t-1})) \\

\cup & \cup && \cup && \cup \\

\cdots & \cdots & \cdots & \cdots & \cdots & \cdots \\

\cup & \cup && \cup && \cup \\

\chi^t(X^{t+1}(X^t\cap Y^{t+k})) & \chi^t(X^t(X^{t-1}\cap Y^{t+k-1})) & \cdots & \chi^t(X^{t-j+1}(X^{t-j}\cap Y^{t+k-j})) & \cdots & \chi^t(X^{t-\ell+1}(X^{t-\ell}\cap Y^{t+k-\ell})) \\

\cup & \cup && \cup && \cup \\

\chi^t(X^{t+1}(X^t\cap Y^{t+k-1})) & \chi^t(X^t(X^{t-1}\cap Y^{t+k-2})) & \cdots & \chi^t(X^{t-j+1}(X^{t-j}\cap Y^{t+k-j-1})) & \cdots & \chi^t(X^{t-\ell+1}(X^{t-\ell}\cap Y^{t+k-\ell-1})) \\

\cup & \cup && \cup && \cup \\

\cdots & \cdots & \cdots & \cdots & \cdots & \cdots \\

\cup & \cup && \cup && \cup \\

\chi^t(X^{t+1}(X^t\cap Y^{t+j})) & \chi^t(X^t(X^{t-1}\cap Y^{t+j-1})) & \cdots & \chi^t(X^{t-j+1}(X^{t-j}\cap Y^{t})) & \cdots & \chi^t(X^{t-\ell+1}(X^{t-\ell}\cap Y^{t+j-\ell})) \\

\cup & \cup && \cup && \cup \\

\chi^t(X^{t+1}(X^t\cap Y^{t+j-1})) & \chi^t(X^t(X^{t-1}\cap Y^{t+j-2})) & \cdots & \chi^t(X^{t-j+1}(X^{t-j}\cap Y^{t-1})) & \cdots & \chi^t(X^{t-\ell+1}(X^{t-\ell}\cap Y^{t+j-\ell-1})) \\

\cup & \cup && \cup && \cup \\

\cdots & \cdots & \cdots & \cdots & \cdots & \cdots \\

\cup & \cup && \cup && \cup \\

\chi^t(X^{t+1}(X^t\cap Y^{t+1})) & \chi^t(X^t(X^{t-1}\cap Y^{t})) & \cdots & \chi^t(X^{t-j+1}(X^{t-j}\cap Y^{t-j+1})) & \cdots & \chi^t(X^{t-\ell+1}(X^{t-\ell}\cap Y^{t-\ell+1})) \\

\cup & \cup && \cup && \cup \\

\chi^t(X^{t+1}(X^t\cap Y^t)) & \chi^t(X^t(X^{t-1}\cap Y^{t-1})) & \cdots & \chi^t(X^{t-j+1}(X^{t-j}\cap Y^{t-j})) & \cdots & \chi^t(X^{t-\ell+1}(X^{t-\ell}\cap Y^{t-\ell})) \\

\cup & \cup && \cup && \cup \\

\chi^t(X^{t+1}(X^t\cap Y^{t-1})) & \chi^t(X^t(X^{t-1}\cap Y^{t-2})) & \cdots & \chi^t(X^{t-j+1}(X^{t-j}\cap Y^{t-j-1})) & \cdots & \chi^t(X^{t-\ell+1}(X^{t-\ell}\cap Y^{t-\ell-1})) \\

\cup & \cup && \cup && \cup \\

\vdots & \vdots & \vdots & \vdots & \vdots & \vdots \\

\vdots & \vdots & \vdots & \vdots & \vdots & \vdots \\

\cup & \cup && \cup && \cup \\

\chi^t(X^{t+1}) & \chi^t(X^t) & \cdots & \chi^t(X^{t-j+1}) & \cdots & \chi^t(X^{t-\ell+1}) \\

\shrtpll & \shrtpll && \shrtpll && \shrtpll \\

1^t  &          &&          &&          \\
\end{array}
\ee
\end{figure}

\newpage

\begin{figure}[tbp]

\be
\label{sminfatt1}
\begin{array}{cccccc}
         &          &&          && B^t      \\

\shrtpll & \shrtpll && \shrtpll && \shrtpll \\

\chi^t(X^{t+1}(X^t)) & \chi^t(X^t(X^{t-1})) & \cdots & \chi^t(X^{t-j+1}(X^{t-j})) & \cdots & \chi^t(X^{t-\ell+1}(X^{t-\ell})) \\

\shrtpll & \shrtpll && \shrtpll && \shrtpll \\

\vdots & \vdots & \vdots & \vdots & \vdots & \vdots \\

\vdots & \vdots & \vdots & \vdots & \vdots & \vdots \\

\shrtpll & \shrtpll && \shrtpll && \shrtpll \\

\chi^t(X^{t+1}(X^t)) & \chi^t(X^t(X^{t-1})) & \cdots & \chi^t(X^{t-j+1}(X^{t-j})) & \cdots & \chi^t(X^{t-\ell+1}(X^{t-\ell})) \\

\shrtpll & \shrtpll && \shrtpll && \shrtpll \\

\chi^t(X^{t+1}(X^t\cap Y^{t+\ell})) & \chi^t(X^t(X^{t-1}\cap Y^{t+\ell-1})) & \cdots & \chi^t(X^{t-j+1}(X^{t-j}\cap Y^{t+\ell-j})) & \cdots & \chi^t(X^{t-\ell+1}(X^{t-\ell}\cap Y^{t})) \\

\cup & \cup && \cup && \cup \\

\chi^t(X^{t+1}(X^t\cap Y^{t+\ell-1})) & \chi^t(X^t(X^{t-1}\cap Y^{t+\ell-2})) & \cdots & \chi^t(X^{t-j+1}(X^{t-j}\cap Y^{t+\ell-j-1})) & \cdots & \chi^t(X^{t-\ell+1}(X^{t-\ell}\cap Y^{t-1})) \\

\cup & \cup && \cup && \shrtpll \\

\cdots & \cdots & \cdots & \cdots & \cdots & \cdots \\

\cup & \cup && \cup && \shrtpll \\

\chi^t(X^{t+1}(X^t\cap Y^{t+k})) & \chi^t(X^t(X^{t-1}\cap Y^{t+k-1})) & \cdots & \chi^t(X^{t-j+1}(X^{t-j}\cap Y^{t+k-j})) & \cdots & \chi^t(X^{t-\ell+1}) \\

\cup & \cup && \cup && \shrtpll \\

\chi^t(X^{t+1}(X^t\cap Y^{t+k-1})) & \chi^t(X^t(X^{t-1}\cap Y^{t+k-2})) & \cdots & \chi^t(X^{t-j+1}(X^{t-j}\cap Y^{t+k-j-1})) & \cdots & \chi^t(X^{t-\ell+1}) \\

\cup & \cup && \cup && \shrtpll \\

\cdots & \cdots & \cdots & \cdots & \cdots & \cdots \\

\cup & \cup && \cup && \shrtpll \\

\chi^t(X^{t+1}(X^t\cap Y^{t+j})) & \chi^t(X^t(X^{t-1}\cap Y^{t+j-1})) & \cdots & \chi^t(X^{t-j+1}(X^{t-j}\cap Y^{t})) & \cdots & \chi^t(X^{t-\ell+1}) \\

\cup & \cup && \cup && \shrtpll \\

\chi^t(X^{t+1}(X^t\cap Y^{t+j-1})) & \chi^t(X^t(X^{t-1}\cap Y^{t+j-2})) & \cdots & \chi^t(X^{t-j+1}(X^{t-j}\cap Y^{t-1})) & \cdots & \chi^t(X^{t-\ell+1}) \\

\cup & \cup && \shrtpll && \shrtpll \\

\cdots & \cdots & \cdots & \cdots & \cdots & \cdots \\

\cup & \cup && \shrtpll && \shrtpll \\

\chi^t(X^{t+1}(X^t\cap Y^{t+1})) & \chi^t(X^t(X^{t-1}\cap Y^{t})) & \cdots & \chi^t(X^{t-j+1}) & \cdots & \chi^t(X^{t-\ell+1}) \\

\cup & \cup && \shrtpll && \shrtpll \\

\chi^t(X^{t+1}(X^t\cap Y^t)) & \chi^t(X^t(X^{t-1}\cap Y^{t-1})) & \cdots & \chi^t(X^{t-j+1}) & \cdots & \chi^t(X^{t-\ell+1}) \\

\cup & \shrtpll && \shrtpll && \shrtpll \\

\chi^t(X^{t+1}(X^t\cap Y^{t-1})) & \chi^t(X^t) & \cdots & \chi^t(X^{t-j+1}) & \cdots & \chi^t(X^{t-\ell+1}) \\

\shrtpll & \shrtpll && \shrtpll && \shrtpll \\

\vdots & \vdots & \vdots & \vdots & \vdots & \vdots \\

\vdots & \vdots & \vdots & \vdots & \vdots & \vdots \\

\shrtpll & \shrtpll && \shrtpll && \shrtpll \\

\chi^t(X^{t+1}) & \chi^t(X^t) & \cdots & \chi^t(X^{t-j+1}) & \cdots & \chi^t(X^{t-\ell+1}) \\

\shrtpll & \shrtpll && \shrtpll && \shrtpll \\

1^t  &          &&          &&          \\
\end{array}
\ee
\end{figure}

\newpage
\begin{figure}[tbp]

\be
\label{sminfatt2}
\begin{array}{cccccc}
         &          &&          && B^t      \\

\shrtpll & \shrtpll && \shrtpll && \shrtpll \\

\chi^t(X^{t+1}(X^t)) & \chi^t(X^t(X^{t-1})) & \cdots & \chi^t(X^{t-j+1}(X^{t-j})) & \cdots & \chi^t(X^{t-\ell+1}(X^{t-\ell})) \\

\shrtpll & \shrtpll && \shrtpll && \shrtpll \\

\chi^t(X^{t+1}(X^t\cap Y^{t+\ell})) & \chi^t(X^t(X^{t-1}\cap Y^{t+\ell-1})) & \cdots & \chi^t(X^{t-j+1}(X^{t-j}\cap Y^{t+\ell-j})) & \cdots & \chi^t(X^{t-\ell+1}(X^{t-\ell}\cap Y^{t})) \\

\cup & \cup && \cup && \cup \\

\chi^t(X^{t+1}(X^t\cap Y^{t+\ell-1})) & \chi^t(X^t(X^{t-1}\cap Y^{t+\ell-2})) & \cdots & \chi^t(X^{t-j+1}(X^{t-j}\cap Y^{t+\ell-j-1})) & \cdots & \chi^t(X^{t-\ell+1}(X^{t-\ell}\cap Y^{t-1})) \\

\cup & \cup && \cup && \shrtpll \\

\cdots & \cdots & \cdots & \cdots & \cdots & \chi^t(X^{t-\ell+1}) \\

\cup & \cup && \cup && \shrtpll \\

\chi^t(X^{t+1}(X^t\cap Y^{t+k})) & \chi^t(X^t(X^{t-1}\cap Y^{t+k-1})) & \cdots & \chi^t(X^{t-j+1}(X^{t-j}\cap Y^{t+k-j})) & \cdots &  \\

\cup & \cup && \cup &&  \\

\chi^t(X^{t+1}(X^t\cap Y^{t+k-1})) & \chi^t(X^t(X^{t-1}\cap Y^{t+k-2})) & \cdots & \chi^t(X^{t-j+1}(X^{t-j}\cap Y^{t+k-j-1})) & \cdots &  \\

\cup & \cup && \cup &&  \\

\cdots & \cdots & \cdots & \cdots &  &  \\

\cup & \cup && \cup &&  \\

\chi^t(X^{t+1}(X^t\cap Y^{t+j})) & \chi^t(X^t(X^{t-1}\cap Y^{t+j-1})) & \cdots & \chi^t(X^{t-j+1}(X^{t-j}\cap Y^{t})) &  &  \\

\cup & \cup && \cup &&  \\

\chi^t(X^{t+1}(X^t\cap Y^{t+j-1})) & \chi^t(X^t(X^{t-1}\cap Y^{t+j-2})) & \cdots & \chi^t(X^{t-j+1}(X^{t-j}\cap Y^{t-1})) &  &  \\

\cup & \cup && \shrtpll &&  \\

\cdots & \cdots & \cdots & \chi^t(X^{t-j+1}) &  &  \\

\cup & \cup && \shrtpll &&  \\

\chi^t(X^{t+1}(X^t\cap Y^{t+1})) & \chi^t(X^t(X^{t-1}\cap Y^{t})) & \cdots &  &   &   \\

\cup & \cup &&   &&   \\

\chi^t(X^{t+1}(X^t\cap Y^t)) & \chi^t(X^t(X^{t-1}\cap Y^{t-1})) & \cdots &   &   &   \\

\cup & \shrtpll &&   &&   \\

\chi^t(X^{t+1}(X^t\cap Y^{t-1})) & \chi^t(X^t) &   &   &   &   \\

\shrtpll & \shrtpll  &&   &&   \\

\chi^t(X^{t+1}) &   &   &   &   &   \\

\shrtpll &   &&   &&   \\

1^t  &   &&   &&   \\
\end{array}
\ee
\end{figure}

\clearpage
\newpage
\vspace{3mm}
{\bf 5.  THE DECOMPOSITION GROUP $(\calr,*)$}
\vspace{3mm}

\vspace{3mm}
{\bf 5.1  The decomposition group}
\vspace{3mm}

Let $\calb$ be a basis of $B$.  Let $\calr$ be the set of tensors
determined by $\calb$.  From Theorem \ref{thm9}, there is a 
bijection $\eta:  \calr\ra B$ with assignment $\eta:  \bmr\mapsto\bmb$
if $\bmb$ is an encoding of $\bmr$ using product (\ref{enctdx}) on the 
generators in $\bmr$.
We define an operation $*$ on $\calr$ by using the operation in $B$.
Let $\bmrdt,\bmrddt\in\calr$.  Let $\eta:  \bmrdt\mapsto\bmbdt$ and 
$\eta:  \bmrddt\mapsto\bmbddt$.  Define an operation $*$ on $\calr$ by
\be
\label{eqno13}
\bmrdt*\bmrddt\rmdef\bmrbr
\ee
if $\eta:  \bmrbr\mapsto\bmbdt\bmbddt$.

\begin{lem}
\label{lem37}
The operation $*$ is well defined.
\end{lem}

\begin{prf}
Let $\bmrgr,\bmrac\in\calr$ such that $\bmrgr=\bmrdt$ and $\bmrac=\bmrddt$.
We have to show that $\bmrdt*\bmrddt=\bmrgr*\bmrac$.  But if $\bmrgr=\bmrdt$,
then $\eta:  \bmrgr\mapsto\bmbdt$, and similarly $\eta:  \bmrac\mapsto\bmbddt$.
Then both $\bmrdt*\bmrddt$ and $\bmrgr*\bmrac$ are determined by $\bmbdt\bmbddt$.
\end{prf}

\begin{thm}
\label{thm31}
The set $\calr$ with operation $*$ forms a group $(\calr,*)$,
and $(\calr,*)\simeq B$ under the bijection $\eta:  \calr\ra B$.
\end{thm}

\begin{prf}
We first show the operation $*$ is associative.  Let 
$\bmr,\bmrdt,\bmrddt\in\calr$.  We need to show
\be
\label{show0}
(\bmr*\bmrdt)*\bmrddt=\bmr*(\bmrdt*\bmrddt).
\ee
Let $\eta:  \bmr\mapsto\bmb$, $\eta:  \bmrdt\mapsto\bmbdt$, and
$\eta:  \bmrddt\mapsto\bmbddt$.  Then (\ref{show0}) is the same as showing
$$
(\bmb\bmbdt)\bmbddt=\bmb(\bmbdt\bmbddt).
$$
But this follows since operation in $B$ is associative.

Let $\bone$ be the identity of $B$.  Let $\eta:  \bone_\calr\mapsto\bone$.
We show $\bone_\calr$ is the identity of $(\calr,*)$.  Let $\bmr\in\calr$.
We need to show $\bone_\calr*\bmr=\bmr$ and $\bmr*\bone_\calr=\bmr$.
But this is the same as showing $\bone\bmb=\bmb$ and $\bmb\bone=\bmb$, 
where $\eta:  \bmr\mapsto\bmb$.  But this follows since $\bone$ is the identity 
of $B$.

Let $\bmr\in\calr$.  We show $\bmr$ has an inverse in $(\calr,*)$.
Let $\eta:  \bmr\mapsto\bmb$.  The group element $\bmb$ has an inverse
$\bmbbr$ in $B$ such that $\bmbbr\bmb=\bone$
and $\bmb\bmbbr=\bone$.  Let $\eta:  \bmrbr\mapsto\bmbbr$.
It follows that $\bmrbr*\bmr=\bone_\calr$ and $\bmr*\bmrbr=\bone_\calr$.

Together these results show that $(\calr,*)$ is a group.
By the definition of operation $*$ given in (\ref{eqno13}),
$(\calr,*)$ is just an isomorphic copy of
$B$ under bijection $\eta$.
\end{prf}
If $\eta:  \bmr\mapsto\bmb$, then $\bmr$ is the decomposition of $\bmb$ into its generators.
For this reason, we call $(\calr,*)$ the {\it decomposition group} of $A$.

We now define a component group on $(\calr,*)$.  Fix time $t$.
Let $\tridnjktarg{0}{0}{t}{\bmrdt},\tridnjktarg{0}{0}{t}{\bmrddt}\in\tridnjktarg{0}{0}{t}{\calr}$.  
Define an operation $\prodjktarg{0}{0}{t}{\cast}$ on set $\tridnjktarg{0}{0}{t}{\calr}$ by
\be
\label{defx}
\tridnjktarg{0}{0}{t}{\bmrdt}\prodjktarg{0}{0}{t}{\cast}\tridnjktarg{0}{0}{t}{\bmrddt}\rmdef\tridnjktarg{0}{0}{t}{\bmrdt*\bmrddt}.
\ee

\begin{lem}
\label{lem32}
Fix time $t$.  The operation $\prodjktarg{0}{0}{t}{\cast}$ on set $\tridnjktarg{0}{0}{t}{\calr}$ 
is well defined.
\end{lem}

\begin{prf}
Let $\tridnjktarg{0}{0}{t}{\bmrgr},\tridnjktarg{0}{0}{t}{\bmrac}\in\tridnjktarg{0}{0}{t}{\calr}$
such that $\tridnjktarg{0}{0}{t}{\bmrgr}=\tridnjktarg{0}{0}{t}{\bmrdt}$ and 
$\tridnjktarg{0}{0}{t}{\bmrac}=\tridnjktarg{0}{0}{t}{\bmrddt}$.
To show the operation is well defined, we need to show
$$
\tridnjktarg{0}{0}{t}{\bmrgr}\prodjktarg{0}{0}{t}{\cast}\tridnjktarg{0}{0}{t}{\bmrac}=
\tridnjktarg{0}{0}{t}{\bmrdt}\prodjktarg{0}{0}{t}{\cast}\tridnjktarg{0}{0}{t}{\bmrddt},
$$
or $\tridnjktarg{0}{0}{t}{\bmrgr*\bmrac}=\tridnjktarg{0}{0}{t}{\bmrdt*\bmrddt}$.
But this is clear from Lemma \ref{lem47}.
\end{prf}

\begin{thm}
\label{thm41}
Fix time $t$.  The set $\tridnjktarg{0}{0}{t}{\calr}$ with operation $\prodjktarg{0}{0}{t}{\cast}$ 
forms a group $\grpjktarg{0}{0}{t}{\calr}{\cast}$.
\end{thm}

\begin{prf}
First we show the operation $\prodjktarg{0}{0}{t}{\cast}$ is associative.
Let $\tridnjktarg{0}{0}{t}{\bmr},\tridnjktarg{0}{0}{t}{\bmrdt},\tridnjktarg{0}{0}{t}{\bmrddt}\in\tridnjktarg{0}{0}{t}{\calr}$.
We need to show
$$
(\tridnjktarg{0}{0}{t}{\bmr}\prodjktarg{0}{0}{t}{\cast}\tridnjktarg{0}{0}{t}{\bmrdt})\prodjktarg{0}{0}{t}{\cast}\tridnjktarg{0}{0}{t}{\bmrddt}
=\tridnjktarg{0}{0}{t}{\bmr}\prodjktarg{0}{0}{t}{\cast}(\tridnjktarg{0}{0}{t}{\bmrdt}\prodjktarg{0}{0}{t}{\cast}\tridnjktarg{0}{0}{t}{\bmrddt}).
$$
But using (\ref{defx}) we have
\begin{align*}
(\tridnjktarg{0}{0}{t}{\bmr}\prodjktarg{0}{0}{t}{\cast}\tridnjktarg{0}{0}{t}{\bmrdt})\prodjktarg{0}{0}{t}{\cast}\tridnjktarg{0}{0}{t}{\bmrddt}
&=\tridnjktarg{0}{0}{t}{\bmr*\bmrdt}\prodjktarg{0}{0}{t}{\cast}\tridnjktarg{0}{0}{t}{\bmrddt} \\
&=\tridnjktarg{0}{0}{t}{(\bmr*\bmrdt)*\,\bmrddt},
\end{align*}
and
\begin{align*}
\tridnjktarg{0}{0}{t}{\bmr}\prodjktarg{0}{0}{t}{\cast}(\tridnjktarg{0}{0}{t}{\bmrdt}\prodjktarg{0}{0}{t}{\cast}\tridnjktarg{0}{0}{t}{\bmrddt})
&=\tridnjktarg{0}{0}{t}{\bmr}\prodjktarg{0}{0}{t}{\cast}\tridnjktarg{0}{0}{t}{\bmrdt*\bmrddt} \\
&=\tridnjktarg{0}{0}{t}{\bmr*\,(\bmrdt*\bmrddt)}.
\end{align*}
Therefore the operation $\prodjktarg{0}{0}{t}{\cast}$ is associative since the operation $*$ in 
group $(\calr,*)$ is associative.

Let $\bone$ be the identity of $(\calr,*)$.  We show
$\tridnjktarg{0}{0}{t}{\bone}$ is the identity of $\grpjktarg{0}{0}{t}{\calr}{\cast}$.
Let $\bmr\in\calr$ and $\tridnjktarg{0}{0}{t}{\bmr}\in\tridnjktarg{0}{0}{t}{\calr}$.  
But using (\ref{defx}) we have
\begin{align*}
\tridnjktarg{0}{0}{t}{\bone}\prodjktarg{0}{0}{t}{\cast}\tridnjktarg{0}{0}{t}{\bmr} &=\tridnjktarg{0}{0}{t}{\bone*\bmr} \\
&=\tridnjktarg{0}{0}{t}{\bmr}
\end{align*}
and
\begin{align*}
\tridnjktarg{0}{0}{t}{\bmr}\prodjktarg{0}{0}{t}{\cast}\tridnjktarg{0}{0}{t}{\bone} &=\tridnjktarg{0}{0}{t}{\bmr*\bone} \\
&=\tridnjktarg{0}{0}{t}{\bmr}.
\end{align*}

Fix $\bmr\in\calr$.  Let $\bmrbr$ be the inverse of $\bmr$ in $(\calr,*)$.
We show $\tridnjktarg{0}{0}{t}{\bmrbr}$ is the inverse of $\tridnjktarg{0}{0}{t}{\bmr}$
in $\grpjktarg{0}{0}{t}{\calr}{\cast}$.  But using (\ref{defx}) we have
\begin{align*}
\tridnjktarg{0}{0}{t}{\bmrbr}\prodjktarg{0}{0}{t}{\cast}\tridnjktarg{0}{0}{t}{\bmr} &=\tridnjktarg{0}{0}{t}{\bmrbr*\bmr} \\
&=\tridnjktarg{0}{0}{t}{\bone}
\end{align*}
and
\begin{align*}
\tridnjktarg{0}{0}{t}{\bmr}\prodjktarg{0}{0}{t}{\cast}\tridnjktarg{0}{0}{t}{\bmrbr} &=\tridnjktarg{0}{0}{t}{\bmr*\bmrbr} \\
&=\tridnjktarg{0}{0}{t}{\bone}.
\end{align*}
Together these results show $\grpjktarg{0}{0}{t}{\calr}{\cast}$ is a group.
\end{prf}

We call groups $\grpjktarg{0}{0}{t}{\calr}{\cast}$, for each $t\in\bmcpz$,
the {\it primary elementary groups} of $(\calr,*)$.

For each $t\in\bmcpz$,
define a map $\phi^t:  \calr\ra\tridnjktarg{0}{0}{t}{\calr}$ by the assignment 
$\phi^t:  \bmr\mapsto\tridnjktarg{0}{0}{t}{\bmr}$.
Note that $\phi^t(\calr)=\tridnjktarg{0}{0}{t}{\calr}$.
Note that $\phi^t$ is {\bf not} a projection since the representatives in $\tridnjktarg{0}{0}{t}{\bmr}$ 
are scattered all over $\bmr$.  Therefore the map $\phi^t$ includes memory of the
sequences $\bmr\in\calr$.  In this sense, it is different from the projection
$\chi^t$ of $B$.

\begin{thm}
\label{homor}
There is a homomorphism from $(\calr,*)$ to $\grpjktarg{0}{0}{t}{\calr}{\cast}$ given 
by the map $\phi^t:  \calr\ra\tridnjktarg{0}{0}{t}{\calr}$, for each $t\in\bmcpz$.
\end{thm}

\begin{prf}
This follows immediately from the definition of operation $\prodjktarg{0}{0}{t}{\cast}$ in (\ref{defx}).
\end{prf}

\begin{cor}
\label{cor36}
There is an isomorphism $\grpjktarg{0}{0}{t}{\calr}{\cast}\simeq B^t$ under the bijection 
$\eta^t:  \tridnjktarg{0}{0}{t}{\calr}\ra B^t$, for each $t\in\bmcpz$.
\end{cor}

\begin{prf}
Let $\bmrdt,\bmrddt\in\calr$.  
Let $\eta:  \bmrdt\mapsto\bmbdt$ and $\eta:  \bmrddt\mapsto\bmbddt$.  Then 
$\eta:  \bmrdt*\bmrddt\mapsto\bmbdt\bmbddt$.  Then from Corollary \ref{cor26},
$\eta^t:  \tridnjktarg{0}{0}{t}{\bmrdt}\mapsto\bdt^t$, 
$\eta^t:  \tridnjktarg{0}{0}{t}{\bmrddt}\mapsto\bddt^t$, and
$\eta^t:  \tridnjktarg{0}{0}{t}{\bmrdt*\bmrddt}\mapsto(\bmbdt\bmbddt)^t$.  In other words,
$\eta^t(\tridnjktarg{0}{0}{t}{\bmrdt})=\bdt^t$, $\eta^t(\tridnjktarg{0}{0}{t}{\bmrddt})=\bddt^t$, and
$\eta^t(\tridnjktarg{0}{0}{t}{\bmrdt*\bmrddt})=\bdt^t\bddt^t$.  But then
$\eta^t(\tridnjktarg{0}{0}{t}{\bmrdt}\prodjktarg{0}{0}{t}{\cast}\tridnjktarg{0}{0}{t}{\bmrddt})=\bdt^t\bddt^t$.
This gives
$$
\eta^t(\tridnjktarg{0}{0}{t}{\bmrdt}\prodjktarg{0}{0}{t}{\cast}\tridnjktarg{0}{0}{t}{\bmrddt})=
\eta^t(\tridnjktarg{0}{0}{t}{\bmrdt})\eta^t(\tridnjktarg{0}{0}{t}{\bmrddt}).
$$
\end{prf}

We can summarize the results of this paper so far by the chain
\be
\label{chain5a}
A\,\,{\stackrel{\simeq}{\lra}}\,\,B\,\,{\stackrel{\simeq}{\ra}}\,\,(\calr,*),
\ee
where $B\simeq (\calr,*)$ under bijection $\eta:  \calr\ra B$.

As an aside, note that under bijection $\eta^t$, 
we have $\eta^t:  \tridnjktarg{0}{0}{t}{\bmr}\mapsto b^t$ if
$\tridnjktarg{0}{0}{t}{\bmr}$ is the \crepc\ of $b^t$.  Assume that
$$
b^t=r_{0,0}^tr_{0,1}^tr_{0,2}^t\cdots r_{j,k}^{t-j}\cdots r_{\ell,\ell}^{t-\ell}.
$$
The \crepc\ of an element $r_{j,k}^{t-j}$ in the \crepc\ of $b^t$ is given by the representatives in
$\tridnjktarg{0}{0}{t}{\bmg^{[t-j,t-j+k]}}$ in $\grpjktarg{0}{0}{t}{\calr}{\cast}$,
where all entries are trivial except for entry $r_{j,k}^{t-j}$.  
We can think of $\tridnjktarg{0}{0}{t}{\bmg^{[t-j,t-j+k]}}$ 
as a special element or ``eigenelement" of $\grpjktarg{0}{0}{t}{\calr}{\cast}$.  
Then we have $\eta^t(\tridnjktarg{0}{0}{t}{\bmg^{[t-j,t-j+k]}})=r_{j,k}^{t-j}$, and
\begin{align*}
\eta^t(\tridnjktarg{0}{0}{t}{\bmr})&=
r_{0,0}^tr_{0,1}^tr_{0,2}^t\cdots r_{j,k}^{t-j}\cdots r_{\ell,\ell}^{t-\ell} \\
&=\eta^t(\tridnjktarg{0}{0}{t}{\bmg^{[t,t]}})
\eta^t(\tridnjktarg{0}{0}{t}{\bmg^{[t,t+1]}})
\eta^t(\tridnjktarg{0}{0}{t}{\bmg^{[t,t+2]}})\cdots
\eta^t(\tridnjktarg{0}{0}{t}{\bmg^{[t-j,t-j+k]}})\cdots
\eta^t(\tridnjktarg{0}{0}{t}{\bmg^{[t-\ell,t-\ell+\ell]}}) \\
&=\eta^t(\tridnjktarg{0}{0}{t}{\bmg^{[t,t]}})\prodjktarg{0}{0}{t}{\cast}
\tridnjktarg{0}{0}{t}{\bmg^{[t,t+1]}})\prodjktarg{0}{0}{t}{\cast}
\tridnjktarg{0}{0}{t}{\bmg^{[t,t+2]}})\prodjktarg{0}{0}{t}{\cast}\cdots
\cdots\prodjktarg{0}{0}{t}{\cast}\tridnjktarg{0}{0}{t}{\bmg^{[t-j,t-j+k]}})\prodjktarg{0}{0}{t}{\cast}\cdots
\cdots\prodjktarg{0}{0}{t}{\cast}\tridnjktarg{0}{0}{t}{\bmg^{[t-\ell,t-\ell+\ell]}})) \\
&=\eta^t(\tridnjktarg{0}{0}{t}{\bmg^{[t,t]}*\bmg^{[t,t+1]}*\bmg^{[t,t+2]}*\cdots*\bmg^{[t-j,t-j+k]}*\cdots*\bmg^{[t-\ell,t-\ell+\ell]}}).
\end{align*}
This shows that any element $\tridnjktarg{0}{0}{t}{\bmr}$ in $\grpjktarg{0}{0}{t}{\calr}{\cast}$
is the product of eigenelements
$$
\tridnjktarg{0}{0}{t}{\bmg^{[t,t]}})\prodjktarg{0}{0}{t}{\cast}
\tridnjktarg{0}{0}{t}{\bmg^{[t,t+1]}})\prodjktarg{0}{0}{t}{\cast}
\tridnjktarg{0}{0}{t}{\bmg^{[t,t+2]}})\prodjktarg{0}{0}{t}{\cast}\cdots
\cdots\prodjktarg{0}{0}{t}{\cast}\tridnjktarg{0}{0}{t}{\bmg^{[t-j,t-j+k]}})\prodjktarg{0}{0}{t}{\cast}\cdots
\cdots\prodjktarg{0}{0}{t}{\cast}\tridnjktarg{0}{0}{t}{\bmg^{[t-\ell,t-\ell+\ell]}})
$$
of $\grpjktarg{0}{0}{t}{\calr}{\cast}$, and any element $\tridnjktarg{0}{0}{t}{\bmr}$ in $\grpjktarg{0}{0}{t}{\calr}{\cast}$
can be written in the form
$$
\tridnjktarg{0}{0}{t}{\bmr}=\tridnjktarg{0}{0}{t}{\bmg^{[t,t]}*\bmg^{[t,t+1]}*\bmg^{[t,t+2]}*\cdots*\bmg^{[t-j,t-j+k]}*\cdots*\bmg^{[t-\ell,t-\ell+\ell]}}.
$$
Then the eigenelements in $\grpjktarg{0}{0}{t}{\calr}{\cast}$ form a \crepc\ 
of any element $\tridnjktarg{0}{0}{t}{\bmr}$ in $\grpjktarg{0}{0}{t}{\calr}{\cast}$.

\vspace{3mm}
{\bf 5.2 The elementary groups}
\vspace{3mm}

We use the following notation.  Fix $t\in\bmcpz$.
Note that \stm\ $\tridnjktarg{0}{0}{t}{\bmr}$ has the triangular form (\ref{rttf}).
We show the subscript indices of representatives of $\tridnjktarg{0}{0}{t}{\bmr}$ in (\ref{rttfind}).
Notice that the columns in (\ref{rttfind}) are numbered 
from $0$ to $\ell$ (left to right), and the rows 
from $0$ to $\ell$ (bottom to top).  Thus column $j$ and row $k$ of (\ref{rttfind}) is index $(j,k)$.
We let index triangle $\tridnjk{0}{0}$ denote
all the indices in (\ref{rttfind}); these are the indices with
lower vertex index $(0,0)$ and upper vertex indices $(0,\ell)$ and $(\ell,\ell)$.
For $0\le j\le\ell$, for $j\le k\le\ell$, we
let {\it index triangle} $\tridnjk{j}{k}$ denote the indices in (\ref{rttfind})
specified by the triangle with lower vertex index $(j,k)$ and upper vertex
indices $(j,\ell)$ and $(j+\ell-k,\ell)$.  In other words, these
are the indices $(m,n)$ where $m,n$ satisfy $k\le n\le\ell$ and $j\le m\le (j+n-k)$.
We now use the index triangle to describe certain triangular subsets
of representatives in $\tridnjktarg{0}{0}{t}{\bmr}$.  
For $\bmr\in\calr$, we let $\tridnjktarg{j}{k}{t}{\bmr}$ be the
representatives in $\tridnjktarg{0}{0}{t}{\bmr}$ in (\ref{rttf}) whose subscript 
indices are specified by the index triangle 
$\tridnjk{j}{k}$.  These are the representatives in (\ref{rttf}) with lower vertex $r_{j,k}^{t-j}$ 
and upper vertices $r_{j,\ell}^{t-j}$ and $r_{j+\ell-k,\ell}^{t-(j+\ell-k)}$.  
Note that for $j=0$ and $k=0$, $\tridnjktarg{j}{k}{t}{\bmr}$ is the same as
previously defined $\tridnjktarg{0}{0}{t}{\bmr}$.
\be
\label{rttfind}
\begin{array}{llllllllll}
  (0,\ell)   & (1,\ell)   & \cdots & \cdots & (j,\ell)   & \cdots & \cdots & \cdots & (\ell-1,\ell)   & (\ell,\ell) \\
  (0,\ell-1) & (1,\ell-1) & \cdots & \cdots & (j,\ell-1) & \cdots & \cdots & \cdots & (\ell-1,\ell-1) & \\
  \vdots & \vdots & \vdots & \vdots & \vdots & \vdots & \vdots & \vdots && \\
  (0,k) & (1,k) & \cdots & \cdots & (j,k) & \cdots & (k,k) &&& \\
  \vdots & \vdots & \vdots & \vdots & \vdots & \vdots &&&& \\
  \cdots & \cdots & \cdots & \cdots & (j,j) &&&&& \\
  \vdots & \vdots & \vdots &&&&&&& \\
  (0,2)   & (1,2) & (2,2) &&&&&&& \\
  (0,1)   & (1,1) &&&&&&&& \\
  (0,0)   &&&&&&&&&
\end{array}
\ee

For $t\in\bmcpz$, for $0\le j\le\ell$, for $j\le k\le\ell$,
let $\tridnjktarg{j}{k}{t}{\calr}$ be the set of all possible triangles
$\tridnjktarg{j}{k}{t}{\bmr}$,
$\tridnjktarg{j}{k}{t}{\calr}\rmdef\{\tridnjktarg{j}{k}{t}{\bmr} : \bmr\in\calr\}$.
Note that for $j=0$ and $k=0$, $\tridnjktarg{j}{k}{t}{\calr}$ is just the 
set $\tridnjktarg{0}{0}{t}{\calr}$ previously defined.

We now examine the structure of group $(\calr,*)$.  We define some ``local"
groups on $(\calr,*)$ which are fundamental to understanding group systems.
We show these groups arise naturally from the generators of $B$.

First we give a general result on products of the form (\ref{exp4}).
Consider any finite group $G$ with normal chain
$$
\bone=G_{-1}\subset G_0\subset G_1\subset\cdots\subset G_m\subset\cdots\subset G_\ell=G,
$$
where each $G_m$ is normal in $G$.  Let $\{g_m\}_{m=0}^\ell$ be any
\compset\ of the normal chain.  Then any $g\in G$ can be written
using a subset of this complete set as a \crepc\
\be
\label{crepg}
g=g_0 g_1\cdots g_m\cdots g_\ell.
\ee

\begin{lem}
\label{lem45}
Let $f,g,h\in G$ and assume $fg=h$.  Let $g$ be as in (\ref{crepg}),
$f=g_0' g_1'\cdots g_m'\cdots g_\ell'$, 
and $h=g_0'' g_1''\cdots g_m''\cdots g_\ell''$.  Consider the product
$fg=h$, or
\be
\label{prodcrepc}
(g_0' g_1'\cdots g_m'\cdots g_\ell')(g_0 g_1\cdots g_m\cdots g_\ell)
=(g_0'' g_1''\cdots g_m''\cdots g_\ell'').
\ee
In the product $fg=h$, the upper coordinates $g_m'',\ldots,g_\ell''$ 
of the \crepc\ of $h$ are solely determined by the upper coordinates 
$g_m',\ldots,g_\ell'$ of the \crepc\ of $f$ and upper coordinates $g_m,\ldots,g_\ell$
of the \crepc\ of $g$, for $m$ such that $0\le m\le\ell$.
\end{lem}

\begin{prf}
Let $m$ be such that $0\le m\le\ell$.  Since $G_{m-1}\lhd G$, we know that
\begin{align*}
(G_{m-1}f)(G_{m-1}g) &=G_{m-1}(fg) \\
 &=G_{m-1}h.
\end{align*}
But $G_{m-1}f=G_{m-1}(g_m'\cdots g_\ell')$,
$G_{m-1}g=G_{m-1}(g_m\cdots g_\ell)$, and 
$G_{m-1}h=G_{m-1}(g_m''\cdots g_\ell'')$.  Therefore
$$
(G_{m-1}(g_m'\cdots g_\ell'))(G_{m-1}(g_m\cdots g_\ell))=G_{m-1}(g_m''\cdots g_\ell''),
$$
or 
$$
G_{m-1}(g_m'\cdots g_\ell')(g_m\cdots g_\ell)=G_{m-1}(g_m''\cdots g_\ell'').
$$
But this last result means $(g_m'\cdots g_\ell')(g_m\cdots g_\ell)$ and
$(g_m''\cdots g_\ell'')$ only differ by an element of $G_{m-1}$.  Then 
the upper coordinates in the \crepc\ of $(g_m'\cdots g_\ell')(g_m\cdots g_\ell)$
must be the same as the upper coordinates $(g_m''\cdots g_\ell'')$.  Therefore
the upper coordinates $(g_m'',\ldots,g_\ell'')$ of $h$ are solely determined 
by the upper coordinates $(g_m',\ldots,g_\ell')$ of $f$ and upper coordinates 
$(g_m,\ldots,g_\ell)$ of $g$.
\end{prf}

\begin{lem}
\label{lem46}
Fix time $t$.  Fix $k$ such that $0\le k\le\ell$.  Let $\bmrdt,\bmrddt\in\calr$.
In the product $\bmrdt*\bmrddt$, the representatives in
$\tridnjktarg{k}{k}{t}{\bmrdt*\bmrddt}$ are uniquely determined by the representatives in
$\tridnjktarg{k}{k}{t}{\bmrdt}$ and $\tridnjktarg{k}{k}{t}{\bmrddt}$.
\end{lem}

\begin{prf}
Equation (\ref{exp4}) in the proof of Lemma \ref{lem47} shows that the \crepc\ of 
$\bdt^t\bddt^t$ is uniquely determined by the products of the \crepc s of $\bdt^t$ and $\bddt^t$.
But this is a product of the form (\ref{prodcrepc}) in Lemma \ref{lem45}.  Then 
applying Lemma \ref{lem45} to this product, 
the upper coordinates $\tridnjktarg{k}{k}{t}{\bmrdt*\bmrddt}$ 
of the \crepc\ of $\bdt^t\bddt^t$ are uniquely determined by the upper coordinates 
$\tridnjktarg{k}{k}{t}{\bmrdt}$ of the \crepc\ of $\bdt^t$ and upper coordinates 
$\tridnjktarg{k}{k}{t}{\bmrddt}$ of the \crepc\ of $\bddt^t$.
\end{prf}

Fix time $t$.  Fix $k$ such that $0\le k\le\ell$.
Let $\tridnjktarg{k}{k}{t}{\bmrdt},\tridnjktarg{k}{k}{t}{\bmrddt}\in\tridnjktarg{k}{k}{t}{\calr}$.
Define an operation 
$\prodjktarg{k}{k}{t}{\cast}$ on $\tridnjktarg{k}{k}{t}{\calr}$ by
\be
\label{opdef0}
\tridnjktarg{k}{k}{t}{\bmrdt}\prodjktarg{k}{k}{t}{\cast}\tridnjktarg{k}{k}{t}{\bmrddt}\rmdef\tridnjktarg{k}{k}{t}{\bmrdt*\bmrddt}.
\ee

\begin{lem}
\label{lem39}
Fix time $t$.  Fix $k$ such that $0\le k\le\ell$.
The operation $\prodjktarg{k}{k}{t}{\cast}$ on $\tridnjktarg{k}{k}{t}{\calr}$ is well defined.
\end{lem}

\begin{prf}
Let $\tridnjktarg{k}{k}{t}{\bmrgr},\tridnjktarg{k}{k}{t}{\bmrac}\in\tridnjktarg{k}{k}{t}{\calr}$
such that $\tridnjktarg{k}{k}{t}{\bmrgr}=\tridnjktarg{k}{k}{t}{\bmrdt}$ and 
$\tridnjktarg{k}{k}{t}{\bmrac}=\tridnjktarg{k}{k}{t}{\bmrddt}$.
To show the operation is well defined, we need to show
$\tridnjktarg{k}{k}{t}{\bmrgr*\bmrac}=\tridnjktarg{k}{k}{t}{\bmrdt*\bmrddt}$.
But this is clear from Lemma \ref{lem46}.
\end{prf}

\begin{thm}
\label{thm51}
Fix time $t$.  Fix $k$ such that $0\le k\le\ell$.
The set $\tridnjktarg{k}{k}{t}{\calr}$ with operation $\prodjktarg{k}{k}{t}{\cast}$ forms a 
group $\grpjktarg{k}{k}{t}{\calr}{\cast}$.
\end{thm}

\begin{prf}
The proof is exactly the same as for Theorem \ref{thm41}.
\end{prf}

We can think of group $\grpjktarg{k}{k}{t}{\calr}{\cast}$ in two different ways.
In view of (\ref{opdef0}), we can think of
group $\grpjktarg{k}{k}{t}{\calr}{\cast}$ as a multiplication
of the upper coordinates $\tridnjktarg{k}{k}{t}{\bmrddt}$
of $\bmrddt^t$ with the upper coordinates $\tridnjktarg{k}{k}{t}{\bmrdt}$
of $\bmrdt^t$ to find the upper coordinates $\tridnjktarg{k}{k}{t}{\bmrdt*\bmrddt}$
of $(\bmrdt*\bmrddt)^t$.  In the second interpretation, the 
group $\grpjktarg{k}{k}{t}{\calr}{\cast}$ is devoid of any meaning
as a multiplication of representatives.  Instead,
we simply think of group $\grpjktarg{k}{k}{t}{\calr}{\cast}$ as a group of arbitrary
elements $\tridnjktarg{k}{k}{t}{\bmr}$ in a set $\tridnjktarg{k}{k}{t}{\calr}$.  The second 
interpretation is the one we use henceforth.

For $0\le j\le k$, index triangles $\tridnjk{j}{k}$ and $\tridnjk{k}{k}$ have
congruent shapes, and we can identify {\it representatives in congruent positions} 
in triangles $\tridnjktarg{j}{k}{t+j}{\bmr}$ and $\tridnjktarg{k}{k}{t+k}{\bmr}$.

\begin{lem}
\label{lem45a}
Fix time $t$.  Fix $k$ such that $0\le k\le\ell$.
Fix $j$ such that $0\le j\le k$.  Fix $\bmr\in\calr$.
Representatives in congruent positions of triangles $\tridnjktarg{j}{k}{t+j}{\bmr}$ 
and $\tridnjktarg{k}{k}{t+k}{\bmr}$ are from the same generator.
\end{lem}

\begin{prf}
Without loss in generality, examine index position $(j,k)$ in triangle
$\tridnjktarg{j}{k}{t+j}{\bmr}$ and position $(k,k)$ in triangle
$\tridnjktarg{k}{k}{t+k}{\bmr}$.  These index positions are congruent.
The representative in index position $(j,k)$ in triangle
$\tridnjktarg{j}{k}{t+j}{\bmr}$ is $r_{j,k}^{(t+j)-j}=r_{j,k}^t$, and
the representative in index position $(k,k)$ in triangle
$\tridnjktarg{k}{k}{t+k}{\bmr}$ is $r_{k,k}^{(t+k)-k}=r_{k,k}^t$.
But representatives $r_{j,k}^t$ and $r_{k,k}^t$ are from the same generator
$\bmg^{[t,t+k]}$.
\end{prf}

\begin{lem}
\label{lem46a}
Fix time $t$.  Fix $k$ such that $0\le k\le\ell$.  
Fix $j$ such that $0\le j\le k$.  Let $\bmrdt,\bmrddt\in\calr$.
In the product $\bmrdt*\bmrddt$, the representatives in
$\tridnjktarg{j}{k}{t+j}{\bmrdt*\bmrddt}$ are uniquely determined by the representatives in
$\tridnjktarg{j}{k}{t+j}{\bmrdt}$ and $\tridnjktarg{j}{k}{t+j}{\bmrddt}$.
\end{lem}

\begin{prf}
Fix time $t$.  Fix $k$ such that $0\le k\le\ell$.  Fix $j$ such that $0\le j\le k$.
Let $\bmrdt,\bmrddt\in\calr$.  From Lemma \ref{lem46}, the representatives in
$\tridnjktarg{k}{k}{t+k}{\bmrdt*\bmrddt}$ are uniquely determined by the representatives
in $\tridnjktarg{k}{k}{t+k}{\bmrdt}$ and $\tridnjktarg{k}{k}{t+k}{\bmrddt}$.
From Lemma \ref{lem45a}, the 
representatives in congruent positions of triangles $\tridnjktarg{j}{k}{t+j}{\bmrdt}$ 
and $\tridnjktarg{k}{k}{t+k}{\bmrdt}$ are from the same generator.  The same holds for
$\tridnjktarg{j}{k}{t+j}{\bmrddt}$ and $\tridnjktarg{k}{k}{t+k}{\bmrddt}$, and for
$\tridnjktarg{j}{k}{t+j}{\bmrdt*\bmrddt}$ and $\tridnjktarg{k}{k}{t+k}{\bmrdt*\bmrddt}$.
Then the representatives in $\tridnjktarg{j}{k}{t+j}{\bmrdt}$ are uniquely
determined by the representatives in $\tridnjktarg{k}{k}{t+k}{\bmrdt}$,
and the same holds for 
$\tridnjktarg{j}{k}{t+j}{\bmrddt}$ and $\tridnjktarg{k}{k}{t+k}{\bmrddt}$, and for
$\tridnjktarg{j}{k}{t+j}{\bmrdt*\bmrddt}$ and $\tridnjktarg{k}{k}{t+k}{\bmrdt*\bmrddt}$.
It follows that the representatives in
$\tridnjktarg{j}{k}{t+j}{\bmrdt*\bmrddt}$ are uniquely determined by the representatives
in $\tridnjktarg{j}{k}{t+j}{\bmrdt}$ and $\tridnjktarg{j}{k}{t+j}{\bmrddt}$.
\end{prf}

Fix time $t$.  Fix $k$ such that $0\le k\le\ell$.  
We now find groups $\grpjktarg{j}{k}{t+j}{\calr}{\cast}$ for $0\le j\le k$.
Let $\tridnjktarg{j}{k}{t+j}{\bmrdt},\tridnjktarg{j}{k}{t+j}{\bmrddt}\in\tridnjktarg{j}{k}{t+j}{\calr}$.
Define an operation $\prodjktarg{j}{k}{t+j}{\cast}$ on $\tridnjktarg{j}{k}{t+j}{\calr}$ by
\be
\label{opdef1}
\tridnjktarg{j}{k}{t+j}{\bmrdt}\prodjktarg{j}{k}{t+j}{\cast}\tridnjktarg{j}{k}{t+j}{\bmrddt}\rmdef\tridnjktarg{j}{k}{t+j}{\bmrdt*\bmrddt}.
\ee

\begin{lem}
\label{lem54}
Fix time $t$.  Fix $k$ such that $0\le k\le\ell$.  Fix $j$ such that $0\le j\le k$.  
The operation $\prodjktarg{j}{k}{t+j}{\cast}$ on $\tridnjktarg{j}{k}{t+j}{\calr}$ 
is well defined.
\end{lem}

\begin{prf}
Let $\tridnjktarg{j}{k}{t+j}{\bmrgr},\tridnjktarg{j}{k}{t+j}{\bmrac}\in\tridnjktarg{j}{k}{t+j}{\calr}$
such that $\tridnjktarg{j}{k}{t+j}{\bmrgr}=\tridnjktarg{j}{k}{t+j}{\bmrdt}$ and 
$\tridnjktarg{j}{k}{t+j}{\bmrac}=\tridnjktarg{j}{k}{t+j}{\bmrddt}$.
To show the operation is well defined, we need to show
$\tridnjktarg{j}{k}{t+j}{\bmrgr*\bmrac}=\tridnjktarg{j}{k}{t+j}{\bmrdt*\bmrddt}$.
But this follows from Lemma \ref{lem46a}. 
\end{prf}

\begin{thm}
\label{thm55}
Fix time $t$.  Fix $k$ such that $0\le k\le\ell$.  Fix $j$ such that $0\le j\le k$.
The set $\tridnjktarg{j}{k}{t+j}{\calr}$ with operation $\prodjktarg{j}{k}{t+j}{\cast}$ forms a 
group $\grpjktarg{j}{k}{t+j}{\calr}{\cast}$.
\end{thm}

\begin{prf}
The proof is similar to Theorem \ref{thm41} and \ref{thm51}.
\end{prf}

We now show the groups $\grpjktarg{j}{k}{t+j}{\calr}{\cast}$ are isomorphic, for $0\le j\le k$.

\begin{thm}
\label{thm56}
Fix time $t$.  Fix $k$ such that $0\le k\le\ell$.  For any $j,j'$
such that $0\le j<j'\le k$, under the 1-1 correspondence of sets
$\tridnjktarg{j}{k}{t+j}{\calr}$ and $\tridnjktarg{j'}{k}{t+j'}{\calr}$ given by
the assignment
\be
\label{assign1}
\tridnjktarg{j}{k}{t+j}{\bmr}\mapsto\tridnjktarg{j'}{k}{t+j'}{\bmr}
\ee
for each $\bmr\in\calr$, there is an isomorphism
$$
\grpjktarg{j}{k}{t+j}{\calr}{\cast}\simeq\grpjktarg{j'}{k}{t+j'}{\calr}{\cast}.
$$
\end{thm}

\begin{prf}
We first show the 1-1 correspondence defined by the assignment (\ref{assign1}) 
is well defined.
Let $\tridnjktarg{j}{k}{t+j}{\bmrdt},\tridnjktarg{j}{k}{t+j}{\bmrddt}\in\tridnjktarg{j}{k}{t+j}{\calr}$
such that $\tridnjktarg{j}{k}{t+j}{\bmrdt}=\tridnjktarg{j}{k}{t+j}{\bmrddt}$.
We have to show that
$\tridnjktarg{j'}{k}{t+j'}{\bmrdt}=\tridnjktarg{j'}{k}{t+j'}{\bmrddt}$. 
But if $\tridnjktarg{j}{k}{t+j}{\bmrdt}=\tridnjktarg{j}{k}{t+j}{\bmrddt}$,
then the generators in $\bmrdt$ and $\bmrddt$ defined by representatives in congruent
positions of $\tridnjktarg{j}{k}{t+j}{\bmrdt}$ and $\tridnjktarg{j}{k}{t+j}{\bmrddt}$
are the same.  But by Lemma \ref{lem45a}, representatives in congruent positions 
of triangles $\tridnjktarg{j}{k}{t+j}{\bmrdt}$ and $\tridnjktarg{j'}{k}{t+j'}{\bmrdt}$ 
are from the same generator.  And the same holds for
$\tridnjktarg{j}{k}{t+j}{\bmrddt}$ and $\tridnjktarg{j'}{k}{t+j'}{\bmrddt}$.
Therefore $\tridnjktarg{j'}{k}{t+j'}{\bmrdt}=\tridnjktarg{j'}{k}{t+j'}{\bmrddt}$. 

We now prove the isomomorphism.
Let $\tridnjktarg{j}{k}{t+j}{\bmrdt},\tridnjktarg{j}{k}{t+j}{\bmrddt}\in\tridnjktarg{j}{k}{t+j}{\calr}$
such that
$\tridnjktarg{j}{k}{t+j}{\bmrdt}\mapsto\tridnjktarg{j'}{k}{t+j'}{\bmrdt}$ and
$\tridnjktarg{j}{k}{t+j}{\bmrddt}\mapsto\tridnjktarg{j'}{k}{t+j'}{\bmrddt}$.
We have to show that
$$
\tridnjktarg{j}{k}{t+j}{\bmrdt}\prodjktarg{j}{k}{t+j}{\cast}\tridnjktarg{j}{k}{t+j}{\bmrddt}\mapsto
\tridnjktarg{j'}{k}{t+j'}{\bmrdt})\prodjktarg{j'}{k}{t+j'}{\cast}\tridnjktarg{j'}{k}{t+j'}{\bmrddt}).
$$
But this is the same as showing
$$
\tridnjktarg{j}{k}{t+j}{\bmrdt*\bmrddt}\mapsto\tridnjktarg{j'}{k}{t+j'}{\bmrdt*\bmrddt}.
$$
\end{prf}

A change of variable in Theorem \ref{thm55} gives the following extension 
of Theorem \ref{thm51}.

\begin{thm}
\label{thm112af}
Fix time $t$.  Fix $k$ such that $0\le k\le\ell$.
The set $\tridnjktarg{j}{k}{t}{\calr}$ with operation $\prodjktarg{j}{k}{t}{\cast}$ forms a 
group $\grpjktarg{j}{k}{t}{\calr}{\cast}$, for $0\le j\le k$.
\end{thm}
We call the groups $\grpjktarg{j}{k}{t}{\calr}{\cast}$, for each $t\in\bmcpz$,
for $0\le k\le\ell$, for $0\le j\le k$, the {\it elementary groups} of $(\calr,*)$.  
We have just shown that any \ellctl\ complete group system $A$ can be decomposed into elementary groups.

\begin{thm}
The decomposition group $(\calr,*)$ of any \ellctl\ complete group system $A$ is a collection of 
elementary groups $\grpjktarg{j}{k}{t}{\calr}{\cast}$ of $(\calr,*)$
for each $t\in\bmcpz$, for $0\le k\le\ell$, for $0\le j\le k$.
\end{thm}

Even though the elementary group $\grpjktarg{j}{k}{t}{\calr}{\cast}$ is
defined using global set $\calr$, the group and group operation is local.
The elementary group $\grpjktarg{j}{k}{t}{\calr}{\cast}$ has no memory of
coordinates outside the set $\tridnjktarg{j}{k}{t}{\calr}$.
The smallest elementary groups are a single point $\grpjktarg{j}{\ell}{t}{\calr}{\cast}$,
for $0\le j\le\ell$, for each $t\in\bmcpz$.  The largest elementary groups,
or primary elementary groups, $\grpjktarg{0}{0}{t}{\calr}{\cast}$, 
are isomorphic to $B^t$, for each $t\in\bmcpz$.

We have just shown in Theorem \ref{thm56} that certain shifts of elementary groups
are isomorphic to each other.  We now give some homomorphism properties of elementary groups.
From the definition of operation $\prodjktarg{j}{k}{t}{\cast}$ in elementary group
$\grpjktarg{j}{k}{t}{\calr}{\cast}$, the following homomorphism is clear.

\begin{thm}
\label{homo1}
There is a homomorphism from $(\calr,*)$ to any 
elementary group $\grpjktarg{j}{k}{t}{\calr}{\cast}$
given by the map $\phi_{j,k}^t:  \calr\ra\tridnjktarg{j}{k}{t}{\calr}$ with assignment
$\phi_{j,k}^t:  \bmr\mapsto\tridnjktarg{j}{k}{t}{\bmr}$.
\end{thm}

The following result shows there is a homomorphism among the elementary groups 
$\grpjktarg{j}{k}{t}{\calr}{\cast}$.

\begin{thm}
\label{homo2}
Fix time $t$.  Fix $k$ such that $0\le k\le\ell$ and $j$ such that $0\le j\le k$.
Fix $m$ such that $k\le m\le\ell$ and $n$ such that $j\le n\le j+m-k$.  Then 
for any $\bmr\in\calr$, $\tridnjktarg{m}{n}{t}{\bmr}$ is a subtriangle of
$\tridnjktarg{j}{k}{t}{\bmr}$.  Then there is a homomorphism from group
$\grpjktarg{j}{k}{t}{\calr}{\cast}$ to group $\grpjktarg{m}{n}{t}{\calr}{\cast}$ defined by
the projection map $\ovphi:  \tridnjktarg{j}{k}{t}{\calr}\ra\tridnjktarg{m}{n}{t}{\calr}$ 
given by the assignment $\ovphi:  \tridnjktarg{j}{k}{t}{\bmr}\mapsto\tridnjktarg{m}{n}{t}{\bmr}$.
\end{thm}

\begin{prf}
We have to show $\ovphi$ is well defined.  Consider 
$\tridnjktarg{j}{k}{t}{\bmrht}\in\tridnjktarg{j}{k}{t}{\calr}$
such that $\tridnjktarg{j}{k}{t}{\bmrht}=\tridnjktarg{j}{k}{t}{\bmr}$.
But then $\tridnjktarg{m}{n}{t}{\bmrht}=\tridnjktarg{m}{n}{t}{\bmr}$ 
and so $\ovphi(\tridnjktarg{j}{k}{t}{\bmrht})=\ovphi(\tridnjktarg{j}{k}{t}{\bmr})$.
Therefore $\ovphi$ is well defined.

Let $\tridnjktarg{j}{k}{t}{\bmrdt},\tridnjktarg{j}{k}{t}{\bmrddt}\in\tridnjktarg{j}{k}{t}{\calr}$.  
Consider the projections 
$\ovphi:  \tridnjktarg{j}{k}{t}{\bmrdt}\mapsto\tridnjktarg{m}{n}{t}{\bmrdt}$ 
and $\ovphi:  \tridnjktarg{j}{k}{t}{\bmrddt}\mapsto\tridnjktarg{m}{n}{t}{\bmrddt}$.  
We want to show
$$
\ovphi(\tridnjktarg{j}{k}{t}{\bmrdt}\cast_{j,k}^t\tridnjktarg{j}{k}{t}{\bmrddt})=
\ovphi(\tridnjktarg{j}{k}{t}{\bmrdt})\cast_{m,n}^t \ovphi(\tridnjktarg{j}{k}{t}{\bmrddt}).
$$
But
\begin{align*}
\ovphi(\tridnjktarg{j}{k}{t}{\bmrdt}\cast_{j,k}^t\tridnjktarg{j}{k}{t}{\bmrddt}) 
&=\ovphi(\tridnjktarg{j}{k}{t}{\bmrdt*\bmrddt}) \\
&=\tridnjktarg{m}{n}{t}{\bmrdt*\bmrddt} \\
&=\tridnjktarg{m}{n}{t}{\bmrdt}\cast_{m,n}^t\tridnjktarg{m}{n}{t}{\bmrddt} \\
&=\ovphi(\tridnjktarg{j}{k}{t}{\bmrdt})\cast_{m,n}^t \ovphi(\tridnjktarg{j}{k}{t}{\bmrddt}).
\end{align*}
\end{prf}

We now give an example of the preceding theorem.  Fix time $t$.  Fix $k$ such that $0\le k\le\ell$.
Let $\ovphi$ be the projection map $\ovphi:  \tridnjktarg{0}{0}{t}{\calr}\ra\tridnjktarg{k}{k}{t}{\calr}$ 
given by the assignment $\ovphi:  \tridnjktarg{0}{0}{t}{\bmr}\mapsto\tridnjktarg{k}{k}{t}{\bmr}$.
Then $\ovphi$ is a homomorphism from group $\grpjktarg{0}{0}{t}{\calr}{\cast}$ to group 
$\grpjktarg{k}{k}{t}{\calr}{\cast}$.  The kernel $K$ of the homomorphism is the 
normal group defined on subset
$\{\bmr\in\calr: \tridnjktarg{k}{k}{t}{\bmr}=\tridnjktarg{k}{k}{t}{\bone}\}$
with operation $\prodjktarg{0}{0}{t}{\cast}$.  Then we have
$$
\grpjktarg{0}{0}{t}{\calr}{\cast}/K\simeq\grpjktarg{k}{k}{t}{\calr}{\cast}
$$
by the first homomorphism theorem.  But from Theorem \ref{thm29}, we know the group 
$\grpjktarg{0}{0}{t}{\calr}{\cast}$ is isomorphic to $B^t$ under the bijection
$\eta^t:  \tridnjktarg{0}{0}{t}{\calr}\ra B^t$ 
given by assignment $\eta^t:  \tridnjktarg{0}{0}{t}{\bmr}\mapsto b^t$,
where the representatives in $\tridnjktarg{0}{0}{t}{\bmr}$ form a \crepc\ of $b^t$ using (\ref{enctd}).
Then it is easy to see that if $\bmr\in K$, then
$\eta^t:  \tridnjktarg{0}{0}{t}{\bmr}\mapsto b^t$ where $b^t\in X_{k-1}^t$.
Then the group $K$ in $\grpjktarg{0}{0}{t}{\calr}{\cast}$
corresponds to group $X_{k-1}^t$ in $B^t$.  This gives
$$
\grpjktarg{0}{0}{t}{\calr}{\cast}/K\simeq B^t/X_{k-1}^t
$$
and so $\grpjktarg{k}{k}{t}{\calr}{\cast}\simeq B^t/X_{k-1}^t$.

\newpage
{\bf 6.  THE GENERATOR GROUP $(\calu,\circ)$}
\vspace{3mm}

\vspace{3mm}
{\bf 6.1  The generator group}
\vspace{3mm}

Fix $\bmr\in\calr$.  Let $\bmu$ be the tensor formed from $\bmr$ by the sequence of inputs of $\bmr$,
\be
\label{input3}
\ldots,\bmr_0^t,\bmr_0^{t-1},\ldots,
\ee
or the first columns of the sequence $\ldots,\tridnjktarg{0}{0}{t}{\bmr},\tridnjktarg{0}{0}{t-1}{\bmr},\ldots$ 
of \stms\ of $\bmr$.  In other words, $\bmu$ is given by the tensor
\be
\label{ugttf}
\begin{array}{llllllll}
        &  r_{0,\ell}^t   & r_{0,\ell}^{t-1}   & \cdots & r_{0,\ell}^{t-j}   & \cdots & r_{0,\ell}^{t-\ell}   & \\
        &  r_{0,\ell-1}^t & r_{0,\ell-1}^{t-1} & \cdots & r_{0,\ell-1}^{t-j} & \cdots & r_{0,\ell-1}^{t-\ell} & \\
        &  \cdots         & \cdots             &        & \cdots             &        & \cdots                & \\
        &  r_{0,k}^t      & r_{0,k}^{t-1}      & \cdots & r_{0,k}^{t-j}      & \cdots & r_{0,k}^{t-\ell}      & \\
        &  r_{0,k-1}^t    & r_{0,k-1}^{t-1}    & \cdots & r_{0,k-1}^{t-j}    & \cdots & r_{0,k-1}^{t-\ell}    & \\
\cdots  &  \cdots         & \cdots             & \cdots & \cdots             & \cdots & \cdots & \cdots       \\
        &  r_{0,j}^t      & r_{0,j}^{t-1}      & \cdots & r_{0,j}^{t-j}      & \cdots & r_{0,j}^{t-\ell}      & \\
        &  r_{0,j-1}^t    & r_{0,j-1}^{t-1}    & \cdots & r_{0,j-1}^{t-j}    & \cdots & r_{0,j-1}^{t-\ell}    & \\
        &  \cdots         & \cdots             &        & \cdots             &        & \cdots                & \\
        &  r_{0,1}^t      & r_{0,1}^{t-1}      & \cdots & r_{0,1}^{t-j}      & \cdots & r_{0,1}^{t-\ell}      & \\
        &  r_{0,0}^t      & r_{0,0}^{t-1}      & \cdots & r_{0,0}^{t-j}      & \cdots & r_{0,0}^{t-\ell}      &
\end{array}
\ee   
We write the tensor $\bmu$ using the same time reverse ordering as $\bmr$.

We can think of this construction in a slightly different way.
Previously we constructed a tensor set $\calr$.  A tensor $\bmr\in\calr$ is shown by the
sequence of columns of generators in the middle rows of (\ref{smgen}).
We now replace each generator $\bmg^{[t,t+k]}$ in $\bmr\in\calr$ 
with a unique identifier $r_{0,k}^t$, where
$\bmg^{[t,t+k]}\mapsto r_{0,k}^t$ in the bijection $[\Lambda_{0,k}^t]\ra R_{0,k}^t$. 
As remarked previously, $r_{0,k}^t$ has no meaning as a representative or group 
element as this point; in this sense we can think of $r_{0,k}^t$ as 
a {\it generator label} for $\bmg^{[t,t+k]}$.  Under the assignment $\bmg^{[t,t+k]}\mapsto r_{0,k}^t$
for $0\le k\le\ell$, for each $t\in\bmcpz$, a tensor $\bmr\in\calr$, as shown by the
middle rows of (\ref{smgen}), becomes a tensor $\bmu$ as shown in (\ref{ugttf}).
Let $\calu$ be the set of tensors $\bmu$ obtained from $\calr$ this way.
This gives the following.

\begin{lem}
There is a bijection $\beta:  \calr\ra\calu$ given by the
assignment $\beta:  \bmr\mapsto\bmu$ where each generator
$\bmg^{[t,t+k]}$ in $\bmr$ is replaced by a single \glab\ $r_{0,k}^t$ in $\bmu$, where
$\bmg^{[t,t+k]}\mapsto r_{0,k}^t$ in the bijection $[\Lambda_{0,k}^t]\ra R_{0,k}^t$.
\end{lem}

Under the bijection $\beta: \calr\ra\calu$
with assignment $\beta: \bmr\ra\bmu$, the sequence (\ref{input3}) of first columns 
of sequence $\ldots,\tridnjktarg{0}{0}{t}{\bmr},\tridnjktarg{0}{0}{t-1}{\bmr},\ldots$ of $\bmr$ is mapped 
to the sequence of columns of $\bmu$.
But from Lemma \ref{nmbrlem}, if $C$ is complete, we know that the set of sequences (\ref{input3})
of first columns of the sequence $\ldots,\tridnjktarg{0}{0}{t}{\bmr},\tridnjktarg{0}{0}{t-1}{\bmr},\ldots$ 
of \stms\ of $\bmr\in\calr$ is just the Cartesian product (\ref{nmbrin}).
Then under the bijection $\beta: \calr\ra\calu$, 
set $\calu$ is just the Cartesian product (\ref{nmbrin}),
\be
\label{input4}
\bigotimes_{t=+\infty}^{t=-\infty} \bigotimes_{0\le k\le\ell} R_{0,k}^t.
\ee

We now develop a compressed version of $(\calr,*)$ called $(\calu,\circ)$.
The operation $*$ in $(\calr,*)$ determines an operation $\circ$ on
$\calu$.  Let $\bmudt,\bmuddt\in\calu$.  Let $\beta:  \bmrdt\mapsto\bmudt$ and
$\beta:  \bmrddt\mapsto\bmuddt$.  Define an operation $\circ$ on $\calu$ by
\be
\label{eqno23}
\bmudt\circ\bmuddt\rmdef\bmubr
\ee
if $\beta:  \bmrdt*\bmrddt\mapsto\bmubr$.

\begin{lem}
The operation $\circ$ is well defined.
\end{lem}

\begin{prf}
The proof is similar to the proof of Lemma \ref{lem37}.
\end{prf}

\begin{thm}
\label{thm88a}
The set $\calu$ with operation $\circ$ forms a group $(\calu,\circ)$,
and $(\calr,*)\simeq(\calu,\circ)$ under the bijection 
$\beta:  \calr\ra\calu$.
\end{thm}

\begin{prf}
The proof is similar to the proof of Theorem \ref{thm31}.
\end{prf}

We know that each $\bmr\in\calr$ in group $(\calr,*)$ corresponds to a sequence of generators.
We see that $\bmu\in\calu$ in group $(\calu,\circ)$ demonstrates this sequence 
using a single generator label in place of each generator.  For this reason we call 
$(\calu,\circ)$ the {\it generator group} of $A$.  Note that $(\calu,\circ)$
is a group but it is not a group system in a natural way.

Collecting Theorems \ref{thm1}, \ref{thm31}, and \ref{thm88a}, we have the following result.

\begin{thm}
\label{homo7}
There is an isomorphism from $A$ to $(\calu,\circ)$ given by the composition map
$\xi\rmdef\beta\bullet\eta^{-1}\bullet\alpha^{-1}:  A \ra\calu$,
where isomorphism $A\simeq B$ is given by the bijection $\alpha^{-1}:  A\ra B$
in Theorem \ref{thm1}, isomorphism $B\simeq (\calr,*)$ is given by the bijection $\eta^{-1}:  B\ra\calr$
in Theorem \ref{thm31}, and isomorphism $(\calr,*)\simeq(\calu,\circ)$ is given by the bijection 
$\beta:  \calr\ra\calu$ in Theorem \ref{thm88a}.
\end{thm}
Then we can summarize the results of this paper so far by the chain
\be
\label{chain7a}
\begin{array}{lllllll}
A   & {\stackrel{\simeq}{\lra}} & B   & {\stackrel{\simeq}{\ra}}   & (\calr,*) & {\stackrel{\simeq}{\ra}}   & (\calu,\circ)    
\end{array}
\ee
where $B\simeq (\calr,*)$ and $(\calr,*)\simeq (\calu,\circ)$ under bijections
$\eta^{-1}:  B\ra\calr$ and $\beta:  \calr\ra\calu$, respectively.

We now define a component group on $(\calu,\circ)$ isomorphic to component group
$\grpjktarg{0}{0}{t}{\calr}{\cast}$ on $(\calr,*)$.  We first define an
{\it input matrix} $\tridnjktarg{0}{0}{t}{\bmu}$ which is congruent
in shape to \stm\ $\tridnjktarg{0}{0}{t}{\bmr}$ in $\bmr$.  Given $\bmu$ in (\ref{ugttf}),
we define $\tridnjktarg{0}{0}{t}{\bmu}$ to be the triangle in $\bmu$ with lower
vertex $r_{0,0}^t$ and upper vertices $r_{0,\ell}^t$ and $r_{0,\ell}^{t-\ell}$,
as shown in (\ref{rcttf}).  Note that all the entries in \stm\ $\tridnjktarg{0}{0}{t}{\bmr}$
are at the same time $t$, but in \inpm\ $\tridnjktarg{0}{0}{t}{\bmu}$, 
all entries in the same column are at the same time, but entries in different columns are
at different times.
\be
\label{rcttf}
\begin{array}{llllllllll}
  r_{0,\ell}^t   & r_{0,\ell}^{t-1}   & \cdots & \cdots & r_{0,\ell}^{t-j}   & \cdots & \cdots        & \cdots & r_{0,\ell}^{t-\ell+1}   & r_{0,\ell}^{t-\ell} \\
  r_{0,\ell-1}^t & r_{0,\ell-1}^{t-1} & \cdots & \cdots & r_{0,\ell-1}^{t-j} & \cdots & \cdots        & \cdots & r_{0,\ell-1}^{t-\ell+1} & \\
  \vdots         & \vdots             & \vdots & \vdots & \vdots             & \vdots & \vdots        & \vdots && \\
  r_{0,k}^t      & r_{0,k}^{t-1}      & \cdots & \cdots & r_{0,k}^{t-j}      & \cdots & r_{0,k}^{t-k} &&& \\
  \vdots         & \vdots             & \vdots & \vdots & \vdots             & \vdots &&&& \\
  \cdots         & \cdots             & \cdots & \cdots & r_{0,j}^{t-j}      &&&&& \\
  \vdots         & \vdots             & \vdots &&&&&&& \\
  r_{0,2}^t      & r_{0,2}^{t-1}      & r_{0,2}^{t-2} &&&&&&& \\
  r_{0,1}^t      & r_{0,1}^{t-1}      &&&&&&&& \\
  r_{0,0}^t                           &&&&&&&&&
\end{array}
\ee  
Note that we can think of tensor $\bmu$ in (\ref{ugttf}) as a sequence of
\inpms\ $\ldots,\tridnjktarg{0}{0}{t}{\bmu},\tridnjktarg{0}{0}{t-1}{\bmu},\ldots$ which overlap
to form $\bmu$.

Let $\beta:  \calr\ra\calu$.  Under the assignment $\beta:  \bmr\mapsto\bmu$,
for each $t\in\bmcpz$, a \stm\ $\tridnjktarg{0}{0}{t}{\bmr}$ given in (\ref{rttf}) 
becomes an \inpm\ $\tridnjktarg{0}{0}{t}{\bmu}$ shown in (\ref{rcttf}).
We now compare the \stm\ $\tridnjktarg{0}{0}{t}{\bmr}$ given in (\ref{rttf}) 
with the \inpm\ $\tridnjktarg{0}{0}{t}{\bmu}$ given in (\ref{rcttf}).
For each $j$, $0\le j\le\ell$, and each $k$, $j\le k\le\ell$, the representative 
$r_{j,k}^{t-j}$ in $\tridnjktarg{0}{0}{t}{\bmr}$ in (\ref{rttf}) is from the generator
$\bmg^{[t-j,t-j+k]}=\ldots,1^{t-j-1},r_{0,k}^{t-j},\ldots,r_{j,k}^{t-j},\ldots,r_{k,k}^{t-j},1^{t-j+k+1},\ldots$.
We replace the representative $r_{j,k}^{t-j}$ in $\tridnjktarg{0}{0}{t}{\bmr}$ in (\ref{rttf})
with \glab\ $r_{0,k}^{t-j}$ in $\tridnjktarg{0}{0}{t}{\bmu}$ in (\ref{rcttf}), where
$r_{0,k}^{t-j}$ is the leading nontrivial entry in the same generator.
Let $\tridnjktarg{0}{0}{t}{\calu}$ be the set of \inpms\ $\tridnjktarg{0}{0}{t}{\bmu}$
obtained from the set of \stms\ $\tridnjktarg{0}{0}{t}{\bmr}$ in $\tridnjktarg{0}{0}{t}{\calr}$
in this way.  Then we have a mapping $\beta^t: \tridnjktarg{0}{0}{t}{\calr}\ra\tridnjktarg{0}{0}{t}{\calu}$ 
with assignment $\beta^t: \tridnjktarg{0}{0}{t}{\bmr}\mapsto\tridnjktarg{0}{0}{t}{\bmu}$,
where $\tridnjktarg{0}{0}{t}{\bmu}$ is obtained from $\tridnjktarg{0}{0}{t}{\bmr}$
as just described.  This gives the following.

\begin{thm}
\label{thm66}
For each $t\in\bmcpz$, there is a bijection 
$\beta^t:  \tridnjktarg{0}{0}{t}{\calr}\ra\tridnjktarg{0}{0}{t}{\calu}$ given by assignment 
$\beta^t:  \tridnjktarg{0}{0}{t}{\bmr}\mapsto\tridnjktarg{0}{0}{t}{\bmu}$, where each
entry $r_{j,k}^{t-j}$ in $\tridnjktarg{0}{0}{t}{\bmr}$ is replaced by
entry $r_{0,k}^{t-j}$ in $\tridnjktarg{0}{0}{t}{\bmu}$
for each $j$, $0\le j\le\ell$, and each $k$, $j\le k\le\ell$.
\end{thm}

\begin{prf}
Clearly the map $\beta^t$ is well defined.  Let $\bmr,\bmrht\in\calr$ such that
$\beta:  \bmr\mapsto\bmu$ and $\beta:  \bmrht\mapsto\bmuht$.  If
$\tridnjktarg{0}{0}{t}{\bmr}=\tridnjktarg{0}{0}{t}{\bmrht}$, then we must have
$\tridnjktarg{0}{0}{t}{\bmu}=\tridnjktarg{0}{0}{t}{\bmuht}$.

Next we show $\beta^t$ is a bijection.
Consider the \stms\ $\tridnjktarg{0}{0}{t}{\bmr}$ in set $\tridnjktarg{0}{0}{t}{\calr}$.  
Fix $j$ such that $0\le j\le\ell$.  Fix $k$ such that $j\le k\le\ell$.  The sets
of representatives $\{r_{j,k}^{t-j}\}$ in $\tridnjktarg{0}{0}{t}{\bmr}$ in set 
$\tridnjktarg{0}{0}{t}{\calr}$ are from
generators $\{\bmg^{[t-j,t-j+k]}\}$ in $[\Lambda_{j,k}^t]$.  From Theorem \ref{thm25a}, we have a bijection
$[\Lambda_{j,k}^t]\ra \, \{r_{j,k}^{t-j}\}$.  Also from Theorem \ref{thm25a}, we have a bijection
$[\Lambda_{j,k}^t]\ra \, \{r_{0,k}^{t-j}\}$.  This gives a bijection
$\{r_{j,k}^{t-j}\}\ra\{r_{0,k}^{t-j}\}$.  Therefore the assignment
$\beta^t$ from $\tridnjktarg{0}{0}{t}{\bmr}$
to $\tridnjktarg{0}{0}{t}{\bmu}$ as described above is 1-1.  By definition,
$\tridnjktarg{0}{0}{t}{\calu}$ is the image of the map $\beta^t$ on
$\tridnjktarg{0}{0}{t}{\calr}$.  Therefore $\beta^t$ is a bijection 
$\tridnjktarg{0}{0}{t}{\calr}\ra\tridnjktarg{0}{0}{t}{\calu}$.
\end{prf}

The triangles $\tridnjktarg{0}{0}{t}{\bmr}$ in sequence 
$\ldots,\tridnjktarg{0}{0}{t}{\bmr},\tridnjktarg{0}{0}{t-1}{\bmr},\ldots$ in $\bmr$
do not overlap and share the same representatives, but triangles $\tridnjktarg{0}{0}{t}{\bmu}$ in sequence 
$\ldots,\tridnjktarg{0}{0}{t}{\bmu},\tridnjktarg{0}{0}{t-1}{\bmu},\ldots$ in $\bmu$ do overlap 
and share the same \glabs, unless $\ell=0$.
If $\beta:  \bmr\mapsto\bmu$, under the bijection 
$\beta^t:  \tridnjktarg{0}{0}{t}{\calr}\ra\tridnjktarg{0}{0}{t}{\calu}$ for each 
$t\in\bmcpz$, a sequence of triangles 
$\ldots,\tridnjktarg{0}{0}{t}{\bmr},\tridnjktarg{0}{0}{t-1}{\bmr},\ldots$
of $\bmr$ becomes a sequence of triangles
$\ldots,\tridnjktarg{0}{0}{t}{\bmu},\tridnjktarg{0}{0}{t-1}{\bmu},\ldots$, which then
overlap to form $\bmu$.  This gives the following.

\begin{cor}
\label{cor66}
Let $\beta:  \calr\ra\calu$ with assignment $\beta:  \bmr\mapsto\bmu$.
Given $\bmr\in\calr$, we can find $\bmu\in\calu$ by finding   
$\beta^t:  \tridnjktarg{0}{0}{t}{\bmr}\mapsto\tridnjktarg{0}{0}{t}{\bmu}$
for each $t\in\bmcpz$, and then overlapping the sequence of triangles
$\ldots,\tridnjktarg{0}{0}{t}{\bmu},\tridnjktarg{0}{0}{t-1}{\bmu},\ldots$ to find $\bmu$ as in (\ref{ugttf}).
Given $\bmu\in\calu$, we can find $\bmr\in\calr$ such that $\beta^{-1}:  \bmu\mapsto\bmr$
by finding $(\beta^t)^{-1}:  \tridnjktarg{0}{0}{t}{\bmu}\mapsto\tridnjktarg{0}{0}{t}{\bmr}$
for each $t\in\bmcpz$, and then constructing $\bmr$ as the sequence
$$
\ldots,\tridnjktarg{0}{0}{t}{\bmr},\tridnjktarg{0}{0}{t-1}{\bmr},\ldots.
$$
\end{cor}

The following corollary follows immediately from Theorem \ref{thm66}.

\begin{cor}
\label{cor67}
Let $\beta:  \calr\ra\calu$ with assignment $\beta:  \bmr\mapsto\bmu$.  
The \glabs\ in $\tridnjktarg{0}{0}{t}{\bmu}$ are uniquely determined
by the generator representatives in $\tridnjktarg{0}{0}{t}{\bmr}$.
\end{cor}

\begin{lem}
\label{lem47a}
Fix time $t$.  Let $\bmudt,\bmuddt\in\calu$.  In the product $\bmudt\circ\bmuddt$, the \glabs\ in 
$\tridnjktarg{0}{0}{t}{\bmudt\circ\bmuddt}$ are uniquely determined by the \glabs\ in
$\tridnjktarg{0}{0}{t}{\bmudt}$ and $\tridnjktarg{0}{0}{t}{\bmuddt}$.
\end{lem}

\begin{prf}
Combine Lemma \ref{lem47} with Corollary \ref{cor67}.
\end{prf}

For each $t\in\bmcpz$,
define a map $\theta^t:  \calu\ra\tridnjktarg{0}{0}{t}{\calu}$ by the assignment $\theta^t:  \bmu\mapsto\tridnjktarg{0}{0}{t}{\bmu}$.
Note that $\theta^t(\calu)=\tridnjktarg{0}{0}{t}{\calu}$.  Fix time $t$.
Let $\tridnjktarg{0}{0}{t}{\bmudt},\tridnjktarg{0}{0}{t}{\bmuddt}\in\tridnjktarg{0}{0}{t}{\calu}$.
Define an operation $\prodjktarg{0}{0}{t}{\ccirc}$ on set $\tridnjktarg{0}{0}{t}{\calu}$ by
\be
\label{defy}
\tridnjktarg{0}{0}{t}{\bmudt}\prodjktarg{0}{0}{t}{\ccirc}\tridnjktarg{0}{0}{t}{\bmuddt}\rmdef\tridnjktarg{0}{0}{t}{\bmudt\circ\bmuddt}.
\ee

\begin{lem}
\label{lem62}
Fix time $t$.  The operation $\prodjktarg{0}{0}{t}{\ccirc}$ on set $\tridnjktarg{0}{0}{t}{\calu}$ 
is well defined.
\end{lem}

\begin{prf}
The proof is similar to Lemma \ref{lem32}, using Lemma \ref{lem47a} in place of Lemma \ref{lem47}.
\end{prf}

\begin{thm}
\label{thm63}
Fix time $t$.  The set $\tridnjktarg{0}{0}{t}{\calu}$ with operation $\prodjktarg{0}{0}{t}{\ccirc}$ 
forms a group $\grpjktarg{0}{0}{t}{\calu}{\ccirc}$.
\end{thm}

\begin{prf}
The proof is similar to the proof of Theorem \ref{thm41}.
\end{prf}

We call the groups $\grpjktarg{0}{0}{t}{\calu}{\ccirc}$, for each $t\in\bmcpz$,
the {\it primary elementary groups} of $(\calu,\circ)$.

\begin{thm}
\label{homou}
There is a homomorphism from $(\calu,\circ)$ to $\grpjktarg{0}{0}{t}{\calu}{\ccirc}$ given by the map
$\theta^t:  \calu\ra\tridnjktarg{0}{0}{t}{\calu}$, for each $t\in\bmcpz$.
\end{thm}

\begin{prf}
This follows immediately from the definition of operation $\prodjktarg{0}{0}{t}{\ccirc}$ in (\ref{defy}).
\end{prf}

\begin{cor}
\label{cor65}
For each $t\in\bmcpz$,
there is an isomorphism $\grpjktarg{0}{0}{t}{\calr}{\cast}\simeq\grpjktarg{0}{0}{t}{\calu}{\ccirc}$ 
under the bijection $\beta^t:  \tridnjktarg{0}{0}{t}{\calr}\ra\tridnjktarg{0}{0}{t}{\calu}$ with
assignment $\beta^t:  \tridnjktarg{0}{0}{t}{\bmr}\ra\tridnjktarg{0}{0}{t}{\bmu}$
if $\beta:  \bmr\mapsto\bmu$ under the bijection $\calr\ra\calu$.
\end{cor}

\begin{prf}
Let $\bmrdt,\bmrddt\in\calr$.
Let $\beta:  \bmrdt\mapsto\bmudt$ and $\beta:  \bmrddt\mapsto\bmuddt$ under the
bijection $\calr\ra\calu$.  Then 
$\tridnjktarg{0}{0}{t}{\bmrdt}\mapsto\tridnjktarg{0}{0}{t}{\bmudt}$ and
$\tridnjktarg{0}{0}{t}{\bmrddt}\mapsto\tridnjktarg{0}{0}{t}{\bmuddt}$.
We have to show that
$$
\tridnjktarg{0}{0}{t}{\bmrdt}\prodjktarg{0}{0}{t}{\cast}\tridnjktarg{0}{0}{t}{\bmrddt}\mapsto
\tridnjktarg{0}{0}{t}{\bmudt}\prodjktarg{0}{0}{t}{\ccirc}\tridnjktarg{0}{0}{t}{\bmuddt}.
$$
But this is the same as showing
\be
\label{assign3a}
\tridnjktarg{0}{0}{t}{\bmrdt*\bmrddt}\mapsto\tridnjktarg{0}{0}{t}{\bmudt\circ\bmuddt}.
\ee
But if $\beta:  \bmrdt\mapsto\bmudt$ and $\beta:  \bmrddt\mapsto\bmuddt$, then
$\bmrdt*\bmrddt\mapsto\bmudt\circ\bmuddt$, and so (\ref{assign3a}) holds.
\end{prf}

\vspace{3mm}
{\bf 6.2  The elementary groups}
\vspace{3mm}

The tensor $\bmu$ in (\ref{ugttf}) is indexed by ordered pairs of the form $(k,t)$,
where subscript $k$ satisfies $0\le k\le\ell$ and superscript $t\in\bmcpz$.  Then tensor $\bmu$ gives
an {\it index tensor} $\bmn$ of ordered pairs $(k,t)$, as shown in (\ref{top}).
\be
\label{top}
\begin{array}{llllllll}
        &  (\ell,t)   & (\ell,t-1)   & \cdots & (\ell,t-j)   & \cdots & (\ell,t-\ell)   & \\
        &  (\ell-1,t) & (\ell-1,t-1) & \cdots & (\ell-1,t-j) & \cdots & (\ell-1,t-\ell) & \\
        &  \cdots     & \cdots       &        & \cdots       &        & \cdots          & \\
        &  (k,t)      & (k,t-1)      & \cdots & (k,t-j)      & \cdots & (k,t-\ell)      & \\
        &  (k-1,t)    & (k-1,t-1)    & \cdots & (k-1,t-j)    & \cdots & (k-1,t-\ell)    & \\
\cdots  &  \cdots     & \cdots       & \cdots & \cdots       & \cdots & \cdots          & \cdots  \\
        &  (j,t)      & (j,t-1)      & \cdots & (j,t-j)      & \cdots & (j,t-\ell)      & \\
        &  (j-1,t)    & (j-1,t-1)    & \cdots & (j-1,t-j)    & \cdots & (j-1,t-\ell)    & \\
        &  \cdots     & \cdots       &        & \cdots       &        & \cdots          & \\
        &  (1,t)      & (1,t-1)      & \cdots & (1,t-j)      & \cdots & (1,t-\ell)      & \\
        &  (0,t)      & (0,t-1)      & \cdots & (0,t-j)      & \cdots & (0,t-\ell)      &
\end{array}
\ee   
We use a similar notation to describe triangular subsets in index tensor $\bmn$ that was
previously used for tensor $\bmr\in\calr$.  For $0\le k\le\ell$ and $t\in\bmcpz$,
let index triangle $\tridnjktarg{0}{k}{t}{\bmn}$ be the ordered pairs $(k,t)$ in (\ref{top})
specified by the triangle with lower vertex $(k,t)$ and upper vertices 
$(\ell,t)$ and $(\ell,t-(\ell-k))$.

For $0\le k\le\ell$ and $t\in\bmcpz$, and for each $\bmu\in\calu$, 
we let $\tridnjktarg{0}{k}{t}{\bmu}$ be the \glabs\
in $\bmu$ in (\ref{ugttf}) whose indices are specified by the index triangle 
$\tridnjktarg{0}{k}{t}{\bmn}$ in (\ref{top}).  These are the \glabs\ with lower vertex $r_{0,k}^t$ 
and upper vertices $r_{0,\ell}^t$ and $r_{0,\ell}^{t-(\ell-k)}$ in (\ref{ugttf}).
Note that for $k=0$, $\tridnjktarg{0}{k}{t}{\bmu}$ is the same as $\tridnjktarg{0}{0}{t}{\bmu}$
previously defined in (\ref{rcttf}).
For $0\le k\le\ell$ and $t\in\bmcpz$, let $\tridnjktarg{0}{k}{t}{\calu}$ 
be the set of all possible triangles $\tridnjktarg{0}{k}{t}{\bmu}$,
$\tridnjktarg{0}{k}{t}{\calu}\rmdef\{\tridnjktarg{0}{k}{t}{\bmu} : \bmu\in\calu\}$.
Note that for $k=0$, $\tridnjktarg{0}{k}{t}{\calu}$ is just the set $\tridnjktarg{0}{0}{t}{\calu}$
previously defined.

Let $\beta:  \bmr\mapsto\bmu$ under the bijection $\calr\ra\calu$.
In Corollary \ref{cor67}, we have seen that the \glabs\ in $\tridnjktarg{0}{0}{t}{\bmu}$
are uniquely determined by the generator representatives in $\tridnjktarg{0}{0}{t}{\bmr}$.
The same thing is true of $\tridnjktarg{0}{k}{t}{\bmu}$ and $\tridnjktarg{0}{k}{t}{\bmr}$,
for the same reason.  This gives the following.

\begin{lem}
\label{lem104}
Fix time $t$.  Fix $k$ such that $0\le k\le\ell$.
Let $\beta:  \bmr\mapsto\bmu$ under the bijection $\calr\ra\calu$.
The \glabs\ in $\tridnjktarg{0}{k}{t}{\bmu}$ are uniquely determined
by the generator representatives in $\tridnjktarg{0}{k}{t}{\bmr}$.
\end{lem}

\begin{lem}
\label{lem60}
Fix time $t$.  Fix $k$ such that $0\le k\le\ell$.  Let $\bmudt,\bmuddt\in\calu$.
In the product $\bmudt\circ\bmuddt$, the \glabs\ in
$\tridnjktarg{0}{k}{t}{\bmudt\circ\bmuddt}$ are uniquely determined 
by the \glabs\ in
$\tridnjktarg{0}{k}{t}{\bmudt}$ and $\tridnjktarg{0}{k}{t}{\bmuddt}$.
\end{lem}

\begin{prf}
Combine Lemma \ref{lem46a} with Lemma \ref{lem104}.
\end{prf}

For each $t\in\bmcpz$, for $0\le k\le\ell$,
let $\tridnjktarg{0}{k}{t}{\calu}$ be the set of all possible triangles
$\tridnjktarg{0}{k}{t}{\bmu}$,
$\tridnjktarg{0}{k}{t}{\calu}\rmdef\{\tridnjktarg{0}{k}{t}{\bmu} : \bmu\in\calu\}$.
We now find groups on sets $\tridnjktarg{0}{k}{t}{\calu}$.
Fix time $t$.  Fix $k$ such that $0\le k\le\ell$.
Let $\tridnjktarg{0}{k}{t}{\bmudt},\tridnjktarg{0}{k}{t}{\bmuddt}\in\tridnjktarg{0}{k}{t}{\calu}$.
Define an operation $\prodjktarg{0}{k}{t}{\ccirc}$ on $\tridnjktarg{0}{k}{t}{\calu}$ by
\be
\label{opdef8}
\tridnjktarg{0}{k}{t}{\bmudt}\prodjktarg{0}{k}{t}{\ccirc}\tridnjktarg{0}{k}{t}{\bmuddt}
\rmdef\tridnjktarg{0}{k}{t}{\bmudt\circ\bmuddt}.
\ee

\begin{lem}
\label{lem74a}
Fix time $t$.  Fix $k$ such that $0\le k\le\ell$.
The operation $\prodjktarg{0}{k}{t}{\ccirc}$ on $\tridnjktarg{0}{k}{t}{\calu}$ 
is well defined.
\end{lem}

\begin{prf}
The proof is similar to Lemma \ref{lem54}, using Lemma \ref{lem60} in place of Lemma \ref{lem46a}.
\end{prf}

\begin{thm}
Fix time $t$.  Fix $k$ such that $0\le k\le\ell$.
The set $\tridnjktarg{0}{k}{t}{\calu}$ with operation $\prodjktarg{0}{k}{t}{\ccirc}$ forms a 
group $\grpjktarg{0}{k}{t}{\calu}{\ccirc}$.
\end{thm}

\begin{prf}
The proof is similar to Theorem \ref{thm41} and \ref{thm51}.
\end{prf}

We call the groups $\grpjktarg{0}{k}{t}{\calu}{\ccirc}$,
for each $t\in\bmcpz$, for $0\le k\le\ell$, the {\it elementary groups} of $(\calu,\circ)$.

\begin{thm}
The generator group $(\calu,\circ)$ of any \ellctl\ complete group system $A$ is a collection of 
elementary groups $\grpjktarg{0}{k}{t}{\calu}{\ccirc}$ of $(\calu,\circ)$
for each $t\in\bmcpz$, for $0\le k\le\ell$.
\end{thm}
Because the elementary groups of $(\calu,\circ)$ are indexed by $0\le k\le\ell$,
we say generator group $(\calu,\circ)$ is an $(\ell+1)$-{\it depth} generator group.


We now show an elementary group of $(\calu,\circ)$ is isomorphic to an elementary group
of $(\calr,*)$.  There is a bijection $\beta:  \calr\ra\calu$ given by the assignment 
$\beta:  \bmr\mapsto\bmu$.  Fix time $t$.  Fix $k$ such that $0\le k\le\ell$.
Define a 1-1 correspondence $\beta_{0,k}^t$ between sets 
$\tridnjktarg{0}{k}{t}{\calr}$ and $\tridnjktarg{0}{k}{t}{\calu}$ by
the assignment
$$
\beta_{0,k}^t:  \tridnjktarg{0}{k}{t}{\bmr}\mapsto\tridnjktarg{0}{k}{t}{\bmu}
$$
if $\beta:  \bmr\mapsto\bmu$ under the bijection $\calr\ra\calu$.

\begin{lem}
Fix time $t$.  Fix $k$ such that $0\le k\le\ell$.  The 1-1 correspondence $\beta_{0,k}^t$
is well defined.
\end{lem}

\begin{prf}
Let $\tridnjktarg{0}{k}{t}{\bmrdt},\tridnjktarg{0}{k}{t}{\bmrddt}\in\tridnjktarg{0}{k}{t}{\calr}$
such that $\tridnjktarg{0}{k}{t}{\bmrdt}=\tridnjktarg{0}{k}{t}{\bmrddt}$.
Let $\beta:  \bmrdt\mapsto\bmudt$ and $\beta:  \bmrddt\mapsto\bmuddt$.
We have to show that $\tridnjktarg{0}{k}{t}{\bmudt}=\tridnjktarg{0}{k}{t}{\bmuddt}$.
But this follows from Lemma \ref{lem104}.
\end{prf}

\begin{thm}
\label{thm82}
Fix time $t$.  For $k$ such that $0\le k\le\ell$, we have 
\be
\grpjktarg{0}{k}{t}{\calr}{\cast}\simeq\grpjktarg{0}{k}{t}{\calu}{\ccirc},
\ee
under the 1-1 correspondence $\beta_{0,k}^t$ between sets 
$\tridnjktarg{0}{k}{t}{\calr}$ and $\tridnjktarg{0}{k}{t}{\calu}$ given by
the assignment
$$
\beta_{0,k}^t:  \tridnjktarg{0}{k}{t}{\bmr}\mapsto\tridnjktarg{0}{k}{t}{\bmu}
$$
if $\beta:  \bmr\mapsto\bmu$ under the bijection $\calr\ra\calu$.
\end{thm}

\begin{prf}
The proof is similar to the proof of Corollary \ref{cor65}.
\end{prf}

Since an elementary group $\grpjktarg{0}{k}{t}{\calr}{\cast}$ of $(\calr,*)$ is
isomorphic to an elementary group $\grpjktarg{j}{k}{t+j}{\calr}{\cast}$ of $(\calr,*)$,
for $0\le j\le k$, we have the following corollary of Theorem \ref{thm82}.

\begin{cor}
\label{cor82}
Fix time $t$.  For $k$ such that $0\le k\le\ell$, for $j$ such that $0\le j\le k$, we have 
\be
\grpjktarg{j}{k}{t+j}{\calr}{\cast}\simeq\grpjktarg{0}{k}{t}{\calr}{\cast}\simeq\grpjktarg{0}{k}{t}{\calu}{\ccirc},
\ee
under the 1-1 correspondence $\beta_{0,k}^t$ between sets 
$\tridnjktarg{0}{k}{t}{\calr}$ and $\tridnjktarg{0}{k}{t}{\calu}$ given in Theorem \ref{thm82},
and the 1-1 correspondence between sets 
$\tridnjktarg{j}{k}{t+j}{\calr}$ and $\tridnjktarg{0}{k}{t}{\calr}$ given in Theorem \ref{thm56}.
\end{cor}

We now give some homomorphism properties of elementary groups.
From the definition of operation $\prodjktarg{0}{k}{t}{\ccirc}$ in elementary group
$\grpjktarg{0}{k}{t}{\calu}{\ccirc}$, the following homomorphism is clear.

\begin{thm}
\label{homo3}
Fix $t\in\bmcpz$.  Fix $k$ such that $0\le k\le\ell$.
There is a homomorphism from $(\calu,\circ)$ to any 
elementary group $\grpjktarg{0}{k}{t}{\calu}{\ccirc}$
given by the map $\theta_{0,k}^t:  \calu\ra\tridnjktarg{0}{k}{t}{\calu}$ with assignment
$\theta_{0,k}^t:  \bmu\mapsto\tridnjktarg{0}{k}{t}{\bmu}$.
\end{thm}

The following result shows there is a homomorphism among the elementary groups 
$\grpjktarg{0}{k}{t}{\calu}{\ccirc}$.  The proof is similar to Theorem \ref{homo2}
with $\ovtheta$ in place of $\ovphi$.

\begin{thm}
\label{homo4}
Fix $t\in\bmcpz$.  Fix $k$ such that $0\le k\le\ell$.  Fix
$m$ such that $k\le m\le\ell$ and $n$ such that $0\le n\le m$.  Then 
for any $\bmu\in\calu$, $\tridnjktarg{0}{m}{t-n}{\bmu}$ is a subtriangle of
$\tridnjktarg{0}{k}{t}{\bmu}$.  Then there is a 
homomorphism from group $\grpjktarg{0}{k}{t}{\calu}{\ccirc}$ to group
$\grpjktarg{0}{m}{t-n}{\calu}{\ccirc}$ defined by the projection map 
$\ovtheta:  \tridnjktarg{0}{k}{t}{\calu}\ra\tridnjktarg{0}{m}{t-n}{\calu}$
given by the assignment $\ovtheta:  \tridnjktarg{0}{k}{t}{\bmu}\mapsto\tridnjktarg{0}{m}{t-n}{\bmu}$.
\end{thm}

We now summarize some of the results on generator group $(\calu,\circ)$.
The group $(\calu,\circ)$ collapses any generator $\bmg^{[t,t+k]}$
in $\bmr\in\calr$ to a single \glab\ $r_{0,k}^t$ in $\bmu\in\calu$.
The group $(\calu,\circ)$ collapses any sequence of triangles 
$\tridnjktarg{j}{k}{t+j}{\calr}$ in $(\calr,*)$, for $0\le j\le k$, to a single triangle
$\tridnjktarg{0}{k}{t}{\calu}$ in $(\calu,\circ)$.
The group $(\calu,\circ)$ collapses any sequence of elementary groups
$\grpjktarg{j}{k}{t+j}{\calr}{\cast}$ in $(\calr,*)$, for $0\le j\le k$, to a single
isomorphic elementary group $\grpjktarg{0}{k}{t}{\calu}{\ccirc}$ in $(\calu,\circ)$.
Except for the alphabet labels $a^t$ of branches in $B^t$, the generator group contains 
all the information that $B$ contains, just in a different and more revealing form.

The last result in this subsection is needed in Subsections 6.4 and 7.3.

\begin{lem}
\label{lem95a}
The product operation $\bmudt\circ\bmuddt$ is equivalent to the evaluation 
of
\be
\label{eq100a}
\tridnjktarg{0}{0}{t}{\bmudt}\prodjktarg{0}{0}{t}{\ccirc}\tridnjktarg{0}{0}{t}{\bmuddt}.
\ee
for $t=+\infty$ to $-\infty$.
\end{lem}

\begin{prf}
Let $\bmu\in\calu$.  Then $\bmu$ is the same as the evaluation of 
$\tridnjktarg{0}{0}{t}{\bmu}$ for $t=+\infty$ to $-\infty$.  Let $\bmudt,\bmuddt\in\calu$.  
Consider the product $\bmudt\circ\bmuddt$.
Then $\bmudt\circ\bmuddt$ is the same as the evaluation of 
$\tridnjktarg{0}{0}{t}{\bmudt\circ\bmuddt}$ for $t=+\infty$ to $-\infty$.  
But from (\ref{opdef8}) we have 
$$
\tridnjktarg{0}{0}{t}{\bmudt\circ\bmuddt}=
\tridnjktarg{0}{0}{t}{\bmudt}\prodjktarg{0}{0}{t}{\ccirc}\tridnjktarg{0}{0}{t}{\bmuddt}
$$
for each $t\in\bmcpz$.  Then $\bmudt\circ\bmuddt$ is the same as the evaluation of 
(\ref{eq100a}) for $t=+\infty$ to $-\infty$.
\end{prf}

We have seen from Lemma \ref{lem74a} that the operation $\prodjktarg{0}{k}{t}{\ccirc}$ 
on set $\tridnjktarg{0}{k}{t}{\calu}$ is well defined.  This means that if
$\tridnjktarg{0}{k}{t}{\bmudt},\tridnjktarg{0}{k}{t}{\bmugr}\in\tridnjktarg{0}{k}{t}{\calu}$
such that $\tridnjktarg{0}{k}{t}{\bmugr}=\tridnjktarg{0}{k}{t}{\bmudt}$, and 
$\tridnjktarg{0}{k}{t}{\bmuddt},\tridnjktarg{0}{k}{t}{\bmuac}\in\tridnjktarg{0}{k}{t}{\calu}$
such that $\tridnjktarg{0}{k}{t}{\bmuac}=\tridnjktarg{0}{k}{t}{\bmuddt}$, then
$\tridnjktarg{0}{k}{t}{\bmugr\circ\bmuac}=\tridnjktarg{0}{k}{t}{\bmudt\circ\bmuddt}$.
Thus the operation in elementary group $\grpjktarg{0}{k}{t}{\calu}{\ccirc}$ just
depends on the elements in set $\tridnjktarg{0}{k}{t}{\calu}$, and is otherwise 
independent of the remaining portion of $\calu$.  Then we can think of the collection of
elementary groups $\grpjktarg{0}{k}{t}{\calu}{\ccirc}$, for $0\le k\le\ell$, for $t\in\bmcpz$, 
as a decomposition of $(\calu,\circ)$ into local groups since each group
$\grpjktarg{0}{k}{t}{\calu}{\ccirc}$ is wholely defined by a local set
$\tridnjktarg{0}{k}{t}{\calu}$.  As shown in Lemma \ref{lem95a},
the fact that we can recover the global operation
$\circ$ in $(\calu,\circ)$ just from the collection of elementary groups means
$(\calu,\circ)$ is completely specified by local operations on local sets.  Similarly we can
think of the elementary groups of $(\calr,*)$ as a local decomposition of
$(\calr,*)$.

\vspace{3mm}
{\bf 6.3  Sequences of elementary groups}
\vspace{3mm}

We have just seen that the generator group 
$(\calu,\circ)$ of any \ellctl\ complete group system $A$ is a collection of 
elementary groups $\grpjktarg{0}{k}{t}{\calu}{\ccirc}$ of $(\calu,\circ)$
for each $t\in\bmcpz$, for $0\le k\le\ell$.  In this subsection, we show that 
finite or infinite sequences of elementary groups in the generator group form a group
with properties similar to an elementary group.  In particular, we analyze the
structure of the group which is the intersection of two elementary groups.

We first define a sequence of ordered pairs $(k,t)$ in index tensor $\bmn$.
Let $\bmt$ be a sequence of times $\bmt=\ldots,t,t',\ldots$, where times $t,t'\in\bmcpz$.
The sequence $\bmt=\ldots,t,t',\ldots$ is written in reverse time order,
so that $t'<t$ as in (\ref{top}).  The sequence $\bmt$ may be the integers $\bmcpz$
or a finite or infinite subset of $\bmcpz$, in reverse time order.
Let $\bmk$ be a sequence of integers $\ldots,k,k',\ldots$ with each integer $k$
satisfying $0\le k\le\ell$.  The {\it index sequence} $(\bmk,\bmt)$ is a sequence
of ordered pairs $(k,t)$ in $\bmn$, defined by
$$
(\bmk,\bmt)\rmdef\ldots,(k,t),(k',t'),\ldots,
$$
such that each $k$ in sequence $\bmk$ is paired with a $t$ in sequence $\bmt$ and vice versa.

We next consider a sequence of triangles in $\bmu$ associated with the 
index sequence $(\bmk,\bmt)$ in $\bmn$.  Fix $\bmu$ in $\calu$.  Consider the sequence of triangles
\be
\label{seq1}
\ldots,\tridnjktarg{0}{k}{t}{\bmu},\tridnjktarg{0}{k'}{t'}{\bmu},\ldots
\ee
specified by $(\bmk,\bmt)$.
Define $\squaritarg{\bmk}{\bmt}{\bmu}$ to be the
sequence in (\ref{seq1}).  Note that triangles in $\squaritarg{\bmk}{\bmt}{\bmu}$
may overlap.  Define $\Box_\bmk^\bmt(\calu)$ to be the set of sequences 
$\{\squaritarg{\bmk}{\bmt}{\bmu}:  \bmu\in\calu\}$.

Let $\squaritarg{\bmk}{\bmt}{\bmudt},\squaritarg{\bmk}{\bmt}{\bmuddt}\in\Box_\bmk^\bmt(\calu)$.
Define an operation $\proditarg{\bmk}{\bmt}{\oplus}$ on 
$\squaritarg{\bmk}{\bmt}{\calu}$ by
\be
\label{defn2}
\squaritarg{\bmk}{\bmt}{\bmudt}\proditarg{\bmk}{\bmt}{\oplus}\squaritarg{\bmk}{\bmt}{\bmuddt}
\rmdef\squaritarg{\bmk}{\bmt}{\bmudt\circ\bmuddt}.
\ee
Note that the operation $\proditarg{\bmk}{\bmt}{\oplus}$ in (\ref{defn2}) can be evaluated as
\begin{align}
\nonumber
\squaritarg{\bmk}{\bmt}{\bmudt}\proditarg{\bmk}{\bmt}{\oplus}\squaritarg{\bmk}{\bmt}{\bmuddt}
&=\squaritarg{\bmk}{\bmt}{\bmudt\circ\bmuddt} \\
\nonumber
&=\ldots,\tridnjktarg{0}{k}{t}{\bmudt\circ\bmuddt},\tridnjktarg{0}{k'}{t'}{\bmudt\circ\bmuddt},\ldots \\
\label{evaln2}
&=\ldots,\tridnjktarg{0}{k}{t}{\bmudt}\prodjktarg{0}{k}{t}{\ccirc}\tridnjktarg{0}{k}{t}{\bmuddt},
\tridnjktarg{0}{k'}{t'}{\bmudt}\prodjktarg{0}{k'}{t'}{\ccirc}\tridnjktarg{0}{k'}{t'}{\bmuddt},\ldots
\end{align}
for each $\squaritarg{\bmk}{\bmt}{\bmudt},\squaritarg{\bmk}{\bmt}{\bmuddt}\in\Box_\bmk^\bmt(\calu)$,
and the last line is easy to evaluate using groups $\grpjktarg{0}{k}{t}{\calu}{\ccirc}$ isomorphic 
to elementary groups.

\begin{lem}
The operation $\proditarg{\bmk}{\bmt}{\oplus}$ is well defined.
\end{lem}

\begin{prf}
Let $\squaritarg{\bmk}{\bmt}{\bmudt},\squaritarg{\bmk}{\bmt}{\bmuddt}\in\Box_\bmk^\bmt(\calu)$.
Let $\squaritarg{\bmk}{\bmt}{\bmugr},\squaritarg{\bmk}{\bmt}{\bmuac}\in\Box_\bmk^\bmt(\calu)$
such that $\squaritarg{\bmk}{\bmt}{\bmugr}=\squaritarg{\bmk}{\bmt}{\bmudt}$ and 
$\squaritarg{\bmk}{\bmt}{\bmuac}=\squaritarg{\bmk}{\bmt}{\bmuddt}$.
To show the operation is well defined, we need to show
$$
\squaritarg{\bmk}{\bmt}{\bmugr}\proditarg{\bmk}{\bmt}{\oplus}\squaritarg{\bmk}{\bmt}{\bmuac}
=\squaritarg{\bmk}{\bmt}{\bmudt}\proditarg{\bmk}{\bmt}{\oplus}\squaritarg{\bmk}{\bmt}{\bmuddt},
$$
or $\squaritarg{\bmk}{\bmt}{\bmugr\circ\bmuac}=\squaritarg{\bmk}{\bmt}{\bmudt\circ\bmuddt}$.
From (\ref{evaln2}), this is the same as showing
\begin{multline*}
\ldots,\tridnjktarg{0}{k}{t}{\bmugr}\prodjktarg{0}{k}{t}{\ccirc}\tridnjktarg{0}{k}{t}{\bmuac},
\tridnjktarg{0}{k'}{t'}{\bmugr}\prodjktarg{0}{k'}{t'}{\ccirc}\tridnjktarg{0}{k'}{t'}{\bmuac},\ldots \\
=\ldots,\tridnjktarg{0}{k}{t}{\bmudt}\prodjktarg{0}{k}{t}{\ccirc}\tridnjktarg{0}{k}{t}{\bmuddt},
\tridnjktarg{0}{k'}{t'}{\bmudt}\prodjktarg{0}{k'}{t'}{\ccirc}\tridnjktarg{0}{k'}{t'}{\bmuddt},\ldots,
\end{multline*}
or
\begin{multline}
\ldots,\tridnjktarg{0}{k}{t}{\bmugr}\prodjktarg{0}{k}{t}{\ccirc}\tridnjktarg{0}{k}{t}{\bmuac}
=\tridnjktarg{0}{k}{t}{\bmudt}\prodjktarg{0}{k}{t}{\ccirc}\tridnjktarg{0}{k}{t}{\bmuddt}, \\
\label{evaln3}
\tridnjktarg{0}{k'}{t'}{\bmugr}\prodjktarg{0}{k'}{t'}{\ccirc}\tridnjktarg{0}{k'}{t'}{\bmuac}
=\tridnjktarg{0}{k'}{t'}{\bmudt}\prodjktarg{0}{k'}{t'}{\ccirc}\tridnjktarg{0}{k'}{t'}{\bmuddt},\ldots.
\end{multline}
But if $\squaritarg{\bmk}{\bmt}{\bmugr}=\squaritarg{\bmk}{\bmt}{\bmudt}$, then
$$
\ldots,\tridnjktarg{0}{k}{t}{\bmugr},\tridnjktarg{0}{k'}{t'}{\bmugr},\ldots=
\ldots,\tridnjktarg{0}{k}{t}{\bmudt},\tridnjktarg{0}{k'}{t'}{\bmudt},\ldots,
$$
or
\be
\label{evaln4}
\ldots,\tridnjktarg{0}{k}{t}{\bmugr}=\tridnjktarg{0}{k}{t}{\bmudt},
\tridnjktarg{0}{k'}{t'}{\bmugr}=\tridnjktarg{0}{k'}{t'}{\bmudt},\ldots.
\ee
Similarly, if $\squaritarg{\bmk}{\bmt}{\bmuac}=\squaritarg{\bmk}{\bmt}{\bmuddt}$,
then
\be
\label{evaln5}
\ldots,\tridnjktarg{0}{k}{t}{\bmuac}=\tridnjktarg{0}{k}{t}{\bmuddt},
\tridnjktarg{0}{k'}{t'}{\bmuac}=\tridnjktarg{0}{k'}{t'}{\bmuddt},\ldots.
\ee
From Lemma \ref{lem74a}, the operation $\prodjktarg{0}{k}{t}{\ccirc}$ 
on $\tridnjktarg{0}{k}{t}{\calu}$ is well defined.  Then using
(\ref{evaln4}) and (\ref{evaln5}), we have verified (\ref{evaln3}).
\end{prf}

\begin{thm}
\label{thm92}
The set $\squaritarg{\bmk}{\bmt}{\calu}$ with operation $\proditarg{\bmk}{\bmt}{\oplus}$
forms a group $\squargrparg{\bmk}{\bmt}{\calu}{\oplus}$.
\end{thm}

\begin{prf}
The proof is essentially the same as Theorem \ref{thm41}.  We recapitulate this proof now.
First we show the operation $\proditarg{\bmk}{\bmt}{\oplus}$ is associative.
Let $\squaritarg{\bmk}{\bmt}{\bmu},\squaritarg{\bmk}{\bmt}{\bmudt},\squaritarg{\bmk}{\bmt}{\bmuddt}\in\squaritarg{\bmk}{\bmt}{\calu}$.
We need to show
$$
(\squaritarg{\bmk}{\bmt}{\bmu}\proditarg{\bmk}{\bmt}{\oplus}\squaritarg{\bmk}{\bmt}{\bmudt})\proditarg{\bmk}{\bmt}{\oplus}\squaritarg{\bmk}{\bmt}{\bmuddt}
=\squaritarg{\bmk}{\bmt}{\bmu}\proditarg{\bmk}{\bmt}{\oplus}(\squaritarg{\bmk}{\bmt}{\bmudt}\proditarg{\bmk}{\bmt}{\oplus}\squaritarg{\bmk}{\bmt}{\bmuddt}).
$$
But using (\ref{defx}) we have
\begin{align*}
(\squaritarg{\bmk}{\bmt}{\bmu}\proditarg{\bmk}{\bmt}{\oplus}\squaritarg{\bmk}{\bmt}{\bmudt})\proditarg{\bmk}{\bmt}{\oplus}\squaritarg{\bmk}{\bmt}{\bmuddt}
&=\squaritarg{\bmk}{\bmt}{\bmu\circ\bmudt}\proditarg{\bmk}{\bmt}{\oplus}\squaritarg{\bmk}{\bmt}{\bmuddt} \\
&=\squaritarg{\bmk}{\bmt}{(\bmu\circ\bmudt)\circ\,\bmuddt},
\end{align*}
and
\begin{align*}
\squaritarg{\bmk}{\bmt}{\bmu}\proditarg{\bmk}{\bmt}{\oplus}(\squaritarg{\bmk}{\bmt}{\bmudt}\proditarg{\bmk}{\bmt}{\oplus}\squaritarg{\bmk}{\bmt}{\bmuddt})
&=\squaritarg{\bmk}{\bmt}{\bmu}\proditarg{\bmk}{\bmt}{\oplus}\squaritarg{\bmk}{\bmt}{\bmudt\circ\bmuddt} \\
&=\squaritarg{\bmk}{\bmt}{\bmu\circ\,(\bmudt\circ\bmuddt)}.
\end{align*}
Therefore the operation $\proditarg{\bmk}{\bmt}{\oplus}$ is associative since the operation $\circ$ in 
group $(\calu,\circ)$ is associative.

Let $\bone$ be the identity of $(\calu,\circ)$.  We show
$\squaritarg{\bmk}{\bmt}{\bone}$ is the identity of $\squargrparg{\bmk}{\bmt}{\calu}{\oplus}$.
Let $\bmu\in\calu$ and $\squaritarg{\bmk}{\bmt}{\bmu}\in\squaritarg{\bmk}{\bmt}{\calu}$.  
But using (\ref{defx}) we have
\begin{align*}
\squaritarg{\bmk}{\bmt}{\bone}\proditarg{\bmk}{\bmt}{\oplus}\squaritarg{\bmk}{\bmt}{\bmu} 
&=\squaritarg{\bmk}{\bmt}{\bone\circ\bmu} \\
&=\squaritarg{\bmk}{\bmt}{\bmu}
\end{align*}
and
\begin{align*}
\squaritarg{\bmk}{\bmt}{\bmu}\proditarg{\bmk}{\bmt}{\oplus}\squaritarg{\bmk}{\bmt}{\bone} 
&=\squaritarg{\bmk}{\bmt}{\bmu\circ\bone} \\
&=\squaritarg{\bmk}{\bmt}{\bmu}.
\end{align*}

Fix $\bmu\in\calu$.  Let $\bmubr$ be the inverse of $\bmu$ in $(\calu,\circ)$.
We show $\squaritarg{\bmk}{\bmt}{\bmubr}$ is the inverse of $\squaritarg{\bmk}{\bmt}{\bmu}$
in $\squargrparg{\bmk}{\bmt}{\calu}{\oplus}$.  But using (\ref{defx}) we have
\begin{align*}
\squaritarg{\bmk}{\bmt}{\bmubr}\proditarg{\bmk}{\bmt}{\oplus}\squaritarg{\bmk}{\bmt}{\bmu} 
&=\squaritarg{\bmk}{\bmt}{\bmubr\circ\bmu} \\
&=\squaritarg{\bmk}{\bmt}{\bone}
\end{align*}
and
\begin{align*}
\squaritarg{\bmk}{\bmt}{\bmu}\proditarg{\bmk}{\bmt}{\oplus}\squaritarg{\bmk}{\bmt}{\bmubr} 
&=\squaritarg{\bmk}{\bmt}{\bmu\circ\bmubr} \\
&=\squaritarg{\bmk}{\bmt}{\bone}.
\end{align*}
Together these results show $\squargrparg{\bmk}{\bmt}{\calu}{\oplus}$ is a group.
\end{prf}

We have the following analogue of Theorems \ref{homo1} and \ref{homo3}.

\begin{thm}
\label{homo5}
There is a homomorphism from $(\calu,\circ)$ to $\squargrparg{\bmk}{\bmt}{\calu}{\oplus}$
given by the map $\omega_\bmk^\bmt:  \calu\ra\squaritarg{\bmk}{\bmt}{\calu}$ with assignment
$\omega_\bmk^\bmt:  \bmu\mapsto\squaritarg{\bmk}{\bmt}{\bmu}$.
\end{thm}

\begin{prf}
This follows immediately from the definition of operation $\proditarg{\bmk}{\bmt}{\oplus}$.
\end{prf}

Consider a second sequence of indices $\bmk'=\ldots,(0,n),(0,n'),\ldots$ at 
the same times $\bmt=\ldots,t,t',\ldots$ as for $\bmk$, such that
indices $\ldots,n,n',\ldots$ and $\ldots,k,k',\ldots$ are related by
$0\le n\le k$ at each time $t\in\bmt$.  Fix $\bmu$ in $\calu$.
Then triangle $\tridnjktarg{0}{n}{t}{\bmu}$ is wholely contained
in triangle $\tridnjktarg{0}{k}{t}{\bmu}$ at each time $t\in\bmt$.
We can construct a group $\squargrparg{\bmk'}{\bmt}{\calu}{\oplus}$
for $\bmk'$ and $\bmt$ in the same way as for $\bmk$ and $\bmt$.
The following shows there is a homomorphism between the two groups.

\begin{thm}
\label{homo6}
There is a homomorphism from $\squargrparg{\bmk}{\bmt}{\calu}{\oplus}$ to 
$\squargrparg{\bmk'}{\bmt}{\calu}{\oplus}$ given by the projection map 
$\ovomega:  \squaritarg{\bmk}{\bmt}{\calu}\ra\squaritarg{\bmk'}{\bmt}{\calu}$
with assignment $\ovomega:  \squaritarg{\bmk}{\bmt}{\bmu}\mapsto\squaritarg{\bmk'}{\bmt}{\bmu}$.
\end{thm}

\begin{prf}
The proof is an analogue of the proof of Theorem \ref{homo2} and Theorem \ref{homo4} with assignment
$\ovomega:  \squaritarg{\bmk}{\bmt}{\bmu}\mapsto\squaritarg{\bmk'}{\bmt}{\bmu}$
in place of assignments $\ovphi:  \tridnjktarg{j}{k}{t}{\bmr}\mapsto\tridnjktarg{m}{n}{t}{\bmr}$
and $\ovtheta:  \tridnjktarg{0}{k}{t}{\bmu}\mapsto\tridnjktarg{0}{m}{t-n}{\bmu}$,
respectively.
\end{prf}

We now show there is a homomorphism from $A$ to $\squargrparg{\bmk}{\bmt}{\calu}{\oplus}$.
Combining Theorem \ref{homo7} and Theorem \ref{homo5} gives the following.

\begin{thm}
\label{homo9}
There is a homomorphism from $A$ to $\squargrparg{\bmk}{\bmt}{\calu}{\oplus}$
given by the composition map
$\omega_\bmk^\bmt\bullet\xi:  A\ra\squaritarg{\bmk}{\bmt}{\calu}$,
where isomorphism $A\simeq(\calu,\circ)$ is given by the bijection 
$\xi:  A\ra\calu$ in Theorem \ref{homo7}, and the
homomorphism from $(\calu,\circ)$ to $\squargrparg{\bmk}{\bmt}{\calu}{\oplus}$
is given by the map $\omega_\bmk^\bmt:  \calu\ra\squaritarg{\bmk}{\bmt}{\calu}$
in Theorem \ref{homo5}.
\end{thm}
We use the homomorphism in Theorem \ref{homo9} to study the structure of block group systems
in Section 8.2.

The sequences in $\squaritarg{\bmk}{\bmt}{\calu}$ at times $t\in\bmt$ are a subset of
the Cartesian product
\be
\label{seqtwo}
\cdots\times\tridnjktarg{0}{k}{t}{\calu}\times\tridnjktarg{0}{k'}{t'}{\calu}\cdots
\ee
at times $t\in\bmt$.  We denote this (usually proper) subset of (\ref{seqtwo})
by the notation
\be
\label{seqtwoa}
\squaritarg{\bmk}{\bmt}{\calu}=\cdots\Join\tridnjktarg{0}{k}{t}{\calu}\Join\tridnjktarg{0}{k'}{t'}{\calu}\Join\cdots.
\ee
We say $\squaritarg{\bmk}{\bmt}{\calu}$ is the {\it trellis product}
of the sets $\ldots,\tridnjktarg{0}{k}{t}{\calu},\tridnjktarg{0}{k'}{t'}{\calu},\ldots$.

From (\ref{evaln2}),
note that $\squargrparg{\bmk}{\bmt}{\calu}{\oplus}$ has component groups isomorphic to 
$$
\ldots,\grpjktarg{0}{k}{t}{\calu}{\ccirc},\grpjktarg{0}{k'}{t'}{\calu}{\ccirc},\ldots
$$
at times $t\in\bmt$.
We indicate this by writing $\squargrparg{\bmk}{\bmt}{\calu}{\oplus}$ as
\be
\label{cmpgrps}
\squargrparg{\bmk}{\bmt}{\calu}{\oplus}=\cdots\Join\grpjktarg{0}{k}{t}{\calu}{\ccirc}
\Join\grpjktarg{0}{k'}{t'}{\calu}{\ccirc}\Join\cdots.
\ee
We say $\squargrparg{\bmk}{\bmt}{\calu}{\oplus}$ is the {\it trellis product}
of the component groups in (\ref{cmpgrps}) (not necessarily a semidirect or direct product group).

The groups $\grpjktarg{0}{k}{t}{\calu}{\ccirc}$ and $\grpjktarg{0}{k'}{t'}{\calu}{\ccirc}$
in (\ref{cmpgrps}) are elementary groups of $(\calu,\circ)$.  Consider any pair of
elementary groups $\grpjktarg{0}{k}{t}{\calu}{\ccirc}$ and $\grpjktarg{0}{k'}{t'}{\calu}{\ccirc}$
in $(\calu,\circ)$.  We know this pair of elementary groups forms the trellis product
\be
\label{tpg1}
\grpjktarg{0}{k}{t}{\calu}{\ccirc}\Join\grpjktarg{0}{k'}{t'}{\calu}{\ccirc}.
\ee
There are three cases to consider in (\ref{tpg1}).
First, the groups in (\ref{tpg1}) may completely overlap, in which case (\ref{tpg1})
simply reduces to the larger of the two elementary groups.  Second they may have a nontrivial
intersection, in which case their intersection is another elementary
group.  Third they may be disjoint.  Without loss of generality we consider the last two cases.

In the language of Hall \cite{MH}, (\ref{tpg1}) is a subdirect product group.
The structure of a subdirect product group is given by Theorem 5.5.1
of Hall \cite{MH}.  We restate Theorem 5.5.1 \cite{MH} for the subdirect product group here.

\begin{thm}
\label{thm130}
Let $G^*$ be the subdirect product of the groups $G_1$ and $G_2$, and let $H_1$
and $H_2$ be the subgroups of $G_1$ and $G_2$, respectively, of elements of one
factor occurring in $G^*$ with the identity of the other factor.
Then $H_1$ is normal in $G_1$ and $H_2$ is normal in $G_2$, and there is an isomorphism
between the factor groups $G_1/H_1\simeq K\simeq G_2/H_2$ such that $(g_1,g_2)$,
$g_1\in G_1$, $g_2\in G_2$, is an element of $G^*$ if and only if $g_1$ and $g_2$
have the same image $k$ in the homomorphisms $G_1\ra K$, $G_2\ra K$.
\end{thm}
$G^*$ is a subgroup of the direct product group $G_1\times G_2$, and
the elements $(g_1,g_2)$ of $G^*$ are a subset of the Cartesian product
$G_1\bigotimes G_2$.

We now apply Theorem \ref{thm130} to our case.  Let $G_1$ be the group 
$\grpjktarg{0}{k}{t}{\calu}{\ccirc}$ and
let $G_2$ be the group $\grpjktarg{0}{k'}{t'}{\calu}{\ccirc}$.
Let $G^*$ be the subdirect product group (\ref{tpg1}).
Let $\bone_\calu$ be the identity of $(\calu,\circ)$.
Let $\bone_{G_1}$ be the identity of $G_1$ and $\bone_{G_2}$ be the identity of $G_2$.
We have 
\begin{align*}
H_1 &=\{\tridnjktarg{0}{k}{t}{\bmu}:  \bmu\in\calu\text{ and }\tridnjktarg{0}{k'}{t'}{\bmu}
=\tridnjktarg{0}{k'}{t'}{\bone_\calu}\} \\
    &=\{\tridnjktarg{0}{k}{t}{\bmu}:  \bmu\in\calu\text{ and }\tridnjktarg{0}{k'}{t'}{\bmu}
=\bone_{G_2}\}.
\end{align*}
And similarly
\begin{align*}
H_2 &=\{\tridnjktarg{0}{k'}{t'}{\bmu}:  \bmu\in\calu\text{ and }\tridnjktarg{0}{k}{t}{\bmu}
=\tridnjktarg{0}{k}{t}{\bone_\calu}\} \\
    &=\{\tridnjktarg{0}{k'}{t'}{\bmu}:  \bmu\in\calu\text{ and }\tridnjktarg{0}{k}{t}{\bmu}
=\bone_{G_1}\}.
\end{align*}

We now apply the Hall theorem to the third case when the two groups are disjoint.
Since the groups are disjoint and we may choose the columns of $\bmu$ independently of one another,
we have $H_1=G_1$ and $H_2=G_2$.  Then $G^*=G_1\bigotimes G_2$, and $G^*$ is the direct product
$G_1\times G_2$.  In other words,
$$
\grpjktarg{0}{k}{t}{\calu}{\ccirc}\Join\grpjktarg{0}{k'}{t'}{\calu}{\ccirc}
=\grpjktarg{0}{k}{t}{\calu}{\ccirc}\times\grpjktarg{0}{k'}{t'}{\calu}{\ccirc}.
$$

We now give the structure of pairs of elementary groups that intersect.
Consider the elementary groups $G_1$ and $G_2$
again but now assume their sets of elements $\tridnjktarg{0}{k}{t}{\calu}$
and $\tridnjktarg{0}{k'}{t'}{\calu}$ intersect.  We have asssumed
$t'<t$.  Then the intersection set is a triangle set
$\tridnjktarg{0}{k-(t-t')}{t'}{\calu}$.  There is an elementary group
$\grpjktarg{0}{k-(t-t')}{t'}{\calu}{\cast}$ defined on this set.  Call
this group $G_3$ and let $\bone_{G_3}$ be the identity of $G_3$.  
The $\tridnjktarg{0}{k}{t}{\bmu}$ in $G_1$ that are in $H_1$ are determined
by the intersection triangle $\tridnjktarg{0}{k-(t-t')}{t'}{\bmu}$.  We have
\begin{align}
\nonumber
H_1 &=\{\tridnjktarg{0}{k}{t}{\bmu}:  \bmu\in\calu\text{ and }\tridnjktarg{0}{k'}{t'}{\bmu}
=\tridnjktarg{0}{k'}{t'}{\bone_\calu}\} \\
\nonumber
    &=\{\tridnjktarg{0}{k}{t}{\bmu}:  \bmu\in\calu\text{ and }\tridnjktarg{0}{k-(t-t')}{t'}{\bmu}
\label{sd1}
=\tridnjktarg{0}{k-(t-t')}{t'}{\bone_\calu}\} \\
    &=\{\tridnjktarg{0}{k}{t}{\bmu}:  \bmu\in\calu\text{ and }\tridnjktarg{0}{k-(t-t')}{t'}{\bmu}
=\bone_{G_3}\}.
\end{align}
And similarly
\begin{align}
\nonumber
H_2 &=\{\tridnjktarg{0}{k'}{t'}{\bmu}:  \bmu\in\calu\text{ and }\tridnjktarg{0}{k}{t}{\bmu}
=\tridnjktarg{0}{k}{t}{\bone_\calu}\} \\
\nonumber
    &=\{\tridnjktarg{0}{k'}{t'}{\bmu}:  \bmu\in\calu\text{ and }\tridnjktarg{0}{k-(t-t')}{t'}{\bmu}
=\tridnjktarg{0}{k-(t-t')}{t'}{\bone_\calu}\} \\
\label{sd2}
    &=\{\tridnjktarg{0}{k'}{t'}{\bmu}:  \bmu\in\calu\text{ and }\tridnjktarg{0}{k-(t-t')}{t'}{\bmu}
=\bone_{G_3}\}.
\end{align}
Then $H_1$ is normal in $G_1$ and $H_2$ is normal in $G_2$, and $K$ in Theorem \ref{thm130}
is isomorphic to $G_3$.  There is an isomorphism between the factor groups $G_1/H_1\simeq G_3\simeq G_2/H_2$ 
such that $(g_1,g_2)$, $g_1\in G_1$, $g_2\in G_2$, is an element of $G^*$ if and only if $g_1$ and $g_2$
have the same image $g_3$ in the homomorphisms $G_1\ra G_3$, $G_2\ra G_3$ given by the 
isomorphism $G_1/H_1\simeq G_3\simeq G_2/H_2$.

\vspace{3mm}
{\bf 6.4  Harmonic theory (normal chains of the generator group)}
\vspace{3mm}

In this subsection, we use a sequence slightly more general than an index sequence,
called a boundary sequence.  A boundary sequence need not have all its ordered pairs in $\bmn$.
We use a sequence of boundary sequences at time epochs
$0,1,\ldots,\tau,\tau+1,\ldots$to ``fill'' all index pairs in index 
tensor $\bmn$.  At each time epoch $\tau$, we use the boundary sequence 
$(\bmk^\tau,\bmt^\tau)$ at time $\tau$ to fill one more index pair in $\bmn$.
We now describe this approach in more detail.

A {\it boundary sequence} $(\bmk^\tau,\bmt^\tau)$ at time epoch $\tau$,
$$
(\bmk^\tau,\bmt^\tau)\rmdef\ldots,(k',t+1),(k,t),(k'',t-1),\ldots,
$$
is a paired sequence of ordered pairs $(k,t)$ where time sequence $\bmt^\tau$
is the integers $\bmcpz$ written in reverse time order, $\ldots,t+1,t,t-1,\ldots$,
and $\bmk^\tau$ is a sequence of integers 
$\ldots,k,k',k'',\ldots$ with each integer $k$ satisfying $-1\le k\le\ell$.
We can ``fill'' $\bmn$ from bottom row to top row using a sequence of boundary sequences 
at time epochs $0,1,\ldots,\tau,\tau+1,\ldots$,
\be
\label{fs}
(\bmk^0,\bmt^0),(\bmk^1,\bmt^1),\ldots,(\bmk^\tau,\bmt^\tau),(\bmk^{\tau+1},\bmt^{\tau+1}),\ldots.
\ee
The initial boundary sequence is
$$
(\bmk^0,\bmt^0)\rmdef\ldots,(-1,t+1),(-1,t),(-1,t-1),\ldots,(-1,t'),\ldots.
$$
The initial boundary sequence does not have any ordered pairs in $\bmn$.
We define $(\bmk^1,\bmt^1)$ to be the same boundary sequence as $(\bmk^0,\bmt^0)$
except that for some time $t'$, an ordered pair $(-1,t')$ in $(\bmk^0,\bmt^0)$
is replaced by ordered pair $(0,t')$ in $(\bmk^1,\bmt^1)$,
$$
(\bmk^1,\bmt^1)\rmdef\ldots,(-1,t+1),(-1,t),(-1,t-1),\ldots,(0,t'),\ldots.
$$
The ordered pair $(0,t')$ is an element of index tensor $\bmn$, but the remaining ordered pairs
in $(\bmk^1,\bmt^1)$ are not elements of $\bmn$.  We call ordered pair $(0,t')$ a 
{\it filled ordered pair} in $\bmn$; the remaining ordered pairs in $\bmn$ are
{\it unfilled ordered pairs}.  In general, for a boundary sequence
\be
\label{fop1}
(\bmk^\tau,\bmt^\tau)\rmdef\ldots,(k',t+1),(k,t),(k'',t-1),\ldots,(k''',t''),\ldots,
\ee
at time epoch $\tau$, we define a boundary sequence $(\bmk^{\tau+1},\bmt^{\tau+1})$
at time epoch $\tau+1$ to be the same as $(\bmk^\tau,\bmt^\tau)$
except that for some time $t''$, there is an ordered pair $(k''',t'')$ in $(\bmk^\tau,\bmt^\tau)$
where $-1\le k'''<\ell$, that is replaced by ordered pair $(k'''+1,t'')$ in $(\bmk^{\tau+1},\bmt^{\tau+1})$, 
\be
\label{fop2}
(\bmk^{\tau+1},\bmt^{\tau+1})\rmdef\ldots,(k',t+1),(k,t),(k'',t-1),\ldots,(k'''+1,t''),\ldots.
\ee
Then $(k'''+1,t'')$ is a filled ordered pair in $\bmn$.
Then at time epoch $\tau+1$, we have filled
the ordered pairs $(0,t'),\ldots,(k'''+1,t'')$ in $\bmn$.
In general, at each time epoch $\tau+1$, we fill one more ordered pair in $\bmn$ than
at time epoch $\tau$.  If we continue in this way, 
we eventually fill all ordered pairs in $\bmn$.  We call such a sequence
in (\ref{fs}) a {\it filling sequence} of $\bmn$.

Any filling sequence of $\bmn$ uniquely
determines a sequence of filled ordered pairs in $\bmn$.  For example, (\ref{fs}) determines
the sequence $(0,t'),\ldots,(k'''+1,t''),\ldots$ of filled ordered pairs.
And conversely any sequence of filled ordered pairs of $\bmn$ uniquely
determines a filling sequence of $\bmn$.  Therefore we can specify any filling sequence
directly, or indirectly through its sequence of filled ordered pairs.

The boundary sequence $(\bmk^\tau,\bmt^\tau)$ at time epoch $\tau$ has the property that 
for each $t\in\bmcpz$, for each ordered pair $(k,t)$ in $(\bmk^\tau,\bmt^\tau)$, 
all ordered pairs $(\kdt,t)$ in $\bmn$ where $-1<\kdt\le k$ have been filled by the sequence 
$(\bmk^0,\bmt^0),(\bmk^1,\bmt^1),\ldots,(\bmk^\tau,\bmt^\tau)$, and all ordered pairs 
$(\kddt,t)$ in $\bmn$ where $k<\kddt\le\ell$ have not been filled by the sequence
$(\bmk^0,\bmt^0),(\bmk^1,\bmt^1),\ldots,(\bmk^\tau,\bmt^\tau)$.
Thus $(\bmk^\tau,\bmt^\tau)$ serves as a boundary between filled and unfilled ordered pairs in $\bmn$.
In addition, the boundary sequence $(\bmk^\tau,\bmt^\tau)$ has the property that for each $t\in\bmcpz$,
the filled ordered pairs $(\kdt,t)$ in column $t$ of $\bmn$, where $-1<\kdt\le k$, appear in the 
sequence $(\bmk^0,\bmt^0),(\bmk^1,\bmt^1),\ldots,(\bmk^\tau,\bmt^\tau)$ in the order
$(0,t),\ldots,(\kdt,t),\ldots,(k,t)$ of ascending $k$.

We now subject the filling sequence to another constraint.  At each time epoch $\tau$,
for each time $t\in\bmcpz$, for all ordered pairs $(k,t)$ in the boundary 
sequence $(\bmk^\tau,\bmt^\tau)$ such that $k<\ell$, we require that triangle 
$\tridnjktarg{0}{k+1}{t}{\bmn}$ contains only unfilled ordered pairs in $\bmn$.
We call such a boundary sequence a {\it normal boundary sequence}, and a filling sequence (\ref{fs})
where all boundary sequences are normal, a {\it normal filling sequence}.

We can use any normal filling sequence (\ref{fs}) to ``fill'' the index tensor $\bmn$.
There are many many normal filling sequences of $\bmn$.  We now give 4 
important examples of normal filling sequences of $\bmn$.  We specify the filling
sequence by the sequence of filled ordered pairs.  If we go left to right
in (\ref{top}), the time index decreases.  We call this {\it backward time}.
If we go right to left, the time index increases.  We call this {\it forward time}.
For the first example of a sequence of filled ordered pairs, we go up the columns in (\ref{top}),
from left to right.  This gives the sequence of filled ordered pairs $(k,t)$,
\be
\label{tdndsbt}
\ldots,(0,t),(1,t),(2,t),\ldots,(j,t),\ldots,(\ell,t),
              (0,t-1),(1,t-1),(2,t-1),\ldots,(j,t-1),\ldots,(\ell,t-1),\ldots.
\ee
It can be easily verified that this gives a normal filling sequence.
We call this the {\it time domain normal filling sequence of $\bmn$} in {\it backward time}.
The reason for this name will be apparent at the end of the subsection.
For the second example of a sequence of filled ordered pairs,
we go up the diagonals in (\ref{top}), from right to left.
This gives the sequence of filled ordered pairs $(k,t)$,
$$
\ldots,(0,t),(1,t-1),(2,t-2),\ldots,(j,t-j),\ldots,(\ell,t-\ell),
              (0,t+1),(1,t),(2,t-1),\ldots,(j,t-j+1),\ldots,(\ell,t-\ell+1),\ldots.
$$
We call this the time domain normal filling sequence of $\bmn$ in {\it forward time}.
Note that we go up the columns in backward time and up the diagonals in forward time
because of the triangular shape of $\tridnjktarg{0}{0}{t}{\bmu}$.
Next we go row by row in (\ref{top}), from bottom row to top row and left to right.
This gives the sequence of filled ordered pairs $(k,t)$,
\be
\label{sdndsbt}
\ldots,(0,t),(0,t-1),\ldots,\ldots,(1,t),(1,t-1),\ldots,(2,t),(2,t-1),
\ldots,(j,t),(j,t-1),\ldots,(\ell,t),(\ell,t-1),\ldots.
\ee
We call this the {\it spectral domain normal filling sequence of $\bmn$} in {\it backward time}.
Finally we may go row by row in (\ref{top}), from bottom row to top row and right to left.
This gives the sequence of filled ordered pairs $(k,t)$,
$$
\ldots,(0,t-1),(0,t),\ldots,(1,t-1),(1,t),\ldots,(2,t-1),(2,t),
\ldots,(j,t-1),(j,t),\ldots,(\ell,t-1),(\ell,t),\ldots.
$$
We call this the spectral domain normal filling sequence of $\bmn$ in {\it forward time}.

We can restate the condition to have a normal boundary sequence in a different way.
Let a boundary sequence $(\bmk^\tau,\bmt^\tau)$ be given by the sequence of
ordered pairs $\ldots,(k,t),\ldots$.
Let $(\bmk_q^\tau,\bmt_q^\tau)$ be the subsequence of ordered pairs $(k,t)$ of $(\bmk^\tau,\bmt^\tau)$
such that $k<\ell$.  Let $(\bmk_{q+}^\tau,\bmt_q^\tau)$ be the subsequence of ordered pairs
obtained from $(\bmk_q^\tau,\bmt_q^\tau)$ by replacing each ordered pair $(k,t)$ in 
$(\bmk_q^\tau,\bmt_q^\tau)$ with ordered pair $(k^+,t)$ in $(\bmk_{q+}^\tau,\bmt_q^\tau)$, 
where $k^+=k+1$.  Then the condition on $(\bmk^\tau,\bmt^\tau)$ to be a normal boundary 
sequence is that for each ordered pair $(k^+,t)$ in paired sequence $(\bmk_{q+}^\tau,\bmt_q^\tau)$, 
the triangle $\tridnjktarg{0}{k^+}{t}{\bmn}$ 
contains only unfilled ordered pairs in $\bmn$.

\begin{lem}
Let $(\bmk^\tau,\bmt^\tau)$ be a normal boundary sequence at any time epoch $\tau$.
The set of triangles
\be
\label{set}
\cup_{t\in\bmt_q^\tau} \tridnjktarg{0}{k^+}{t}{\bmn},
\ee
indexed by paired sequence $(\bmk_{q+}^\tau,\bmt_q^\tau)$, is the set of all unfilled ordered pairs
in $\bmn$ at time epoch $\tau$.
\end{lem}

\begin{prf}
By definition of normal boundary sequence, we know that each triangle in union set (\ref{set}) only 
contains unfilled ordered pairs.  Therefore the union set only contains 
unfilled ordered pairs.  It remains to prove that union set (\ref{set}) contains any
unfilled ordered pair.  Any ordered pair $(k,t)$ in $(\bmk^\tau,\bmt^\tau)$ is filled.
There are two cases to consider.  Suppose $(k,t)=(\ell,t)$.  Then $(\ell,t)$ is filled
in $\bmn$, and all ordered pairs in column $t$ of $\bmn$ are filled.
Suppose $k<\ell$.  Then $t\in\bmt^\tau$.  
And any unfilled ordered pair $(k',t)$ in column $t$ of $\bmn$ must satisfy $k<k'\le\ell$.
Then $(k',t)$ is contained in triangle $\tridnjktarg{0}{k+1}{t}{\bmn}$ where $t\in\bmt^\tau$.
Then $(k',t)$ is contained in union set (\ref{set}).
\end{prf}

Let $(\bmk^\tau,\bmt^\tau)$ be a normal boundary sequence.  We define a subset of $\calu$
using the paired sequence $(\bmk_{q+}^\tau,\bmt_q^\tau)$.
Define $\capkcalu{\tau}$ to be the subset of sequences 
$\{\bmu\in\calu:  \squaritarg{\bmk_{q+}^\tau}{\bmt_q^\tau}{\bmu}=\squaritarg{\bmk_{q+}^\tau}{\bmt_q^\tau}{\bone}\}$,
where $\bone$ is the identity of $(\calu,\circ)$.  Note that $\capkcalu{\tau}$
is a subset of $\calu$, whereas $\squaritarg{\bmk_{q+}^\tau}{\bmt_q^\tau}{\bmu}$ is a sequence
of triangles
$$
\ldots,\tridnjktarg{0}{k^+}{t}{\bmu},\tridnjktarg{0}{(k')^+}{t'}{\bmu},\ldots
$$
of $\bmu\in\calu$, as in (\ref{seq1}).

\begin{lem}
\label{lem81}
Let
\be
\label{nfs}
(\bmk^0,\bmt^0),(\bmk^1,\bmt^1),\ldots,(\bmk^\tau,\bmt^\tau),(\bmk^{\tau+1},\bmt^{\tau+1}),\ldots
\ee
be a normal filling sequence of $\bmn$.  At each time epoch $\tau$, we have
$\capkcalu{\tau}\subset\capkcalu{\tau+1}$.
\end{lem}

\begin{prf}
Consider any normal boundary sequence $(\bmk^\tau,\bmt^\tau)$ at any time epoch $\tau$.
For any time epoch $\tau$, the entries in the sequence of triangles
\be
\label{seq11}
\ldots,\tridnjktarg{0}{k^+}{t}{\bmn},\tridnjktarg{0}{(k')^+}{t'}{\bmn},\ldots
\ee
of set (\ref{set}), indexed by paired sequence $(\bmk_{q+}^\tau,\bmt_q^\tau)$,
are in 1-1 correspondence with the entries in the sequence of triangles 
\be
\label{seq2}
\ldots,\tridnjktarg{0}{k^+}{t}{\bone},\tridnjktarg{0}{(k')^+}{t'}{\bone},\ldots
\ee
of $\squaritarg{\bmk_{q+}^\tau}{\bmt_q^\tau}{\bone}$.  But the entries in the 
sequence of triangles (\ref{seq11}) are the unfilled ordered pairs in $\bmn$ at
time epoch $\tau$.  Since the unfilled ordered pairs in $\bmn$ at
time epoch $\tau+1$ are contained in the unfilled ordered pairs in $\bmn$ at
time epoch $\tau$, we have the entries in the sequence of triangles of
$\squaritarg{\bmk_{q+}^{\tau+1}}{\bmt_q^{\tau+1}}{\bone}$ are contained in 
the entries in the sequence of triangles (\ref{seq2}) of 
$\squaritarg{\bmk_{q+}^\tau}{\bmt_q^\tau}{\bone}$.  Then 
$\capkcalu{\tau}\subset\capkcalu{\tau+1}$. 
\end{prf}

\begin{thm}
For any normal boundary sequence $(\bmk^\tau,\bmt^\tau)$ at any time epoch $\tau$, 
$(\capkcalu{\tau},\circ)$ is a normal subgroup of $(\calu,\circ)$.
\end{thm}

\begin{prf}
Let $\bmu\in\capkcalu{\tau}$ and let $\bmubr\in\calu$.  We have to show that 
$\bmubr\circ\bmu\circ\bmubr^{-1}$ is an element of $\capkcalu{\tau}$.
From Lemma \ref{lem95a}, the evaluation of $\bmubr\circ\bmu\circ\bmubr^{-1}$ is 
equivalent to the evaluation 
\be
\label{fop3}
\tridnjktarg{0}{0}{t}{\bmubr}\prodjktarg{0}{0}{t}{\ccirc}\tridnjktarg{0}{0}{t}{\bmu}
\prodjktarg{0}{0}{t}{\ccirc}\tridnjktarg{0}{0}{t}{\bmubr^{-1}}
\ee
for each $t\in\bmcpz$.  Then to show that 
$\bmubr\circ\bmu\circ\bmubr^{-1}\in\capkcalu{t}$, it is sufficient to show that
\be
\label{fop4}
\tridnjktarg{0}{k^+}{t}{\bmubr}\prodjktarg{0}{k^+}{t}{\ccirc}\tridnjktarg{0}{k^+}{t}{\bmu}
\prodjktarg{0}{k^+}{t}{\ccirc}\tridnjktarg{0}{k^+}{t}{\bmubr^{-1}}=\tridnjktarg{0}{k^+}{t}{\bone}
\ee
for each $t\in\bmt_q^\tau$.  But by definition, 
$\tridnjktarg{0}{k^+}{t}{\bmu}=\tridnjktarg{0}{k^+}{t}{\bone}$ for each $t\in\bmt_q^\tau$.
And by properties of elementary groups, we have that 
$$
\tridnjktarg{0}{k^+}{t}{\bmubr}\prodjktarg{0}{k^+}{t}{\ccirc}\tridnjktarg{0}{k^+}{t}{\bone}
\prodjktarg{0}{k^+}{t}{\ccirc}\tridnjktarg{0}{k^+}{t}{\bmubr^{-1}}=\tridnjktarg{0}{k^+}{t}{\bone}
$$
for each $t\in\bmt_q^\tau$.  This proves (\ref{fop4}).  
\end{prf}

Since $\capkcalu{\tau}$ is a normal subroup of $(\calu,\circ)$ for any normal boundary sequence 
$(\bmk^\tau,\bmt^\tau)$ at any time epoch $\tau$, then from Lemma \ref{lem81}, for any normal 
filling sequence (\ref{nfs}) of $\bmn$ we may construct a normal chain
\be
\label{nc1}
\bone=\capkcalu{0}\subset\capkcalu{1}\subset\cdots\subset\capkcalu{\tau}\subset\capkcalu{\tau+1}\subset\cdots\subset(\calu,\circ).
\ee
Note that the first term $\capkcalu{0}$ is the identity $\bone$ of $(\calu,\circ)$.
We say (\ref{nc1}) is the {\it normal chain of a normal filling sequence} of $\bmn$.
This gives the following.

\begin{thm}
Any normal filling sequence of $\bmn$ gives a normal chain (\ref{nc1}) of $(\calu,\circ)$.
\end{thm}
There are many many normal chains of $(\calu,\circ)$ because there are many many normal 
filling sequences of $\bmn$.

Each normal chain (\ref{nc1}) of $(\calu,\circ)$ gives a \cdc\ of $(\calu,\circ)$.
We now find \creps\ of the \cdc.  First we find a \crep\ of the quotient group 
$\capkcalu{\tau+1}/\capkcalu{\tau}$.  A \crep\ of $\capkcalu{\tau+1}/\capkcalu{\tau}$ is an element of 
$\capkcalu{\tau+1}$ that is not an element of $\capkcalu{\tau}$.

We say an element $\bmu\in\calu$ is a {\it generator} $\bmu_{g,k}^t$ of $(\calu,\circ)$ if $\bmu_{g,k}^t$
contains one and only one nontrivial \glab\ $r_{0,k}^t$ for some $k$ such that $0\le k\le\ell$ 
and some time $t\in\bmcpz$.  As in (\ref{fop1}) and (\ref{fop2}),
suppose $(k'''+1,t'')$ is the filled ordered pair
of $(\bmk^{\tau+1},\bmt^{\tau+1})$ that is not filled in $(\bmk^\tau,\bmt^\tau)$.
Then generator $\bmu_{g,k'''+1}^{t''}$ of $(\calu,\circ)$ is an element of 
$\capkcalu{\tau+1}$ that is not an element of $\capkcalu{\tau}$.  This gives the following.

\begin{thm}
The generators of $(\calu,\circ)$ form a \crepc\ of any \cdc\ of $(\calu,\circ)$
formed by the normal chain (\ref{nc1}) of $(\calu,\circ)$ of any normal filling 
sequence (\ref{nfs}) of $\bmn$.
\end{thm}

Given a generator $\bmg^{[t,t+k]}$ in $B$, there is a corresponding generator 
$\bmg_A^{[t,t+k]}$ in $A$, given by the bijection $\alpha:  B\ra A$ in Theorem \ref{thm1}.
Then there is a bijection $\alpha:  B\ra A$ between 
the set of generators $\bmg^{[t,t+k]}$ in $B$ and the set of generators 
$\bmg_A^{[t,t+k]}$ in $A$.  Under the bijection $\eta^{-1}:  B\ra\calr$ in Theorem \ref{thm31}, 
the generator $\bmg^{[t,t+k]}$ in $B$ gives a tensor $\bmr\in\calr$ containing just one
nontrivial generator $\bmg^{[t,t+k]}$.  Further,  under the bijection $\beta:  \calr\ra\calu$ 
in Theorem \ref{thm88a}, the tensor $\bmr\in\calr$ containing just one
nontrivial generator $\bmg^{[t,t+k]}$ gives a generator 
$\bmu_{g,k}^t$ in $(\calu,\circ))$.  Then there is a bijection $\beta\bullet\eta^{-1}:  B\ra\calu$
between the set of generators $\bmg^{[t,t+k]}$ in $B$ and the set of generators 
$\bmu_{g,k}^t$ in $(\calu,\circ)$.

Thus there is a bijection $\eta\bullet\beta^{-1}:  \calu\ra B$ which takes the set of generators 
$\bmu_{g,k}^t$ in $(\calu,\circ)$ to the set of generators $\bmg^{[t,t+k]}$ in $B$, and a 
bijection $\alpha\bullet\eta\bullet\beta^{-1}:  \calu\ra A$ which takes the set of generators 
$\bmu_{g,k}^t$ in $(\calu,\circ)$ to the set of generators $\bmg_A^{[t,t+k]}$ in $A$.
Furthermore, the bijection $\eta\bullet\beta^{-1}$ is an isomorphism between between
$(\calu,\circ)$ and $B$,
and the bijection $\alpha\bullet\eta\bullet\beta^{-1}$ is an isomorphism between
$(\calu,\circ)$ and $A$.
Thus we can reconstruct the set of elements $B$ and $A$ using the \creps, i.e, generators, of 
the \cdc\ of $(\calu,\circ)$ obtained from any normal filling sequence of $\bmn$.

If we use generators of $(\calu,\circ)$ obtained from the normal chain of the time domain
normal filling sequence of $\bmn$ in backward time (\ref{tdndsbt}), 
we obtain the time domain encoder of $B$ 
in (\ref{enctdc}).  If we use generators of $(\calu,\circ)$ obtained from the normal chain of the 
spectral domain normal filling sequence of $\bmn$ in backward time (\ref{sdndsbt}), 
we obtain the Forney-Trott spectral domain encoder of $A$ 
in (\ref{encftd}).  Thus the time domain encoder of $B$ and $A$ discussed in Section 3
(see also \cite{KM5}) and the spectral domain encoder of $A$ in \cite{FT} can be more easily
obtained just from the normal chain of their respective normal filling sequence of $\bmn$.
In addition, there are many other encoders and expansions of $B$ and $A$ that can be obtained from
the normal chain of many other normal filling sequences of $\bmn$.

In the last chapter of his Ph.D. thesis on
group codes \cite{TR1}, Mitchell Trott mentions that ``... surely there must be
an orthogonal decomposition..." of group codes.  We believe that the generators
in a \crepc\ of the generator group partially fulfill the goal 
of an orthogonal decomposition.  However, note that in general the product of two generators
in a \crepc\ may give more than two generators unless the generator group is a direct
product group.

The generator group has perfect symmetry with respect to time.
But when we attempt to ``observe'' the generator group by using a \crepc\ to find a group element
in the group system, 
the symmetry is ``broken''.  For example, the time domain encoder in forward time and backward time
may give different group elements unless the generator group is abelian.  In the same way, any two
normal chains of the generator group may give different group elements unless the generator group 
is abelian.

\vspace{3mm}
{\bf 6.5  Recovery of group system $A$ from generator group $(\calu,\circ)$}
\vspace{3mm}

We first restate and prove Theorem \ref{thmone}, the {\it first homomorphism theorem for group systems}.

\begin{thm}
\label{fhgs}
Consider any group $\msfcpg$.  Suppose there is a homomorphism $p^t:  \msfcpg\ra \msfcpg^t$ from $\msfcpg$
to a group $\msfcpg^t$ for each $t\in\bmcpz$.  In general group $\msfcpg^t$ may be different for each 
$t\in\bmcpz$.  Define the direct product group $(\msfcpg_\amalg,\plus)$ 
by $(\msfcpg_\amalg,\plus)\rmdef\cdots\times \msfcpg^t\times \msfcpg^{t+1}\times\cdots$.
There is a homomorphism $p:  G\ra G_\amalg$, from $\msfcpg$ to the direct product group 
$(\msfcpg_\amalg,\plus)$, defined by
\be
\label{eqthm37}
p(\msfg)\rmdef\ldots,p^t(\msfg),p^{t+1}(\msfg),\ldots.
\ee
Define $\msfg_\amalg\rmdef\ldots,p^t(\msfg),p^{t+1}(\msfg),\ldots$.  Then 
$p:  G\ra G_\amalg$ with assignment $p:  \msfg\mapsto\msfg_\amalg$.  Then
$$
\msfcpg/\msfcpg_K\simeq\imp,
$$
where group $\imp$ is the image of the homomorphism $p$, and where 
$\msfcpg_K$ is the kernel of the homomorphism $p$.  We have $\imp$ is a group system 
defined by a componentwise operation in $\msfcpg^t$ for each $t\in\bmcpz$.  Lastly we have 
$\msfcpg\simeq\imp$ \ifof\ the kernel $\msfcpg_K$ of the homomorphism $p$ is the identity.
\end{thm}

\begin{prf}
Since there is a homomorphism $p^t:  \msfcpg\ra \msfcpg^t$ from $\msfcpg$ to a group 
$\msfcpg^t$ for each $t\in\bmcpz$, we must have $p^t(\msfgdt\msfgddt)=p^t(\msfgdt)p^t(\msfgddt)$ 
for each $t\in\bmcpz$.  Then
\begin{align*}
p(\msfgdt\msfgddt) &=\ldots,p^t(\msfgdt\msfgddt),p^{t+1}(\msfgdt\msfgddt),\ldots \\
                   &=\ldots,p^t(\msfgdt)p^t(\msfgddt),p^{t+1}(\msfgdt)p^{t+1}(\msfgddt),\ldots \\
                   &=p(\msfgdt)\plus p(\msfgddt).
\end{align*}
Then there is a homomorphism $p$ from $\msfcpg$ to the direct product group $(\msfcpg_\amalg,\plus)$.
We have $\msfcpg/\msfcpg_K\simeq\imp$ from the first homomorphism theorem.
We have $\imp$ is a group system since the global operation $\plus$ in $\imp$ is defined 
by a componentwise operation in $\msfcpg^t$ for each $t\in\bmcpz$.
\end{prf}
We refer to (\ref{eqthm37}) by the notation $p=\ldots,p^t,p^{t+1},\ldots$.

We now ask whether we can reverse the chain in (\ref{chain7a}), i.e., starting with a generator
group $(\calu,\circ)$, can we recover $A$.  We now show that we can recover $A$ directly 
from group $(\calu,\circ)$ using the \fhgs.

\begin{thm}
\label{thm71}
The \fhgs\ constructs a group system $\imfu$ with component group
$A^t$ from an $(\ell+1)$-depth generator group $(\calu,\circ)$ using a homomorphism $f_u$.
\end{thm}

\begin{prf}
The composition $\alpha^t\bullet\eta^t\bullet(\beta^t)^{-1}\bullet\theta^t$ is a map 
$\calu\ra\tridnjktarg{0}{0}{t}{\calu}\ra\tridnjktarg{0}{0}{t}{\calr}\ra B^t\ra A^t$ given by the assignment 
$\alpha^t\bullet\eta^t\bullet(\beta^t)^{-1}\bullet\theta^t:  \bmu\mapsto\tridnjktarg{0}{0}{t}{\bmu}\mapsto\tridnjktarg{0}{0}{t}{\bmr}\mapsto b^t\mapsto a^t$.
From Theorem \ref{homou}, we know there is a homomorphism from $(\calu,\circ)$ to 
$\grpjktarg{0}{0}{t}{\calu}{\ccirc}$ given by the map $\theta^t$; from Corollary \ref{cor65}, we know there is an isomorphism
from $\grpjktarg{0}{0}{t}{\calu}{\ccirc}$ to $\grpjktarg{0}{0}{t}{\calr}{\cast}$ given by the map $(\beta^t)^{-1}$; 
from Corollary \ref{cor36}, we know there is an isomorphism
from $\grpjktarg{0}{0}{t}{\calr}{\cast}$ to $B^t$ given by the map $\eta^t$; and from Theorem \ref{thm1},
we know there is a homomorphism from $B^t$ to $A^t$ given by the map $\alpha^t$.  
Let $f_u^t\rmdef\alpha^t\bullet\eta^t\bullet(\beta^t)^{-1}\bullet\theta^t$.  
Then $f_u^t:  \calu\ra A^t$ is a
homomorphism from $(\calu,\circ)$ to $A^t$ for each $t\in\bmcpz$.  Consider the Cartesian product 
$A_\amalg\rmdef\cdots\times A^t\times A^{t+1}\times\cdots$ (note here $A^t$ is interpreted
as a set).  Define the direct product group
$(A_\amalg,+)$ by $(A_\amalg,+)\rmdef\cdots\times A^t\times A^{t+1}\times\cdots$.
Then from Theorem \ref{fhgs}, using $(\calu,\circ)$ for information group $\msfcpg$
and alphabet group $A^t$ for $\msfcpg^t$, $t\in\bmcpz$, 
there is a homomorphism $f_u:  \calu\ra A_\amalg$, from $(\calu,\circ)$
to the direct product group $(A_\amalg,+)$, defined by
$$
f_u(\bmu)\rmdef\ldots,f_u^t(\bmu),f_u^{t+1}(\bmu),\ldots.
$$
Define
\begin{align*}
\bma_\amalg\rmdef &\ldots,f_u^t(\bmu),f_u^{t+1}(\bmu),\ldots \\
            = &\ldots,a^t,a^{t+1},\ldots.
\end{align*}
Then $f_u:  \calu\ra A_\amalg$
with assignment $f_u:  \bmu\mapsto\bma_\amalg$.
We can think of $\bma_\amalg$ as a ``sliding block" mapping of $\bmu$.  Applying the first
homomorphism theorem for groups, we have
$$
(\calu,\circ)/(\calu,\circ)_K\simeq\imfu,
$$
where group $\imfu$ is the image of the homomorphism $f_u$, and where 
$(\calu,\circ)_K$ is the kernel of the homomorphism $f_u$.  Since group $\imfu$ is a
subgroup of the direct product group $(A_\amalg,+)$, then $\imfu$ is a
group system where global operation $+$ defines the componentwise 
operation in group $A^t$ for each $t\in\bmcpz$.
\end{prf}

\begin{thm}
\label{thm72}
The homomorphism $f_u$ is a bijection $f_u:  \calu\ra A$.  Then $\imfu=A$.
The group system $\imfu$ is \ellctl\ and complete.
\end{thm}

\begin{prf}
Fix $\bmu\in\calu$.  Let
$$
\bmr_\amalg\rmdef\ldots,\tridnjktarg{0}{0}{t}{\bmr},\tridnjktarg{0}{0}{t-1}{\bmr},\ldots,
$$
where $\tridnjktarg{0}{0}{t}{\bmr}$ is given by the assignment $(\beta^t)^{-1}\bullet\theta^t:  \bmu\mapsto\tridnjktarg{0}{0}{t}{\bmu}\mapsto\tridnjktarg{0}{0}{t}{\bmr}$ 
for each $t\in\bmcpz$ of Theorem \ref{thm71}.
From Corollary \ref{cor66}, the assignment $(\beta^t)^{-1}\bullet\theta^t:  \bmu\mapsto\tridnjktarg{0}{0}{t}{\bmu}\mapsto\tridnjktarg{0}{0}{t}{\bmr}$ 
for each $t\in\bmcpz$ is the same as the assignment 
$\beta^{-1}:  \bmu\mapsto\bmr_\amalg$ of bijection $\beta^{-1}:  \calu\ra\calr$.  
Then $\bmr_\amalg\in\calr$.  Let
$$
\bmb_\amalg\rmdef\ldots,b^t,b^{t+1},\ldots,
$$
where $b^t$ is given by the assignment
$\eta^t:  \tridnjktarg{0}{0}{t}{\bmr}\mapsto b^t$ for each $t\in\bmcpz$ of Theorem \ref{thm71}.  
From Corollary \ref{cor26}, the assignment 
$\eta^t:  \tridnjktarg{0}{0}{t}{\bmr}\mapsto b^t$ for each $t\in\bmcpz$
is the same as the assignment $\eta:  \bmr_\amalg\mapsto\bmb_\amalg$ of 
bijection $\eta:  \calr\ra B$.  Then $\bmb_\amalg\in B$.  Let
$$
\bma_\amalg'\rmdef\ldots,a^t,a^{t+1},\ldots,
$$
where $a^t$ is given by the assignment $\alpha^t:  b^t\mapsto a^t$ for each $t\in\bmcpz$
of Theorem \ref{thm71}.  From Theorem \ref{thm1}, 
the assignment $\alpha^t:  b^t\mapsto a^t$ for each $t\in\bmcpz$
is the same as the assignment
$\alpha:  \bmb_\amalg\mapsto\bma_\amalg'$ of the bijection $\alpha:  B\ra A$.
Then $\bma_\amalg'\in A$.  Then the assignment 
$\alpha^t\bullet\eta^t\bullet(\beta^t)^{-1}\bullet\theta^t:  \bmu\mapsto\tridnjktarg{0}{0}{t}{\bmu}\mapsto\tridnjktarg{0}{0}{t}{\bmr}\mapsto b^t\mapsto a^t$
for each $t\in\bmcpz$ of Theorem \ref{thm71} is the same as the assignment 
$\alpha\bullet\eta\bullet\beta^{-1}:  \bmu\mapsto\bmr_\amalg\mapsto\bmb_\amalg\mapsto\bma_\amalg'$ 
of the bijection $\alpha\bullet\eta\bullet\beta^{-1}:  \calu\ra\calr\ra B\ra A$.
But the map $\alpha^t\bullet\eta^t\bullet(\beta^t)^{-1}\bullet\theta^t$ is the same as
the map $f_u^t$.  Then the bijection $\alpha\bullet\eta\bullet\beta^{-1}:  \calu\ra A$
is the same as map $f_u$, since $f_u=\ldots,f_u^t,f_u^{t+1},\ldots$.  Then
$f_u(\bmu)=\bma_\amalg=\bma_\amalg'$, and $f_u$ is a bijection $f_u:  \calu\ra A$.
Then $\imfu=A$.

Since $\imfu=A$, the group system $\imfu$ is \ellctl\ and complete.
\end{prf}
These results give the chain shown in (\ref{chain7b}).  As shown 
we can recover $A$ directly from group $(\calu,\circ)$ using the \fhgs.
\be
\label{chain7b}
\begin{array}{lllllll}
\da & \la                       & \la & \la                       & \la       & {\stackrel{=}{\la}}        & \imfu                                  \\
\da &                           &     &                           &           &                            & \ua\,f_u=\ldots,f_u^t,f_u^{t+1},\ldots \\
A   & {\stackrel{\simeq}{\lra}} & B   & {\stackrel{\simeq}{\ra}}  & (\calr,*) & {\stackrel{\simeq}{\ra}}   & (\calu,\circ)    
\end{array}
\ee

\begin{cor}
\label{cor80}
We have $(\calu,\circ)\simeq\imfu=A$.
\end{cor}

\begin{prf}
The kernel of the homomorphism $f_u$ is the identity so $(\calu,\circ)\simeq\imfu=A$.
\end{prf}

We can abbreviate chain (\ref{chain7b}) as shown in (\ref{chain7bb}).
\be
\label{chain7bb}
\begin{array}{lllll}
\da & {\stackrel{=}{\la}}  & \imfu                                  \\
\da &                           & \ua\,f_u=\ldots,f_u^t,f_u^{t+1},\ldots \\
A   & {\stackrel{\simeq}{\ra}}  & (\calu,\circ)    
\end{array}
\ee

\vspace{3mm}
{\bf 6.6  Construction of any group system $C$ from generator group $(\calu,\circ)$}
\vspace{3mm}

In Subsection 6.5, we showed how we could reverse the steps from $A$ to $(\calu,\circ)$
to recover $A$ from $(\calu,\circ)$.  In this subsection, we assume that we are given
the generator group $(\calu,\circ)$ of some group system $A$, and we show how to construct
all other group systems $C$ that have generator group $(\calu,\circ)$ as an input group.  First we
describe a special \ellctl\ complete group system that we can always construct
from any $(\ell+1)$-depth generator group $(\calu,\circ)$.
For each $t\in\bmcpz$, we simply let the alphabet group $A^t$ be $\grpjktarg{0}{0}{t}{\calu}{\ccirc}$.
From Theorem \ref{homou}, we know there is a homomorphism from $(\calu,\circ)$ to 
$\grpjktarg{0}{0}{t}{\calu}{\ccirc}$ given by the map $\theta^t$, for each $t\in\bmcpz$.  Consider the Cartesian product 
$$
U_\amalg\rmdef\cdots\times\tridnjktarg{0}{0}{t}{\calu}\times\tridnjktarg{0}{0}{t+1}{\calu}\times\cdots.
$$
Define the direct product group $(U_\amalg,\ovcirc)$ by
$$
(U_\amalg,\ovcirc)\rmdef\cdots\times\grpjktarg{0}{0}{t}{\calu}{\ccirc}\times\grpjktarg{0}{0}{t+1}{\calu}{\ccirc}\times\cdots.
$$
Then from Theorem \ref{fhgs}, using $(\calu,\circ)$ for information group $\msfcpg$
and the primary elementary group $\grpjktarg{0}{0}{t}{\calu}{\ccirc}$ for $\msfcpg^t$, $t\in\bmcpz$, 
there is a homomorphism $\theta:  \calu\ra U_\amalg$, from $(\calu,\circ)$
to the direct product group $(U_\amalg,\ovcirc)$, defined by
$$
\theta(\bmu)\rmdef\ldots,\theta^t(\bmu),\theta^{t+1}(\bmu),\ldots.
$$
Define
\begin{align*}
\bmu_s\rmdef &\ldots,\theta^t(\bmu),\theta^{t+1}(\bmu),\ldots \\
           = &\ldots,\tridnjktarg{0}{0}{t}{\bmu},\tridnjktarg{0}{0}{t+1}{\bmu},\ldots.
\end{align*}
Then $\theta:  \calu\ra U_\amalg$ with assignment $\theta:  \bmu\mapsto\bmu_s$.
We can think of $\bmu_s$ as the sequence of \inpms\ $\tridnjktarg{0}{0}{t}{\bmu}$ of $\bmu$, 
now written in conventional time order and not overlapped.  Then
$$
(\calu,\circ)/(\calu,\circ)_K\simeq\imtheta,
$$
where group $\imtheta$ is the image of the homomorphism $\theta$, and where 
$(\calu,\circ)_K$ is the kernel of the homomorphism $\theta$.  Since group $\imtheta$ is a
subgroup of the direct product group $(U_\amalg,\ovcirc)$, then $\imtheta$ is a
group system where global operation $\ovcirc$ is defined by the componentwise 
operation $\prodjktarg{0}{0}{t}{\ccirc}$ in group $\grpjktarg{0}{0}{t}{\calu}{\ccirc}$
for each $t\in\bmcpz$.  We denote group $\imtheta$ by $(U_s,\ovcirc)$,
where $U_s$ is the subset of the Cartesian product $\calu_\amalg$ determined by $\bmu\in\calu$, or
equivalently the subset of $\calu_\amalg$ defined by $\imtheta$.
We call $(U_s,\ovcirc)$ the {\it generator group system} of $(\calu,\circ)$.
We can summarize the preceding discussion as follows.  

\begin{thm}
\label{thm83}
The \fhgs\ constructs a group system $(U_s,\ovcirc)$ with component group
$\grpjktarg{0}{0}{t}{\calu}{\ccirc}$ from an $(\ell+1)$-depth generator group $(\calu,\circ)$ 
using a homomorphism $\theta$, where $\theta=\ldots,\theta^t,\theta^{t+1},\ldots$,
and $\theta^t$ is a homomorphism $\theta^t:  (\calu,\circ)\mapsto\grpjktarg{0}{0}{t}{\calu}{\ccirc}$, 
for each $t\in\bmcpz$.  The homomorphism $\theta$ is a bijection.  
We have $(\calu,\circ)\simeq\imtheta=(U_s,\ovcirc)$ under the assignment
$\theta: \bmu\mapsto\bmu_s$ given by the bijection $\theta: \calu\ra U_s$.
\end{thm}

\begin{prf}
For each $\bmu_s\in U_s$, there can be only one
$\bmu\in\calu$ such that $\theta:  \bmu\ra\bmu_s$ because the sequence
$\bmu_s=\ldots,\tridnjktarg{0}{0}{t}{\bmu},\tridnjktarg{0}{0}{t+1}{\bmu},\ldots$ defines a unique $\bmu$.  Then
$\theta:  \calu\ra U_s$ given by the assignment $\theta:  \bmu\ra\bmu_s$ is a bijection.
Then the kernel $(\calu,\circ)_K$ is the identity and $(\calu,\circ)\simeq\imtheta=(U_s,\ovcirc)$.
\end{prf}
We can summarize the construction in Theorem \ref{thm83} as shown in chain (\ref{chain7f}).
\be
\label{chain7f}
\begin{array}{lll}
\da             & {\stackrel{=}{\la}} & \imtheta                                         \\
\da             &                     & \ua\,\theta=\ldots,\theta^t,\theta^{t+1},\ldots  \\
(U_s,\ovcirc)   &                     & (\calu,\circ)
\end{array}
\ee
The chain (\ref{chain7f}) forms a linear system with input group $(\calu,\circ)$,
homorphism $\theta$, and output group $(U_s,\ovcirc)$.  We say a linear system is {\it invertible} 
if the homorphism is a bijection.  Since the homorphism $\theta$
in (\ref{chain7f}) is a bijection, then the linear 
system in (\ref{chain7f}) is invertible.  If the linear system in (\ref{chain7f}) is
invertible, then the input $\bmu\in\calu$ can be discovered from the output
$\bmu_s\in U_s$.

\begin{lem}
\label{lem98}
Let $\Xi\rmdef\{\bmu_{g,k}^t:  0\le k\le\ell,t\in\bmcpz\}$ be the set of generators in $(\calu,\circ)$.
The sequences in set $\theta(\Xi)$ are generators of $(U_s,\ovcirc)$ and form a basis $\calb$ of
$(U_s,\ovcirc)$.
\end{lem}

\begin{prf}
Let $\bmu$ be any element in $(\calu,\circ)$.  Then $\theta(\bmu)$ is an element in $(U_s,\ovcirc)$.

We now show any element in $\theta(\Xi)$ is a generator of $(U_s,\ovcirc)$.  
Fix $k$ such that $0\le k\le\ell$.  Fix generator  $\bmu_{g,k}^t$ in $\Xi$.
Under the bijection $\theta$ of the \fhgs\, shown in (\ref{chain7f}), 
nontrivial \glab\ $r_{0,k}^t$ in $\bmu_{g,k}^t$ lies in $(k+1)$ 
primary elementary groups $\grpjktarg{0}{0}{t}{\calu}{\ccirc},\grpjktarg{0}{0}{t+1}{\calu}{\ccirc},\ldots,\grpjktarg{0}{0}{t+k}{\calu}{\ccirc}$,
and therefore generator $\bmu_{g,k}^t$ becomes a sequence $\bmu_s^{[t,t+k]}$ of span $k+1$ in $U_s$.
To show $\bmu_s^{[t,t+k]}$ is a generator, we have to show $\bmu_s^{[t,t+k]}$ is a \crep\
of the time domain granule (\ref{qgx}).  It is sufficient to show $\bmu_s^{[t,t+k]}$ is a \crep\
of (\ref{qgx1}).  But $\bmu_s^{[t,t+k]}$ is a sequence of span $k+1$ in $U_s$.  Therefore it is a member
of the numerator of (\ref{qgx1}) but not of either term in the denominator.  Therefore $\bmu_s^{[t,t+k]}$ 
is a \crep\ of (\ref{qgx1}).

We have shown $\bmu_s^{[t,t+k]}$ is a generator of $(U_s,\ovcirc)$ for $0\le k\le\ell$ and $t\in\bmcpz$.
Therefore $\theta(\Xi)$ is a basis $\calb$ of $(U_s,\ovcirc)$.
\end{prf}

We see that the \fhgs\ constructs generators in $(U_s,\ovcirc)$ from generators in $(\calu,\circ)$.
There is a bijection between the generators of group system $(U_s,\ovcirc)$
and the generators in $(\calu,\circ)$ given by the restriction of $\theta$ to the
generators of $(\calu,\circ)$.

\begin{thm}
\label{thm81}
The group system $(U_s,\ovcirc)=\imtheta$ constructed by Theorem \ref{thm83} is \ellctl\ and complete.
\end{thm}

\begin{prf}
Since the generator group $(\calu,\circ)$ is $(\ell+1)$-depth, there is at least one
generator $\bmu_{g,\ell}^t$ in $(\calu,\circ)$ which has a nontrivial \glab\ $r_{0,\ell}^t$.  
Under the bijection $\theta$ of the \fhgs\, shown in (\ref{chain7f}),
$r_{0,\ell}^t$ lies in $(\ell+1)$ primary elementary groups 
$\grpjktarg{0}{0}{t}{\calu}{\ccirc},\grpjktarg{0}{0}{t+1}{\calu}{\ccirc},\ldots,\grpjktarg{0}{0}{t+\ell}{\calu}{\ccirc}$,
and therefore generator $\bmu_{g,\ell}^t$ becomes a sequence $\bmu_s^{[t,t+\ell]}$ of span
$\ell+1$ in $U_s$.  From the same argument used to show $\bmu_s^{[t,t+k]}$ is a generator
of span $k+1$ in the proof of Lemma \ref{lem98}, we know $\bmu_s^{[t,t+\ell]}$ is a generator 
of span $\ell+1$.  Therefore $(U_s,\ovcirc)$ is \ellctl.

The group system $(U_s,\ovcirc)$ is complete since it is determined by component groups
$\grpjktarg{0}{0}{t}{\calu}{\ccirc}$ in $(\calu,\circ)$, which have no global constraints.
\end{prf}

\begin{lem}
\label{lem95b}
The generator group of $(U_s,\ovcirc)$ is $(\calu,\circ)$.
\end{lem}

\begin{prf}
Under the bijection $\theta^{-1}: U_s\ra\calu$, the group system $(U_s,\ovcirc)$
collapses to the generator group, which is $(\calu,\circ)$.
\end{prf}

Theorem \ref{thm83} shows that we can always construct a special \ellctl\ complete group system 
from any $(\ell+1)$-depth generator group $(\calu,\circ)$, namely $(U_s,\ovcirc)$.
We now show how to construct all other group systems $C$ that have the generator group $(\calu,\circ)$
of an \ellctl\ complete group system $A$ as an input group.

\begin{thm}
\label{thm84}
{\bf (Construction)}
We construct a group system $C\rmdef\imhu$ with component group
$A^t$ from an $(\ell+1)$-depth generator group $(\calu,\circ)$ using a homomorphism $h_u$.
From Theorem \ref{homou}, we know there is a homomorphism from $(\calu,\circ)$ to 
$\grpjktarg{0}{0}{t}{\calu}{\ccirc}$ given by the map $\theta^t$, for each $t\in\bmcpz$.  
For each $t\in\bmcpz$, let map $\mu^t:  \tridnjktarg{0}{0}{t}{\calu}\ra A^t$ be
a homomorphism from $\grpjktarg{0}{0}{t}{\calu}{\ccirc}$ to $A^t$ given by the assignment $\tridnjktarg{0}{0}{t}{\bmu}\mapsto a^t$.  
Let $h_u^t\rmdef\mu^t\bullet\theta^t$.  Then $h_u^t:  \calu\ra A^t$ is a
homomorphism from $(\calu,\circ)$ to $A^t$ for each $t\in\bmcpz$.  Consider the Cartesian product 
$A_\amalg\rmdef\cdots\times A^t\times A^{t+1}\times\cdots$.  Define the direct product group
$(A_\amalg,+)$ by $(A_\amalg,+)\rmdef\cdots\times A^t\times A^{t+1}\times\cdots$.
Then from Theorem \ref{fhgs}, using $(\calu,\circ)$ for information group $\msfcpg$
and alphabet group $A^t$ for $\msfcpg^t$, $t\in\bmcpz$, 
there is a homomorphism $h_u:  \calu\ra A_\amalg$, from $(\calu,\circ)$
to the direct product group $(A_\amalg,+)$, defined by
$$
h_u(\bmu)\rmdef\ldots,h_u^t(\bmu),h_u^{t+1}(\bmu),\ldots.
$$
Define
\begin{align*}
\bma_\amalg\rmdef &\ldots,h_u^t(\bmu),h_u^{t+1}(\bmu),\ldots \\
            = &\ldots,a^t,a^{t+1},\ldots.
\end{align*}
Then $h_u:  \calu\ra A_\amalg$
with assignment $h_u:  \bmu\mapsto\bma_\amalg$.  Applying the first
homomorphism theorem for groups, we have
$$
(\calu,\circ)/(\calu,\circ)_K\simeq\imhu,
$$
where group $\imhu$ is the image of the homomorphism $h_u$, and where 
$(\calu,\circ)_K$ is the kernel of the homomorphism $h_u$.  Since group $\imhu$ is a
subgroup of the direct product group $(A_\amalg,+)$, then $\imhu$ is a
group system where global operation $+$ defines the componentwise 
operation in group $A^t$ for each $t\in\bmcpz$.  If $h_u$ is a bijection, 
we have $(\calu,\circ)\simeq\imhu=C$.
\end{thm}

\begin{prf}
The proof follows directly from the \fhgs.
\end{prf}
We can summarize the construction in Theorem \ref{thm84} as shown in chain (\ref{chain7e}).
\be
\label{chain7e}
\begin{array}{lll}
\da & {\stackrel{=}{\la}} & \imhu                                       \\
\da &                     & \ua\,h_u=\ldots,h_u^t,h_u^{t+1},\ldots      \\
C   &                     & (\calu,\circ)
\end{array}
\ee
The linear system in (\ref{chain7e}) is invertible if homomorphism $h_u$ is a bijection.

We now use chain (\ref{chain7f}) to show that the construction in chain (\ref{chain7e}) 
can be split into two steps, as shown in chain (\ref{chain7c}).
The group $(\calu,\circ)$, and the map $\theta=\ldots,\theta^t,\theta^{t+1},\ldots$ and group
$\imtheta=(U_s,\ovcirc)$ previously found are shown in (\ref{chain7c}).  In Theorem \ref{thm84}, 
we have seen that $h_u^t$ is the composition $h_u^t=\mu^t\bullet\theta^t$ 
for each $t\in\bmcpz$.  We define the map 
$\mu:  U_s\ra C$ shown in (\ref{chain7c}) by $\mu\rmdef\ldots,\mu^t,\mu^{t+1},\ldots$.  
Then $h_u$ in Theorem \ref{thm84}
is the composition $h_u=\mu\bullet\theta$ of the maps $\theta$ and $\mu$ in (\ref{chain7c}).
Then (\ref{chain7c}) gives a two step way of implementing chain (\ref{chain7e}).  We first construct
the generator group system $(U_s,\ovcirc)$ from $(\calu,\circ)$ using homomorphism $\theta$,
and then construct a homomorphism $\mu^t$ from component group $\grpjktarg{0}{0}{t}{\calu}{\ccirc}$ 
to an alphabet group $A^t$, for each $t\in\bmcpz$; this gives group system $C$.
Note that all the information needed to construct $(U_s,\ovcirc)$ is contained in $(\calu,\circ)$.
In the sense of this two step construction, we can regard the generator group system $(U_s,\ovcirc)$
as a``parent" group system for all other group systems formed from the generator group $(\calu,\circ)$
by the \fhgs.
\be
\label{chain7c}
\begin{array}{lll}
 \da & {\stackrel{\mu=\ldots,\mu^t,\mu^{t+1},\ldots}{\la}} & (U_s,\ovcirc)                                    \\
 \da &                                                     & \ua\,\theta=\ldots,\theta^t,\theta^{t+1},\ldots  \\
 C   &                                                     & (\calu,\circ)           
\end{array}
\ee

We confirm that we can recover $A$ as well from the two step construction shown in chain (\ref{chain7c}).

\begin{thm}
All \ellctl\ complete group systems $A$ can be found from chain (\ref{chain7c}), starting from
the class of all $(\ell+1)$-depth generator groups $(\calu,\circ)$.
\end{thm}

\begin{prf}
The homomorphism $f_u^t=\alpha^t\bullet\eta^t\bullet(\beta^t)^{-1}\bullet\theta^t$ 
for each $t\in\bmcpz$ used in Theorem \ref{thm71} to construct $A$ from its generator
group is the composition $h_u^t=\mu^t\bullet\theta^t$
for each $t\in\bmcpz$ used in Theorem \ref{thm84}, where 
$\mu^t=\alpha^t\bullet\eta^t\bullet(\beta^t)^{-1}$ for each $t\in\bmcpz$.
\end{prf}

The two step construction shown in chain (\ref{chain7c}) gives the following.

\begin{thm}
\label{thm85}
The group system $C=\imhu$ constructed by Theorem \ref{thm84} is $l$-controllable and complete,
where $l\le\ell$.
\end{thm}

\begin{prf}
Consider any $\bmadt_\amalg,\bmaddt_\amalg\in\imhu$.  Fix any $t\in\bmcpz$.
We show there exists $\bma_\amalg\in\imhu$ such that
$\chi^{(-\infty,t)}(\bma_\amalg)=\chi^{(-\infty,t)}(\bmadt_\amalg)$ and 
$\chi^{[t+\ell,+\infty)}(\bma_\amalg)=\chi^{[t+\ell,+\infty)}(\bmaddt_\amalg)$;
this means $\imhu$ is $l$-controllable where $l\le\ell$.  Let $\bmudt_s\in U_s$
such that $\mu:  \bmudt_s\mapsto\bmadt_\amalg$, and let $\bmuddt_s\in U_s$
such that $\mu:  \bmuddt_s\mapsto\bmaddt_\amalg$.  Since $(U_s,\ovcirc)$ is
\ellctl, there exists $\bmu_s\in U_s$ such that
$\chi^{(-\infty,t)}(\bmu_s)=\chi^{(-\infty,t)}(\bmudt_s)$ and 
$\chi^{[t+\ell,+\infty)}(\bmu_s)=\chi^{[t+\ell,+\infty)}(\bmuddt_s)$.
Then let $\mu:  \bmu_s\mapsto\bma_\amalg$.  But then there exists
$\bma_\amalg\in\imhu$ such that
$\chi^{(-\infty,t)}(\bma_\amalg)=\chi^{(-\infty,t)}(\bmadt_\amalg)$ and 
$\chi^{[t+\ell,+\infty)}(\bma_\amalg)=\chi^{[t+\ell,+\infty)}(\bmaddt_\amalg)$.

The group system $C=\imhu$ is complete since it is determined by component groups
$\grpjktarg{0}{0}{t}{\calu}{\ccirc}$ in $(\calu,\circ)$, which have no global constraints.
\end{prf}

We now give an example to show that the inequality $l<\ell$ in Theorem \ref{thm85} is necessary.

\begin{ex}{1}
First we give an example of a complete \ellctl\ group system $A$.  
Let $A$ be the group system formed by the identity and the sequence
$\ldots,0,0,1,1,0,0,\ldots$, where the first 1 is at time $t$.
The alphabet group $A^t$ is isomorphic to $\bmcpz_2$ for $t$ and $t+1$.  Elsewhere the
alphabet group is the identity.  This group system is 1-controllable.
The group system $A$ is isomorphic to $\bmcpz_2$.
There are two generator sequences, the identity $\bone$ and the sequence 
$\bmg_a^{[t,t+1]}\rmdef\ldots,0,0,1,1,0,0,\ldots$.  Now we find the 
$2$-depth generator group $(\calu,\circ)$ of $A$.  Using (\ref{ugttf}),
there are two tensors $\bmu\in\calu$, the identity 
\be
\label{ugttf1}
\begin{array}{lllllllll}
\cdots  & 1_{0,1}^{t+1}    &  1_{0,1}^t      & 1_{0,1}^{t-1}      & \cdots & 1_{0,1}^{t-j}      & \cdots & 1_{0,1}^{t-\ell}  & \cdots  \\
\cdots  & 1_{0,0}^{t+1}    &  1_{0,0}^t      & 1_{0,0}^{t-1}      & \cdots & 1_{0,0}^{t-j}      & \cdots & 1_{0,0}^{t-\ell}  & \cdots  
\end{array}
\ee   
and the generator $\bmu_{g,1}^t$,
\be
\label{ugttf2}
\begin{array}{lllllllll}
\cdots  & 1_{0,1}^{t+1}    &  r_{0,1}^t  & 1_{0,1}^{t-1}      & \cdots & 1_{0,1}^{t-j}      & \cdots & 1_{0,1}^{t-\ell}  & \cdots  \\
\cdots  & 1_{0,0}^{t+1}    &  1_{0,0}^t  & 1_{0,0}^{t-1}      & \cdots & 1_{0,0}^{t-j}      & \cdots & 1_{0,0}^{t-\ell}  & \cdots  
\end{array}
\ee   
It is clear that the $2$-depth generator group $(\calu,\circ)$ of $A$ is isomorphic to $\bmcpz_2$.

We can recover $A$ from $2$-depth generator group $(\calu,\circ)$ as in Subsection 6.5.  
In fact $A$ is just isomorphic to
$(U_s,\ovcirc)$, the generator group system of $(\calu,\circ)$, where
alphabet group $\grpjktarg{0}{0}{t}{\calu}{\ccirc}$ is simply isomorphic to $A^t$ for each $t\in\bmcpz$.
The homomorphism $\theta$ is a bijection.

Now consider obtaining any group system $C$ from $(U_s,\ovcirc)$ as in (\ref{chain7c}).
As an example,
we let $A^t$ be isomorphic to $\bmcpz_2$ as before but choose $A^{t+1}$ to be the identity.  
Then the homomorphism $\mu^{t+1}$ from $\grpjktarg{0}{0}{t+1}{\calu}{\ccirc}$ to $A^{t+1}$ maps $\bmcpz_2$ 
to the identity.  We obtain the group system $C'$ formed by the identity and the sequence
$\ldots,0,0,1,0,0,0,\ldots$, where the first 1 is at time $t$.  The group system
$C'$ is isomorphic to $\bmcpz_2$.  This group system is 0-controllable.
The homomorphism $h_u$ is a bijection.

We have $(\calu,\circ)\simeq A\simeq C'$, where $A$ is 1-controllable and $C'$ is 0-controllable.  
This proves that the inequality $l<\ell$ in Theorem \ref{thm85} is necessary, even when 
$h_u$ is a bijection.  Since $h_u$ is a bijection, the linear system formed by input 
group $(\calu,\circ)$, homomorphism $h_u$, and output group $C'$ is still invertible even 
though $C'$ is 0-controllable.

Since $C'$ is 0-controllable, there must be a $1$-depth generator group $(\calu',\circ')$
which can be used to construct $C'$ using the \fhgs, where the homomorphism $h_u'$ 
is a bijection.  In fact, the tensors $\bmu'$ in $\calu'$ are just
$$
\begin{array}{lllllllll}
\cdots  & 1_{0,0}^{t+1}    &  1_{0,0}^t   & 1_{0,0}^{t-1}      & \cdots & 1_{0,0}^{t-j}   & \cdots & 1_{0,0}^{t-\ell}  & \cdots  
\end{array}
$$
and
$$
\begin{array}{lllllllll}
\cdots  & 1_{0,0}^{t+1}    &  r_{0,0}^t  & 1_{0,0}^{t-1}      & \cdots & 1_{0,0}^{t-j}    & \cdots & 1_{0,0}^{t-\ell}  & \cdots
\end{array}
$$
and $(\calu',\circ')$ is isomorphic to $\bmcpz_2$.
Thus $C'$ can be constructed from two generator groups,
$2$-depth generator group $(\calu,\circ)$ and $1$-depth generator group $(\calu',\circ')$.
Although $C'$ can be constructed from $(\calu,\circ)$, $(\calu,\circ)$ is not
the generator group of $C'$.  The generator group of $C'$ is the unique group constructed
from the canonical realization of $C'$, and this is the $1$-depth generator group $(\calu',\circ')$.
\end{ex}

In construction chain (\ref{chain7e}), the \fhgs\ constructs an $l$-controllable complete group system 
$C$ from an $(\ell+1)$-depth generator group $(\calu,\circ)$ using a homomorphism $h_u$,
where $l\le\ell$.  We have seen from the example above that we may have $l<\ell$.
We now show how to construct all \ellctl\ complete group systems $A$ up to isomorphism from
the set of all $(\ell+1)$-depth generator groups $(\calu,\circ)$ and avoid
the case $l<\ell$.

By definition, any $(\ell+1)$-depth generator group $(\calu,\circ)$ is the generator group of
some \ellctl\ complete group system $A$.  Therefore we can recover the set of all
\ellctl\ complete group systems $A$ from the set of all $(\ell+1)$-depth generator groups $(\calu,\circ)$.
Let $\{(\calu,\circ)\}$ be the set of all $(\ell+1)$-depth generator groups $(\calu,\circ)$.
We divide the set $\{(\calu,\circ)\}$ into classes $[(\calu,\circ)]$ 
such that $(\calu,\circ)$ and $(\calu',\circ')$ are in the same class if
$(\calu,\circ)\simeq(\calu',\circ')$.  Let $(\calu,\circ)$ be a representative in $[(\calu,\circ)]$.
The following lemma is clear.

\begin{lem}
The class $[(\calu,\circ)]$ forms an equivalence class of $\{(\calu,\circ)\}$.  The set
$\{(\calu,\circ)\}$ can be divided into equivalence classes $[(\calu,\circ)]$.
\end{lem}
  
\begin{lem}
\label{lem125x}
If two group systems $A$ and $A'$ have generator groups $(\calu,\circ)$ and $(\calu',\circ')$
in the same equivalence class $[(\calu,\circ)]$, then $A$ and $A'$ are isomorphic.
\end{lem}

\begin{prf}
Since $(\calu,\circ)$ and $(\calu',\circ')$ are in the
same equivalence class $[(\calu,\circ)]$, then $(\calu,\circ)$ and $(\calu',\circ')$ are 
isomorphic.  Since $(\calu,\circ)\simeq A$ and $(\calu',\circ')\simeq A'$, then $A$ and $A'$ 
are isomorphic.
\end{prf}

\begin{lem}
\label{lem126x}
Fix any equivalence class $[(\calu,\circ)]$.  Pick any generator group $(\calu',\circ')$
in $[(\calu,\circ)]$.  Then $(\calu',\circ')$ is the generator group of an \ellctl\ complete
generator group system $(U_s',\ovcirc')$.
\end{lem}

\begin{prf}
Fix any equivalence class $[(\calu,\circ)]$.  Pick any generator group $(\calu',\circ')$
in $[(\calu,\circ)]$.  Form the \ellctl\ complete generator group system $(U_s',\ovcirc')$ 
of $(\calu',\circ')$.  From Lemma \ref{lem95b}, the generator group of $(U_s',\ovcirc')$ 
is $(\calu',\circ')$.
\end{prf}

\begin{thm}
\label{thm129x}
The equivalence classes $[(\calu',\circ')]$, where $(\calu',\circ')$ is the generator group of
an \ellctl\ complete generator group system $(U_s',\ovcirc')$, divide the set $\{(\calu,\circ)\}$.  
Then any \ellctl\ complete group system $A$ has a generator group $(\calu,\circ)$ 
in one and only one equivalence class $[(\calu',\circ')]$,
and $A$ is isomorphic to $(U_s',\ovcirc')$.
\end{thm}

\begin{prf}
From Lemma \ref{lem126x}, each equivalence class $[(\calu,\circ)]$ has a representative 
$(\calu',\circ')$, where $(\calu',\circ')$ is the generator group of
an \ellctl\ complete generator group system $(U_s',\ovcirc')$.
Then the equivalence classes $[(\calu',\circ')]$,
for each generaator group system $(U_s',\ovcirc')$, divide the set $\{(\calu,\circ)\}$.  
Then any \ellctl\ complete group system $A$
has a generator group $(\calu,\circ)$ in one and only one equivalence class $[(\calu',\circ')]$.
Since $(\calu,\circ)$ and $(\calu',\circ')$ are in the same equivalence class, by Lemma \ref{lem125x}
$A$ is isomorphic to $(U_s',\ovcirc')$.
\end{prf}

We now use Theorem \ref{thm129x} and chain (\ref{chain7f}) 
of Theorem \ref{thm83} to find all \ellctl\ complete group systems $A$ up to isomorphism from
the set of all $(\ell+1)$-depth generator groups $(\calu,\circ)$.  
To construct all \ellctl\ complete group systems 
$A$ up to isomorphism, we divide the the set of all $(\ell+1)$-depth generator groups into equivalence classes
$[(\calu,\circ)]$.  Pick one representative $(\calu,\circ)$ from each equivalence class.
Construct the \ellctl\ complete generator group system $(U_s,\ovcirc)$ of $(\calu,\circ)$ 
using chain (\ref{chain7f}) of Theorem \ref{thm83}.
The set of all generator group systems $(U_s,\ovcirc)$ obtained this way, one for each equivalence class
$[(\calu,\circ)]$, is the set of all \ellctl\ complete group systems $A$ up to isomorphism.

In this subsection we have shown how to construct all \ellctl\ complete group systems $A$
that have a generator group as an input group.  Some of the difficulties with the framework
were pointed out in Example 1.  In the next section, we discuss an 
easy way to find all generator groups.  We rectify some of the difficulties in Example 1
by giving a new notion of isomorphism for group systems.

\newpage
{\bf 7.  THE ELEMENTARY SYSTEM}
\vspace{3mm}

\vspace{3mm}
{\bf 7.1  The elementary system}
\vspace{3mm}

Fix $k$ such that $0\le k\le\ell$.  Fix $t\in\bmcpz$.  
Let set $V_k^t$ be a collection of elements $v_{k}^t$, which may be
integers or any arbitrary objects; besides this, there is no requirement
on set $V_k^t$.  Consider the Cartesian product $\bigotimes_{t=+\infty}^{-\infty}V_k^t$.
Let $\bmv_k$ be a sequence in the Cartesian product, where
\be
\bmv_k\rmdef\ldots,v_{k}^t,v_{k}^{t-1},\ldots
\ee
Consider the set $\calv$, which is the double Cartesian product
\be
\label{dprod1}
\bigotimes_{0\le k\le\ell}\bigotimes_{t=+\infty}^{-\infty}V_k^t.
\ee
An element $\bmv\in\calv$ is a selection of one sequence $\bmv_k$ from 
each Cartesian product $\bigotimes_{t=+\infty}^{-\infty}V_k^t$,
for $0\le k\le\ell$.  We call $\calv$ the {\it elementary set}.

An element $\bmv\in\calv$ can be thought of as a column of sequences $\bmv_k$, or
\be
\label{decomp}
\bmv=
\left(
\begin{array}{l}
\bmv_\ell \\
\bmv_{\ell-1} \\
\vdots \\
\bmv_k \\
\vdots \\
\bmv_2 \\
\bmv_1 \\
\bmv_0 \\
\end{array}
\right)
\ee
We can see the element $\bmv$ in (\ref{decomp}) is in the same form as
$\bmu$ in (\ref{ugttf}) by simply replacing $v_{k}^t$ by $r_{0,k}^t$.
Because $\bmv$ in (\ref{decomp}) has the same form as $\bmu$ in (\ref{ugttf}),
we can use the same triangle notation $\tridnjktarg{0}{k}{t}{\bmv}$ and
$\tridnjktarg{0}{k}{t}{\calv}$ for $\bmv$ and $\calv$, for $0\le k\le\ell$, 
for $t\in\bmcpz$, as $\tridnjktarg{0}{k}{t}{\bmu}$ and 
$\tridnjktarg{0}{k}{t}{\calu}$ for $\bmu$ and $\calu$.
Note that the entry in index position $(0,k)$ in $\tridnjktarg{0}{k}{t}{\bmv}$
is $v_{k}^t$.

Given an element $\bmw\in\calv$, we can use the triangle notation
$\tridnjktarg{0}{k}{t}{\bmw}$ to find the decomposition of
$\bmw$ into the form (\ref{decomp}) as follows.

\begin{alg} 
\label{alg2}

\vspace{2mm}
\noindent Let $\bmw\in\calv$.  We decompose $\bmw$ as follows. \\
\noindent {\bf DO} \\
\noindent {\bf FOR} $k=\ell$ to $0$ (counting in order), \\
\noindent {\bf FOR} $t=+\infty$ to $-\infty$, \\
let $w_{k}^t$ be the entry in index position $(0,k)$ in $\tridnjktarg{0}{k}{t}{\bmw}$. \\
{\bf ENDFOR} \\
\noindent Set $\bmw_k=\ldots,w_{k}^t,w_{k}^{t-1},\ldots$. \\
{\bf ENDFOR} \\
\noindent Then
$$
\bmw=(\bmw_\ell,\bmw_{\ell-1},\ldots,\bmw_k,\ldots,\bmw_2,\bmw_1,\bmw_0)^T.
$$
{\bf ENDDO}
\end{alg}

The {\it elementary list} $\call$ is an infinite collection of groups defined 
on triangular subsets $\tridnjktarg{0}{k}{t}{\calv}$ of the elementary set $\calv$. 
The groups in the elementary list $\call$ are
$$
\{\grpjktarg{0}{k}{t}{\calv}{\odot}:  0\le k\le\ell,t\in\bmcpz\}.
$$
The {\it $(\ell+1)$-depth elementary system} $\cale$ is an elementary set $\calv$,
the elementary list $\call$, and a homomorphism condition.
The homomorphism condition is that for each $k$ such that $0\le k<\ell$, for each
$t\in\bmcpz$, there is a homomorphism from $\grpjktarg{0}{k}{t}{\calv}{\odot}$ to 
$\grpjktarg{0}{k+1}{t}{\calv}{\odot}$ and $\grpjktarg{0}{k+1}{t-1}{\calv}{\odot}$
under the projection map from set $\tridnjktarg{0}{k}{t}{\calv}$ to set
$\tridnjktarg{0}{k+1}{t}{\calv}$ and set $\tridnjktarg{0}{k+1}{t-1}{\calv}$
given by the assignment $\tridnjktarg{0}{k}{t}{\bmv}\mapsto\tridnjktarg{0}{k+1}{t}{\bmv}$
and $\tridnjktarg{0}{k}{t}{\bmv}\mapsto\tridnjktarg{0}{k+1}{t-1}{\bmv}$.

The $(\ell+1)$-depth elementary system $\cale$ is nested.  For example,
the list $\{\grpjktarg{0}{k}{t}{\calv}{\odot}:  k=\ell,t\in\bmcpz\}$
forms a 1-depth elementary system $\cale_\ell$.  For this trivial case,
the homomorphism condition is vacuous.  The list
$\{\grpjktarg{0}{k}{t}{\calv}{\odot}:  \ell-1\le k\le\ell,t\in\bmcpz\}$
forms a 2-depth elementary system $\cale_{\ell-1}$.  In general the following holds.
 
\begin{thm}
The list $\{\grpjktarg{0}{k}{t}{\calv}{\odot}:  \ell-m+1\le k\le\ell,t\in\bmcpz\}$
forms an $m$-depth elementary system $\cale_{\ell-m+1}$.
\end{thm}

In Subsection 7.2 we show that we can form an $(\ell+1)$-depth global group $(\calv,\cdot)$ 
from any $(\ell+1)$-depth elementary system $\cale$ with elementary set $\calv$.  
Then in Subsection 7.3, we show how to construct \ellctl\ complete group systems
from a given $(\ell+1)$-depth global group $(\calv,\cdot)$.  We show that we can start with
the $(\ell+1)$-depth elementary system $\cale_A$ of any \ellctl\ complete group system $A$
and use the global group formed from $\cale_A$ to obtain an isomorphic copy of the generator group
$(\calu,\circ)$ of $A$ and hence recover $A$.  We define a new notion of group isomorphism
and show how to start with equivalence classes of all $(\ell+1)$-depth elementary systems $\cale$
and directly find all \ellctl\ complete group systems $A$ up to this new group isomorphism.  
Finally in Subsection 7.4, we give a brief discussion of how to construct all $(\ell+1)$-depth 
elementary systems $\cale$.

\vspace{3mm}
{\bf 7.2  The global group $(\calv,\cdot)$}
\vspace{3mm}

We now define a global operation $\cdot$ on the elementary set
$\calv$ using the infinite collection of groups in the elementary system $\cale$.
We show this forms a group $(\calv,\cdot)$.  We say $(\calv,\cdot)$ is the {\it global group}
of elementary system $\cale$.

To motivate the definition of the group operation $\cdot$, we first give
an example of how to evaluate the operation $\prodjktarg{0}{0}{t}{\odot}$ 
in elementary group $\grpjktarg{0}{0}{t}{\calv}{\odot}$.
For this example, we deal with time reversed sequences
of the form $\ldots,t,t-1,\ldots,t-n,\ldots$, and denote
the time interval $t,t-1,\ldots,t-n$ in a time reversed sequence by $[t,t-n]$.
Fix $t\in\bmcpz$.  Let $\bmvdt,\bmvddt\in\calv$.  Consider the product operation
\be
\label{eexp21}
\tridnjktarg{0}{0}{t}{\bmvdt}\prodjktarg{0}{0}{t}{\odot}\tridnjktarg{0}{0}{t}{\bmvddt}
\ee
in group $\grpjktarg{0}{0}{t}{\calv}{\odot}$.  Let
$$
\tridnjktarg{0}{0}{t}{\bmvbr}=
\tridnjktarg{0}{0}{t}{\bmvdt}\prodjktarg{0}{0}{t}{\odot}\tridnjktarg{0}{0}{t}{\bmvddt}.
$$
We evaluate (\ref{eexp21}) by a recursive step procedure, for steps 
$k=\ell,\ldots,0$ in order.  At each step $k$, we use the elementary groups
$\grpjktarg{0}{k}{t-j}{\calv}{\odot}$, for $0\le j\le k$, that are contained
in $\grpjktarg{0}{0}{t}{\calv}{\odot}$.  First for step $k=\ell$, we evaluate
$$
\tridnjktarg{0}{\ell}{t-j}{\bmvdt}\prodjktarg{0}{\ell}{t-j}{\odot}\tridnjktarg{0}{\ell}{t-j}{\bmvddt}
=\tridnjktarg{0}{\ell}{t-j}{\bmvbr}
$$
in elementary group $\grpjktarg{0}{\ell}{t-j}{\calv}{\odot}$, for $0\le j\le\ell$.
This gives the finite sequence of elements
$$
\bmvbr_\ell^{[t,t-\ell]}\rmdef (\vbr_{\ell}^t,\vbr_{\ell}^{t-1},\ldots,\vbr_{\ell}^{t-k},\vbr_{\ell}^{t-(k+1)},\ldots,\vbr_{\ell}^{t-(\ell-1)},\vbr_{\ell}^{t-\ell}),
$$
which are just the sequence of triangles
$$
\tridnjktarg{0}{\ell}{t}{\bmvbr},\tridnjktarg{0}{\ell}{t-1}{\bmvbr},\ldots,\tridnjktarg{0}{\ell}{t-k}{\bmvbr},\tridnjktarg{0}{\ell}{t-(k+1)}{\bmvbr},\ldots,\tridnjktarg{0}{\ell}{t-(\ell-1)}{\bmvbr},\tridnjktarg{0}{\ell}{t-\ell}{\bmvbr}
$$
in elementary groups
\begin{multline*}
\grpjktarg{0}{\ell}{t}{\calv}{\odot},\grpjktarg{0}{\ell}{t-1}{\calv}{\odot},\ldots,\grpjktarg{0}{\ell}{t-k}{\calv}{\odot},\grpjktarg{0}{\ell}{t-(k+1)}{\calv}{\odot}, \\
\ldots,\grpjktarg{0}{\ell}{t-(\ell-1)}{\calv}{\odot},\grpjktarg{0}{\ell}{t-\ell}{\calv}{\odot}.
\end{multline*}

Next, for step $k=\ell-1$, we evaluate
$$
\tridnjktarg{0}{\ell-1}{t-j}{\bmvdt}\prodjktarg{0}{\ell-1}{t-j}{\odot}\tridnjktarg{0}{\ell-1}{t-j}{\bmvddt}
=\tridnjktarg{0}{\ell-1}{t-j}{\bmvbr}
$$
in elementary group $\grpjktarg{0}{\ell-1}{t-j}{\calv}{\odot}$, for $0\le j\le\ell-1$.
For each $j$ such that $0\le j\le\ell-1$, there is a homomorphism from
$\grpjktarg{0}{\ell-1}{t-j}{\calv}{\odot}$ to 
$\grpjktarg{0}{\ell}{t-j}{\calv}{\odot}$ and $\grpjktarg{0}{\ell}{t-j-1}{\calv}{\odot}$
under the projection map from set $\tridnjktarg{0}{\ell-1}{t-j}{\calv}$ to set
$\tridnjktarg{0}{\ell}{t-j}{\calv}$ and set $\tridnjktarg{0}{\ell}{t-j-1}{\calv}$.
Then the evaluation of
$\tridnjktarg{0}{\ell}{t-j}{\bmvdt}\prodjktarg{0}{\ell}{t-j}{\odot}\tridnjktarg{0}{\ell}{t-j}{\bmvddt}$
and
$\tridnjktarg{0}{\ell}{t-j-1}{\bmvdt}\prodjktarg{0}{\ell}{t-j-1}{\odot}\tridnjktarg{0}{\ell}{t-j-1}{\bmvddt}$
is already done in the previous step $k=\ell$.  Then the only new element we 
need to find in $\tridnjktarg{0}{\ell-1}{t-j}{\bmvbr}$ is $\vbr_{\ell-1}^{t-j}$.
Then in step $k=\ell-1$, we just need to find the elements in the finite sequence
$$
\bmvbr_{\ell-1}^{[t,t-(\ell-1)]}\rmdef (\vbr_{\ell-1}^t,\vbr_{\ell-1}^{t-1},\ldots,\vbr_{\ell-1}^{t-k},\vbr_{\ell-1}^{t-(k+1)},\ldots,\vbr_{\ell-1}^{t-(\ell-1)}).
$$
To this point, we have found the finite sequences 
$\bmvbr_\ell^{[t,t-\ell]},\bmvbr_{\ell-1}^{[t,t-(\ell-1)]}$.
The elements in the column of finite sequences
$(\bmvbr_\ell^{[t,t-\ell]},\bmvbr_{\ell-1}^{[t,t-(\ell-1)]})^T$, 
shown as the first two rows of (\ref{ftseq}), form the sequence of overlapping triangles
$$
\tridnjktarg{0}{\ell-1}{t}{\bmvbr},\tridnjktarg{0}{\ell-1}{t-1}{\bmvbr},\ldots,\tridnjktarg{0}{\ell-1}{t-k}{\bmvbr},\tridnjktarg{0}{\ell-1}{t-(k+1)}{\bmvbr},\ldots,\tridnjktarg{0}{\ell-1}{t-(\ell-1)}{\bmvbr}
$$
in elementary groups
\begin{multline*}
\grpjktarg{0}{\ell-1}{t}{\calv}{\odot},\grpjktarg{0}{\ell-1}{t-1}{\calv}{\odot},\ldots,\grpjktarg{0}{\ell-1}{t-k}{\calv}{\odot},\grpjktarg{0}{\ell-1}{t-(k+1)}{\calv}{\odot}, \\
\ldots,\grpjktarg{0}{\ell-1}{t-(\ell-1)}{\calv}{\odot}.
\end{multline*}

In general, let $k+1$ be such that $0\le k+1<\ell$.  
Assume we have found the finite sequences
$\bmvbr_\ell^{[t,t-\ell]},\bmvbr_{\ell-1}^{[t,t-(\ell-1)]},\ldots,\bmvbr_{k+1}^{[t,t-(k+1)]}$,
and assume the elements in the column of finite sequences
$(\bmvbr_\ell^{[t,t-\ell]},\bmvbr_{\ell-1}^{[t,t-(\ell-1)]},\ldots,\bmvbr_{k+1}^{[t,t-(k+1)]})^T$, 
shown as the first four rows of (\ref{ftseq}), form the sequence of overlapping triangles
$$
\tridnjktarg{0}{k+1}{t}{\bmvbr},\tridnjktarg{0}{k+1}{t-1}{\bmvbr},\ldots,\tridnjktarg{0}{k+1}{t-k}{\bmvbr},\tridnjktarg{0}{k+1}{t-(k+1)}{\bmvbr}
$$
in elementary groups
$$
\grpjktarg{0}{k+1}{t}{\calv}{\odot},\grpjktarg{0}{k+1}{t-1}{\calv}{\odot},\ldots,\grpjktarg{0}{k+1}{t-k}{\calv}{\odot},\grpjktarg{0}{k+1}{t-(k+1)}{\calv}{\odot}.
$$
Then for step $k$, we evaluate
$$
\tridnjktarg{0}{k}{t-j}{\bmvdt}\prodjktarg{0}{k}{t-j}{\odot}\tridnjktarg{0}{k}{t-j}{\bmvddt}
=\tridnjktarg{0}{k}{t-j}{\bmvbr}
$$
in elementary group $\grpjktarg{0}{k}{t-j}{\calv}{\odot}$, for $0\le j\le k$.
For each $j$ such that $0\le j\le k$, there is a homomorphism from
$\grpjktarg{0}{k}{t-j}{\calv}{\odot}$ to 
$\grpjktarg{0}{k+1}{t-j}{\calv}{\odot}$ and $\grpjktarg{0}{k+1}{t-j-1}{\calv}{\odot}$
under the projection map from set $\tridnjktarg{0}{k}{t-j}{\calv}$ to set
$\tridnjktarg{0}{k+1}{t-j}{\calv}$ and set $\tridnjktarg{0}{k+1}{t-j-1}{\calv}$.
Then the evaluation of
$\tridnjktarg{0}{k+1}{t-j}{\bmvdt}\prodjktarg{0}{k+1}{t-j}{\odot}\tridnjktarg{0}{k+1}{t-j}{\bmvddt}$
and
$\tridnjktarg{0}{k+1}{t-j-1}{\bmvdt}\prodjktarg{0}{k+1}{t-j-1}{\odot}\tridnjktarg{0}{k+1}{t-j-1}{\bmvddt}$
is already done in the previous step $k+1$.  Then the only new element we 
need to find in $\tridnjktarg{0}{k}{t-j}{\bmvbr}$ is $\vbr_{k}^{t-j}$.
Then in step $k$, we just need to find the elements in the finite sequence
$$
\bmvbr_k^{[t,t-k]}\rmdef (\vbr_{k}^t,\vbr_{k}^{t-1},\ldots,\vbr_{k}^{t-k}).
$$
To this point, we have found the finite sequences 
$\bmvbr_\ell^{[t,t-\ell]},\bmvbr_{\ell-1}^{[t,t-(\ell-1)]},\ldots,\bmvbr_{k+1}^{[t,t-(k+1)]},\bmvbr_k^{[t,t-k]}$.
The elements in the column of finite sequences
$(\bmvbr_\ell^{[t,t-\ell]},\bmvbr_{\ell-1}^{[t,t-(\ell-1)]},\ldots,\bmvbr_{k+1}^{[t,t-(k+1)]},\bmvbr_k^{[t,t-k]})^T$, 
shown as the first five rows of (\ref{ftseq}), form the sequence of overlapping triangles
$$
\tridnjktarg{0}{k}{t}{\bmvbr},\tridnjktarg{0}{k}{t-1}{\bmvbr},\ldots,\tridnjktarg{0}{k}{t-k}{\bmvbr}
$$
in elementary groups
$$
\grpjktarg{0}{k}{t}{\calv}{\odot},\grpjktarg{0}{k}{t-1}{\calv}{\odot},\ldots,\grpjktarg{0}{k}{t-k}{\calv}{\odot}.
$$

This process ends with step $k=0$.  At this point, we have found the finite sequences
$\bmvbr_\ell^{[t,t-\ell]},\bmvbr_{\ell-1}^{[t,t-(\ell-1)]},\ldots,\bmvbr_{k+1}^{[t,t-(k+1)]},\bmvbr_k^{[t,t-k]},\ldots,\bmvbr_1^{[t,t-1]},\bmvbr_0^{[t,t]}$.
The elements in the column of finite sequences
\be
\label{ftseq}
\begin{array}{lllllllll}
\bmvbr_\ell^{[t,t-\ell]}&=(\vbr_{\ell}^t&,\vbr_{\ell}^{t-1}&,\ldots&,\vbr_{\ell}^{t-k}&,\vbr_{\ell}^{t-(k+1)}&,\ldots&,\vbr_{\ell}^{t-(\ell-1)}&,\vbr_{\ell}^{t-\ell}) \\
\bmvbr_{\ell-1}^{[t,t-(\ell-1)]}&=(\vbr_{\ell-1}^t&,\vbr_{\ell-1}^{t-1}&,\ldots&,\vbr_{\ell-1}^{t-k}&,\vbr_{\ell-1}^{t-(k+1)}&,\ldots&,\vbr_{\ell-1}^{t-(\ell-1)}) &\\
\vdots &&&&&&&&\\
\bmvbr_{k+1}^{[t,t-(k+1)]}&=(\vbr_{k+1}^t&,\vbr_{k+1}^{t-1}&,\ldots&,\vbr_{k+1}^{t-k}&,\vbr_{k+1}^{t-(k+1)}) &&&\\
\bmvbr_k^{[t,t-k]}&=(\vbr_{k}^t&,\vbr_{k}^{t-1}&,\ldots&,\vbr_{k}^{t-k}) &&&&\\
\vdots &&&&&&&&\\
\bmvbr_1^{[t,t-1]}&=(\vbr_{1}^t&,\vbr_{1}^{t-1}) &&&&&&\\
\bmvbr_0^{[t,t]}&=(\vbr_{0}^t) &&&&&&&\\
\end{array}
\ee
form the single triangle $\tridnjktarg{0}{0}{t}{\bmvbr}$ in group
$\grpjktarg{0}{0}{t}{\calv}{\odot}$.  This example has given an evaluation of
product operation (\ref{eexp21}) in group $\grpjktarg{0}{0}{t}{\calv}{\odot}$.

Based on the example just given, we now define a group operation $\cdot$ on set 
$\calv$ with the following algorithm, using groups in the elementary system.

\begin{alg} 
\label{alg3} 

\vspace{2mm}
\noindent Let $\bmvdt,\bmvddt\in\calv$.  We evaluate $\bmvdt\cdot\bmvddt$ as follows. \\
\noindent {\bf DO} \\
\noindent {\bf FOR} $k=\ell$ to $0$ (counting in order), \\
\noindent {\bf FOR} $t=+\infty$ to $-\infty$, \\
evaluate
\be
\label{subop3}
\tridnjktarg{0}{k}{t}{\bmvdt}\prodjktarg{0}{k}{t}{\odot}\tridnjktarg{0}{k}{t}{\bmvddt},
\ee
\noindent and let $\vbr_{k}^t$ be the entry in index position $(0,k)$ in the product. \\
{\bf ENDFOR} \\
\noindent Set $\bmvbr_k=\ldots,\vbr_{k}^t,\vbr_{k}^{t-1},\ldots$. \\
{\bf ENDFOR} \\
\noindent Then $\bmvdt\cdot\bmvddt\rmdef\bmvbr$, where
$$
\bmvbr=(\bmvbr_\ell,\bmvbr_{\ell-1},\ldots,\bmvbr_k,\ldots,\bmvbr_2,\bmvbr_1,\bmvbr_0)^T.
$$
{\bf ENDDO}
\end{alg}
Notice that this algorithm is similar to the decomposition algorithm, 
Algorithm \ref{alg2}.  In addition, we see this algorithm is an extension
of the evaluation of a product operation for a single elementary group
$\grpjktarg{0}{0}{t}{\calv}{\odot}$ given in the preceding example,
with the finite sequence $\bmvbr_k^{[t,t-k]}$ replaced by the infinite
sequence $\bmvbr_k$.

\vspace{4mm}
\begin{lem}
\label{lem94}
We have that $\bmvdt\cdot\bmvddt\in\calv$ and the operation $\bmvdt\cdot\bmvddt$ 
is well defined.
\end{lem}

\begin{prf}
It is clear that $\vbr_{k}^t$ in Algorithm \ref{alg3} is an element in $V_k^t$.  Then
the sequence $\bmvbr_k$ is an element in the Cartesian product 
$\bigotimes_{t=+\infty}^{-\infty}V_k^t$.  Then $\bmvbr\in\calv$.

Let $\bmvgr,\bmvac\in\calv$ and suppose $\bmvgr=\bmvdt$ and $\bmvac=\bmvddt$.
Then it is clear by the evaluation of $\bmvgr\cdot\bmvac$ and $\bmvdt\cdot\bmvddt$
using Algorithm \ref{alg3} that $\bmvgr\cdot\bmvac=\bmvdt\cdot\bmvddt$.
\end{prf}

We now show $(\calv,\cdot)$ forms a group.  First we use the groups in the
elementary system to construct two elements of $\calv$, an element $\bone_\calv$ 
that is a candidate to be the identity of $(\calv,\cdot)$, and an
element that is a candidate to be an inverse in $(\calv,\cdot)$.  In the
construction, we reverse the decomposition procedure in Algorithm \ref{alg2}.  

We first construct $\bone_\calv$.  Fix $k$ such that $0\le k\le\ell$, 
and consider the sequence
\be
\label{sdk1}
\ldots,\tridnjktarg{0}{k}{t}{\bone_{0,k}^t},\tridnjktarg{0}{k}{t-1}{\bone_{0,k}^{t-1}},\ldots
\ee
of identities of groups
$$
\ldots,\grpjktarg{0}{k}{t}{\calv}{\odot},\grpjktarg{0}{k}{t-1}{\calv}{\odot},\ldots.
$$
Let $1_k^t$ be the single entry in index position $(0,k)$ in
$\tridnjktarg{0}{k}{t}{\bone_{0,k}^t}$ in (\ref{sdk1}).  We know $1_k^t$ is an 
element in $V_k^t$.  Then 
$$
\bone_{k}\rmdef\ldots,1_k^t,1_k^{t-1},\ldots
$$
is a sequence of elements in the Cartesian product 
$\bigotimes_{t=+\infty}^{-\infty}V_k^t$.  Define $\bone_\calv$ by
$$
\bone_\calv\rmdef(\bone_\ell,\bone_{\ell-1},\ldots,\bone_k,\ldots,\bone_2,\bone_1,\bone_0)^T.
$$
Since each $\bone_k$ is a sequence of elements in the Cartesian product 
$\bigotimes_{t=+\infty}^{-\infty}V_k^t$, then $\bone_\calv$ is an element
in $\calv$.  We will use the following lemma to show that $\bone_\calv$ 
is the identity of $(\calv,\cdot)$.

\begin{lem}
\label{lem99}
For each $k$ such that $0\le k\le\ell$, for each $t\in\bmcpz$, the following statement is true:  
the identity $\tridnjktarg{0}{k}{t}{\bone_{0,k}^t}$ of group $\grpjktarg{0}{k}{t}{\calv}{\odot}$
is $\tridnjktarg{0}{k}{t}{\bone_\calv}$.
\end{lem}

\begin{prf}
We prove this by induction.  Fix $t\in\bmcpz$.  First consider $k=\ell$.
The identity of group $\grpjktarg{0}{\ell}{t}{\calv}{\odot}$ is $\tridnjktarg{0}{\ell}{t}{\bone_{0,\ell}^t}$.
But $\tridnjktarg{0}{\ell}{t}{\bone_{0,\ell}^t}$ is the single element $1_\ell^t$, 
and $1_\ell^t$ is an element in the sequence $\bone_\ell$ of $\bone_\calv$.
Therefore $\tridnjktarg{0}{\ell}{t}{\bone_{0,\ell}^t}$ is the same as $\tridnjktarg{0}{\ell}{t}{\bone_\calv}$.
Therefore the statement is true for $k=\ell$.

In general, assume the statement is true for $k+1$, for $0<k+1\le\ell$; 
we show it is true for $k$.  We want to show that
the identity $\tridnjktarg{0}{k}{t}{\bone_{0,k}^t}$ of group $\grpjktarg{0}{k}{t}{\calv}{\odot}$
is $\tridnjktarg{0}{k}{t}{\bone_\calv}$.
There is a homomorphism from group $\grpjktarg{0}{k}{t}{\calv}{\odot}$ to
group $\grpjktarg{0}{k+1}{t}{\calv}{\odot}$ and group 
$\grpjktarg{0}{k+1}{t-1}{\calv}{\odot}$ given by the projection of set
$\tridnjktarg{0}{k}{t}{\calv}$ to set $\tridnjktarg{0}{k+1}{t}{\calv}$ and set
$\tridnjktarg{0}{k+1}{t-1}{\calv}$.  Therefore
the projection of the identity $\tridnjktarg{0}{k}{t}{\bone_{0,k}^t}$ of group
$\grpjktarg{0}{k}{t}{\calv}{\odot}$ is the identity 
$\tridnjktarg{0}{k+1}{t}{\bone_{0,k+1}^t}$ of group $\grpjktarg{0}{k+1}{t}{\calv}{\odot}$ and 
the identity $\tridnjktarg{0}{k+1}{t-1}{\bone_{0,k+1}^{t-1}}$ of group $\grpjktarg{0}{k+1}{t-1}{\calv}{\odot}$.
But by assumption
$\tridnjktarg{0}{k+1}{t}{\bone_{0,k+1}^t}$ is the same as $\tridnjktarg{0}{k+1}{t}{\bone_\calv}$
and $\tridnjktarg{0}{k+1}{t-1}{\bone_{0,k+1}^{t-1}}$ is the same as $\tridnjktarg{0}{k+1}{t-1}{\bone_\calv}$.
Then except possibly for the element in index position $(0,k)$, $\tridnjktarg{0}{k}{t}{\bone_{0,k}^t}$ 
is the same as $\tridnjktarg{0}{k}{t}{\bone_\calv}$.
Now the element in index position $(0,k)$ in $\tridnjktarg{0}{k}{t}{\bone_{0,k}^t}$ is $1_k^t$,
and $1_k^t$ is an element in the sequence $\bone_k$ of $\bone_\calv$.
Therefore $\tridnjktarg{0}{k}{t}{\bone_{0,k}^t}$ is the same as $\tridnjktarg{0}{k}{t}{\bone_\calv}$.
Therefore the statement is true for $k$.
\end{prf}

Fix any $\bmv\in\calv$.  We now construct an element $\bmv^{-1}\in\calv$ 
that is a candidate to be the inverse of $\bmv$ in $(\calv,\cdot)$.
The procedure is very similar to constructing $\bone_\calv$.
Fix $k$ such that $0\le k\le\ell$, and consider the sequence
\be
\label{sdk2}
\ldots,\tridnjktarg{0}{k}{t}{\bme_{0,k}^t},\tridnjktarg{0}{k}{t-1}{\bme_{0,k}^{t-1}},\ldots
\ee
of inverses of the sequence of elements
$$
\ldots,\tridnjktarg{0}{k}{t}{\bmv},\tridnjktarg{0}{k}{t-1}{\bmv},\ldots
$$
in the sequence of groups
$$
\ldots,\grpjktarg{0}{k}{t}{\calv}{\odot},\grpjktarg{0}{k}{t-1}{\calv}{\odot},\ldots.
$$
Let $e_k^t$ be the single entry in index position $(0,k)$ in
$\tridnjktarg{0}{k}{t}{\bme_{0,k}^t}$ in (\ref{sdk2}).  We know $e_k^t$ is an 
element in $V_k^t$.  Then 
$$
\bme_k\rmdef\ldots,e_k^t,e_k^{t-1},\ldots
$$
is a sequence of elements in the Cartesian product 
$\bigotimes_{t=+\infty}^{-\infty}V_k^t$.  Define $\bmv^{-1}$ by
$$
\bmv^{-1}\rmdef(\bme_\ell,\bme_{\ell-1},\ldots,\bme_k,\ldots,\bme_2,\bme_1,\bme_0)^T.
$$
Since each $\bme_k$ is a sequence of elements in the Cartesian product 
$\bigotimes_{t=+\infty}^{-\infty}V_k^t$, then $\bmv^{-1}$ is an element
in $\calv$.  We will use the following lemma to show that $\bmv^{-1}$ 
is the inverse of $\bmv$ in $(\calv,\cdot)$.

\begin{lem}
\label{lem100}
For each $k$ such that $0\le k\le\ell$, for each $t\in\bmcpz$, the following statement is true:  
the inverse $\tridnjktarg{0}{k}{t}{\bme_{0,k}^t}$ of element $\tridnjktarg{0}{k}{t}{\bmv}$ in
group $\grpjktarg{0}{k}{t}{\calv}{\odot}$ is $\tridnjktarg{0}{k}{t}{\bmv^{-1}}$.
\end{lem}
 
\begin{prf}
We prove this by induction.  Fix $t\in\bmcpz$.  First consider $k=\ell$.
The inverse of element $\tridnjktarg{0}{\ell}{t}{\bmv}$ in
group $\grpjktarg{0}{\ell}{t}{\calv}{\odot}$ is $\tridnjktarg{0}{\ell}{t}{\bme_{0,\ell}^t}$.
But $\tridnjktarg{0}{\ell}{t}{\bme_{0,\ell}^t}$ is the single element $e_\ell^t$, 
and $e_\ell^t$ is an element in the sequence $\bme_\ell$ of $\bmv^{-1}$.
Therefore $\tridnjktarg{0}{\ell}{t}{\bme_{0,\ell}^t}$ is the same as $\tridnjktarg{0}{\ell}{t}{\bmv^{-1}}$.
Therefore the statement is true for $k=\ell$.

In general, assume the statement is true for $k+1$, for $0<k+1\le\ell$; 
we show it is true for $k$.  We want to show that
the inverse $\tridnjktarg{0}{k}{t}{\bme_{0,k}^t}$ of element $\tridnjktarg{0}{k}{t}{\bmv}$ in
group $\grpjktarg{0}{k}{t}{\calv}{\odot}$ is $\tridnjktarg{0}{k}{t}{\bmv^{-1}}$.
There is a homomorphism from group $\grpjktarg{0}{k}{t}{\calv}{\odot}$ to
group $\grpjktarg{0}{k+1}{t}{\calv}{\odot}$ and group 
$\grpjktarg{0}{k+1}{t-1}{\calv}{\odot}$ given by the projection of set
$\tridnjktarg{0}{k}{t}{\calv}$ to set $\tridnjktarg{0}{k+1}{t}{\calv}$ and set
$\tridnjktarg{0}{k+1}{t-1}{\calv}$.  Therefore
the projection of the inverse $\tridnjktarg{0}{k}{t}{\bme_{0,k}^t}$ of $\tridnjktarg{0}{k}{t}{\bmv}$ in
group $\grpjktarg{0}{k}{t}{\calv}{\odot}$ is the inverse 
$\tridnjktarg{0}{k+1}{t}{\bme_{0,k+1}^t}$ of $\tridnjktarg{0}{k+1}{t}{\bmv}$ 
in group $\grpjktarg{0}{k+1}{t}{\calv}{\odot}$ and 
the inverse $\tridnjktarg{0}{k+1}{t-1}{\bme_{0,k+1}^{t-1}}$ of $\tridnjktarg{0}{k+1}{t-1}{\bmv}$ 
in group $\grpjktarg{0}{k+1}{t-1}{\calv}{\odot}$.  But by assumption
$\tridnjktarg{0}{k+1}{t}{\bme_{0,k+1}^t}$ is the same as $\tridnjktarg{0}{k+1}{t}{\bmv^{-1}}$
and $\tridnjktarg{0}{k+1}{t-1}{\bme_{0,k+1}^{t-1}}$ is the same as $\tridnjktarg{0}{k+1}{t-1}{\bmv^{-1}}$.
Then except possibly for the element in index position $(0,k)$, $\tridnjktarg{0}{k}{t}{\bme_{0,k}^t}$ 
is the same as $\tridnjktarg{0}{k}{t}{\bmv^{-1}}$.
Now the element in index position $(0,k)$ in $\tridnjktarg{0}{k}{t}{\bme_{0,k}^t}$ is $e_k^t$,
and $e_k^t$ is an element in the sequence $\bme_k$ of $\bmv^{-1}$.
Therefore $\tridnjktarg{0}{k}{t}{\bme_{0,k}^t}$ is the same as $\tridnjktarg{0}{k}{t}{\bmv^{-1}}$.
Therefore the statement is true for $k$.
\end{prf}
Since from Lemma \ref{lem99}, $\tridnjktarg{0}{k}{t}{\bone_\calv}$ is the identity 
of group $\grpjktarg{0}{k}{t}{\calv}{\odot}$, we have that
$$
\tridnjktarg{0}{k}{t}{\bmv^{-1}}\prodjktarg{0}{k}{t}{\odot}\tridnjktarg{0}{k}{t}{\bmv}
=\tridnjktarg{0}{k}{t}{\bone_\calv}
$$
and
$$
\tridnjktarg{0}{k}{t}{\bmv}\prodjktarg{0}{k}{t}{\odot}\tridnjktarg{0}{k}{t}{\bmv^{-1}}
=\tridnjktarg{0}{k}{t}{\bone_\calv}.
$$

Using Lemmas \ref{lem99} and \ref{lem100}, we can prove the following.

\begin{thm}
The set $\calv$ with operation $\cdot$ forms a group $(\calv,\cdot)$.
\end{thm}

\begin{prf}
First we show the operation $\cdot$ is associative.
Let $\bmv,\bmvdt,\bmvddt\in\calv$.  We need to show
\be
\label{eva00}
(\bmv\cdot\bmvdt)\cdot\bmvddt
=\bmv\cdot(\bmvdt\cdot\bmvddt).
\ee
To find the left hand side of (\ref{eva00}), for each $k$ from $k=\ell$ to
$0$, we first evaluate
\be
\label{eval11}
(\tridnjktarg{0}{k}{t}{\bmv}\prodjktarg{0}{k}{t}{\odot}\tridnjktarg{0}{k}{t}{\bmvdt})\prodjktarg{0}{k}{t}{\odot}\tridnjktarg{0}{k}{t}{\bmvddt}
\ee
for $t=+\infty$ to $-\infty$.
And to find the right hand side of (\ref{eva00}), for each $k$ from $k=\ell$ to
$0$, we first evaluate
\be
\label{eval12}
\tridnjktarg{0}{k}{t}{\bmv}\prodjktarg{0}{k}{t}{\odot}(\tridnjktarg{0}{k}{t}{\bmvdt}\prodjktarg{0}{k}{t}{\odot}\tridnjktarg{0}{k}{t}{\bmvddt})
\ee
for $t=+\infty$ to $-\infty$.  But we know group
$\grpjktarg{0}{k}{t}{\calv}{\odot}$ is associative so (\ref{eval11}) is the 
same as (\ref{eval12}).  This means the left hand side and right hand side
of (\ref{eva00}) evaluate to the same element in $\calv$.

We show $\bone_\calv$ is the identity of $(\calv,\cdot)$.  Let $\bmv\in\calv$.  
We need to show $\bone_\calv\cdot\bmv=\bmv$ and $\bmv\cdot\bone_\calv=\bmv$.
First let $\bmvbr\rmdef\bone_\calv\cdot\bmv$.  We use Algorithm \ref{alg3} to evaluate
$\bone_\calv\cdot\bmv$ and find $\bmvbr$.  From Algorithm \ref{alg3}, 
to find $\bone_\calv\cdot\bmv$, for each $k$ from $k=\ell$ to $0$, we first evaluate
\be
\label{eval15}
\tridnjktarg{0}{k}{t}{\bone_\calv}\prodjktarg{0}{k}{t}{\odot}\tridnjktarg{0}{k}{t}{\bmv}
\ee
for $t=+\infty$ to $-\infty$.  But we know from Lemma \ref{lem99},
$\tridnjktarg{0}{k}{t}{\bone_\calv}$ is the identity of group
$\grpjktarg{0}{k}{t}{\calv}{\odot}$.  Then (\ref{eval15}) reduces to
\be
\label{eval16}
\tridnjktarg{0}{k}{t}{\bone_\calv}\prodjktarg{0}{k}{t}{\odot}\tridnjktarg{0}{k}{t}{\bmv}
=\tridnjktarg{0}{k}{t}{\bmv}.
\ee
Continuing with Algorithm \ref{alg3},
the entry in index position $(0,k)$ in the product $\tridnjktarg{0}{k}{t}{\bmv}$
is $v_{k}^t$.  Set $\bmvbr_k=\ldots,v_{k}^t,v_{k}^{t-1},\ldots$.
Then $\bone_\calv\cdot\bmv=\bmvbr$, where
$$
\bmvbr=(\bmvbr_\ell,\bmvbr_{\ell-1},\ldots,\bmvbr_k,\ldots,\bmvbr_2,\bmvbr_1,\bmvbr_0)^T.
$$
But comparing $\bmvbr$ with $\bmv$ in (\ref{decomp}) shows that $\bmvbr=\bmv$ and so
$\bone_\calv\cdot\bmv=\bmv$.  A similar argument shows that $\bmv\cdot\bone_\calv=\bmv$.

Let $\bmv\in\calv$.  We show $\bmv^{-1}$ is the inverse of $\bmv$ in $(\calv,\cdot)$.
We need to show $\bmv^{-1}\cdot\bmv=\bone_\calv$ and $\bmv\cdot\bmv^{-1}=\bone_\calv$.
First let $\bmvbr\rmdef\bmv^{-1}\cdot\bmv$.  We use Algorithm \ref{alg3} to evaluate
$\bmv^{-1}\cdot\bmv$ and find $\bmvbr$.  From Algorithm \ref{alg3}, 
to find $\bmv^{-1}\cdot\bmv$, for each $k$ from $k=\ell$ to $0$, we first evaluate
\be
\label{eval17}
\tridnjktarg{0}{k}{t}{\bmv^{-1}}\prodjktarg{0}{k}{t}{\odot}\tridnjktarg{0}{k}{t}{\bmv}
\ee
for $t=+\infty$ to $-\infty$.  But we know from Lemma \ref{lem100},
$\tridnjktarg{0}{k}{t}{\bmv^{-1}}$ is the inverse of
$\tridnjktarg{0}{k}{t}{\bmv}$ in $\grpjktarg{0}{k}{t}{\calv}{\odot}$.
Then (\ref{eval17}) reduces to
\be
\label{eval18}
\tridnjktarg{0}{k}{t}{\bmv^{-1}}\prodjktarg{0}{k}{t}{\odot}\tridnjktarg{0}{k}{t}{\bmv}
=\tridnjktarg{0}{k}{t}{\bone_\calv}.
\ee
Continuing with Algorithm \ref{alg3},
the entry in index position $(0,k)$ in the product $\tridnjktarg{0}{k}{t}{\bone_\calv}$
is $1_{k}^t$.  Set $\bmvbr_k=\ldots,1_{k}^t,1_{k}^{t-1},\ldots$.
Then $\bmv^{-1}\cdot\bmv=\bmvbr$, where
$$
\bmvbr=(\bmvbr_\ell,\bmvbr_{\ell-1},\ldots,\bmvbr_k,\ldots,\bmvbr_2,\bmvbr_1,\bmvbr_0)^T.
$$
But since $\bmvbr_k=\bone_k$ then $\bmvbr=\bone_\calv$ and so
$\bmv^{-1}\cdot\bmv=\bone_\calv$.  A similar argument shows that $\bmv\cdot\bmv^{-1}=\bone_\calv$.

Together these results show $(\calv,\cdot)$ is a group.
\end{prf}

We call the group $(\calv,\cdot)$ formed from an $(\ell+1)$-depth elementary system
$\cale$ an {\it $(\ell+1)$-depth global group}.
The groups $\grpjktarg{0}{k}{t}{\calv}{\odot}$ in the elementary system are
the {\it elementary groups} of $(\calv,\cdot)$.  We have just shown the following.

\begin{thm}
Any $(\ell+1)$-depth elementary system $\cale$ forms an $(\ell+1)$-depth 
global group $(\calv,\cdot)$ by the procedure just described.
\end{thm}

The $(\ell+1)$-depth global group is nested.  For example,
the top row of an $(\ell+1)$-depth global group $(\calv,\cdot)$ is a
1-depth global group $(\calv,\cdot)_\ell$.  The top two rows
of an $(\ell+1)$-depth global group $(\calv,\cdot)$ is a
2-depth global group $(\calv,\cdot)_{\ell-1}$.
In general the following holds.

\begin{thm}
The top $m$ rows of an $(\ell+1)$-depth global group $(\calv,\cdot)$ form an
$m$-depth global group $(\calv,\cdot)_{\ell-m+1}$.
\end{thm}

To summarize, in Subsection 7.2 we have constructed the chain
\be
\cale\ra(\calv,\cdot),
\ee
where $\cale$ is an $(\ell+1)$-depth elementary system, and $(\calv,\cdot)$ is an
$(\ell+1)$-depth global group formed from $\cale$.  We finish this subsection with
the following result used in the next subsection.

\begin{lem}
\label{lem95}
The product operation $\bmvdt\cdot\bmvddt$ is equivalent to the evaluation 
of
\be
\label{eq100}
\tridnjktarg{0}{0}{t}{\bmvdt}\prodjktarg{0}{0}{t}{\odot}\tridnjktarg{0}{0}{t}{\bmvddt}.
\ee
for $t=+\infty$ to $-\infty$.
\end{lem}

\begin{prf}
The product operation $\bmvdt\cdot\bmvddt$ defined in Algorithm \ref{alg3} is the same as the 
evaluation of $\tridnjktarg{0}{0}{t}{\bmvdt}\prodjktarg{0}{0}{t}{\odot}\tridnjktarg{0}{0}{t}{\bmvddt}$
by the method given in the example at the beginning of Subsection 7.2 (see (\ref{eexp21})),
for $t=+\infty$ to $-\infty$.
\end{prf}

\begin{thm}
\label{thm11282}
The $(\ell+1)$-depth global group $(\calv,\cdot)$ formed from the $(\ell+1)$-depth elementary system
$\cale$ is uniquely determined by $\cale$.
\end{thm}

\begin{prf}
$\cale$ contains the elementary set $\calv$.
From Lemma \ref{lem95}, the global operation $\cdot$ in $(\calv,\cdot)$ is uniquely determined
by the elementary groups $\{\grpjktarg{0}{0}{t}{\calu}{\ccirc}:  t\in\bmcpz\}$ in the elementary
list of $\cale$.
\end{prf}

\vspace{3mm}
{\bf 7.3  Construction of any group system $C$ from the elementary system}
\vspace{3mm}

We now show that any \ellctl\ complete group system $A$ can be reduced 
to an elementary system $\cale_A$.

\begin{thm}
\label{thm99}
The $(\ell+1)$-depth generator group $(\calu,\circ)$ of any \ellctl\ complete group system $A$ contains an 
$(\ell+1)$-depth elementary system $\cale_A$ with elementary set $\calu$ and elementary list
$\{\grpjktarg{0}{k}{t}{\calu}{\ccirc}:  0\le k\le\ell,t\in\bmcpz\}$.

\end{thm}

\begin{prf}
First we show that $\calu$ can be considered to be an elementary set.
We have seen that $\calu$ is just the double Cartesian product 
(\ref{input4}).  Comparing set $\calu$ in (\ref{input4}) to set $\calv$
in (\ref{dprod1}), we see the double product is interchanged and there is a
bijection between the two sets provided there is a bijection between the
sets $R_{0,k}^t$ and $V_k^t$, for $0\le k\le\ell$, for each $t\in\bmcpz$.
Then $\calu$ is an elementary set.  Lastly, from Theorem \ref{homo4}, 
the elementary groups $\grpjktarg{0}{k}{t}{\calu}{\ccirc}$ of $(\calu,\circ)$, 
for $0\le k\le\ell$ and $t\in\bmcpz$, satisfy the homomorphism condition of groups in an 
$(\ell+1)$-depth elementary system.  We define the elementary system $\cale_A$
to be the elementary set $\calu$ and elementary list
$\{\grpjktarg{0}{k}{t}{\calu}{\ccirc}:  0\le k\le\ell,t\in\bmcpz\}$.
\end{prf}
We call $\cale_A$ the {\it elementary system of} $A$.
Essentially the elementary system $\cale_A$ of $A$ is just the generator group $(\calu,\circ)$ of $A$
stripped of its global operation $\circ$.
Then we can summarize the results of this paper so far by the chain
\be
\label{chain8a}
\begin{array}{lllllllll}
 \da & \la                        & \la   & \la                        & \la         & {\stackrel{=}{\la}}      & \imfu         &      & \\
 \da &                            &       &                            &             &                          & \ua\,f_u      &      & \\
 A   & {\stackrel{\simeq}{\lra}}  & B     & {\stackrel{\simeq}{\ra}}   & (\calr,*)   & {\stackrel{\simeq}{\ra}} & (\calu,\circ) & \ra  & \cale_A        
\end{array}
\ee
where $(\calr,*)$ is a decomposition group; $(\calu,\circ)$ is a generator group; and $\cale_A$ is an 
elementary system.  In the remainder of this subsection, we ask whether we can reverse 
the chain in (\ref{chain8a}), i.e., can we recover $A$ from $\cale_A$.
Then we define a new notion of isomorphism for group systems and show how to obtain
all \ellctl\ complete group systems $C$ up to this new isomorphism from the set of all
$(\ell+1)$-depth elementary systems $\cale$. 

\begin{thm}
\label{thm120}
We know the generator group $(\calu,\circ)$ of any \ellctl\ complete group system $A$ forms an 
$(\ell+1)$-depth elementary system $\cale_A$ with elementary set $\calu$.
Form the $(\ell+1)$-depth global group $(\calu,\star)$ of $\cale_A$.
The global operation $\bmudt\star\bmuddt$ in $(\calu,\star)$ is the same as the global 
operation $\bmudt\circ\bmuddt$ in $(\calu,\circ)$.  Therefore there is an isomorphism 
$(\calu,\star)\simeq (\calu,\circ)$ under the 1-1 correspondence $\calu=\calu$ 
given by the assignment $\bmu=\bmu$.
\end{thm}

\begin{prf}
From Theorem \ref{thm99}, we have seen that $\cale_A$ is an elementary system
with elementary set $\calu$ and elementary groups $\grpjktarg{0}{k}{t}{\calu}{\ccirc}$.
We use $\cale_A$ to define a global group $(\calu,\star)$.  Let $\bmudt,\bmuddt\in\calu$.  
From Lemma \ref{lem95}, the product operation $\bmudt\star\bmuddt$ in global group 
$(\calu,\star)$ is equivalent to the evaluation of
\be
\label{subop11}
\tridnjktarg{0}{0}{t}{\bmudt}\prodjktarg{0}{0}{t}{\ccirc}\tridnjktarg{0}{0}{t}{\bmuddt},
\ee
for $t=+\infty$ to $-\infty$.  From Lemma \ref{lem95a},
the product operation $\bmudt\circ\bmuddt$ in generator group $(\calu,\circ)$ 
is equivalent to the evaluation of
\be
\label{subop12}
\tridnjktarg{0}{0}{t}{\bmudt}\prodjktarg{0}{0}{t}{\ccirc}\tridnjktarg{0}{0}{t}{\bmuddt}
\ee
for $t=+\infty$ to $-\infty$.  We see that (\ref{subop11}) and 
(\ref{subop12}) are the same.  Therefore $\bmudt\star\bmuddt$ and $\bmudt\circ\bmuddt$ 
are the same.  Therefore there is an isomorphism $(\calu,\star)\simeq (\calu,\circ)$ under the
1-1 correspondence $\calu=\calu$ given by the assignment $\bmu=\bmu$.
\end{prf}

For $\bmudt,\bmuddt\in\calu$, the global operation $\bmudt\star\bmuddt$ in $(\calu,\star)$ 
is the same as the global operation $\bmudt\circ\bmuddt$ in $(\calu,\circ)$.  This is a
little stronger condition than isomorphism, and we say that $(\calu,\star)$ is
{\it essentially identical} to $(\calu,\circ)$, written $(\calu,\star)\eid (\calu,\circ)$.

\begin{lem}
The global group $(\calu,\star)$ of $\cale_A$ is essentially identical to the generator group 
$(\calu,\circ)$ of $A$, $(\calu,\star)\eid (\calu,\circ)$.
\end{lem}

We have just constructed the chain
\be
\label{chain8b}
\cale_A\ra(\calu,\star),
\ee
where $\cale_A$ is an $(\ell+1)$-depth elementary system, and $(\calu,\star)$ is an
$(\ell+1)$-depth global group.  We can incorporate the chain (\ref{chain8b}) into chain
(\ref{chain8a}) as follows.
\be
\label{chain8c}
\begin{array}{lllllllll}
 \da & \la                        & \la   & \la                        & \la         & {\stackrel{=}{\la}}      & \imfu         &     &             \\
 \da &                            &       &                            &             &                          & \ua\,f_u      &     &             \\
 A   & {\stackrel{\simeq}{\lra}}  & B     & {\stackrel{\simeq}{\ra}}   & (\calr,*)   & {\stackrel{\simeq}{\ra}} & (\calu,\circ) & \ra & \cale_A     \\       
     &                            &       &                            &             &                          & \uda\,\eid    &     & \uda =      \\
     &                            &       &                            &             &                          & (\calu,\star) & \la & \cale_A
\end{array}
\ee
This gives the following.
\begin{thm}
\label{thm116}
We may recover any \ellctl\ complete group system $A$ from the $(\ell+1)$-depth 
elementary system $\cale_A$ of $A$ using the chain (\ref{zchain8c}).
\be
\label{zchain8c}
\begin{array}{lllll}
\da & {\stackrel{=}{\la}}        & \imfu         &      &         \\
\da &                            & \ua\,f_u      &      &         \\
A   &                            & (\calu,\star) & \la  & \cale_A    
\end{array}
\ee
\end{thm}

\begin{prf}
Since $(\calu,\circ)\eid (\calu,\star)$, we can easily recover the generator group 
$(\calu,\circ)$ of $A$ from the global group $(\calu,\star)$ of $\cale_A$.
Then we can recover $A$ from $(\calu,\circ)$
using the homomorphism $f_u$ in the \fhgs\ as done previously in Subsection 6.5.
\end{prf}

In fact, we can recover $A$ directly from $(\calu,\star)$ using the \fhgs.
The homomorphism $f_u$ in the top half of chain (\ref{chain8c})
just uses the primary elementary groups $\grpjktarg{0}{0}{t}{\calu}{\ccirc}$ of $(\calu,\circ)$
for each $t\in\bmcpz$.  But the latter groups are already available in $(\calu,\star)$.
In fact these groups are available in $\cale_A$, so $A$ can be recovered
directly from the elementary system $\cale_A$ as well.

So far we have shown that the set of all elementary systems $\cale_A$
of all \ellctl\ complete group systems $A$, or $\{\cale_A\}$, is contained in the set of all 
$(\ell+1)$-depth elementary systems $\cale$, $\{\cale\}$, or $\{\cale_A\}\subset\{\cale\}$.
We now show that we can construct at least one \ellctl\ complete group system $A$ from
any $(\ell+1)$-depth elementary system $\cale$.

Assume we are given any $(\ell+1)$-depth elementary system $\cale$.  First find the
$(\ell+1)$-depth global group $(\calv,\cdot)$ of $\cale$.  We construct an 
\ellctl\ complete group system from global group $(\calv,\cdot)$.
In Theorem \ref{thm83} at the beginning of Subsection 6.6, 
we showed that we could always construct a special
\ellctl\ complete group system $(U_s,\ovcirc)$ from any $(\ell+1)$-depth generator group 
$(\calu,\circ)$ of $A$.
The construction of $(U_s,\ovcirc)$ uses the \fhgs\ with $(\calu,\circ)$ as an input group,
as summarized in chain (\ref{chain7f}).  But the \fhgs\ can accept any group as input,
and there is no essential difference between any $(\ell+1)$-depth global group $(\calv,\cdot)$ and any 
$(\ell+1)$-depth generator group $(\calu,\circ)$.  Therefore,
if we replace $(\ell+1)$-depth generator group $(\calu,\circ)$ in Theorem \ref{thm83} with 
$(\ell+1)$-depth global group $(\calv,\cdot)$ as an input, we may use the exact same approach 
of Theorem \ref{thm83} and chain (\ref{chain7f}) of Subsection 6.6
to construct a special \ellctl\ complete group system
$(V_s,\ovcdot)$ from global group $(\calv,\cdot)$.  We give this construction now.

We change the prelude to Theorem \ref{thm83} in Subsection 6.6 to accommodate $(\calv,\cdot)$
instead of $(\calu,\circ)$.  For each $t\in\bmcpz$, we simply let the alphabet group $A^t$ 
be $\grpjktarg{0}{0}{t}{\calv}{\odot}$ instead of $\grpjktarg{0}{0}{t}{\calu}{\ccirc}$.
For each $t\in\bmcpz$, define a map $\theta_v^t:  \calv\ra\tridnjktarg{0}{0}{t}{\calv}$ with assignment
$\theta_v^t:  \bmv\ra\tridnjktarg{0}{0}{t}{\bmv}$.  Using Lemma \ref{lem95}, the map
$\theta_v^t$ is a homomorphism from $(\calv,\cdot)$ to $\grpjktarg{0}{0}{t}{\calv}{\odot}$.  
Consider the Cartesian product 
$$
V_\amalg\rmdef\cdots\times\tridnjktarg{0}{0}{t}{\calv}\times\tridnjktarg{0}{0}{t+1}{\calv}\times\cdots.
$$
Define the direct product group $(V_\amalg,\ovcdot)$ by
$$
(V_\amalg,\ovcdot)\rmdef\cdots\times\grpjktarg{0}{0}{t}{\calv}{\odot}\times\grpjktarg{0}{0}{t+1}{\calv}{\odot}\times\cdots.
$$
Then from Theorem \ref{fhgs}, using $(\calv,\cdot)$ for information group $\msfcpg$
and the primary elementary group $\grpjktarg{0}{0}{t}{\calv}{\odot}$ for $\msfcpg^t$, $t\in\bmcpz$, 
there is a homomorphism $\theta_v:  \calv\ra V_\amalg$, from $(\calv,\cdot)$
to the direct product group $(V_\amalg,\ovcdot)$, defined by
$$
\theta_v(\bmv)\rmdef\ldots,\theta_v^t(\bmv),\theta_v^{t+1}(\bmv),\ldots.
$$
Define
\begin{align*}
\bmv_s\rmdef &\ldots,\theta_v^t(\bmv),\theta_v^{t+1}(\bmv),\ldots \\
           = &\ldots,\tridnjktarg{0}{0}{t}{\bmv},\tridnjktarg{0}{0}{t+1}{\bmv},\ldots.
\end{align*}
Then $\theta_v:  \calv\ra V_\amalg$ with assignment $\theta_v:  \bmv\mapsto\bmv_s$.
We can think of $\bmv_s$ as the sequence of triangles $\tridnjktarg{0}{0}{t}{\bmv}$ of $\bmv$, 
now written in conventional time order and not overlapped.  Then
$$
(\calv,\cdot)/(\calv,\cdot)_K\simeq\imtheta_v,
$$
where group $\imtheta_v$ is the image of the homomorphism $\theta_v$, and where 
$(\calv,\cdot)_K$ is the kernel of the homomorphism $\theta_v$.  Since group $\imtheta_v$ is a
subgroup of the direct product group $(V_\amalg,\ovcdot)$, then $\imtheta_v$ is a
group system where global operation $\ovcdot$ is defined by the componentwise 
operation $\prodjktarg{0}{0}{t}{\odot}$ in group $\grpjktarg{0}{0}{t}{\calv}{\odot}$ 
for each $t\in\bmcpz$.  We denote group $\imtheta_v$ by $(V_s,\ovcdot)$,
where $V_s$ is the subset of the Cartesian product $\calv_\amalg$ determined by $\bmv\in\calv$, or
equivalently the subset of $\calv_\amalg$ defined by $\imtheta_v$.
We call $(V_s,\ovcdot)$ the {\it global group system} of $(\calv,\cdot)$.  

In the same manner as for Theorem \ref{thm83}, the homomorphism $\theta_v$ is a bijection.  
We have $(\calv,\cdot)\simeq\imtheta_v=(V_s,\ovcdot)$ under the assignment
$\theta_v: \bmv\mapsto\bmv_s$ given by the bijection $\theta_v: \calv\ra V_s$.
In the same manner as for Theorem \ref{thm81},
the group system $(V_s,\ovcdot)=\imtheta_v$ is \ellctl\ and complete.
And in the same manner as for Lemma \ref{lem95b}, the generator group of $(V_s,\ovcdot)$
is $(\calv,\cdot)$.

Then chain (\ref{chain7f}) of Theorem \ref{thm83} is modified as shown in
chain (\ref{ychain7f}).
\be
\label{ychain7f}
\begin{array}{lll}
\da             & {\stackrel{=}{\la}} & \imtheta_v                                             \\
\da             &                     & \ua\,\theta_v=\ldots,\theta_v^t,\theta_v^{t+1},\ldots  \\
(V_s,\ovcdot)   &                     & (\calv,\cdot)
\end{array}
\ee
We can incorporate the $(\ell+1)$-depth elementary system $\cale$ into chain (\ref{ychain7f})
as shown in chain (\ref{zchain8e}).
\be
\label{zchain8e}
\begin{array}{lllll}
\da             & {\stackrel{=}{\la}} & \imtheta_v        &      &        \\
\da             &                     & \ua\,\theta_v     &      &        \\
(V_s,\ovcdot)   &                     & (\calv,\cdot)     & \la  & \cale    
\end{array}
\ee
We can summarize these results as follows.

\begin{thm}
\label{thm115}
Given any $(\ell+1)$-depth elementary system $\cale$, we may always use the chain (\ref{zchain8e})
to construct an \ellctl\ complete group system $(V_s,\ovcdot)$ from the $(\ell+1)$-depth 
global group $(\calv,\cdot)$ of $\cale$.
The homomorphism $\theta_v$ is a bijection, and we have $(\calv,\cdot)\simeq\imtheta_v=(V_s,\ovcdot)$.
The generator group of $(V_s,\ovcdot)$ is $(\calv,\cdot)$,
and the elementary system $\cale_{(V_s,\ovcdot)}$ of $(V_s,\ovcdot)$ is $\cale$.  
\end{thm}

\begin{thm}
\label{thm119}
The set of all elementary systems $\cale_A$
of all \ellctl\ complete group systems $A$, or $\{\cale_A\}$, is the same as the set of all 
$(\ell+1)$-depth elementary systems $\cale$, $\{\cale\}$, or $\{\cale_A\}=\{\cale\}$.
\end{thm}

\begin{prf}
Theorem \ref{thm116} shows that $\{\cale_A\}\subset\{\cale\}$.  The group system
$(V_s,\ovcdot)$ of Theorem \ref{thm115} is an \ellctl\ complete group system $A$.
Then Theorem \ref{thm115} shows that given any $(\ell+1)$-depth elementary system $\cale$,
we can find an \ellctl\ complete group system $A$ whose elementary system 
$\cale_A$ is $\cale$; then $\{\cale\}\subset\{\cale_A\}$.
\end{prf}

Since $\{\cale_A\}=\{\cale\}$, we can henceforth assume that any elementary system $\cale$
has an elementary set which uses notation $\calu$, an elementary list which uses notation
$\{\grpjktarg{0}{k}{t}{\calu}{\ccirc}:  0\le k\le\ell,t\in\bmcpz\}$, and a
global group which uses notation $(\calu,\star)$.  As previously mentioned,
the entries $r_{0,k}^t$ in $\calu$ are to be regarded as abstract labels
rather than representatives.  Then again
we can use Theorem \ref{thm83} and chain (\ref{chain7f}) of Subsection 6.6
to construct a special \ellctl\ complete group system
$(U_s,\ovstar)$ from global group $(\calu,\star)$, the {\it global group system}
of $(\calu,\star)$, after replacing group $(\calu,\circ)$ with group $(\calu,\star)$ and 
group $(U_s,\ovcirc)$ with group $(U_s,\ovstar)$ in Theorem \ref{thm83}.  
The elementary system $\cale$ can be incorporated into chain (\ref{chain7f}) 
as shown in chain (\ref{zchain7f}).
\be
\label{zchain7f}
\begin{array}{lllll}
\da             & {\stackrel{=}{\la}} & \imtheta      &      &          \\
\da             &                     & \ua\,\theta   &      &          \\
(U_s,\ovstar)   &                     & (\calu,\star) & \la  & \cale
\end{array}
\ee
Chain (\ref{zchain7f}) is essentially just a repeat of chain (\ref{zchain8e}) with $(\calu,\star)$
in place of $(\calv,\cdot)$ and  $(U_s,\ovstar)$ in place of $(V_s,\ovcdot)$.
We have the following restatement of Theorem \ref{thm115} using chain (\ref{zchain7f})
in place of chain (\ref{zchain8e}).

\begin{thm}
\label{thm115a}
Given any $(\ell+1)$-depth elementary system $\cale$, we may always use the chain (\ref{zchain7f})
to construct an \ellctl\ complete group system $(U_s,\ovstar)$ from the $(\ell+1)$-depth 
global group $(\calu,\star)$ of $\cale$.
The homomorphism $\theta$ is a bijection, and we have $(\calu,\star)\simeq\imtheta=(U_s,\ovstar)$.
The generator group of $(U_s,\ovstar)$ is $(\calu,\star)$,
and the elementary system $\cale_{(U_s,\ovstar)}$ of $(U_s,\ovstar)$ is $\cale$.  
\end{thm}
The chain (\ref{zchain7f}) forms a linear system with input group $(\calu,\star)$,
homorphism $\theta$, and output group $(U_s,\ovstar)$.  Since homorphism $\theta$ is a bijection, 
then the linear system in (\ref{zchain7f}) is invertible.

We have seen in Theorem \ref{thm84} of Subsection 6.6
how to construct all group systems $C$ that have the generator 
group $(\calu,\circ)$ of an \ellctl\ complete group system $A$ as an input group.
The elementary groups of $(\calu,\circ)$ are the
same as the elementary groups of $(\calu,\star)$, and so the 
global group $(\calu,\star)$ of $\cale$ is essentially identical to
the generator group $(\calu,\circ)$ of $A$.  Therefore we can reuse Theorem \ref{thm84} 
and chain (\ref{chain7e}) of Subsection 6.6 to construct all group systems $C$ that have the global
group $(\calu,\star)$ of $\cale$ as input group, after simply replacing group $(\calu,\circ)$
with group $(\calu,\star)$ in Theorem \ref{thm84}.
The elementary system $\cale$ can be incorporated into chain (\ref{chain7e}) 
as shown in chain (\ref{zchain7e}).
\be
\label{zchain7e}
\begin{array}{lllll}
\da & {\stackrel{=}{\la}}        & \imhu         &      &        \\
\da &                            & \ua\,h_u      &      &        \\
C   &                            & (\calu,\star) & \la  & \cale
\end{array}
\ee
In chain (\ref{zchain7e}), $\cale$ is any $(\ell+1)$-depth elementary system, $(\calu,\star)$ is an
$(\ell+1)$-depth global group formed from $\cale$, and $h_u$ is a homomorphism in the \fhgs\
given in Theorem \ref{thm84}.  The linear system in (\ref{zchain7e}) of input group $(\calu,\star)$,
homorphism $h_u$, and output group $C$ is invertible if homomorphism $h_u$ is a bijection.

In general, any group system $C$ constructed from chain (\ref{chain7e}) or
chain (\ref{zchain7e}) is $l$-controllable, 
where $l\le\ell$, as discussed in Theorem \ref{thm85} of Subsection 6.6.
We have seen from Example 1 that we may have $l<\ell$.
In Subsection 6.6 we showed how to avoid the case $l<\ell$
and search for all \ellctl\ complete group systems $C$
up to isomorphism by starting from equivalence classes of the generator group.
We now give a similar result and show how to
search for all \ellctl\ complete group systems $C$
up to a new definition of isomorphism by starting from equivalence classes of all 
$(\ell+1)$-depth global groups $(\calu,\star)$ of any $(\ell+1)$-depth elementary system $\cale$.

The definition of group system isomorphism given in Section 2 is just the definition of
isomorphism for finite groups applied to group systems.  But the definition of isomorphism of 
finite groups is somewhat defective for group systems since it
does not include the notion of time used in group systems.  We have seen in Example 1
of Subsection 6.6 some of the problems that this can cause.  We now give a new definition 
of group system isomorphism including the notion of time that may be more appropriate than the
definition for finite groups.  We say that two group systems $A$ and $A'$ are isomorphic
in this new definition if their elementary systems $\cale_A$ and $\cale_{A'}$ are 
essentially the same.  Then we give a systematic way to search
for all \ellctl\ complete group systems $A$ up to this new isomorphism.

Consider an elementary system $\cale$ defined on elementary set $\calu$ and 
elementary list $\call$, and an elementary system $\caleht$ defined on elementary set $\caluht$ and 
elementary list $\callht$.  We use notation $U_k^t$ for $\calu$ in place of $V_k^t$ for
$\calv$, and similarly $\cpuht_k^t$ for $\caluht$.  We say two elementary systems $\cale$ and $\caleht$ are 
{\it list isomorphic} under bijection $\lambda$, and write $\cale\stackrel{l}{\simeq}\caleht$, 
if there is a bijection $\lambda:  \calu\ra\caluht$ formed by bijections 
$\lambda_k^t:  U_k^t\ra\cpuht_k^t$, for $0\le k\le\ell$ and $t\in\bmcpz$, and if the groups 
$\grpjktarg{0}{k}{t}{\calu}{\ccirc}$ and $\grpjktarg{0}{k}{t}{\caluht}{\ccircht}$
in their respective elementary lists $\call$ and $\callht$ are isomorphic,
for $0\le k\le\ell$ and $t\in\bmcpz$, under the bijection $\lambda:  \calu\ra\caluht$.

We say two $(\ell+1)$-depth generator groups $(\calu,\circ)$ and $(\caluht,\circht)$
are {\it list isomorphic} under a bijection $\lambda:  \calu\ra\caluht$
if they contain list isomorphic elementary systems under bijection $\lambda$.  

\begin{lem}
\label{lem119}
If two $(\ell+1)$-depth generator groups $(\calu,\circ)$ and $(\caluht,\circht)$ are list isomorphic
under a bijection $\lambda:  \calu\ra\caluht$, then they are isomorphic under bijection $\lambda$.
\end{lem}

\begin{prf}
If two $(\ell+1)$-depth generator groups $(\calu,\circ)$ and $(\caluht,\circht)$ are list isomorphic
under a bijection $\lambda:  \calu\ra\caluht$, then we know $\grpjktarg{0}{0}{t}{\calu}{\ccirc}$ and 
$\grpjktarg{0}{0}{t}{\caluht}{\ccircht}$ are isomorphic for each $t\in\bmcpz$,
under bijection $\lambda$.  Let $\bmudt,\bmuddt\in\calu$.  Let $\lambda:  \bmudt\mapsto\bmudtht$ and 
$\lambda:  \bmuddt\mapsto\bmuddtht$.  Then we have
\be
\label{eqn105}
\tridnjktarg{0}{0}{t}{\bmudt}\prodjktarg{0}{0}{t}{\ccirc}\tridnjktarg{0}{0}{t}{\bmuddt}
=\tridnjktarg{0}{0}{t}{\bmudtht}\prodjktarg{0}{0}{t}{\ccircht}\tridnjktarg{0}{0}{t}{\bmuddtht}
\ee
for $t=+\infty$ to $-\infty$.  But by Lemma \ref{lem95a}, the product operation 
$\bmudt\circ\bmuddt$ is equivalent to the evaluation of
$$
\tridnjktarg{0}{0}{t}{\bmudt}\prodjktarg{0}{0}{t}{\ccirc}\tridnjktarg{0}{0}{t}{\bmuddt}
$$
for $t=+\infty$ to $-\infty$.  Then (\ref{eqn105}) is equivalent to the product evaluation
$$
\bmudt\circ\bmuddt=\bmudtht\,\circht\,\bmuddtht.
$$
This means $(\calu,\circ)$ and $(\caluht,\circht)$ are isomorphic under bijection $\lambda$.
\end{prf}
A list isomorphism is more restrictive than an isomorphism.  For example, 
the $(\ell+1)$-depth generator group $(\calu,\circ)$ formed by the identity and generator $\bmu_{g,\ell}^t$, 
and the $(\ell+1)$-depth generator group $(\calu',\circ')$ formed by the identity and generator 
$\bmu_{g,\ell}^{t+1}$, are isomorphic but not list isomorphic.  Although the two generator groups
are isomorphic, they are not isomorphic under a bijection $\lambda$ formed by a list isomorphism.

\begin{lem}
The generator group $(\calu,\circ)$ of an \ellctl\ complete group system $A$ is unique,
up to list isomorphism.
\end{lem}

\begin{prf}
The generator group $(\calu,\circ)$ of an \ellctl\ complete group system $A$ is wholely
determined by chain (\ref{chain7a}).
Clearly we may choose any elements for the sets $R_{0,k}^t$, for $0\le k\le\ell$, 
for each $t\in\bmcpz$, for the generator group $(\calu,\circ)$ of $A$.
\end{prf}
By the same approach the elementary system $\cale_A$ of an \ellctl\ complete 
group system $A$ is unique, up to list isomorphism.

We say two group systems are {\it list isomorphic} \ifof\ their generator groups are 
list isomorphic for some bijection $\lambda$.  The definition of list isomorphism is useful 
because of the following.

\begin{lem}
\label{lem120}
If two group systems $A$ and $\hat{A}$ are list isomorphic under bijection $\lambda$,
then they are isomorphic and both are $l$-controllable, for some integer $l$.
\end{lem}

\begin{prf}
If two \ellctl\ complete group systems $A$ and $\hat{A}$ are list isomorphic,
then their generator groups $(\calu,\circ)$ and $(\caluht,\circht)$ are list isomorphic 
for some bijection $\lambda$.  Then from Lemma \ref{lem119}, $(\calu,\circ)$ and 
$(\caluht,\circht)$ are isomorphic under bijection $\lambda$.  Since $(\calu,\circ)\simeq A$
and $(\caluht,\circht)\simeq\hat{A}$, then $A$ and $\hat{A}$ are isomorphic.

The controllability of $A$ and $\hat{A}$ is completely determined by the elementary system.
\end{prf}

Note that we can construct a group system $A$ from a global group $(\calu,\star)$, but
the generator group of $A$ may not contain an elementary system that is list isomorphic
to the elementary system of $(\calu,\star)$.  This occurs when the generator group of
$A$ is not the same as the global group.  A similar situation is encountered in Example 1,
where an $(\ell+1)$-depth generator group is used to construct an 
$l$-controllable group system with $l<\ell$.

Let $\{\cale\}$ be the set of all $(\ell+1)$-depth elementary systems $\cale$.
We divide the set $\{\cale\}$ into classes $[\cale]_l$ 
such that $\cale$ and $\caleht$ are in the same class if
$\cale\stackrel{l}{\simeq}\caleht$.  Let $\cale$ be a representative in $[\cale]_l$.
The following result is clear.

\begin{lem}
The class $[\cale]_l$ forms an equivalence class of $\{\cale\}$.  The set $\{\cale\}$
can be divided into equivalence classes $[\cale]_l$.
\end{lem}
  
\begin{lem}
\label{lem125}
If two group systems $A$ and $A'$ have elementary systems $\cale$ and $\cale'$ in the
same equivalence class $[\cale]_l$, then $A$ and $A'$ are list isomorphic.
\end{lem}

\begin{prf}
Let $(\calu,\circ)$ be the generator group of $A$ and let $\cale$ be the elementary system
of $(\calu,\circ)$.  Let $(\calu',\circ')$ be the generator group of $A'$ and let $\cale'$ 
be the elementary system of $(\calu',\circ')$.  Since $\cale$ and $\cale'$ are in the
same equivalence class $[\cale]_l$, then $\cale$ and $\cale'$ are list isomorphic
under a bijection $\lambda:  \calu\ra\calu'$.  Then by definition, $(\calu,\circ)$
and $(\calu',\circ')$ are list isomorphic under bijection $\lambda:  \calu\ra\calu'$.
Then again by definition, $A$ and $A'$ are list isomorphic.
\end{prf}

\begin{lem}
\label{lem126}
Fix any equivalence class $[\cale]_l$.  Pick any elementary system $\cale'$ in $[\cale]_l$.
Then $\cale'$ is the elementary system $\cale_{(U_s,\ovstar)}$ of an \ellctl\ complete global 
group system $(U_s,\ovstar)$.
\end{lem}

\begin{prf}
Fix any equivalence class $[\cale]_l$.  Pick any elementary system $\cale'$ in $[\cale]_l$.
Form the $(\ell+1)$-depth global group $(\calu,\star)$ of $\cale'$.
Form the \ellctl\ complete global group system $(U_s,\ovstar)$ of $(\calu,\star)$.  From
Theorem \ref{thm115a}, the generator group of $(U_s,\ovstar)$ is $(\calu,\star)$,
and the elementary system $\cale_{(U_s,\ovstar)}$ of $(U_s,\ovstar)$ is $\cale'$.
\end{prf}

\begin{thm}
\label{thm129}
The equivalence classes $[\cale_{(U_s,\ovstar)}]_l$, where $(U_s,\ovstar)$ is an 
\ellctl\ complete global group system, divide the set $\{\cale\}$.  
Then any \ellctl\ complete group system $A$ has an elementary system $\cale_A$ 
in one and only one equivalence class $[\cale_{(U_s,\ovstar)}]_l$,
and $A$ is list isomorphic to $(U_s,\ovstar)$.
\end{thm}

\begin{prf}
From Lemma \ref{lem126}, each equivalence class $[\cale]_l$ has a representative 
$\cale_{(U_s,\ovstar)}$, where $(U_s,\ovstar)$ is an \ellctl\ complete global group system.  
Then the equivalence classes $[\cale_{(U_s,\ovstar)}]_l$,
for each global group system $(U_s,\ovstar)$, divide the set $\{\cale\}$.  
Then any \ellctl\ complete group system $A$
has an elementary system $\cale_A$ in one and only one equivalence class $[\cale_{(U_s,\ovstar)}]_l$.
Since $\cale_A$ and $\cale_{(U_s,\ovstar)}$ are in the same equivalence class, by Lemma \ref{lem125}
$A$ is list isomorphic to $(U_s,\ovstar)$.
\end{prf}

In similar manner as discussed in Subsection 6.6 for generator groups, 
we can now use Theorem \ref{thm129} and chain (\ref{zchain7f})
of Theorem \ref{thm115a} to find all \ellctl\ complete group systems $A$ up to list 
isomorphism from the set of all $(\ell+1)$-depth elementary systems $\cale$.  
To construct all \ellctl\ complete group systems $A$ up to list isomorphism, 
we divide the the set of all $(\ell+1)$-depth elementary systems into equivalence 
classes $[\cale]_l$.  Pick one representative $\cale$ from each equivalence class.
Find its $(\ell+1)$-depth global group $(\calu,\star)$.  Construct the \ellctl\ complete 
global group system $(U_s,\ovstar)$ of $(\calu,\star)$ using chain (\ref{zchain7f})
of Theorem \ref{thm115a}.  The set of all global group systems 
$(U_s,\ovstar)$ obtained this way, one for each equivalence class $[\cale]_l$,
is the set of all \ellctl\ complete group systems $A$ up to list isomorphism.

\vspace{3mm}
{\bf 7.4  Construction of any elementary system $\cale$}
\vspace{3mm}

In the previous subsection, we discussed how to construct all \ellctl\ complete group systems $C$
from all $(\ell+1)$-depth elementary systems $\cale$.
We now give a brief discussion of how to construct all $(\ell+1)$-depth elementary systems $\cale$.
We first discuss how to construct a single elementary system $\cale$.
Since an elementary system $\cale$ is nested, to construct an $(\ell+1)$-depth 
elementary system $\cale$, we first construct a 1-depth elementary system 
$\cale_\ell=\{\grpjktarg{0}{k}{t}{\calu}{\ccirc}:  k=\ell,t\in\bmcpz\}$;
then a 2-depth elementary system 
$\cale_{\ell-1}=\{\grpjktarg{0}{k}{t}{\calu}{\ccirc}:  \ell-1\le k\le\ell,t\in\bmcpz\}$,
where there is a homomorphism from groups $\grpjktarg{0}{\ell-1}{t}{\calu}{\ccirc}$
in $\cale_{\ell-1}$ to groups
$\grpjktarg{0}{\ell}{t}{\calu}{\ccirc}$ and $\grpjktarg{0}{\ell}{t-1}{\calu}{\ccirc}$
in $\cale_\ell$ for each $t\in\bmcpz$; and continue on.  In this way,
we obtain a sequence of elementary systems
$\cale_\ell,\cale_{\ell-1},\ldots,\cale_m,\ldots,\cale_1,\cale_0=\cale$
which ends in $\cale_0=\cale$, where
$\cale_m=\{\grpjktarg{0}{k}{t}{\calu}{\ccirc}:  m\le k\le\ell,t\in\bmcpz\}$.

We may construct the sequence 
$\cale_\ell,\cale_{\ell-1},\ldots,\cale_m,\ldots,\cale_1,\cale_0$
in the same way as we evaluated (\ref{eexp21}) at the beginning of
Subsection 7.2.  Thus assume we have found the partial sequence
$\cale_\ell,\cale_{\ell-1},\ldots,\cale_{m+1}$ for some $m$, $0<m\le\ell$.
We show how to find $\cale_m$.  To find $\cale_m$ we have to find groups
$\grpjktarg{0}{m}{t}{\calu}{\ccirc}$ for each $t\in\bmcpz$, such that there is a 
homomorphism from $\grpjktarg{0}{m}{t}{\calu}{\ccirc}$ to 
$\grpjktarg{0}{m+1}{t}{\calu}{\ccirc}$ and $\grpjktarg{0}{m+1}{t-1}{\calu}{\ccirc}$
in the elementary list, under the projection map from set $\tridnjktarg{0}{m}{t}{\calu}$ 
to sets $\tridnjktarg{0}{m+1}{t}{\calu}$ and $\tridnjktarg{0}{m+1}{t-1}{\calu}$.  The elementary groups 
$\grpjktarg{0}{m+1}{t}{\calu}{\ccirc}$ and $\grpjktarg{0}{m+1}{t-1}{\calu}{\ccirc}$
intersect and form the trellis product group
\be
\label{tpg2}
\grpjktarg{0}{m+1}{t}{\calu}{\ccirc}\Join\grpjktarg{0}{m+1}{t-1}{\calu}{\ccirc}
\ee
discussed following Theorem \ref{thm130}.  Then Theorem \ref{homo6} shows there is 
a homomorphism from $\grpjktarg{0}{m}{t}{\calu}{\ccirc}$ to trellis product group (\ref{tpg2}).  Thus
$\grpjktarg{0}{m}{t}{\calu}{\ccirc}$ is an extension of (\ref{tpg2}).  Note that set
$\tridnjktarg{0}{m}{t}{\calu}$ is the same as set 
$\tridnjktarg{0}{m+1}{t}{\calu}\Join\tridnjktarg{0}{m+1}{t-1}{\calu}$
except for the addition of elements at index position $(0,m)$.
This approach gives the construction of group $\grpjktarg{0}{m}{t}{\calu}{\ccirc}$.
Continuing in this way, we finally obtain $\cale_0=\cale$.  In Subsection 8.3, 
we give the precise details of the construction of $\grpjktarg{0}{m}{t}{\calu}{\ccirc}$
for a particular example of a group system.   More construction details can be found in v2-v6
of \cite{KM6}.

To construct all $(\ell+1)$-depth elementary systems $\cale$, we just iterate
the above approach, first constructing all 1-depth elementary systems, then
all 2-depth elementary systems for each of the 1-depth elementary systems, and so on.

\newpage
{\bf 8.  THE GENERATOR GROUP AND BLOCK CODES}
\vspace{3mm}

In Subsection 6.3 we showed that finite or infinite sequences of elementary
groups of generator group $(\calu,\circ)$ form a group.  Theorem \ref{homo9} showed
there was a homomorphism from $A$ to any sequence of elementary groups in its generator group.
In Section 8, we define block codes over groups and use elementary groups in the generator group
to study their structure.  In Subsection 8.1, we use all possible combinations of nontrivial
elementary groups in $(\calu,\circ)$ that form a group 
to study the structure of all block codes with $\ell=2$.
In Subsection 8.2, we give an example of a homomorphism from the $(8,4,4)$ extended Hamming code,
where $\ell=3$, to an elementary group in its generator group.
Finally, since the nontrivial \glabs\ have a pyramid shape in $\calu$, the nontrivial
elementary groups form a pyramid, and so the
construction of a block group system using its elementary system
is particularly simple.  In Subsection 8.3, we give an example of this
construction for the $(8,4,4)$ extended Hamming code.

\vspace{3mm}
{\bf 8.1  The block code}
\vspace{3mm}

In this subsection, we only consider \ellctl\ group systems $A$ which are nontrivial
on a finite time interval $[t,t+\ell+\nu]$, $\nu\ge 0$, but are trivial outside this
time interval.  In this case we say $A$ is a {\it block group system}.  Note that 
a block group system $A$ as considered here is defined on $t\in\bmcpz$ 
but is isomorphic to a set of finite sequences defined 
on $[t,t+\ell+\nu]$ with a componentwise group addition.  This latter structure is 
usually referred to in engineering literature and coding literature as a linear block code.
We only study the most important case of a block group system and linear
block code, the case $\nu=0$.  The general case for $\nu>0$ is a straightforward
extension of the case for $\nu=0$.

Assume an \ellctl\ block group system $A$ is nontrivial on $[t,t+\ell]$.  
The work of Forney and Trott shows that any block group system can be
decomposed into a set of generators which can be arranged in a group trellis 
\cite{FT}.  The generators of span $\ell+1$ must be of the form 
$\bmg^{[t,t+\ell]}$.  By definition there must be at least one nontrivial 
generator of span $\ell+1$.  If there are generators of span $\ell$, they must
be of the form $\bmg^{[t,t+\ell-1]}$ and $\bmg^{[t+1,t+\ell]}$.  
In general, generators of span $k+1$, $0\le k\le\ell$, must be of the form
$\bmg^{[t+n,t+n+k]}$, for $0\le n\le\ell-k$.  Knowing the decomposition into
generators, we can describe the form of the set $\calu$.  The sets $R_{0,k}^{t'}$ 
of \glabs\ which may be nontrivial in set $\calu$
are shown in (\ref{blkuttf}); all \glabs\ outside this triangle shape 
or pyramid shape must be trivial.  
Any tensor $\bmu\in\calu$, such as shown in (\ref{ugttf}), can only have
nontrivial \glabs\ in the sets shown in (\ref{blkuttf}); all other
\glabs\ outside this pyramid shape must be the identity.
Since the nontrivial \glabs\ have a pyramid shape in $\calu$, the 
construction of a block group system using its elementary system
is particularly simple.  The pyramid shape (\ref{blkuttf}) 
for a specific $A$ is called the {\it generator pyramid} of $A$.
For any specific block group system $A$, some of the \glab\ sets $R_{0,k}^{t'}$
shown in (\ref{blkuttf}) may be the identity set.

\be
\label{blkuttf}
\begin{array}{llllllllllll}
                   &                     &                    &                    &        &        &                    &        &        & \cdots             &  R_{0,\ell}^t   & \cdots \\
                   &                     &                    &                    &        &        &                    &        & \cdots & R_{0,\ell-1}^{t+1} &  R_{0,\ell-1}^t & \cdots \\
                   &                     &                    &                    &        &        &                    &        & \vdots & \vdots             &  \vdots         &        \\
                   &                     &                    &                    &        & \cdots & R_{0,k}^{t+\ell-k} & \cdots & \cdots & R_{0,k}^{t+1}      &  R_{0,k}^t      & \cdots \\
                   &                     &                    &                    &        & \vdots & \vdots             & \vdots & \vdots & \vdots             &  \vdots         &        \\
                   &                     & \cdots             & R_{0,2}^{t+\ell-2} & \cdots & \cdots & R_{0,2}^{t+\ell-k} & \cdots & \cdots & R_{0,2}^{t+1}      &  R_{0,2}^t      & \cdots \\
                   & \cdots              & R_{0,1}^{t+\ell-1} & R_{0,1}^{t+\ell-2} & \cdots & \cdots & R_{0,1}^{t+\ell-k} & \cdots & \cdots & R_{0,1}^{t+1}      &  R_{0,1}^t      & \cdots \\
\cdots             & R_{0,0}^{t+\ell}    & R_{0,0}^{t+\ell-1} & R_{0,0}^{t+\ell-2} & \cdots & \cdots & R_{0,0}^{t+\ell-k} & \cdots & \cdots & R_{0,0}^{t+1}      &  R_{0,0}^t      & \cdots
\end{array}
\ee

In the next example, we use the results on $\squargrparg{\bmk}{\bmt}{\calu}{\oplus}$
in Subsection 6.3 to study the additive structure of a block group system $A$ in terms of 
its generator pyramid.  If the block group system is nontrivial on $[t,t+\ell]$, 
the generator pyramid is confined to $[t,t+\ell]$, and the most interesting sequences 
$\bmk$ and $\bmt$ in $\squargrparg{\bmk}{\bmt}{\calu}{\oplus}$ are those 
for which $\bmt\subset [t,t+\ell]$.

\begin{ex}{2}
In this example, we study the block group system $A$ for the case $\ell=2$.
Then the generator pyramid (\ref{blkuttf}) becomes (\ref{blkuttf1}).  We identify all nontrivial
groups of the form $\squargrparg{\bmk}{\bmt}{\calu}{\oplus}$.  First we
find the simplest groups by considering sequences $\bmk$ and $\bmt$ of only one term.
These are the elementary groups $\grpjktarg{0}{k}{t'}{\calu}{\ccirc}$ on triangles 
$\tridnjktarg{0}{k}{t'}{\calu}$ for some $k$ such that $0\le k\le\ell$ and some $t'\in [t,t+2]$.  
From (\ref{blkuttf1}), the only sets of the form $\tridnjktarg{0}{k}{t'}{\calu}$
in the block group system $A$ with nontrivial elements are
$\tridnjktarg{0}{2}{t}{\calu}$, $\tridnjktarg{0}{1}{t}{\calu}$, $\tridnjktarg{0}{1}{t+1}{\calu}$,
$\tridnjktarg{0}{0}{t}{\calu}$, $\tridnjktarg{0}{0}{t+1}{\calu}$, and $\tridnjktarg{0}{0}{t+2}{\calu}$.
These are listed in the first column of Table \ref{pgrp1}.
Using (\ref{blkuttf1}), the sets of \glabs\ in set $\tridnjktarg{0}{k}{t'}{\calu}$ 
are shown in the second column of Table \ref{pgrp1}.  Under the assignment
$\bmg^{[t,t+k]}\mapsto r_{0,k}^t$ given by the bijection $[\Lambda_{0,k}^t]\ra R_{0,k}^t$, a \glab\
$r_{0,k}^t$ in the second column corresponds to a generator $\bmg^{[t,t+k]}$ in $A$ whose first nontrivial
component is $r_{0,k}^t$.  The sets $\squaritarg{\bmk}{\bmt}{\calu}$ 
in column one are associated with the groups
$\grpjktarg{0}{2}{t}{\calu}{\ccirc}$,
$\grpjktarg{0}{1}{t}{\calu}{\ccirc}$,
$\grpjktarg{0}{1}{t+1}{\calu}{\ccirc}$,
$\grpjktarg{0}{0}{t}{\calu}{\ccirc}$,
$\grpjktarg{0}{0}{t+1}{\calu}{\ccirc}$, and
$\grpjktarg{0}{0}{t+2}{\calu}{\ccirc}$, respectively,
shown in the third column.  The elementary groups in the third column of
Table \ref{pgrp1} are shown in (\ref{blkuttf2}), stacked
according to the \glabs\ in (\ref{blkuttf1}) that define them.
As an example, set $\tridnjktarg{0}{1}{t}{\calu}$
is shown in the second row of the first column.  The \glabs\ in 
$\tridnjktarg{0}{1}{t}{\calu}$ are shown in the second column; 
these are the set of triples $(r_{0,1}^t,r_{0,2}^t,1_{0,2}^{t-1})$
in the Cartesian product of the sets of \glabs\ 
$R_{0,1}^t\times R_{0,2}^t\times\{1_{0,2}^{t-1}\}$, where $\{1_{0,2}^{t-1}\}$ 
is the trivial set just containing the identity \glab.  
The set $\tridnjktarg{0}{1}{t}{\calu}$
is associated with group $\grpjktarg{0}{1}{t}{\calu}{\ccirc}$ in the third column.

\be
\label{blkuttf1}
\begin{array}{lllll}
        &                & \cdots        &  R_{0,2}^t & \cdots \\
        & \cdots         & R_{0,1}^{t+1} &  R_{0,1}^t & \cdots \\
 \cdots & R_{0,0}^{t+2}  & R_{0,0}^{t+1} &  R_{0,0}^t & \cdots
\end{array}
\ee

\begin{table}[h]

\begin{tabular}{|l|l|l|} \hline
Set $\squaritarg{\bmk}{\bmt}{\calu}$ & Sets of \glabs\ in $\squaritarg{\bmk}{\bmt}{\calu}$ & Group $\squargrparg{\bmk}{\bmt}{\calu}{\oplus}$ \\   \hline 
$\tridnjktarg{0}{2}{t}{\calu}$ & $R_{0,2}^t$ & $E_{0,2}^t\rmdef \grpjktarg{0}{2}{t}{\calu}{\ccirc}$  \\   \hline 
$\tridnjktarg{0}{1}{t}{\calu}$ & $R_{0,1}^t\times R_{0,2}^t\times\{1_{0,2}^{t-1}\}$ & $E_{0,1}^t\rmdef \grpjktarg{0}{1}{t}{\calu}{\ccirc}$  \\   \hline       
$\tridnjktarg{0}{1}{t+1}{\calu}$ & $R_{0,1}^{t+1}\times\{1_{0,2}^{t+1}\}\times R_{0,2}^t$ & $E_{0,1}^{t+1}\rmdef \grpjktarg{0}{1}{t+1}{\calu}{\ccirc}$  \\   \hline       
$\tridnjktarg{0}{0}{t}{\calu}$ & $R_{0,0}^t\times R_{0,1}^t\times R_{0,2}^t\times\{1_{0,1}^{t-1}\}\times\{1_{0,2}^{t-1}\}\times\{1_{0,2}^{t-2}\}$ & $E_{0,0}^t\rmdef \grpjktarg{0}{0}{t}{\calu}{\ccirc}$  \\   \hline       
$\tridnjktarg{0}{0}{t+1}{\calu}$ & $R_{0,0}^{t+1}\times R_{0,1}^{t+1}\times\{1_{0,2}^{t+1}\}\times R_{0,1}^t\times R_{0,2}^t\times\{1_{0,2}^{t-1}\}$ & $E_{0,0}^{t+1}\rmdef \grpjktarg{0}{0}{t+1}{\calu}{\ccirc}$  \\   \hline       
$\tridnjktarg{0}{0}{t+2}{\calu}$ & $R_{0,0}^{t+2}\times\{1_{0,1}^{t+2}\}\times\{1_{0,2}^{t+2}\}\times R_{0,1}^{t+1}\times\{1_{0,2}^{t+1}\}\times R_{0,2}^t$ & $E_{0,0}^{t+2}\rmdef \grpjktarg{0}{0}{t+2}{\calu}{\ccirc}$  \\   \hline       
\end{tabular}

\caption{Nontrivial elementary groups of the block group system $A$ for $\ell=2$.}
\label{pgrp1}

\end{table}

\be
\label{blkuttf2}
\begin{array}{lll}
                &               &  E_{0,2}^t \\
                & E_{0,1}^{t+1} &  E_{0,1}^t \\
 E_{0,0}^{t+2}  & E_{0,0}^{t+1} &  E_{0,0}^t 
\end{array}
\ee

We now find the remaining sets in (\ref{blkuttf1}) that involve index 
sequences $\bmk$ and $\bmt$ of two or more terms.  Index sequences 
$\bmk$ and $\bmt$ of two or more terms
form multiple triangles of the form $\tridnjktarg{0}{k}{t'}{\calu}$ 
in $\squaritarg{\bmk}{\bmt}{\calu}$.
These form a trellis product of sets shown in the first column of Table \ref{pgrp2}.
For this example, all cases of multiple triangles in $\squaritarg{\bmk}{\bmt}{\calu}$
overlap and form polygon shapes.  The nontrivial sets of
\glabs\ in polygon shape $\squaritarg{\bmk}{\bmt}{\calu}$ are shown in the second column 
of Table \ref{pgrp2}.  The group $\squargrparg{\bmk}{\bmt}{\calu}{\oplus}$
that corresponds to $\squaritarg{\bmk}{\bmt}{\calu}$ in column one is shown in the third column 
of Table \ref{pgrp2}.  For example $\tridnjktarg{0}{1}{t+1}{\calu}$ and $\tridnjktarg{0}{1}{t}{\calu}$ 
form the trellis product $\tridnjktarg{0}{1}{t+1}{\calu}\Join\tridnjktarg{0}{1}{t}{\calu}$ shown in 
the first row of the first column of Table \ref{pgrp2};
$\tridnjktarg{0}{1}{t+1}{\calu}\Join\tridnjktarg{0}{1}{t}{\calu}$
overlap and form the polygon shape whose \glabs\ are from the Cartesian product of sets
$R_{0,1}^{t+1}\times\{1_{0,2}^{t+1}\}\times R_{0,1}^t\times R_{0,2}^t\times\{1_{0,2}^{t-1}\}$,
a union of the sets of \glabs\ from $\tridnjktarg{0}{1}{t+1}{\calu}$ and 
$\tridnjktarg{0}{1}{t}{\calu}$.  The nontrivial sets of \glabs\ in the 
polygon shape are shown in the second column of Table \ref{pgrp2}.  
The group corresponding to $\squaritarg{\bmk}{\bmt}{\calu}$,
the trellis product $E_{0,1}^{t+1}\Join E_{0,1}^t$, is shown in the third column.

Tables \ref{pgrp1} and \ref{pgrp2} give all nontrivial groups
$\squargrparg{\bmk}{\bmt}{\calu}{\oplus}$ up to isomorphism.
The groups in Tables \ref{pgrp1} and \ref{pgrp2} may have other
descriptions which are isomorphic.  For example, 
$E_{0,0}^{t+2}\Join E_{0,0}^{t+1}\Join E_{0,1}^t$ is isomorphic to
group $E_{0,0}^{t+2}\Join E_{0,0}^{t+1}$ listed in Table \ref{pgrp2}.

\begin{table}[h]

\begin{tabular}{|l|l|l|} \hline
Set $\squaritarg{\bmk}{\bmt}{\calu}$ & Nontrivial sets of \glabs\ in $\squaritarg{\bmk}{\bmt}{\calu}$ & Group $\squargrparg{\bmk}{\bmt}{\calu}{\oplus}$ \\   \hline 
$\tridnjktarg{0}{1}{t+1}{\calu}\Join\tridnjktarg{0}{1}{t}{\calu}$  & $R_{0,1}^{t+1}\times R_{0,1}^t\times R_{0,2}^t$ & $E_{0,1}^{t+1}\Join E_{0,1}^t$  \\   \hline 
$\tridnjktarg{0}{1}{t+1}{\calu}\Join\tridnjktarg{0}{0}{t}{\calu}$  & $R_{0,1}^{t+1}\times R_{0,0}^t\times R_{0,1}^t\times R_{0,2}^t$ & $E_{0,1}^{t+1}\Join E_{0,0}^t$  \\   \hline       
$\tridnjktarg{0}{0}{t+1}{\calu}\Join\tridnjktarg{0}{0}{t}{\calu}$  & $R_{0,0}^{t+1}\times R_{0,1}^{t+1}\times R_{0,0}^t\times R_{0,1}^t\times R_{0,2}^t$ & $E_{0,0}^{t+1}\Join E_{0,0}^t$  \\   \hline       
$\tridnjktarg{0}{0}{t+2}{\calu}\Join\tridnjktarg{0}{1}{t}{\calu}$  & $R_{0,0}^{t+2}\times R_{0,1}^{t+1}\times R_{0,1}^t\times R_{0,2}^t$ & $E_{0,0}^{t+2}\Join E_{0,1}^t$  \\   \hline       
$\tridnjktarg{0}{0}{t+2}{\calu}\Join\tridnjktarg{0}{0}{t}{\calu}$  & $R_{0,0}^{t+2}\times R_{0,1}^{t+1}\times R_{0,0}^t\times R_{0,1}^t\times R_{0,2}^t$ & $E_{0,0}^{t+2}\Join E_{0,0}^t$  \\   \hline       
$\tridnjktarg{0}{0}{t+2}{\calu}\Join\tridnjktarg{0}{0}{t+1}{\calu}$  & $R_{0,0}^{t+2}\times R_{0,0}^{t+1}\times R_{0,1}^{t+1}\times R_{0,1}^t\times R_{0,2}^t$ & $E_{0,0}^{t+2}\Join E_{0,0}^{t+1}$  \\   \hline       
$\tridnjktarg{0}{0}{t+2}{\calu}\Join\tridnjktarg{0}{0}{t+1}{\calu}\Join\tridnjktarg{0}{0}{t}{\calu}$  & $R_{0,0}^{t+2}\times R_{0,0}^{t+1}\times R_{0,1}^{t+1}\times R_{0,0}^t\times R_{0,1}^t\times R_{0,2}^t$ & $E_{0,0}^{t+2}\Join E_{0,0}^{t+1}\Join E_{0,0}^t$  \\   \hline       
\end{tabular}

\caption{Remaining nontrivial sets and groups of the block group system $A$ for $\ell=2$.}
\label{pgrp2}

\end{table}

We have the following results.

1.  $E_{0,2}^t$ is a subgroup of every group in (\ref{blkuttf2}).  Alternatively, 
in the top row of (\ref{blkuttf1}), the set $R_{0,2}^t$ is contained in 
every group.

2.  In the bottom row of (\ref{blkuttf1}), each set $R_{0,0}^{t+n}$, 
for $0\le n\le 2$, is only contained in one group in (\ref{blkuttf2}), $E_{0,0}^{t+n}$.

3.  The groups in (\ref{blkuttf2}) contain other groups in (\ref{blkuttf2}) as
subgroups.  Each group $E_{0,m}^{t+n}$ in (\ref{blkuttf2}) is defined on a 
triangle set $\tridnjktarg{0}{m}{t+n}{\calu}$.  The subgroups of group $E_{0,m}^{t+n}$ 
in (\ref{blkuttf2}) can be found from triangle set $\tridnjktarg{0}{m}{t+n}{\calu}$.
For example group $E_{0,0}^t$ in (\ref{blkuttf2}) has subgroups 
$E_{0,2}^t,E_{0,1}^t$ in (\ref{blkuttf2}).

4.  The triangle sets of any two groups in (\ref{blkuttf2}) intersect in
a triangle set which defines another group in (\ref{blkuttf2}).
The intersection of these two groups is a subgroup of each group.

5.  If any group in (\ref{blkuttf2}) is a subgroup of another group, the
projection map from the latter to the former defines a homomorphism.
This is an application of Theorem \ref{homo6}.

6.  The trellis product of any two groups in (\ref{blkuttf2}) at different times
is a group.  For example, $E_{0,1}^{t+1}\Join E_{0,1}^t$ is a group.  This is an
application of Theorem \ref{thm92} and (\ref{cmpgrps}).  
These groups are shown in Table \ref{pgrp2}.

7.  The trellis product of any three groups in (\ref{blkuttf2}) at different times
is a group.  For example, $E_{0,0}^{t+2}\Join E_{0,1}^{t+1}\Join E_{0,0}^t$ is a group.
This is an application of Theorem \ref{thm92} and (\ref{cmpgrps}).  This group is shown 
in Table \ref{pgrp2}.  Any other trellis product of three groups at different times
is isomorphic to $E_{0,0}^{t+2}$ or a trellis product of two groups shown 
in Table \ref{pgrp2}.

8.  Any collection of groups in (\ref{blkuttf2}) is a group.  For each time $t+n$, 
$n\in\{0,1,2\}$, there is some largest group $E_{0,m}^{t+n}$
which contains any other groups $E_{0,m'}^{t+n}$ at the same time as subgroups.
Then the trellis product of the largest groups at different times is a group.
This is an application of Theorem \ref{thm92} and (\ref{cmpgrps}).
Therefore any collection of groups is isomorphic to a group listed in 
Table \ref{pgrp1} or Table \ref{pgrp2}.

9.  Theorem \ref{homo9} shows there is a homomorphism from the block group 
system $A$ to the trellis product of any groups in (\ref{blkuttf2}).

10.  Since each \glab\ in (\ref{blkuttf1}) determines a generator,
we can use (\ref{blkuttf2}) to study the additive structure 
of a block group system $A$ in terms of its generators.  It follows that
the group $E_{0,2}^t$ is isomorphic to the group of 
all generators of span 3 (including the identity generator).  
The group $E_{0,1}^{t+1}\Join E_{0,1}^t$ is isomorphic to the
group of all pairs of generators of span 2 and 3.  The group 
$E_{0,0}^{t+2}\Join E_{0,1}^{t+1}\Join E_{0,0}^t$ is isomorphic to the 
group of all triples of generators of span 1, 2, and 3.  This last 
group is isomorphic to the block group system.

11.  If any trellis product of groups in (\ref{blkuttf2}) is a subgroup of 
another trellis product of groups in (\ref{blkuttf2}), the projection map from 
the latter to the former defines a homomorphism.
This is an application of Theorem \ref{homo6}.

The results in this example are not surprising; they are merely an extension to 
a block group system of what happens for a single group.  For a single group, 
the set of cosets of highest order \creps\ in a \crepc\ forms a quotient group.  
Next the set of cosets of all pairs of a
highest order and second highest order \crep\ forms a quotient group, 
and so on.  In the block group system, the set of \glabs\ of longest span generators forms an 
elementary group.  There are two sets of second longest span generators, 
those starting at time $t$ and those starting at time $t+1$.  The set 
of all pairs of \glabs\ of a longest span generator with a second longest span generator starting 
at time $t$ forms an elementary group.  Also the set of all pairs of \glabs\ of a longest span generator 
with a second longest span generator starting at time $t+1$ forms an elementary group.
The trellis product of these two groups also forms a group, 
the group of all pairs of \glabs\ of a longest and second longest span generator.  
The intersection of these two groups is also a group, the elementary group of \glabs\
of longest span generators.  And so on.  The groups in Table \ref{pgrp1}, Table \ref{pgrp2}, 
and (\ref{blkuttf2}) give a simple and direct explanation of the group structure of a block 
group system in terms of its \glabs\ and then by inference generators.  
\end{ex}

\vspace{3mm}
{\bf 8.2  Homomorphisms}
\vspace{3mm}

We now give an example to show that Theorem \ref{homo9} holds for the
binary $(8,4,4)$ extended Hamming code or first order Reed-Muller code.
This homomorphism gives information about the structure of the block code.

\begin{ex}{3}
The theory of group systems developed in \cite{FT} can be applied to block codes.
We study the binary $(8,4,4)$ extended Hamming code or first order Reed-Muller code
used as Example 4 in \cite{FT} (p.~1503 and 1510).  This code is linear over the alphabet
group $A^t=\bmcpz_2\times\bmcpz_2$.  By the convention used here, 
we assume the corresponding block group system $A$ has infinite extent but is only nontrivial 
over $[t,t+\ell]$; for this code $\ell=3$.  The codewords of $A$ are shown in the first
column of Table \ref{pgrp4}.  The four elements of $A^t$ in parentheses are the
codewords of the $(8,4,4)$ block code over $[t,t+3]$; the ellipses are trivial half infinite sequences
of the identity of $A^t$.  

From \cite{FT}, we can choose granule representatives or generators of $A$ as follows:
\be
\label{ga7}
\begin{array}{ll}
[\Gamma^{[t,t+3]}]   &=\{\ldots,(00,00,00,00),\ldots;\ldots,(10,10,10,10),\ldots\}, \\\relax
[\Gamma^{[t,t+1]}]   &=\{\ldots,(00,00,00,00),\ldots;\ldots,(11,11,00,00),\ldots\}, \\\relax
[\Gamma^{[t+1,t+2]}] &=\{\ldots,(00,00,00,00),\ldots;\ldots,(00,11,11,00),\ldots\}, \\\relax
[\Gamma^{[t+2,t+3]}] &=\{\ldots,(00,00,00,00),\ldots;\ldots,(00,00,11,11),\ldots\}.
\end{array}
\ee
We label the nontrivial generators in granules $[\Gamma^{[t',t'+k]}]$ of (\ref{ga7}) 
by $\bmg_a^{[t',t'+k]}$.  To make the example clearer,
we label the identity path $\bone$ in $A$ corresponding to generators $\ldots,(00,00,00,00),\ldots$ in 
granules $[\Gamma^{[t',t'+k]}]$ of (\ref{ga7}) by $\bone_a^{[t',t'+k]}$.
Therefore we use the following labels for generators in the granules (\ref{ga7}).
\be
\label{ga7x}
\begin{array}{lll}
[\Gamma^{[t,t+3]}]   &=\{\ldots,(00,00,00,00),\ldots;\ldots,(10,10,10,10),\ldots\} & \ra\{\bone_a^{[t,t+3]},\bmg_a^{[t,t+3]}\}, \\\relax
[\Gamma^{[t,t+1]}]   &=\{\ldots,(00,00,00,00),\ldots;\ldots,(11,11,00,00),\ldots\} & \ra\{\bone_a^{[t,t+1]},\bmg_a^{[t,t+1]}\}, \\\relax
[\Gamma^{[t+1,t+2]}] &=\{\ldots,(00,00,00,00),\ldots;\ldots,(00,11,11,00),\ldots\} & \ra\{\bone_a^{[t+1,t+2]},\bmg_a^{[t+1,t+2]}\}, \\\relax
[\Gamma^{[t+2,t+3]}] &=\{\ldots,(00,00,00,00),\ldots;\ldots,(00,00,11,11),\ldots\} & \ra\{\bone_a^{[t+2,t+3]},\bmg_a^{[t+2,t+3]}\}.
\end{array}
\ee
Finally there are no nontrivial generators of span 3 or span 1 in $A$.  
We indicate the identity $\bone$ in $A$ corresponding to
trivial generators for these cases as $\bone_a^{[t,t+2]},\bone_a^{[t+1,t+3]}$ for span 3,
and $\bone_a^{[t,t]},\bone_a^{[t+1,t+1]},\bone_a^{[t+2,t+2]},\bone_a^{[t+3,t+3]}$ for span 1.

We can decompose the codewords of $A$ in column 1 of Table \ref{pgrp4} into the 
generators of $A$ given in (\ref{ga7x}); this decomposition is
shown in the second column of Table \ref{pgrp4}.  We do not include the trivial generators 
of span 3 or span 1 in the second column.

Using the generators in $A$, we can determine the \glabs\ in $(\calu,\circ)$.  
Each codeword in $A$ is decomposed into 4 generators
in $A$ in the second column.  The 4 generators in the second column give
4 \glabs\ in the third column.  The 4 \glabs\ determine a $\bmu\in\calu$,
where the remaining \glabs\ are trivial.
Then Table \ref{pgrp4} gives a bijection between the sixteen codewords of $A$
and the sixteen possible choices of $\bmu\in\calu$ determined by the 4 \glabs\ in 
the third column.

Looked at in another way, from (\ref{blkuttf}) the sets $R_{0,k}^{t'}$ of \glabs\ 
which may be nontrivial in set $\calu$ are 
shown in (\ref{blkuttf3}); all \glabs\ outside this triangle shape 
or pyramid shape must be the identity.  
The \glab\ of a nontrivial generator $\bmg_a^{[t',t'+k]}$ is indicated by $r_{0,k}^{t'}$ in $R_{0,k}^{t'}$, 
and the \glab\ of an identity generator $\bone_a^{[t',t'+k]}$ is indicated by $1_k^{t'}$.
Since there are no nontrivial generators
of span 3 or span 1, there are two rows of identities in (\ref{blkuttf3});
these are trivial sets just containing the identity \glab.  
The array (\ref{blkuttf3}) shows there are 4 nontrivial sets of \glabs,
$R_{0,1}^{t+2}=\{1_{0,1}^{t+2},r_{0,1}^{t+2}\}$, $R_{0,1}^{t+1}=\{1_{0,1}^{t+1},r_{0,1}^{t+1}\}$,
$R_{0,1}^t=\{1_{0,1}^t,r_{0,1}^t\}$, and $R_{0,3}^t=\{1_{0,3}^t,r_{0,3}^t\}$.  
Then (\ref{blkuttf3}) is the generator pyramid of $A$.
\be
\label{blkuttf3}
\begin{array}{cccccc}
        &                   &                  & \cdots             &  R_{0,3}^t     & \cdots \\
        &                   & \cdots           & \{1_{0,2}^{t+1}\}  &  \{1_{0,2}^t\} & \cdots \\
        & \cdots            & R_{0,1}^{t+2}    & R_{0,1}^{t+1}      &  R_{0,1}^t     & \cdots \\
 \cdots & \{1_{0,0}^{t+3}\} & \{1_{0,0}^{t+2}\}& \{1_{0,0}^{t+1}\}  &  \{1_{0,0}^t\} & \cdots
\end{array}
\ee
There are 16 possible choices of an element from each of the 4 nontrivial sets.  These 16 choices are shown 
in the third column of Table \ref{pgrp4}.

We can identify all nontrivial groups of the form $\squargrparg{\bmk}{\bmt}{\calu}{\oplus}$ 
in the same way as for Example 2.  The simplest groups are found
by considering sequences $\bmk$ and $\bmt$ of only one term.  These are the 
elementary groups $E_{0,k}^{t'}\rmdef\grpjktarg{0}{1}{t'}{\calu}{\ccirc}$ 
on triangles $\tridnjktarg{0}{1}{t'}{\calu}$ for some $k$ such that $0\le k\le\ell$ and $t'\in [t,t+3]$.
These simplest groups are shown in (\ref{blkuttf4}), stacked according to the 
\glabs\ in (\ref{blkuttf3}) that define them.
An example is shown in Table \ref{pgrp3}, the group
$E_{0,1}^{t+1}=\grpjktarg{0}{1}{t+1}{\calu}{\ccirc}$.  From (\ref{blkuttf3})
the Cartesian product of sets of \glabs\ in $\tridnjktarg{0}{1}{t+1}{\calu}$ is
$R_{0,1}^{t+1}\times\{1_{0,2}^{t+1}\}\times\{1_{0,3}^{t+1}\}\times\{1_{0,2}^t\}\times R_{0,3}^t\times\{1_{0,3}^{t-1}\}$.
Then group $E_{0,1}^{t+1}$ is the elementary group of all pairs of \glabs\ of a longest 
generator of span 4 starting at time $t$ with a generator of span 2 starting 
at time $t+1$.  

\begin{table}[h]

\begin{tabular}{|l|l|l|} \hline
Set $\squaritarg{\bmk}{\bmt}{\calu}$ & Sets of \glabs\ in $\squaritarg{\bmk}{\bmt}{\calu}$ & Group $\squargrparg{\bmk}{\bmt}{\calu}{\oplus}$ \\   \hline 
$\tridnjktarg{0}{1}{t+1}{\calu}$ & $R_{0,1}^{t+1}\times\{1_{0,2}^{t+1}\}\times\{1_{0,3}^{t+1}\}\times\{1_{0,2}^t\}\times R_{0,3}^t\times\{1_{0,3}^{t-1}\}$ & $E_{0,1}^{t+1}=\grpjktarg{0}{1}{t+1}{\calu}{\ccirc}$  \\   \hline       
\end{tabular}

\caption{An elementary group of the binary $(8,4,4)$ extended Hamming code.}
\label{pgrp3}

\end{table}

\be
\label{blkuttf4}
\begin{array}{llll}
                 &                &               &  E_{0,3}^t  \\
                 &                & E_{0,2}^{t+1} &  E_{0,2}^t  \\
                 & E_{0,1}^{t+2}  & E_{0,1}^{t+1} &  E_{0,1}^t  \\
 E_{0,0}^{t+3}   & E_{0,0}^{t+2}  & E_{0,0}^{t+1} &  E_{0,0}^t
\end{array}
\ee

Theorem \ref{homo9} shows there is a homomorphism from $A$ to 
$\squargrparg{\bmk}{\bmt}{\calu}{\oplus}$.
We now verify Theorem \ref{homo9} holds for the 
example in Table \ref{pgrp3}.  For the example in Table \ref{pgrp3}, $\bmk$ and $\bmt$ 
are the finite sequence $(0,k)=(0,1)$ at time $t+1$.  Then
$\squargrparg{\bmk}{\bmt}{\calu}{\oplus}=\grpjktarg{0}{1}{t+1}{\calu}{\ccirc}$.
We now show there is a homomorphism from $A$ to 
$\grpjktarg{0}{1}{t+1}{\calu}{\ccirc}$.  The bijection 
$\xi:  A\ra\calu$ in Theorem \ref{homo9} is the mapping 
in Table \ref{pgrp4} between codewords of $A$ in the first column and elements 
$\bmu\in\calu$ determined by the 4 \glabs\ in the third column.  
We now show the map $\omega_\bmk^\bmt:  \calu\ra\tridnjktarg{0}{1}{t+1}{\calu}$
is indicated by the map from the third column to the fourth.  From (\ref{blkuttf3})
and  Table \ref{pgrp3}, the only sets of \glabs\ which may be nontrivial 
in $\tridnjktarg{0}{1}{t+1}{\calu}$ are $R_{0,1}^{t+1}$ and $R_{0,3}^t$, 
where $R_{0,1}^{t+1}=\{1_{0,1}^{t+1},r_{0,1}^{t+1}\}$
and $R_{0,3}^t=\{1_{0,3}^t,r_{0,3}^t\}$.
Note that Table \ref{pgrp4} is divided into four quartets of codewords,
where the 4 \glabs\ in each quartet in the third column share the same value
in $R_{0,1}^{t+1}\times R_{0,3}^t$ in the fourth column.
Each quartet of 4 \glabs\ in the third column gives 4 elements
$\bmu\in\calu$.  Each shared value in $R_{0,1}^{t+1}\times R_{0,3}^t$ in the fourth column gives
one element $\tridnjktarg{0}{1}{t+1}{\bmu}\in\tridnjktarg{0}{1}{t+1}{\calu}$.
Therefore the map $\omega_\bmk^\bmt:  \calu\ra\tridnjktarg{0}{1}{t+1}{\calu}$
is given by the map from each quartet of 
$\bmu\in\calu$ determined by \glabs\ in the third column to each single element
$\tridnjktarg{0}{1}{t+1}{\bmu}\in\tridnjktarg{0}{1}{t+1}{\calu}$
determined by their shared value in $R_{0,1}^{t+1}\times R_{0,3}^t$
in the fourth column.

It remains to verify the assignment 
$\omega_\bmk^\bmt\bullet\xi:  A\ra\tridnjktarg{0}{1}{t+1}{\calu}$ 
gives a homomorphism from $A$ to $\grpjktarg{0}{1}{t+1}{\calu}{\ccirc}$.  
To show this, it is sufficient to show the 4 quartets of codewords in the first column 
of Table \ref{pgrp4} form a quotient group with the first quartet 
$$
\{\ldots,(00,00,00,00),\ldots;\ldots,(00,00,11,11),\ldots;\ldots,(11,11,00,00),\ldots;\ldots,(11,11,11,11),\ldots\}
$$
as a normal subgroup.  But it is easy to show the above quartet is a subgroup of $A$ and
also normal.
\end{ex}

\begin{table}[h]

\begin{tabular}{|l|l|l|l|} \hline
Codewords in    & Decomposition into generators                                         & Generator labels                          & Generator labels in Cartesian                    \\ 
$A$             & of $A$ of span 4 and span 2                                           & for generators of                         & product of nontrivial sets of                    \\
                &                                                                       & span 4 and span 2                         & \glabs\ in $\tridnjktarg{0}{1}{t+1}{\calu}$,     \\
                &                                                                       &                                           & $R_{0,1}^{t+1}\times R_{0,3}^t$                  \\  \hline
$\ldots,(00,00,00,00),\ldots$ & $\bone_a^{[t,t+3]},\bone_a^{[t,t+1]},\bone_a^{[t+1,t+2]},\bone_a^{[t+2,t+3]}$ & $1_{0,3}^t,1_{0,1}^t,1_{0,1}^{t+1},1_{0,1}^{t+2}$   & $(1_{0,1}^{t+1},1_{0,3}^t)$    \\
$\ldots,(00,00,11,11),\ldots$ & $\bone_a^{[t,t+3]},\bone_a^{[t,t+1]},\bone_a^{[t+1,t+2]},\bmg_a^{[t+2,t+3]}$  & $1_{0,3}^t,1_{0,1}^t,1_{0,1}^{t+1},r_{0,1}^{t+2}$   &     \\
$\ldots,(11,11,00,00),\ldots$ & $\bone_a^{[t,t+3]},\bmg_a^{[t,t+1]},\bone_a^{[t+1,t+2]},\bone_a^{[t+2,t+3]}$  & $1_{0,3}^t,r_{0,1}^t,1_{0,1}^{t+1},1_{0,1}^{t+2}$   &     \\
$\ldots,(11,11,11,11),\ldots$ & $\bone_a^{[t,t+3]},\bmg_a^{[t,t+1]},\bone_a^{[t+1,t+2]},\bmg_a^{[t+2,t+3]}$   & $1_{0,3}^t,r_{0,1}^t,1_{0,1}^{t+1},r_{0,1}^{t+2}$   &     \\  \hline

$\ldots,(00,11,11,00),\ldots$ & $\bone_a^{[t,t+3]},\bone_a^{[t,t+1]},\bmg_a^{[t+1,t+2]},\bone_a^{[t+2,t+3]}$  & $1_{0,3}^t,1_{0,1}^t,r_{0,1}^{t+1},1_{0,1}^{t+2}$   & $(r_{0,1}^{t+1},1_{0,3}^t)$    \\
$\ldots,(00,11,00,11),\ldots$ & $\bone_a^{[t,t+3]},\bone_a^{[t,t+1]},\bmg_a^{[t+1,t+2]},\bmg_a^{[t+2,t+3]}$   & $1_{0,3}^t,1_{0,1}^t,r_{0,1}^{t+1},r_{0,1}^{t+2}$   &     \\
$\ldots,(11,00,11,00),\ldots$ & $\bone_a^{[t,t+3]},\bmg_a^{[t,t+1]},\bmg_a^{[t+1,t+2]},\bone_a^{[t+2,t+3]}$   & $1_{0,3}^t,r_{0,1}^t,r_{0,1}^{t+1},1_{0,1}^{t+2}$   &     \\
$\ldots,(11,00,00,11),\ldots$ & $\bone_a^{[t,t+3]},\bmg_a^{[t,t+1]},\bmg_a^{[t+1,t+2]},\bmg_a^{[t+2,t+3]}$    & $1_{0,3}^t,r_{0,1}^t,r_{0,1}^{t+1},r_{0,1}^{t+2}$   &     \\  \hline

$\ldots,(10,10,10,10),\ldots$ & $\bmg_a^{[t,t+3]},\bone_a^{[t,t+1]},\bone_a^{[t+1,t+2]},\bone_a^{[t+2,t+3]}$  & $r_{0,3}^t,1_{0,1}^t,1_{0,1}^{t+1},1_{0,1}^{t+2}$   & $(1_{0,1}^{t+1},r_{0,3}^t)$    \\
$\ldots,(10,10,01,01),\ldots$ & $\bmg_a^{[t,t+3]},\bone_a^{[t,t+1]},\bone_a^{[t+1,t+2]},\bmg_a^{[t+2,t+3]}$   & $r_{0,3}^t,1_{0,1}^t,1_{0,1}^{t+1},r_{0,1}^{t+2}$   &     \\
$\ldots,(01,01,10,10),\ldots$ & $\bmg_a^{[t,t+3]},\bmg_a^{[t,t+1]},\bone_a^{[t+1,t+2]},\bone_a^{[t+2,t+3]}$   & $r_{0,3}^t,r_{0,1}^t,1_{0,1}^{t+1},1_{0,1}^{t+2}$   &     \\
$\ldots,(01,01,01,01),\ldots$ & $\bmg_a^{[t,t+3]},\bmg_a^{[t,t+1]},\bone_a^{[t+1,t+2]},\bmg_a^{[t+2,t+3]}$    & $r_{0,3}^t,r_{0,1}^t,1_{0,1}^{t+1},r_{0,1}^{t+2}$   &     \\  \hline

$\ldots,(10,01,01,10),\ldots$ & $\bmg_a^{[t,t+3]},\bone_a^{[t,t+1]},\bmg_a^{[t+1,t+2]},\bone_a^{[t+2,t+3]}$   & $r_{0,3}^t,1_{0,1}^t,r_{0,1}^{t+1},1_{0,1}^{t+2}$   & $(r_{0,1}^{t+1},r_{0,3}^t)$    \\
$\ldots,(10,01,10,01),\ldots$ & $\bmg_a^{[t,t+3]},\bone_a^{[t,t+1]},\bmg_a^{[t+1,t+2]},\bmg_a^{[t+2,t+3]}$    & $r_{0,3}^t,1_{0,1}^t,r_{0,1}^{t+1},r_{0,1}^{t+2}$   &     \\
$\ldots,(01,10,01,10),\ldots$ & $\bmg_a^{[t,t+3]},\bmg_a^{[t,t+1]},\bmg_a^{[t+1,t+2]},\bone_a^{[t+2,t+3]}$    & $r_{0,3}^t,r_{0,1}^t,r_{0,1}^{t+1},1_{0,1}^{t+2}$   &     \\
$\ldots,(01,10,10,01),\ldots$ & $\bmg_a^{[t,t+3]},\bmg_a^{[t,t+1]},\bmg_a^{[t+1,t+2]},\bmg_a^{[t+2,t+3]}$     & $r_{0,3}^t,r_{0,1}^t,r_{0,1}^{t+1},r_{0,1}^{t+2}$   &     \\  \hline
\end{tabular}

\caption{Mapping of codewords of the binary $(8,4,4)$ extended Hamming code 
to $\tridnjktarg{0}{1}{t+1}{\calu}$ for the homomorphism 
$A\ra\grpjktarg{0}{1}{t+1}{\calu}{\ccirc}$.}
\label{pgrp4}

\end{table}

\vspace{3mm}
{\bf 8.3  Construction}
\vspace{3mm}

In this subsection we give an example of the construction of a block group system $C$
for $\ell=3$.  As remarked in Subsection 7.3, 
since $\{\cale_A\}=\{\cale\}$, we can always assume that any elementary system $\cale$
has an elementary set which uses notation $\calu$, an elementary list which uses notation
$\{\grpjktarg{0}{k}{t}{\calu}{\ccirc}:  0\le k\le\ell,t\in\bmcpz\}$, and a
global group which uses notation $(\calu,\star)$.  Therefore we can use this notation
here in construction as well as analysis.

We do not construct all block group systems for $\ell=3$.  Instead, we
construct all block groups systems $C$ with $\ell=3$ which have a specific generator pyramid,
the generator pyramid shown in (\ref{blkuttf3}).  Further, we assume that $|R_{0,3}^t|=2$ and 
$|R_{0,1}^{t+2}|=|R_{0,1}^{t+1}|=|R_{0,1}^t|=2$ in (\ref{blkuttf3}).  Obviously then, the block group 
systems we construct will include the extended $(8,4,4)$ Hamming code discussed 
in Subsection 8.2.  

In Subsection 7.3 we discussed how to construct all \ellctl\ complete group systems $C$
from all $(\ell+1)$-depth elementary systems $\cale$ by using chain (\ref{zchain7e}) 
and Theorem \ref{thm84}.  In Subsection 7.4 we discussed how to construct all 
$(\ell+1)$-depth elementary systems $\cale$.  We begin chain 
(\ref{zchain7e}) by constructing all possible $(\ell+1)$-depth elementary systems $\cale$ that have the 
generator pyramid (\ref{blkuttf3}).  It is clear that we only need to
construct the elementary groups in $\cale$ with nontrivial entries in the generator pyramid.
We construct the elementary system $\cale$ using a row by row construction as discussed in
Subsection 7.4.  We start with the top row of $\cale$.  There is only one nontrivial
elementary group in the top row, $\grpjktarg{0}{3}{t}{\calu}{\ccirc}$.  Since $|R_{0,3}^t|=2$, we must have
$\grpjktarg{0}{3}{t}{\calu}{\ccirc}\simeq\bmcpz_2$.  Let $(\bmcpz_2)_3^t$ be the group
isomorphic to $\bmcpz_2$ defined on $R_{0,3}^t$.  Then 
$\grpjktarg{0}{3}{t}{\calu}{\ccirc}\simeq(\bmcpz_2)_3^t$.  We are done with the top row;
we move to the second row.  There is a homomorphism from the elementary group 
$\grpjktarg{0}{2}{t}{\calu}{\ccirc}$ to $\grpjktarg{0}{3}{t}{\calu}{\ccirc}$.
Since the set $\tridnjktarg{0}{2}{t}{\calu}$ overlaps two trivial \glab\ sets in
the generator pyramid, the kernel of the homomorphism is the identity.  Then 
$\grpjktarg{0}{2}{t}{\calu}{\ccirc}\simeq\grpjktarg{0}{3}{t}{\calu}{\ccirc}\simeq(\bmcpz_2)_3^t$.  
In the same way we have 
$\grpjktarg{0}{2}{t+1}{\calu}{\ccirc}\simeq\grpjktarg{0}{3}{t}{\calu}{\ccirc}\simeq(\bmcpz_2)_3^t$.
We are done with the second row; we move to the third row.
Consider the elementary group $\grpjktarg{0}{1}{t}{\calu}{\ccirc}$.
The set $\tridnjktarg{0}{1}{t}{\calu}$ overlaps two nontrivial \glab\ sets in
the generator pyramid.  Then $\grpjktarg{0}{1}{t}{\calu}{\ccirc}$ must be a group of
order 4, either $\bmcpz_2\times\bmcpz_2$ or $\bmcpz_4$.  But since there is a 
homomorphism from $\grpjktarg{0}{1}{t}{\calu}{\ccirc}$ to 
$\grpjktarg{0}{2}{t}{\calu}{\ccirc}$,
we must have $\grpjktarg{0}{1}{t}{\calu}{\ccirc}\simeq\bmcpz_2\times\bmcpz_2
\simeq\bmcpz_2\times(\bmcpz_2)_3^t$.  Let $(\bmcpz_2)_1^t$ be the group
isomorphic to $\bmcpz_2$ defined on $R_{0,1}^t$.  Then
$\grpjktarg{0}{1}{t}{\calu}{\ccirc}\simeq(\bmcpz_2)_1^t\times(\bmcpz_2)_3^t$.
In similar fashion, the elementary groups
$\grpjktarg{0}{1}{t+1}{\calu}{\ccirc}$ and $\grpjktarg{0}{1}{t+2}{\calu}{\ccirc}$
are isomorphic to $(\bmcpz_2)_1^{t+1}\times(\bmcpz_2)_3^t$ and
$(\bmcpz_2)_1^{t+2}\times(\bmcpz_2)_3^t$, respectively.
We are done with the third row; we move to the last row.

First consider the group $\grpjktarg{0}{0}{t}{\calu}{\ccirc}$.  There is a 
homomorphism from $\grpjktarg{0}{0}{t}{\calu}{\ccirc}$ to $\grpjktarg{0}{1}{t}{\calu}{\ccirc}$.
Since all the \glab\ sets in set $\tridnjktarg{0}{0}{t}{\calu}$ are trivial except for those in set
$\tridnjktarg{0}{1}{t}{\calu}$, we must have the kernel of the homomorphism is the identity.
Then $\grpjktarg{0}{0}{t}{\calu}{\ccirc}\simeq\grpjktarg{0}{1}{t}{\calu}{\ccirc}\simeq(\bmcpz_2)_1^t\times(\bmcpz_2)_3^t$.
In similar way, we must have 
$\grpjktarg{0}{0}{t+3}{\calu}{\ccirc}\simeq\grpjktarg{0}{1}{t+2}{\calu}{\ccirc}\simeq(\bmcpz_2)_1^{t+2}\times(\bmcpz_2)_3^t$.

We now consider the last two groups in the last row, $\grpjktarg{0}{0}{t+1}{\calu}{\ccirc}$ 
and $\grpjktarg{0}{0}{t+2}{\calu}{\ccirc}$.  First consider
$\grpjktarg{0}{0}{t+1}{\calu}{\ccirc}$.  There is a homomorphism from 
$\grpjktarg{0}{0}{t+1}{\calu}{\ccirc}$ to $\grpjktarg{0}{1}{t+1}{\calu}{\ccirc}$
and $\grpjktarg{0}{1}{t}{\calu}{\ccirc}$.  The groups $\grpjktarg{0}{1}{t+1}{\calu}{\ccirc}$
and $\grpjktarg{0}{1}{t}{\calu}{\ccirc}$ form the trellis product group
\be
\label{eq110}
\grpjktarg{0}{1}{t+1}{\calu}{\ccirc}\Join\grpjktarg{0}{1}{t}{\calu}{\ccirc}.
\ee
We know $\grpjktarg{0}{1}{t+1}{\calu}{\ccirc}\simeq(\bmcpz_2)_1^{t+1}\times(\bmcpz_2)_3^t$ and
$\grpjktarg{0}{1}{t}{\calu}{\ccirc}\simeq(\bmcpz_2)_1^t\times(\bmcpz_2)_3^t$.

We now apply Theorem \ref{thm130}.  Let 
$G_1\rmdef\grpjktarg{0}{1}{t+1}{\calu}{\ccirc}\simeq(\bmcpz_2)_1^{t+1}\times(\bmcpz_2)_3^t$
and $G_2\rmdef\grpjktarg{0}{1}{t}{\calu}{\ccirc}\simeq(\bmcpz_2)_3^t\times(\bmcpz_2)_1^t$.
Let $G^*$ be the subdirect product of $G_1$ and $G_2$ given by (\ref{eq110}).
The elementary groups $\grpjktarg{0}{1}{t+1}{\calu}{\ccirc}$ and $\grpjktarg{0}{1}{t}{\calu}{\ccirc}$
intersect in elementary group
$\grpjktarg{0}{2}{t}{\calu}{\ccirc}\simeq\grpjktarg{0}{3}{t}{\calu}{\ccirc}\simeq(\bmcpz_2)_3^t$.  
Let $G_3\rmdef\grpjktarg{0}{2}{t}{\calu}{\ccirc}\simeq(\bmcpz_2)_3^t$.  Let $1_3^t$ be the identity of
$(\bmcpz_2)_3^t$ and $2_3^t$ be the remaining element of $(\bmcpz_2)_3^t$.  From (\ref{sd1}) and 
(\ref{sd2}), we have $H_1\simeq(\bmcpz_2)_1^{t+1}\times1_3^t$ and $H_2\simeq 1_3^t\times(\bmcpz_2)_1^t$.
We have factor groups 
$$
G_1/H_1\simeq(\bmcpz_2)_1^{t+1}\times(\bmcpz_2)_3^t/(\bmcpz_2)_1^{t+1}\times1_3^t\simeq G_3
$$
and
$$
G_2/H_2\simeq(\bmcpz_2)_3^t\times(\bmcpz_2)_1^t/1_3^t\times(\bmcpz_2)_1^t\simeq G_3.
$$

For the isomorphisms $G_1\simeq(\bmcpz_2)_1^{t+1}\times(\bmcpz_2)_3^t$ and
$G_2\simeq(\bmcpz_2)_3^t\times(\bmcpz_2)_1^t$, the Cartesian product $G_1\bigotimes G_2$
corresponds to 
\be
\label{eq111}
(\bmcpz_2)_1^{t+1}\times(\bmcpz_2)_3^t\bigotimes(\bmcpz_2)_3^t\times(\bmcpz_2)_1^t.
\ee
The elements of $G^*$ are a subset of $G_1\bigotimes G_2$ of the form
$(g_1,g_2)$, where $g_1$ and $g_2$ have the same image $g_3$ in the homomorphisms 
$G_1\ra G_3$, $G_2\ra G_3$ given by the isomorphism $G_1/H_1\simeq G_3\simeq G_2/H_2$.
Then the subset $G^*$ of $G_1\bigotimes G_2$ corresponds to elements
in (\ref{eq111}) of the form
\be
\label{eq112}
(\bmcpz_2)_1^{t+1}\times z_3^t\bigotimes z_3^t\times(\bmcpz_2)_1^t,
\ee
where $z_3^t$ is any element of $(\bmcpz_2)_3^t$.  Since $G^*$ is a subgroup of the direct product
group $G_1\times G_2$, then (\ref{eq112}) means $G^*$ is isomorphic to 
$(\bmcpz_2)_1^{t+1}\times(\bmcpz_2)_3^t\times(\bmcpz_2)_1^t$.  Then we have
\be
\label{tpg3}
\grpjktarg{0}{1}{t+1}{\calu}{\ccirc}\Join\grpjktarg{0}{1}{t}{\calu}{\ccirc}\simeq(\bmcpz_2)_1^{t+1}\times(\bmcpz_2)_3^t\times(\bmcpz_2)_1^t.
\ee
There is a homomorphism from $\grpjktarg{0}{0}{t+1}{\calu}{\ccirc}$ to (\ref{tpg3}).
Since the set of generator labels in $\tridnjktarg{0}{0}{t+1}{\calu}$ but not in
$\tridnjktarg{0}{1}{t+1}{\calu}\Join\tridnjktarg{0}{1}{t}{\calu}$ is trivial, we must have
$$
\grpjktarg{0}{0}{t+1}{\calu}{\ccirc}\simeq(\bmcpz_2)_1^{t+1}\times(\bmcpz_2)_3^t\times(\bmcpz_2)_1^t.
$$
In a similar way, we must have the last group
$$
\grpjktarg{0}{0}{t+2}{\calu}{\ccirc}\simeq(\bmcpz_2)_1^{t+2}\times(\bmcpz_2)_3^t\times(\bmcpz_2)_1^{t+1}.
$$

We can summarize the preceding results by
\be
\label{tpg4}
\begin{array}{ll}
\grpjktarg{0}{0}{t}{\calu}{\ccirc}   &\simeq(\bmcpz_2)_1^t\times(\bmcpz_2)_3^t, \\
\grpjktarg{0}{0}{t+1}{\calu}{\ccirc} &\simeq(\bmcpz_2)_1^{t+1}\times(\bmcpz_2)_3^t\times(\bmcpz_2)_1^t, \\
\grpjktarg{0}{0}{t+2}{\calu}{\ccirc} &\simeq(\bmcpz_2)_1^{t+2}\times(\bmcpz_2)_3^t\times(\bmcpz_2)_1^{t+1}, \\
\grpjktarg{0}{0}{t+3}{\calu}{\ccirc} &\simeq(\bmcpz_2)_1^{t+2}\times(\bmcpz_2)_3^t.
\end{array}
\ee
We have just found all possible nontrivial elementary groups that have
the generator pyramid (\ref{blkuttf3}) with $|R_{0,3}^t|=2$ and 
$|R_{0,1}^{t+2}|=|R_{0,1}^{t+1}|=|R_{0,1}^t|=2$.  $\cale$ is defined by the
elementary groups in (\ref{tpg4}); all other elementary
groups in $\cale$ are trivial.  Since all possible nontrivial elementary groups have been found
in (\ref{tpg4}), there is only one $(\ell+1)$-depth elementary system $\cale$
with the specified generator pyramid, up to list isomorphism.

The $(\ell+1)$-depth elementary system $\cale$ specifies an $(\ell+1)$-depth global group $(\calu,\star)$.
Next we will apply chain (\ref{zchain7e}) and the \fhgs, Theorem \ref{thm84}, to $(\calu,\star)$,
after replacing group $(\calu,\circ)$ in Theorem \ref{thm84} with $(\calu,\star)$.  
Since the formal mechanism of the \fhgs\ only requires knowledge of the elementary groups 
already found in (\ref{tpg4}), there is no need to specify the details of 
the global operation $\star$ in $(\calu,\star)$.  From Theorem \ref{thm84}, there is a map
$\theta^t:  \calu\ra\tridnjktarg{0}{0}{t}{\calu}$ which gives a homomorphism for each $t\in\bmcpz$.
We now find a map $\mu^{t'}:  \tridnjktarg{0}{0}{t'}{\calu}\ra A^{t'}$ which gives a 
homomorphism from each of the four elementary groups $\grpjktarg{0}{0}{t'}{\calu}{\ccirc}$ 
in (\ref{tpg4}) to an alphabet group $A^{t'}$, for $t'\in[t,t+3]$.
We can choose any $A^{t'}$ such that there is a homomorphism from   
$\grpjktarg{0}{0}{t'}{\calu}{\ccirc}$ to $A^{t'}$.  For example, if
$\grpjktarg{0}{0}{t'}{\calu}{\ccirc}\simeq\bmcpz_2\times\bmcpz_2\times\bmcpz_2$, we
can choose $A^{t'}\simeq\bmcpz_2\times\bmcpz_2\times\bmcpz_2$, 
$A^{t'}\simeq\bmcpz_2\times\bmcpz_2$, or $A^{t'}\simeq\bmcpz_2$.
And furthermore, we can mix alphabet groups by choosing different 
$A^{t'}$ for different $t'\in[t,t+3]$.  For each choice of alphabet
group $A^{t'}$, we obtain a group system with alphabet groups
$A^{t'}$, for $t'\in[t,t+3]$.

As a first alphabet choice, we choose the alphabet group
$A^{t'}=\bmcpz_2\times\bmcpz_2$ for each $t'$ in $[t,t+3]$.  The first column of
Table \ref{code1} shows the elementary groups in (\ref{tpg4}).  The second column
shows the nontrivial \glab\ sets used in the elementary groups.  The third column
shows the groups in (\ref{tpg4}) isomorphic to the elementary groups.  The elements 
in these isomorphic groups are essentially the Cartesian product of \glab\ sets in the 
second column.  The fourth column shows one possible homomorphism mapping $\mu^{t'}$ 
from the isomorphic groups in the third column to alphabet group
$A^{t'}=\bmcpz_2\times\bmcpz_2$.  It is easy to verify that
the mappings $\mu^{t'}$ are homomorphisms.  The first quartet of four rows are easily seen to be
an automorphism of $\bmcpz_2\times\bmcpz_2$.  The next octet of eight rows are
a homomorphism from $\bmcpz_2\times\bmcpz_2\times\bmcpz_2$ to $\bmcpz_2\times\bmcpz_2$,
where the normal subgroup is $\{(0,0,0),(0,1,1)\}$.  The third octet is the same
homomorphism as the second octet, and the last quartet is the same as the first.

\begin{table}[h]

\begin{tabular}{|l|l|l|l|} \hline
Elementary                             & Generator label                           & Group isomorphic to                                                 & Homomorphism map                                        \\ 
group                                  & sets used in                              & elementary group                                                    & $\mu^{t'}:  \tridnjktarg{0}{0}{t'}{\calu}\ra A^{t'}$    \\
                                       & elementary                                &                                                                     & where $A^{t'}=\bmcpz_2\times\bmcpz_2$                   \\
                                       & group                                     &                                                                     &                                                         \\  \hline

$\grpjktarg{0}{0}{t}{\calu}{\ccirc}$   & $R_{0,3}^t, R_{0,1}^t$                    & $\simeq(\bmcpz_2)_3^t\times(\bmcpz_2)_1^t$                          & $(0,0)\mapsto(0,0)$ \\
                                       &                                           &                                                                     & $(1,0)\mapsto(1,0)$ \\
                                       &                                           &                                                                     & $(0,1)\mapsto(1,1)$ \\
                                       &                                           &                                                                     & $(1,1)\mapsto(0,1)$ \\  \hline

$\grpjktarg{0}{0}{t+1}{\calu}{\ccirc}$ & $R_{0,3}^t, R_{0,1}^t, R_{0,1}^{t+1}$     & $\simeq(\bmcpz_2)_3^t\times(\bmcpz_2)_1^t\times(\bmcpz_2)_1^{t+1}$  & $(0,0,0)\mapsto(0,0)$ \\
                                       &                                           &                                                                     & $(1,0,0)\mapsto(1,0)$ \\
                                       &                                           &                                                                     & $(0,1,0)\mapsto(1,1)$ \\
                                       &                                           &                                                                     & $(0,0,1)\mapsto(1,1)$ \\
                                       &                                           &                                                                     & $(1,1,0)\mapsto(0,1)$ \\
                                       &                                           &                                                                     & $(1,0,1)\mapsto(0,1)$ \\
                                       &                                           &                                                                     & $(0,1,1)\mapsto(0,0)$ \\
                                       &                                           &                                                                     & $(1,1,1)\mapsto(1,0)$ \\  \hline

$\grpjktarg{0}{0}{t+2}{\calu}{\ccirc}$ & $R_{0,3}^t, R_{0,1}^{t+1}, R_{0,1}^{t+2}$ & $\simeq(\bmcpz_2)_3^t\times(\bmcpz_2)_1^{t+1}\times(\bmcpz_2)_1^{t+2}$  & $(0,0,0)\mapsto(0,0)$ \\
                                       &                                           &                                                                         & $(1,0,0)\mapsto(1,0)$ \\
                                       &                                           &                                                                         & $(0,1,0)\mapsto(1,1)$ \\
                                       &                                           &                                                                         & $(0,0,1)\mapsto(1,1)$ \\
                                       &                                           &                                                                         & $(1,1,0)\mapsto(0,1)$ \\
                                       &                                           &                                                                         & $(1,0,1)\mapsto(0,1)$ \\
                                       &                                           &                                                                         & $(0,1,1)\mapsto(0,0)$ \\
                                       &                                           &                                                                         & $(1,1,1)\mapsto(1,0)$ \\  \hline

$\grpjktarg{0}{0}{t+3}{\calu}{\ccirc}$ & $R_{0,3}^t, R_{0,1}^{t+2}$                & $\simeq(\bmcpz_2)_3^t\times(\bmcpz_2)_1^{t+2}$                          & $(0,0)\mapsto(0,0)$ \\
                                       &                                           &                                                                         & $(1,0)\mapsto(1,0)$ \\
                                       &                                           &                                                                         & $(0,1)\mapsto(1,1)$ \\
                                       &                                           &                                                                         & $(1,1)\mapsto(0,1)$ \\  \hline
\end{tabular}

\caption{Homomorphism from isomorphic group to alphabet group $A^{t'}=\bmcpz_2\times\bmcpz_2$,
$t'\in[t,t+3]$, to construct the binary $(8,4,4)$ extended Hamming code.}
\label{code1}

\end{table}

We now continue the construction in chain (\ref{zchain7e}) using Theorem \ref{thm84}.
We know $\theta^t$ and we have just found $\mu^{t'}$ for $t'\in[t,t+3]$; $\mu^{t'}$ is trivial
elsewhere.  Let $h_u^t\rmdef\mu^t\bullet\theta^t$ for each $t\in\bmcpz$.  
The homomorphisms $h_u^t:  \calu\ra A^t$,
for each $t\in\bmcpz$, define a homomorphism $h_u:  \calu\ra A_\amalg$ from $(\calu,\star)$
to the direct product group $(A_\amalg,+)$.  The image of homomorphism $h_u$, $\imhu$,
is the constructed group system $C$.

Table \ref{code2} demonstrates the codewords of the group system $C$ obtained by the 
homomorphisms in Table \ref{code1}.  The generator sets in the second column of
Table \ref{code1} are various combinations of the 4 sets $R_{0,3}^t, R_{0,1}^t, R_{0,1}^{t+1}, R_{0,1}^{t+2}$,
which are shown in the first column of Table \ref{code2}.  Similarly, the isomorphic groups in
the third column of Table \ref{code1} are various combinations of the 4
groups $(\bmcpz_2)_3^t$, $(\bmcpz_2)_1^t$, $(\bmcpz_2)_1^{t+1}$, $(\bmcpz_2)_1^{t+2}$,
which are shown in the second column of Table \ref{code2}.  There are 16 choices of \glabs\
in the sets $R_{0,3}^t, R_{0,1}^t, R_{0,1}^{t+1}, R_{0,1}^{t+2}$ in the first column of Table \ref{code2}, and 16 choices
of elements in the groups $(\bmcpz_2)_3^t$, $(\bmcpz_2)_1^t$, $(\bmcpz_2)_1^{t+1}$, $(\bmcpz_2)_1^{t+2}$
in the second column of Table \ref{code2}.  For each choice of elements 
in the second column of Table \ref{code2}, we
can use the mappings in the fourth column of Table \ref{code1} to determine an
alphabet letter in $A^{t'}=\bmcpz_2\times\bmcpz_2$, where $t'\in[t,t+3]$.  The alphabet
letters for $t'\in[t,t+3]$ determine a codeword over the interval $[t,t+3]$.  Then the 16 choices of 
elements in the second column of Table \ref{code2} determine 16 codewords, shown in the third column of
Table \ref{code2}.  For example, if we choose the elements $1,0,0,0$ for groups
$(\bmcpz_2)_3^t$, $(\bmcpz_2)_1^t$, $(\bmcpz_2)_1^{t+1}$, $(\bmcpz_2)_1^{t+2}$ 
in the second column of Table \ref{code2}, then 
the mappings in the fourth column of Table \ref{code1} give the generator codeword $10,10,10,10$
in the third column of Table \ref{code2}.

Comparing the first and third column of Table \ref{code2} with the first and third
column of Table \ref{pgrp4} shows that Table \ref{code2} gives the codewords of the 
binary $(8,4,4)$ extended Hamming code.
In fact the choice of homomorphisms in Table \ref{code1} guarantees construction of the
extended Hamming code, since the chosen mappings determine the correct element 
of alphabet group $A^{t'}=\bmcpz_2\times\bmcpz_2$ needed for each possible combination 
of generators of the binary $(8,4,4)$ extended Hamming code at each time $t'$, $t'\in[t,t+3]$.
We know the Hamming code is \ellctl\ for $\ell=3$.

\begin{table}[h]

\begin{tabular}{|l|l|l|} \hline
Choice of generator                                  & Choice of elements in groups                                                           & Codeword   \\ 
labels in sets                                       & $\left((\bmcpz_2)_3^t, (\bmcpz_2)_1^t, (\bmcpz_2)_1^{t+1}, (\bmcpz_2)_1^{t+2}\right)$  &            \\
$R_{0,3}^t, R_{0,1}^t, R_{0,1}^{t+1}, R_{0,1}^{t+2}$ &                                                                                        &            \\  \hline

$1_{0,3}^t,1_{0,1}^t,1_{0,1}^{t+1},1_{0,1}^{t+2}$ & (0,0,0,0) & $\ldots,(00,00,00,00),\ldots$     \\
$1_{0,3}^t,r_{0,1}^t,1_{0,1}^{t+1},1_{0,1}^{t+2}$ & (0,1,0,0) & $\ldots,(11,11,00,00),\ldots$     \\
$1_{0,3}^t,1_{0,1}^t,r_{0,1}^{t+1},1_{0,1}^{t+2}$ & (0,0,1,0) & $\ldots,(00,11,11,00),\ldots$     \\
$1_{0,3}^t,1_{0,1}^t,1_{0,1}^{t+1},r_{0,1}^{t+2}$ & (0,0,0,1) & $\ldots,(00,00,11,11),\ldots$     \\

$1_{0,3}^t,r_{0,1}^t,r_{0,1}^{t+1},1_{0,1}^{t+2}$ & (0,1,1,0) & $\ldots,(11,00,11,00),\ldots$     \\
$1_{0,3}^t,r_{0,1}^t,1_{0,1}^{t+1},r_{0,1}^{t+2}$ & (0,1,0,1) & $\ldots,(11,11,11,11),\ldots$     \\
$1_{0,3}^t,1_{0,1}^t,r_{0,1}^{t+1},r_{0,1}^{t+2}$ & (0,0,1,1) & $\ldots,(00,11,00,11),\ldots$     \\
$1_{0,3}^t,r_{0,1}^t,r_{0,1}^{t+1},r_{0,1}^{t+2}$ & (0,1,1,1) & $\ldots,(11,00,00,11),\ldots$     \\  \hline

$r_{0,3}^t,1_{0,1}^t,1_{0,1}^{t+1},1_{0,1}^{t+2}$ & (1,0,0,0) & $\ldots,(10,10,10,10),\ldots$     \\
$r_{0,3}^t,r_{0,1}^t,1_{0,1}^{t+1},1_{0,1}^{t+2}$ & (1,1,0,0) & $\ldots,(01,01,10,10),\ldots$     \\
$r_{0,3}^t,1_{0,1}^t,r_{0,1}^{t+1},1_{0,1}^{t+2}$ & (1,0,1,0) & $\ldots,(10,01,01,10),\ldots$     \\
$r_{0,3}^t,1_{0,1}^t,1_{0,1}^{t+1},r_{0,1}^{t+2}$ & (1,0,0,1) & $\ldots,(10,10,01,01),\ldots$     \\

$r_{0,3}^t,r_{0,1}^t,r_{0,1}^{t+1},1_{0,1}^{t+2}$ & (1,1,1,0) & $\ldots,(01,10,01,10),\ldots$     \\
$r_{0,3}^t,r_{0,1}^t,1_{0,1}^{t+1},r_{0,1}^{t+2}$ & (1,1,0,1) & $\ldots,(01,01,01,01),\ldots$     \\
$r_{0,3}^t,1_{0,1}^t,r_{0,1}^{t+1},r_{0,1}^{t+2}$ & (1,0,1,1) & $\ldots,(10,01,10,01),\ldots$     \\
$r_{0,3}^t,r_{0,1}^t,r_{0,1}^{t+1},r_{0,1}^{t+2}$ & (1,1,1,1) & $\ldots,(01,10,10,01),\ldots$     \\  \hline
\end{tabular}

\caption{Codewords of the binary $(8,4,4)$ extended Hamming code.}
\label{code2}

\end{table}

As a final step, we check that the homomorphism $h_u$ is a bijection and the 
linear system is invertible.  It is evident from column 2 of Table \ref{code2} that
there are 16 distinct inputs and from column 3 that there are 16 distinct outputs.
Therefore $h_u$ is a bijection and the input can be determined from the output.

It is useful to prove the homomorphism $h_u$ is a bijection in another way.  We show that the kernel of
the homomorphism $h_u$ is trivial.  Note that the first and last quartet
in Table \ref{code1} use the mapping $(0,0)\mapsto(0,0)$, and the second and third octet 
use the mappings $(0,0,0)\mapsto(0,0)$ and $(0,1,1)\mapsto(0,0)$.  The only way to have a 
nontrivial kernel is to use the mapping $(0,0)\mapsto(0,0)$ in the first and
fourth quartet, and the mapping $(0,1,1)\mapsto(0,0)$ in the second and third octet.
But the choice of elements $(0,1,1)$ for
$(\bmcpz_2)_3^t\times(\bmcpz_2)_1^t\times(\bmcpz_2)_1^{t+1}$ in the second octet
means the element in $(\bmcpz_2)_1^t$ is 1, which contradicts the choice of elements 
$(0,0)$ in the first quartet.  And the choice of elements $(0,1,1)$ for
$(\bmcpz_2)_3^t\times(\bmcpz_2)_1^{t+1}\times(\bmcpz_2)_1^{t+2}$ in the third octet
means the element in $(\bmcpz_2)_1^{t+2}$ is 1, which contradicts the choice of elements 
$(0,0)$ in the fourth quartet.  Therefore the kernel of the homomorphism $h_u$ is trivial.

We have constructed the binary $(8,4,4)$ extended Hamming code $C$ using the homomorphisms 
shown in Table \ref{code1}.  We may substitute other homomorphisms to 
$A^{t'}=\bmcpz_2\times\bmcpz_2$ and obtain other group systems.  By a similar
argument to the preceding paragraph, the kernel of the homomorphism $h_u$ is trivial,
and so $h_u$ is a bijection and the linear system is invertible.

As a second alphabet choice, we simply choose an alphabet $A^{t'}$ isomorphic to
$\grpjktarg{0}{0}{t'}{\calu}{\ccirc}$ for each $t'\in[t,t+3]$.  Then we have
\be
\label{tpg6}
\begin{array}{lll}
\grpjktarg{0}{0}{t}{\calu}{\ccirc}   &\simeq(\bmcpz_2)_1^t\times(\bmcpz_2)_3^t                             & \simeq A^t,     \\
\grpjktarg{0}{0}{t+1}{\calu}{\ccirc} &\simeq(\bmcpz_2)_1^t\times(\bmcpz_2)_3^t\times(\bmcpz_2)_1^{t+1}     & \simeq A^{t+1}, \\
\grpjktarg{0}{0}{t+2}{\calu}{\ccirc} &\simeq(\bmcpz_2)_1^{t+2}\times(\bmcpz_2)_3^t\times(\bmcpz_2)_1^{t+1} & \simeq A^{t+2}, \\
\grpjktarg{0}{0}{t+3}{\calu}{\ccirc} &\simeq(\bmcpz_2)_1^{t+2}\times(\bmcpz_2)_3^t                         & \simeq A^{t+3}.
\end{array}
\ee
Since group $\grpjktarg{0}{0}{t'}{\calu}{\ccirc}\simeq A^{t'}$ for each $t'\in[t,t+3]$,
the \fhgs\ constructs the global group system $(U_s,\ovstar)$.  For this code, the first 
three columns of Table \ref{code1} remain the same, but in the fourth column of Table \ref{code1}, 
the homomorphism $\mu^{t'}$ becomes an isomorphism to $A^{t'}$ for each $t'\in[t,t+3]$.  
The first two columns of Table \ref{code2} remain the same,
but the third column gives codewords different from the extended Hamming code.
For example, the choice of elements $(0,0,0,0)$ in the second column gives the identity codeword
$\ldots,(00,000,000,00),\ldots$ in the third column.
We know the global group system $(U_s,\ovstar)$ is \ellctl, the homomorphism $h_u$ is a
bijection, and the linear system is invertible.

We have seen there is only one $(\ell+1)$-depth elementary system $\cale$ up to list isomorphism
having the generator pyramid (\ref{blkuttf3}) with $|R_{0,3}^t|=2$ and 
$|R_{0,1}^{t+2}|=|R_{0,1}^{t+1}|=|R_{0,1}^t|=2$.  From Section 7.3, since there is only one
$(\ell+1)$-depth elementary system $\cale$ up to list isomorphism, there is only one \ellctl\
complete group system up to list isomorphism that can be constructed from $\cale$.  
Therefore all \ellctl\ codes constructed here from $\cale$ are list isomorphic.  
This means for example that the binary $(8,4,4)$ extended
Hamming code is list isomorphic to the global group sytem $(U_s,\ovstar)$, 
and either one may be taken as a representative of the equivalence class $[\cale]_l$
of the list isomorphism.

If we want to find all \ellctl\ complete group systems up to list isomorphism from $\cale$,
then we can just find $(U_s,\ovstar)$ and we are done.  If we want to find all group systems
from $\cale$, then we may find groups systems that are $l$-controllable where $l<\ell$.
In general then, we need to determine $l$.  To do this,
the generators of the code can be found from the time domain granules of the \cdc,
as in Section 3.  Then we need to find a generator of maximum span $l+1$.

\end{document}